\documentclass[aps,a4paper,superscriptaddress,showpacs,preprintnumbers,amsmath,amssymb]{revtex4}

\usepackage{ulem}

\usepackage{psfrag} \usepackage{graphicx} \usepackage{dcolumn}
\usepackage{color} \usepackage{latexsym,amsfonts} \usepackage{bm}
\usepackage{amssymb}
\begin{document}

\title{Radiative corrections of order $O(\alpha E_e/m_N)$\\ to
  Sirlin's radiative corrections of order $O(\alpha/\pi)$, induced by
  \\ the hadronic structure of the neutron}

\author{A. N. Ivanov}\email{ivanov@kph.tuwien.ac.at}
\affiliation{Atominstitut, Technische Universit\"at Wien, Stadionallee
  2, A-1020 Wien, Austria}
\author{R. H\"ollwieser}\email{roman.hoellwieser@gmail.com}
\affiliation{Atominstitut, Technische Universit\"at Wien, Stadionallee
  2, A-1020 Wien, Austria}\affiliation{Department of Physics,
  Bergische Universit\"at Wuppertal, Gaussstr. 20, D-42119 Wuppertal,
  Germany} \author{N. I. Troitskaya}\email{natroitskaya@yandex.ru}
\affiliation{Atominstitut, Technische Universit\"at Wien, Stadionallee
  2, A-1020 Wien, Austria}
\author{M. Wellenzohn}\email{max.wellenzohn@gmail.com}
\affiliation{Atominstitut, Technische Universit\"at Wien, Stadionallee
  2, A-1020 Wien, Austria} \affiliation{FH Campus Wien, University of
  Applied Sciences, Favoritenstra\ss e 226, 1100 Wien, Austria}
\author{Ya. A. Berdnikov}\email{berdnikov@spbstu.ru}\affiliation{Peter
  the Great St. Petersburg Polytechnic University, Polytechnicheskaya
  29, 195251, Russian Federation}

\date{\today}

\begin{abstract}
We investigate the contributions of the hadronic structure of the
neutron to radiative $O(\alpha E_e/m_N)$ corrections (or the inner
$O(\alpha E_e/m_N)$ RC) to the neutron beta decay, where $\alpha$,
$E_e$ and $m_N$ are the fine-structure constant, the electron energy
and the nucleon mass, respectively. We perform the calculation within
the effective quantum field theory of strong low-energy pion-nucleon
interactions described by the linear $\sigma$-model with chiral $SU(2)
\times SU(2)$ symmetry and electroweak hadron-hadron, hadron-lepton
and lepton-lepton interactions for the electron-lepton family with
$SU(2)_L \times U(1)_Y$ symmetry of the Standard Electroweak Theory
(Ivanov {\it et al.}, Phys. Rev. D {\bf 99}, 093006 (2019)). We show
that after renormalization, carried out in accordance with Sirlin's
prescription (Sirlin, Phys. Rev. {\bf 164}, 1767 (1967)), the inner
$O(\alpha E_e/m_N)$ RC are of the order of a few parts of $10^{-5} -
10^{-4}$. This agrees well with the results obtained in (Ivanov {\it
  et al.}, Phys. Rev. D {\bf 99}, 093006 (2019)).
\end{abstract}

\maketitle

\section{Introduction}
\label{sec:introduction}

According to Sirlin \cite{Sirlin1967, Sirlin1978}, the contribution of
the hadronic structure of the neutron to the radiative $O(\alpha/\pi)$
corrections (or the {\it inner} $O(\alpha/\pi)$ RC
\cite{Wilkinson1970}) to the neutron lifetime is a constant,
calculated to leading order (LO) in the large nucleon mass $m_N$
expansion, where $\alpha$ is the fine-structure constant
\cite{PDG2020}.  Because of the divergent contribution, this constant
has been removed by renormalization of the Fermi coupling constant
$G_V$ and the axial coupling constant $g_A$ \cite{Sirlin1967,
  Sirlin1978}. This result has been confirmed by Shann
\cite{Shann1971} for the calculation of the $O(\alpha/\pi)$ RC to the
correlation coefficients of the neutron beta decay with a polarized
neutron and unpolarized electron and proton (see also
\cite{Myhrer2004, Gudkov2006, Ivanov2013}). However, as has been shown
in \cite{Ivanov2017a}, the contributions of the inner
$O(\alpha^2/\pi^2)$ RC to the neutron radiative beta decay should have
a non-trivial dependence on the electron and photon energies, even
these RC are calculated to LO in the large nucleon mass $m_N$
expansion.

Recently \cite{Ivanov2019a}, we have calculated the $O(\alpha
E_e/m_N)$ RC as next-to-leading order (NLO) corrections in the large
nucleon mass $m_N$ expansion to Sirlin's $O(\alpha/\pi)$ RC
\cite{Sirlin1967} (or to the {\it outer} model-independent RC
\cite{Wilkinson1970}), where $E_e$ is an electron energy, We have
carried out the calculation within the effective quantum field theory
of strong and electroweak low-energy interactions L$\sigma$M$\&$SET.
In this theory strong low-energy pion-nucleon interactions are
described by the linear $\sigma$-model (L$\sigma$M) with chiral $SU(2)
\times SU(2)$ symmetry \cite{GellMann1960, Lee1972, Nowak1996}. For
the description of electroweak hadron-hadron, hadron-lepton and
lepton-lepton interactions for the electron-lepton family we have used
the Standard Electroweak Theory (SET) with $SU(2)_L \times U(1)_Y$
symmetry \cite{Weinberg1971}. This effective quantum field theory is
some kind of a hadronized version of the Standard Model (SM)
\cite{PDG2020, DGH2014}. From a gauge invariant set of the Feynman
diagrams with one-photon exchange, where the contribution of strong
low-energy interactions is presented by the axial coupling constant
$g_A$ only (see Fig.\,7 in Ref.\cite{Ivanov2019a}), we have reproduced
outer $O(\alpha/\pi)$ RC \cite{Sirlin1967} and calculated NLO
$O(\alpha E_e/m_N)$ terms. This confirms Sirlin's confidence level for
this kind of $O(\alpha E_e/m_N)$ RC.

We have calculated the contributions of strong low-energy interactions
to $O(\alpha E_e/m_N)$ RC within the L$\sigma$M in the limit
$m_{\sigma} \to \infty$ of the $\sigma$-meson mass. In such a limit
and in the tree-approximation the L$\sigma$M reproduces all results of
the current algebra in the form of effective chiral Lagrangians of
pion-nucleon interactions with non-linear realization of chiral
$SU(2)\times SU(2)$ symmetry and different parametrizations of the
pion-field \cite{Weinberg1967a, Weinberg1968, Weinberg1979}.

For the exponential parametrization of the pion-field the Lagrangian
${\cal L}_{\rm L \sigma M}\big|_{m_{\sigma} \to \infty}$ of the
L$\sigma$M, taken at $m_{\sigma} \to \infty$, reduces to the
Lagrangian of the chiral quantum field theory with the structure of
low-energy interactions agreeing well with Gasser-Leutwyler's chiral
perturbation theory (ChPT) or the heavy baryon chiral perturbation
theory (HB$\chi$PT) \cite{Gasser1984} - \cite{Scherer2011} with chiral
$SU(2)\times SU(2)$ symmetry(see, for example, Ecker
\cite{Ecker1995}).  We denote the Lagrangian of the HB$\chi$PT as
${\cal L}_{\rm HB \chi PT}$. At the tree-level, the Lagrangians ${\cal
  L}_{\rm L \sigma M}\big|_{m_{\sigma} \to \infty}$ and ${\cal L}_{\rm
  HB \chi PT}$ differ only by the value of the {\it bare} axial
coupling constant $g^{(0)}_A$. Indeed, it is $g^{(0)}_A = 1$ in ${\cal
  L}_{\rm L \sigma M}\big|_{m_{\sigma} \to \infty}$ and $g^{(0)}_A
\neq 1$ in ${\cal L}_{\rm HB \chi PT}$ (see also
\cite{Weinberg1967a}).  However, as has been shown in
\cite{Ivanov2019a}, a deviation of the axial coupling constant from
unity $g_A > 1$ can be obtained in the L$\sigma$M in the
one-hadron-loop approximation. We get $g_A > 1$ taking the limit
$m_{\sigma} \to \infty$ and renormalizing the contribution of the
hadronic axial-vector current. In turn, hadron-loop corrections (or
chiral-hadron-loop corrections), calculated in the HB$\chi$PT, lead to
appearance effective low-energy interactions proportional to
low-energy constants (LECs) \cite{Gasser1984} -
\cite{Scherer2011}. These LECs play an important role for the correct
description of the dynamics of low-energy processes within the
HB$\chi$PT \cite{Gasser1984} - \cite{Scherer2011}. Unfortunately, LECs
do not appear in the observables of low-energy processes described by
the L$\sigma$M. Nevertheless, fortunately, it turns out that the LECs
of the HB$\chi$PT do not contribute to the inner $O(\alpha E_e/m_N)$
RC. This should in principle allow us to apply the L$\sigma$M for the
description of strong low-energy interactions in the inner $O(\alpha
E_e/m_N)$ RC. To confirm this, we propose to discuss the studies
carried out by Alvarez {\it et al.}  \cite{Myhrer2005} and Ando {\it
  et al.}  \cite{Myhrer2004}. We will focus our attention on using
them to analyze the applicability of the L$\sigma$M to the computation
of the inner $O(\alpha E_e/m_N)$ RC.

In our study of the inner $O(\alpha/\pi)$ and $O(\alpha E_e/m_N)$ RC
to the neutron beta decay the contributions of strong low-energy
interactions, calculated in the L$\sigma$M, are proportional to
$g^2_{\pi N}$ and $g^2_{\pi N}/m_N$, where $g_{\pi N}$ is the
pion-nucleon coupling constant.  Alvarez {\it et al.}
\cite{Myhrer2005} have analyzed the amplitude of the low-energy $\pi
N$-scattering. They have compared the contributions of the L$\sigma$M
with chiral $SU(2)\times SU(2)$ symmetry, taken in its extended
version - the extended linear $\sigma$-model (EL$\sigma$M), and the
HB$\chi$PT. Recall that the EL$\sigma$M differs from the L$\sigma$M by
the phenomenological local current-current interaction \cite{Lee1972}
and the interaction proportional to the phenomenological parameter
$\varepsilon_3$ \cite{Campbell1979}. The local current-current
phenomenological interaction is intended to introduce $g_A > 1$ into
the L$\sigma$M at the tree-level. The parameter $\varepsilon_3$
defines a correction to the Goldberger-Treiman relation
\cite{Goldberger1958} and leads to a deviation of the divergence of
the hadronic axial-vector current from its canonical form
\cite{GellMann1960}.

As has been shown by Alvarez {\it et al.}  \cite{Myhrer2005}, the
contributions of the EL$\sigma$M and HB$\chi$PT coincide fully at the
tree-approximation (see \cite{Fettes1998, Nowak1996}). Then, taking
results obtained by Alvarez {\it et al.}  \cite{Myhrer2005} in the
limit $m_{\sigma} \to \infty$ and setting $\varepsilon_3 = 0$, one may
show that the difference between the contributions of these two
theories appears only in the terms dependent on LECs. These terms are
proportional to $g^2_{\pi N}/m^2_N$ and $g^2_{\pi N}/m^3_N$,
respectively. Since in the neutron beta decay we compute the
contributions of strong low-energy interactions proportional to
$g^2_{\pi N}$ and $g^2_{\pi N}/m_N$ only, the problematic terms
$g^2_{\pi N}/m^2_N$ and $g^2_{\pi N}/m^3_N$, which can depend on LECs,
do not appear at all.

A correctness of the application of the L$\sigma$M without LECs to the
computation of the inner $O(\alpha E_e/\pi)$ RC can be also confirmed
by the results obtained by Ando {\it et al.}  \cite{Myhrer2004}. They
have studied the $O(\alpha/\pi)$ RC and the $O(E_e/m_N)$ corrections,
caused by weak magnetism and proton recoil, to the neutron beta decay
within the HB$\chi$PT. As has been shown by Ando {\it et al.}
\cite{Myhrer2004}, only two LECs, namely $(\alpha/2\pi)\, e^R_V$ and
$(\alpha/2\pi)\,e^R_A$, are needed for the consistent analyzes of
these corrections. The value of $(\alpha/2\pi)\,e^R_V$ has been fixed
in terms of the inner $O(\alpha/\pi)$ RC, induced by the $\gamma
W^-$-box and calculated by Marciano and Sirlin \cite{Sirlin1986}. In
turn, the difference of LECs $(\alpha/2\pi)(e^R_A - e^R_V)$ has been
removed by renormalization of the axial coupling constant $g_A$.
Then, no LECs proportional to $1/m_N$ have been found in
\cite{Myhrer2004} for the calculation of RC to the neutron lifetime
and correlation coefficients of the neutron beta decay. This should
indicate that the inner $O(\alpha E_e/m_N)$ RC, which we calculate in
this paper within the L$\sigma$M$\&$SET, should not contradict the
results, that can be, in principle, obtained describing strong
low-energy interactions within the HB$\chi$PT.

This paper is addressed to the calculation of the contribution of the
hadronic structure of the neutron to the inner $O(\alpha/\pi)$ and
$O(\alpha E_e/m_N)$ RC to the neutron beta decay. We calculate them in
the two-loop approximation within the effective quantum field theory
L$\sigma$M$\&$SET \cite{Ivanov2019a}.  The complete set of the
two-loop Feynman diagrams is shown in Fig.\,\ref{fig:fig1} -
Fig.\,\ref{fig:fig4}. They are conditioned by one-photon exchange and
the contributions of strong low-energy interactions, which do not
reduce to the axial coupling constant $g_A$ after renormalization
\cite{Ivanov2019a}.  We treat the inner $O(\alpha E_e/m_N)$ RC as
next-to-leading order (NLO) corrections in the large nucleon mass
expansion to the inner $O(\alpha/\pi)$ RC \cite{Sirlin1967}. They can
be observable after the removal of the inner $O(\alpha/\pi)$ RC by
renormalization of the Fermi weak coupling constant and the axial
coupling constant (see \cite{Sirlin1967}).

The paper is organized as follows. In section \ref{sec:comment}, we
describe in outline the approach to the analytical calculation of the
two-loop Feynman diagrams in Fig.\,\ref{fig:fig1} -
Fig.\,\ref{fig:fig4} in agreement with the applicability of the
L$\sigma$M for the calculation of strong low-energy interactions. In
section \ref{sec:amplitude}, we define the general expression for the
contribution of the one-virtual photon exchanges to the amplitude of
the neutron beta decay within the L$\sigma$M$\&$SET. In section
\ref{sec:lifetime}, we present the contributions of the inner
$O(\alpha E_e/m_N)$ RC i) to the amplitude of the neutron beta decay,
ii) to the electron-energy and angular distribution of the neutron
beta decay with unpolarized massive fermions and iii) to the rate of
the neutron beta decay. We show that in the total electron-energy
region $m_e \le E_e \le E_0$ the inner $O(\alpha E_e/m_N)$ RC are of
the order of a few parts of $10^{-5} - 10^{-4}$. In section
\ref{sec:Abschluss}, we discuss the obtained results and perspectives
of further development of the effective quantum field theory of strong
and electroweak low-energy interactions L$\sigma$M$\&$SET, where
strong low-energy interactions are described by the HB$\chi$PT.

In the Supplemental Material in Appendices A, B, C, D, E and F we give
i) the analytical expressions for the Feynman diagrams in
Fig.\,\ref{fig:fig1} - Fig.\,\ref{fig:fig4}, obtained by using the
Lagrangian Eq.(44) in Ref.\cite{Ivanov2019a}, ii) the analysis of
gauge invariance of the Feynman diagrams in Fig.\,\ref{fig:fig1} -
Fig.\,\ref{fig:fig4}, and iii) the analytical calculation of the
Feynman diagrams using the standard procedure
\cite{Feynman1950}-\cite{Smirnov2012}. The numerical values of the
structure constants of the inner $O(\alpha E_e/m_N)$ RC to the
amplitude of the neutron beta decay are evaluated by using Wolfram
Mathematica 12.0.

\vspace{-0.15in}
\section{\bf The approach to analytical
  calculations of the Feynman diagrams in Fig.\,\ref{fig:fig1} -
  Fig.\,\ref{fig:fig4}}
\label{sec:comment}

An important role of strong low-energy interactions in decay processes
has been pointed by Weinberg \cite{Weinberg1957}.  In this connection,
according to Sirlin \cite{Sirlin1978}, the current algebra is a nice
tool for the analysis of contributions of strong low-energy
interactions in the $O(\alpha/\pi)$ RC to semileptonic and leptonic
decays of hadrons. The method of the current algebra is
model-independent. It is based on the use of equal-time commutators of
the hadronic currents and their divergences imposed by the $SU(2)
\times SU(2)$ or $SU(3) \times SU(3)$ symmetries of strong low-energy
interactions \cite{Adler1968, DeAlfaro1973} (see also
\cite{DGH2014}). Indeed, as has been pointed out by Sirlin
\cite{Sirlin1978}:''In fact, a current algebra formulation is probably
our only hope of controlling the effects of the strong interactions in
a clear and logical manner.'' He has shown \cite{Sirlin1978,
  Sirlin1974} that the contributions of strong low-energy interactions
to the inner $O(\alpha/\pi)$ RC have the standard $V - A$ structure
\cite{Feynman1958, Nambu1960, Marshak1969} in the amplitude of the
neutron beta decay.

In our analysis we treat the inner $O(\alpha E_e/m_N)$ RC as NLO
corrections in the large nucleon mass $m_N$ expansion to the inner
$O(\alpha/\pi)$ RC. This causes the inner $O(\alpha E_e/m_N)$ RC to
have the $V - A$ structure as well. To reproduce the inner
$O(\alpha/\pi)$ RC with the $V - A$ structure and to calculate NLO
corrections $O(\alpha E_e/m_N)$ we use the {\it leading logarithmic
  approximation} (LLA) (see, for example, \cite{Czarnecki2002,
  Bissegger2007}) for the analytical calculation of the Feynman
diagrams in Fig.\,\ref{fig:fig1} - Fig.\,\ref{fig:fig4}. As has been
pointed out by Bissegger and Fuhrer \cite{Bissegger2007}, the linear
$\sigma$-model without a nucleon is equivalent to the ChPT by Gasser
and Leutwyler \cite{Gasser1984} in the LLA.  The application of the
LLA to the calculation of the inner $O(\alpha E_e/m_N)$ RC within the
effective field theory L$\sigma$M$\&$SET can also be justified as
follows. It is well-known by example of the bosonization of the
Extended Nambu-Jona-Lasinio (ENJL) model \cite{Bijnens1996} that the
divergent parts of the Feynman diagrams or the counterterms preserve
fully the symmetry of the dynamical quark system. Indeed, the
effective local Lagrangian of the bound quark-antiquark pairs, induced
by the divergent parts of one-quark-loop diagrams, preserves the
symmetry of the ENJL quark model \cite{Bijnens1996} with a local
four-quark interaction, invariant under chiral $SU(2) \times SU(2)$ or
$SU(3) \times SU(3)$ symmetries.

Thus, keeping only the divergent contributions of the Feynman diagrams
in Fig.\,\ref{fig:fig1} - Fig.\,\ref{fig:fig4} we preserve the chiral
$SU(2) \times SU(2)$ symmetry of the L$\sigma$M. The divergent
contributions of the Feynman diagrams in Fig.\,\ref{fig:fig1} -
Fig.\,\ref{fig:fig4}, calculated by following the standard procedure
\cite{Feynman1950} - \cite{Smirnov2012} with the $n$-dimensional
regularization \cite{Kinoshita1974a} - \cite{Smirnov2012}, are
proportional to $\Gamma(2 - n/2)(Q/m^2_N)^{-4+n}/U^{n/2}$. In this
product $Q$ is a function of the Feynman parameters, momenta and
squares of the masses of interacting particles in dependence of the
structure of the Feynman diagram and $U$ is the determinant of the
Feynman diagram depending on the Feynman parameters and the structure
of the Feynman diagram \cite{Kinoshita1974a, Kinoshita1974b,
  Smirnov2004, Smirnov2006, Smirnov2012}. Then, taking the limit $n
\to 4$ and keeping the divergent contributions proportional to
$\Gamma(2 - n/2)$, we get $(\Gamma(2 - n/2) - 2\,{\ell
  n}Q/m^2_N)/U^2$.  Expanding $(\Gamma(2 - n/2) - 2\,{\ell
  n}Q/m^2_N)/U^2$ in powers of $k_n\cdot q/m^2_N$, $k_n \cdot
k_e/m^2_N$ and $m^2_N/M^2_W$ and integrating over the Feynman
parameters we obtain the leading order contributions, determined by
$(\Gamma(2 - n/2) - 2\,{\ell n}Q/m^2_N)/U^2\big|_{q = k_e = 0}$, and
the NLO contributions proportional to $k_n\cdot q/m^2_N$, $k_n\cdot
k_e/m^2_N$ and $m^2_N/M^2_W$. The sum of these contributions preserve
the $V - A$ structure of the amplitude of the neutron beta decay (see
Appendix F).

\vspace{-0.15in}
\section{Generating functional of inner radiative corrections
  in the one-virtual photon exchange approximation}
\label{sec:amplitude}

The general expression for contributions to the amplitude of the
neutron beta decay, taken in the one-virtual photon exchange
approximation with a photon coupled to the hadronic structure of the
neutron, is defined by \cite{Ivanov2019a, Ivanov2018b, Ivanov2018c}
\begin{eqnarray}\label{eq:1}
M(n \to p e^-\bar{\nu}_e)_{\rm st} = \Big\langle {\rm in},
\bar{\nu}_e(\vec{k}_{\bar{\nu}},+ \frac{1}{2}), e^-(\vec{k}_e,
\sigma_e), p(\vec{k}_p,\sigma_p) \Big|{\rm T}e^{\textstyle i\int
  d^4x\,{\cal L}_{\rm L\sigma M \& SET}(x)}\Big| n(\vec{k}_n,
\sigma_n), {\rm in} \Big\rangle_{\rm one-photon-approx.},
\end{eqnarray}
where ${\cal L}_{\rm L\sigma M \& SET}$ is the Lagrangian of the
effective quantum field theory L$\sigma$M$\&$SET defined by Eq.(44) in
Ref.\cite{Ivanov2019a}, and ${\rm T}$ is the time-ordering operator
\cite{Itzykson1980}.

\begin{figure}
 \includegraphics[height=0.31\textheight]{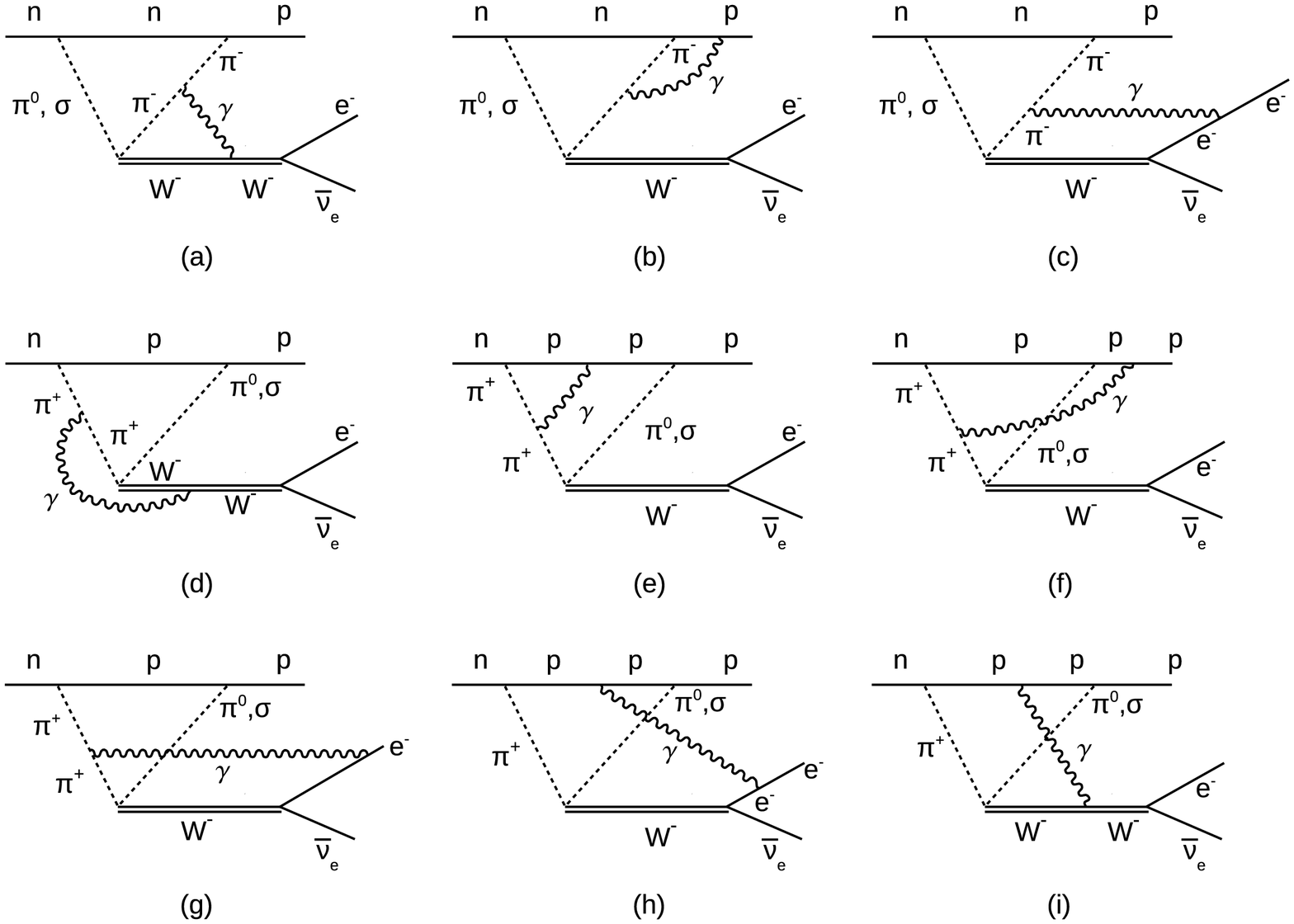} \centering
 \includegraphics[height=0.33\textheight]{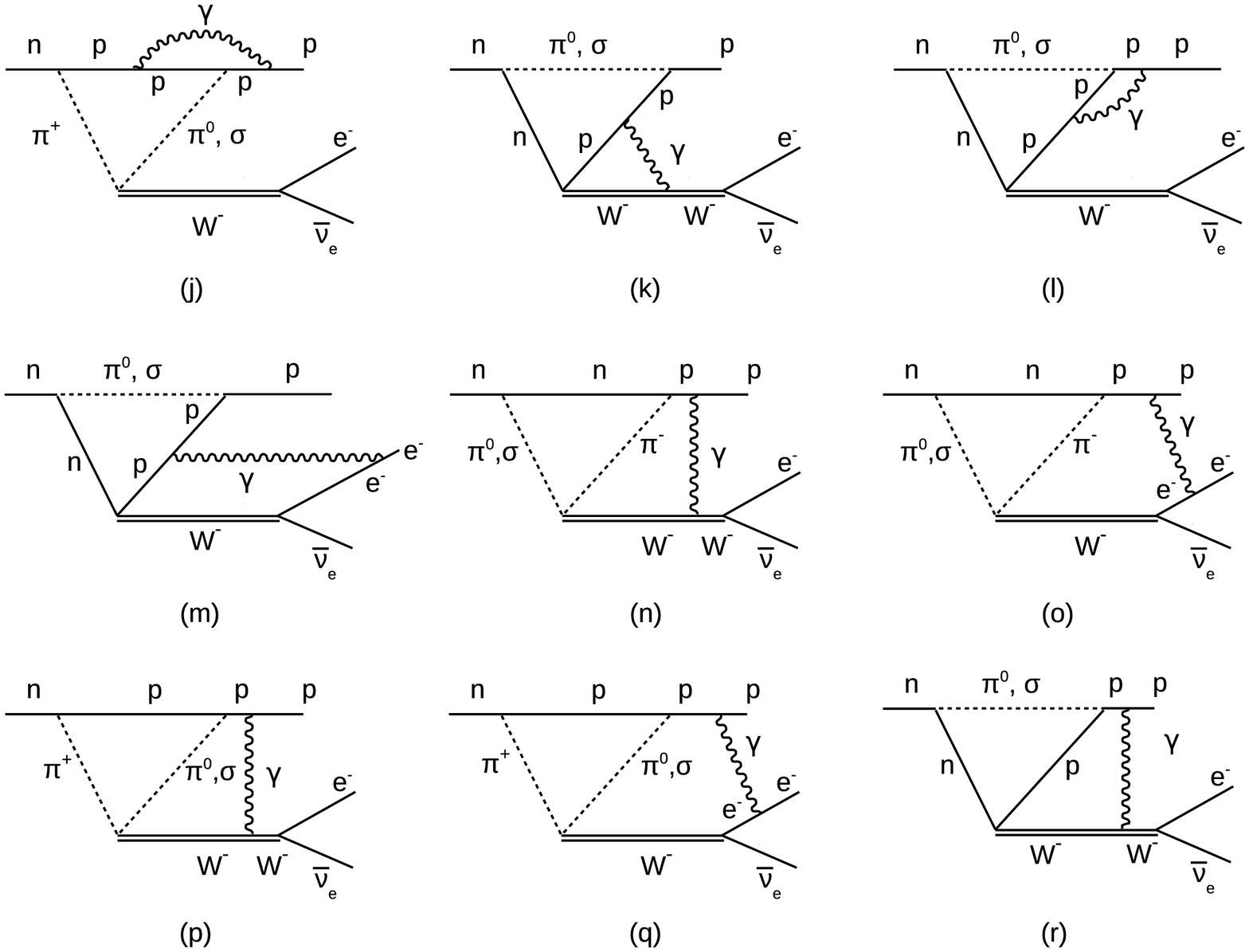}\\ \centering
 \includegraphics[height=0.10\textheight]{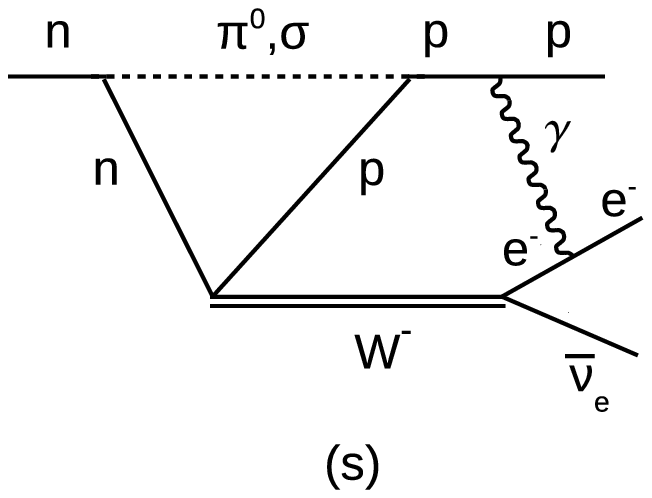}
  \caption{The two-loop Feynman diagrams of the inner RC with a
    virtual photon coupled to the hadronic structure of the neutron,
    charged decay particles and the electroweak $W^-$-boson. }
\label{fig:fig1}
\end{figure}
\begin{figure}
\centering \includegraphics[height=0.23\textheight]{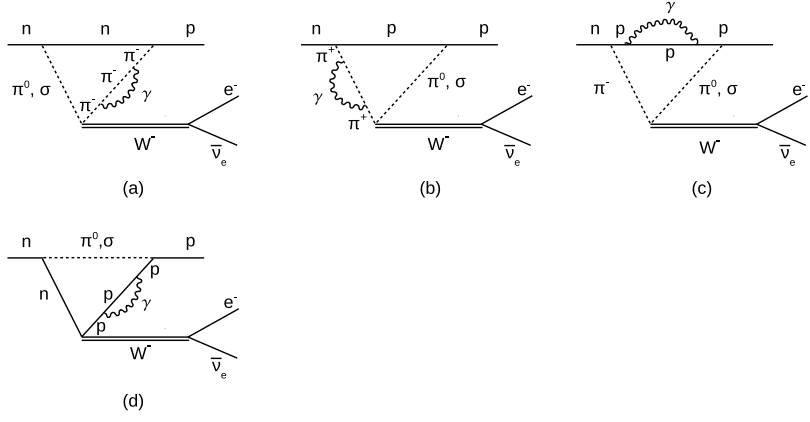}
  \caption{The two-loop Feynman diagrams of the inner RC with
    self-energy corrections to the virtual charged hadrons.}
\label{fig:fig2}
\end{figure}
\begin{figure}
q\centering \includegraphics[height=0.33\textheight]{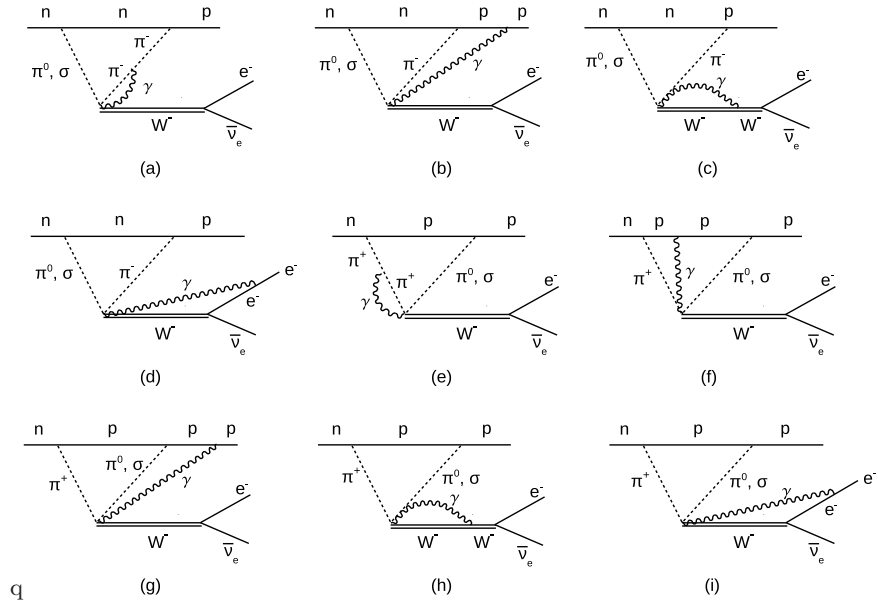}
  \caption{The two-loop Feynman diagrams of the inner RC, induced by
    the interactions of the $\gamma W^-$-pair with the $\pi \pi$- and
    $\pi \sigma$-pairs.}
\label{fig:fig3} 
\end{figure}
\begin{figure}
  \centering \includegraphics[height=0.18\textheight]{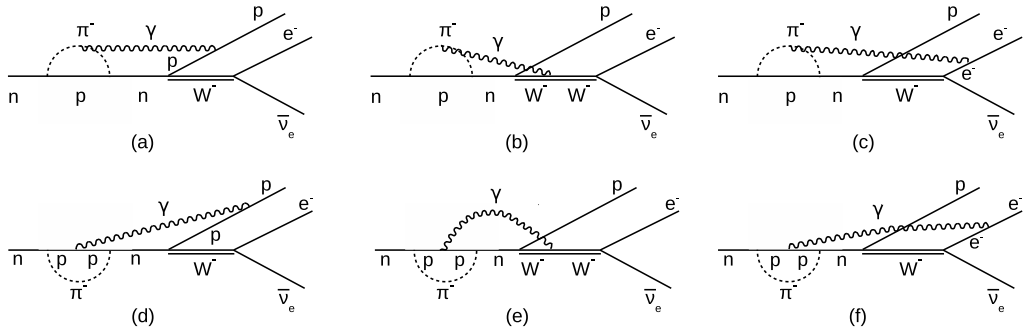}
  \caption{The two-loop Feynman diagrams of the inner RC, induced by a
    virtual photon emitted by virtual hadrons from the self-energy
    hadronic corrections to the neutron.}
\label{fig:fig4} 
\end{figure}

The wave functions of fermions in the initial and
  final states are determined in terms of operators of creation
  (annihilation) \cite{Itzykson1980, Ivanov2018} (see also
  \cite{Ivanov2019a,Ivanov2018b, Ivanov2018c})
\begin{eqnarray}\label{eq:2}
\Big|n(\vec{k}_n, \sigma_n), {\rm in}\Big\rangle &=&
a^{\dagger}_{n,\rm in}(\vec{k}_n, \sigma_n)\Big|0\Big\rangle,
\nonumber\\ \Big\langle {\rm in}, \bar{\nu}_e(\vec{k}_{\bar{\nu}},+
\frac{1}{2}), e^-(\vec{k}_e, \sigma_e), p(\vec{k}_p,\sigma_p)\Big| &=&
\Big\langle 0\Big|b_{\bar{\nu}_e,\rm in}(\vec{k}_{\bar{\nu}},+
\frac{1}{2}) a_{e,\rm in} (\vec{k}_e, \sigma_e) a_{p,\rm
  in}(\vec{k}_p,\sigma_p).
\end{eqnarray}
The operators of creation (annihilation) obey standard anticommutation
relations \cite{Itzykson1980,Ivanov2018}. The required contribution of
the hadronic structure of the neutron to the inner $O(\alpha/\pi)$ RC
appears in the two-loop approximation with one-virtual photon
exchange. It is defined by the Feynman diagrams in
Fig.\,\ref{fig:fig1} - Fig.\ref{fig:fig4}. Since, as has been shown in
\cite{Ivanov2018b, Ivanov2018c}, the RC to the one-pion-pole exchanges
are of the order of $10^{-9}$ (see also \cite{Ivanov2019b}), we omit
them from consideration. The analytical expressions for the Feynman
diagrams in Fig.\,\ref{fig:fig1} - Fig.\ref{fig:fig4} and their
properties under gauge transformation of the virtual photon propagator
are given and investigated in Appendices A, B, C, D, E and F.  For the
calculation of the Feynman diagrams in Fig.\,\ref{fig:fig1} -
Fig.\,\ref{fig:fig4} we have used the standard technique
\cite{Feynman1950} - \cite{Smirnov2012}. The numerical evaluation of
the structure constants of the analytical expressions for the Feynman
diagrams in Fig.\,\ref{fig:fig1} - Fig.\,\ref{fig:fig4} we have
carried out with Wolfram Mathematica 12.0.

\vspace{-0.15in}
\section{The inner $O(\alpha/\pi)$ and
  $O(\alpha E_e/m_N)$ radiative corrections}
\label{sec:lifetime}

The contributions of the inner $O(\alpha/\pi)$ and $O(\alpha E_e/m_N)$
RC are calculated to the amplitude of the neutron beta decay in
Appendices A, B, C, D, E and F.  These corrections
are described by the complete set of the Feynman diagrams in
Fig.\,\ref{fig:fig1} - Fig.\,\ref{fig:fig4}. The analytical
expressions for the Feynman diagrams in Fig.\,\ref{fig:fig1} -
Fig.\,\ref{fig:fig4} we have given in Appendix A. As
we have shown in Appendix B, this set of Feynman
diagrams is gauge invariant. In other words it does not depend on
longitudinal polarization states of a virtual photon. The inner
$O(\alpha E_e/m_N)$ RC are obtained as NLO terms for the inner
$O(\alpha/\pi)$ RC, calculated to LO in the large nucleon mass $m_N$
expansion \cite{Sirlin1967, Sirlin1978}. Following \cite{Sirlin1967}
we have absorbed the terms of order $O(\alpha/\pi)$ by renormalization
of the Fermi weak couping constant $G_V$ and the axial coupling
constant $g_A$. As a result, the contribution of the Feynman diagrams
in Fig.\,\ref{fig:fig1} - Fig.\,\ref{fig:fig4}, calculated to NLO in
the large nucleon mass $m_N$ expansion and the electroweak $W^-$-boson
mass $M_W$ expansion, is given by (see Appendix F)
\begin{eqnarray}\label{eq:3}
 M(n \to p e^- \bar{\nu}_e)^{(\rm NLO)} _{\rm st} &=& -
 \frac{\alpha}{2\pi}\,G_V\Big\{ \Big(G^{(V)}_{\rm st}\, \frac{k_n
   \cdot q}{m^2_N} + H^{(V)}_{\rm st}\, \frac{k_n \cdot k_e}{m^2_N} +
 \big(G^{(W)}_{\rm st} + F^{(W)}_{\rm
   st}\big)\,\frac{m^2_N}{M^2_W}\Big) \Big[\bar{u}_e \gamma^{\mu} (1 -
   \gamma^5) v_{\bar{\nu}}\Big]\,\Big[\bar{u}_p \gamma_{\mu} u_n\Big]
 \nonumber\\ && + \Big( G^{(A)}_{\rm st} \, \frac{k_n \cdot q}{m^2_N }
 + H^{(A)}_{\rm st} \, \frac{k_n \cdot k_e}{m^2_N }+ H^{(W)}_{\rm st}
 \,\frac{m^2_N}{M^2_W}\Big) \Big[\bar{u}_e \gamma^{\mu} (1 - \gamma^5)
   v_{\bar{\nu}}\Big]\, \Big[\bar{u}_p \gamma_{\mu}\gamma^5
   u_n\Big]\Big\},
\end{eqnarray}
where $q = k_p - k_n = - k_e - k_{\bar{\nu}}$ is a 4-momentum transfer
and $E_e/m_N \sim 10^{-3}$ and $m^2_N/M^2_W \sim 10^{-4}$,
respectively.

The contribution of the inner $O(\alpha E_e/m_N)$ RC in
Eq.(\ref{eq:3}) agrees well with the assertion that the inner
$O(\alpha/\pi)$ RC do not depend on the electron energy
\cite{Sirlin1967, Sirlin1978}. The structure constants in
Eq.(\ref{eq:3}) are equal to $G^{(V)}_{\rm st} = - 70.71$,
$H^{(V)}_{\rm st} = 67.75$, $G^{(W)}_{st} = 8.94$, $G^{(A)}_{\rm st} =
41.95$, $H^{(A)}_{\rm st} = - 40.78$, $H^{(W)}_{\rm st} = 2.10$ and
$F^{(W)}_{\rm st} = -1.64$ (see Appendix F).  The
Lorentz structure of Eq.(\ref{eq:3}) is obtained at the neglect of the
contributions of order $O(m_e m_N/M^2_W) \sim O(k_n\cdot q/M^2_W) \sim
O(k_n\cdot k_e/M^2_W) \sim 10^{-7}$ and $O(E^2_0/m^2_N) \sim
O(m^2_{\pi}/M^2_W) \sim 10^{-6}$, respectively. We
would like to emphasize that all structure constants in the matrix
element Eq.(\ref{eq:3}) are induced by the contributions of the first
class currents \cite{Weinberg1958}, which are $G$-even \cite{Lee1956a}
(see also \cite{Ivanov2018}).

\begin{figure}
\includegraphics[height=0.23\textheight]{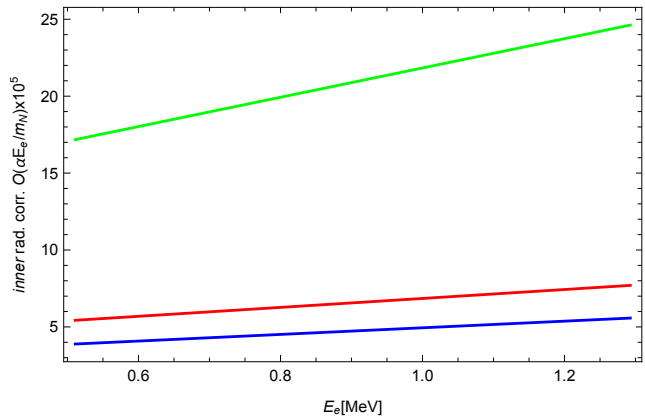}
  \caption{The RC $(\alpha/\pi)g_{\rm st}(E_e)$
    (\textcolor{red}{red}), $(\alpha/\pi)f_{\rm st}(E_e)$
    (\textcolor{blue}{blue}) and $(\alpha/\pi)h_{\rm st}(E_e)$
    (\textcolor{green}{green}), caused by the hadronic structure of
    the neutron, to the electron-energy and angular distribution of
    the neutron beta decay and to the neutron lifetime in the electron
    energy region $m_e \le E_e < E_0$. These are NLO corrections in
    the large nucleon mass $m_N$ expansion to Sirlin's inner RC, which
    have been absorbed by renormalization of the Fermi weak coupling
    constant $G_V$ and the axial coupling constant $g_A$
    \cite{Sirlin1967}}
\label{fig:fig5}
\end{figure}

In the rest frame of the neutron and in the non-relativistic
approximation for the proton the amplitude of the neutron beta decay
with the contribution of the inner $O(\alpha E_e/m_N)$ RC is given by
\begin{eqnarray}\label{eq:4}
 \hspace{-0.30in}M(n \to p e^- \bar{\nu}_e) &=& - 2 m_N G_V
 \Big\{\Big(1 + \frac{\alpha}{2\pi}\,\bar{g}_{\rm
   st}(E_e)\Big)[\varphi^{\dagger}_p \varphi_n][\bar{u}_e \gamma^0 (1
   - \gamma^5) v_{\bar{\nu}}] \nonumber\\ \hspace{-0.30in}&& +
 \Big(g_A + \frac{\alpha}{2\pi}\,\bar{f}_{\rm st}(E_e)\Big)
     [\varphi^{\dagger}_p\, \vec{\sigma}\, \varphi_n] \cdot [\bar{u}_e
       \vec{\gamma}\, (1 - \gamma^5) v_{\bar{\nu}}] + \ldots \Big\},
\end{eqnarray}
where the ellipsis implies the contributions of other terms (see, for
example, \cite{Ivanov2013, Ivanov2019a}), which we do not take into
account here.  The functions $\bar{g}_{\rm st}(E_e)$ and $\bar{f}_{\rm
  st}(E_e)$ are defined in terms of the structure constants as follows
\begin{eqnarray}\label{eq:5}
 \hspace{-0.30in}\bar{g}_{\rm st}(E_e) &=& - G^{(V)}_{\rm
   st}\frac{E_0}{m_N} + \big( G^{(W)}_{\rm st} + F^{(W)}_{\rm
   st}\big)\, \frac{m^2_N}{M^2_W} + H^{(V)}_{\rm st}\frac{E_e}{m_N} =
 0.098 \,\Big(1 + 0.95\,\frac{E_e}{E_0}\Big), \nonumber\\
 \hspace{-0.30in}\bar{f}_{\rm st}(E_e) &=& + G^{(A)}_{\rm
   st}\frac{E_0}{m_N} - H^{(W)}_{\rm st}\frac{m^2_N}{M^2_W} -
 H^{(A)}_{\rm st}\frac{E_e}{m_N} = 0.057\, \Big(1 +
 \frac{E_e}{E_0}\Big),
\end{eqnarray}
where $E_0 = (m^2_n - m^2_p + m^2_e)/2 m_n = 1.2926\, {\rm MeV}$ is
the end-point energy of the electron-energy spectrum \cite{Abele2008,
  Nico2009}. The electron-energy and angular distribution of the
neutron beta decay for unpolarized massive fermions, taking into
account the RC Eq.(\ref{eq:4}), is given by
\begin{eqnarray}\label{eq:6}
\frac{d^5\lambda_n(E_e, \vec{k}_e,
  \vec{k}_{\bar{\nu}})}{d E_e d\Omega_e d\Omega_{\bar{\nu}}} &=& (1 +
3 g^2_A)\,\frac{|G_V|^2}{16\pi^5}\,\Big\{1 + \frac{\alpha}{\pi}\Big(
g_{\rm st}(E_e) + 3\, f_{\rm st}(E_e)\Big) + \Big[a_0 +
  \frac{\alpha}{\pi}\Big( g_{\rm st}(E_e) - f_{\rm st}(E_e) \Big)\Big]
\, \frac{\vec{k}_e \cdot \vec{k}_{\bar{\nu}}}{E_e E_{\bar{\nu}}} +
\ldots\Big\}\nonumber\\ && \times\, \sqrt{E^2_e -
  m^2_e} E_e F(E_e, Z = 1),
\end{eqnarray}
where $d\Omega_e$ and $d\Omega_{\bar{\nu}}$ are infinitesimal solid
angles in the directions of the electron and antineutrino 3-momenta,
$a_0 =(1 - g^2_A)/(1 + 3 g^2_A)$ \cite{Abele2008, Nico2009}, and
$F(E_e, Z = 1)$ is the well-known relativistic Fermi function,
describing electron-proton Coulomb final-state interaction
\cite{Blatt1952, Wilkinson1982}.  The ellipsis denote the
contributions of other terms (see, for example,
\cite{Ivanov2013}). The functions $g_{st}(E_e)$ and $f_{st}(E_e)$ are
related to the functions $\bar{g}_{\rm st}(E_e)$ and $\bar{f}_{\rm
  st}(E_e)$ as follows
\begin{eqnarray}\label{eq:7}
\hspace{-0.3in}g_{st}(E_e) &=& \frac{1}{1 + 3 g^2_A}\, \bar{g}_{\rm
  st}(E_e) = \frac{0.098}{1 + 3 g^2_A}\,\Big(1 + 0.95\,
\frac{E_e}{E_0}\Big) = 0.017 \,\Big(1 + 0.95\,
\frac{E_e}{E_0}\Big),\nonumber\\
\hspace{-0.3in}f_{st}(E_e) &=& \frac{g_A}{1 + 3 g^2_A}\,
\bar{f}_{\rm st}(E_e) = \frac{0.057 g_A}{1 + 3 g^2_A}\,\Big(1 +
\frac{E_e}{E_0}\Big) = 0.012\,\Big(1 +
\frac{E_e}{E_0}\Big). 
\end{eqnarray}
These corrections depend strongly on the axial coupling constant
$g_A$. The numerical values are evaluated for $g_A = 1.2764$
\cite{Abele2018} (see also
\cite{Sirlin2018}). The rate of the neutron beta decay is defined by the
integral
\begin{eqnarray}\label{eq:8}
\hspace{-0.3in}\lambda_n = (1 + 3
g^2_A)\,\frac{|G_V|^2}{\pi^3}\int^{E_0}_{m_e}\,\Big(1 +
\frac{\alpha}{\pi}\, h_{\rm st}(E_e) + \ldots \Big) \, \sqrt{E^2_e -
  m^2_e}\,E_e F(E_e, Z = 1)\,dE_e,
\end{eqnarray}
where the function $h_{\rm st}(E_e)$ is equal to
\begin{eqnarray}\label{eq:9}
\hspace{-0.3in}h_{\rm st}(E_e) = g_{\rm st}(E_e) + 3 f_{\rm st}(E_e) =
0.053\Big(1 + \frac{E_e}{E_0}\Big).
\end{eqnarray}
The functions $(\alpha/\pi)\,g_{\rm st}(E_e)$, $(\alpha/\pi)\,f_{\rm
  st}(E_e)$ and $(\alpha/\pi)\,h_{\rm st}(E_e)$ are calculated at the
neglect of the terms of the order $10^{-6}$. In Fig.\,\ref{fig:fig5} we
plot these functions for $g_A = 1.2764$.
The values of the functions $(\alpha/\pi)g_{\rm st}(E_e)$ are
$(\alpha/\pi)f_{\rm st}(E_e)$ are of the same order and of the order
of a few parts of $10^{-5}$. An increase of the values of the function
$(\alpha/\pi)h_{\rm st}(E_e)$ to a few parts of $10^{-4}$ is caused by
the hadronic axial-vector current. Its contribution to the neutron
lifetime is enhanced by a factor of 3 with respect to the contribution
of the hadronic vector current.

In order to calculate correctly the relative contribution of the inner
$O(\alpha E_e/m_N)$ RC, described by the function $(\alpha/\pi)h_{\rm
  st}(E_e)$, to the neutron lifetime we make a replacement $1 + \ldots
\to \zeta(E_e)$ in Eq.(\ref{eq:8}). We take the correlation function
$\zeta(E_e)$ in the form, calculated in \cite{Ivanov2013}. It contains
a complete set of outer $O(\alpha/\pi)$ RC \cite{Sirlin1967} and the
$O(E_e/m_N)$ corrections, caused by weak magnetism and proton
recoil. In addition the correlation function $\zeta(E_e)$ includes the
contributions of the inner $O(\alpha/\pi)$ RC defined by $\Delta^V_R$
and $\Delta^A_R$. They are induced by the Feynman $\gamma W^-$-box
diagrams and calculated to LO in the large nucleon mass $m_N$
expansion in \cite{Sirlin1986, Sirlin2004, Sirlin2006, Seng2018,
  Seng2018a, Sirlin2019, Hayen2020}. Having integrated over the
electron energy we get
\begin{eqnarray}\label{eq:10}
\hspace{-0.3in}\lambda_n = (1 + 3
g^2_A)\,\frac{|G_V|^2}{\pi^3}\Big(6,136 \times 10^{-2} + 1.18 \times
10^{-5}\Big) \propto 1 + 1.92 \times 10^{-4},
\end{eqnarray}
where the term $1.18 \times 10^{-5}$ is the contribution of the
function $(\alpha/\pi)h_{\rm st}(E_e)$ defining the inner $O(\alpha
E_e/m_N)$ RC in the neutron lifetime.  In turn, the relative
contribution of these RC is of the order $10^{-4}$.

\section{Discussion}
\label{sec:Abschluss}

We have calculated the inner $O(\alpha E_e/m_N)$ RC, induced by the
hadronic structure of the neutron i) to the amplitude of the neutron
beta decay, ii) to the electron-energy and angular distribution of the
neutron beta decay for unpolarized massive fermions and iii) to the
neutron lifetime.  We treat them as NLO terms in the large nucleon
mass $m_N$ expansion to Sirlin's inner $O(\alpha/\pi)$ RC
\cite{Sirlin1967, Sirlin1978}, calculated to LO in the large nucleon
mass $m_N$ expansion.

For the calculation of the inner $O(\alpha/\pi)$ and $O(\alpha
E_e/m_N)$ RC we have used the effective quantum field theory
L$\sigma$M$\&$SET of strong and electroweak low-energy interactions,
proposed in \cite{Ivanov2019a}.  In such an effective quantum field
theory strong low-energy interactions are described by the linear
$\sigma$-model (L$\sigma$M) with chiral $SU(2) \times SU(2)$ symmetry
\cite{GellMann1960}.  In turn, electroweak interactions are described
by the Standard Electroweak Theory (SET) with $SU(2)_L \times U_Y$
symmetry \cite{Weinberg1971, DGH2014} (see also \cite{PDG2020}). The
hadronic and leptonic sectors are represented by the nucleon coupled
to pions and the scalar isoscalar $\sigma$-meson and the
electron-lepton family, respectively.  The application of this
effective quantum field theory to the calculation of the inner
$O(\alpha E_e/m_N)$ RC is well motivated and justified by the results,
obtained in \cite{Ivanov2019a}.

The contributions of the inner $O(\alpha/\pi)$ and $O(\alpha E_e/m_N)$
RC are described in the L$\sigma$M$\&$SET by the two-loop Feynman
diagrams. They are shown in Fig\,\ref{fig:fig1} -
Fig\,\ref{fig:fig4}. We have calculated these Feynman diagrams in the
LLA \cite{Czarnecki2002, Bissegger2007}. This approximation is
justified as follows: i) the linear $\sigma$-model without the nucleon
is equivalent to the ChPT by Gasser and Leutwyler \cite{Gasser1984} in
the LLA (see \cite{Bissegger2007}) , ii) the leading divergent and
finite logarithms preserve the chiral $SU(2) \times SU(2)$ symmetry of
strong low-energy interactions, described by the L$\sigma$M (see also
\cite{Bijnens1996} for bosonization of the ENJL quark model) and iii)
the contributions of the inner $O(\alpha/\pi)$ and $O(\alpha E_e/m_N)$
RC to the amplitude o the neutron beta decay have the standard $V - A$
structure in agreement with \cite{Sirlin1967, Sirlin1978}. Such a $V -
A$ structure of the inner $O(\alpha/\pi)$ RC has been pointed out by
Sirlin \cite{Sirlin1967, Sirlin1978} within the current algebra
approach \cite{Adler1968, DeAlfaro1973}.  The latter has allowed to
remove the inner $O(\alpha/\pi)$ RC by renormalization of the Fermi
weak couping constant $G_V$ and the axial coupling constant $g_A$
\cite{Sirlin1967, Sirlin1978}.

After renormalization we have got a set of the inner $O(\alpha
E_e/m_N)$ RC of the order of a few parts of $10^{-5} - 10^{-4}$. We would
like to emphasize that these corrections are calculated in agreement
with the constraints on the applicability of the L$\sigma$M for the
description of strong low-energy interactions \cite{Myhrer2005} and
\cite{Bissegger2007}. It agrees also with analysis of the RC in the
neutron beta decay, performed in \cite{Myhrer2004} within the
HB$\chi$PT.

The inner $O(\alpha E_e/m_N)$ RC are represented by two functions
$(\alpha/\pi)g_{\rm st}(E_e)$ and $(\alpha/\pi) f_{\rm st}(E_e)$ in
the electron-energy and angular distribution of the neutron beta
decay. In turn, the contribution of the inner $O(\alpha E_e/m_N)$ to
the neutron lifetime is determined by the function $(\alpha/\pi)
h_{\rm st}(E_e)$. It is a linear superposition of the functions
$g_{\rm st}(E_e)$ and $f_{\rm st}(E_e)$, i.e. $h_{\rm st}(E_e) =
g_{\rm st}(E_e) + 3 f_{\rm st}(E_e) = 0.053\, (1 + E_e/E_0)$. We have
plotted the functions $(\alpha/\pi)g_{\rm st}(E_e)$, $(\alpha/\pi)
f_{\rm st}(E_e)$ and $(\alpha/\pi) h_{\rm st}(E_e)$ in
Fig.\,\ref{fig:fig5}. In the electron-energy region $m_e \le E_e \le
E_0$ the numerical values of the functions $(\alpha/\pi)g_{\rm
  st}(E_e)$ and $(\alpha/\pi) f_{\rm st}(E_e)$ are of the order of a
few parts of $10^{-5}$. Nevertheless, the function $(\alpha/\pi)
h_{\rm st}(E_e)$ is of the order of $10^{-4}$ and varies over the
region $1.72 \times 10^{-4} \le (\alpha/\pi)h_{\rm st}(E_e) \le 2.46
\times 10^{-4}$.  The contribution of the function $(\alpha/\pi)
h_{\rm st}(E_e)$, integrated over the phase volume of the neutron beta
decay, is of the order of $10^{-5}$ to the rate of the neutron beta
decay. In turn, its relative contribution is of the order of
$10^{-4}$.

As has been shown in \cite{Ivanov2021c}, the inner $O(\alpha E_e/m_N)$
RC Eq.(\ref{eq:7}) provide the SM theoretical description of the
neutron beta decay at the level of $10^{-5} - 10^{-4}$ together with
i) the $O(\alpha E_e/m_N)$ RC, calculated in \cite{Ivanov2019a}, ii)
the $O(E^2_e/m^2_N) \sim 10^{-5}$ corrections, caused by weak
magnetism and proton recoil \cite{Ivanov2020b}, and iii) Wilkinson's
corrections \cite{Wilkinson1982} (see also \cite{Ivanov2013,
  Ivanov2017w, Ivanov2019w}).  The theoretical accuracy of these
corrections is of a few parts of $10^{-6}$ \cite{Ivanov2021c}. Such a
SM theoretical background of the neutron beta decay should be very
important for experimental searches of interactions beyond the SM
\cite{Paul2009, Abele2016, Bodek2019} with experimental uncertainties
of a few part of $10^{-5}$ and even better.

Of course, a very challenging extension of our approach to the
calculation of the inner $O(\alpha E_e/m_N)$ RC is the use of the
effective quantum field theory of strong and electroweak low-energy
interactions HB$\chi$PT$\&$SET. In this effective theory the hadronic
part is described by the HB$\chi$PT, which is accepted as an effective
low-energy dynamics of QCD (see, for example, Ecker \cite{Ecker1995,
  Ecker1996}). In such an effective theory, we could do calculations
beyond the LLA and take into account a variety of Lorentz structures
that differ from the standard $V - A$ structure. We would like to
emphasize that the problem of the reformulation of the effective
quantum field theory of strong and electroweak low-energy interactions
L$\sigma$M$\&$SET, where the L$\sigma$M is replaced by the HB$\chi$PT,
is not straightforward.  We are planning to devote to the analysis and
solution of this problem our subsequent researches.

\section{Acknowledgements}

We thank Hartmut Abele for discussions stimulating the work under
corrections of the order $10^{-5}$ to the neutron beta decay. We are
grateful to Vladimir Smirnov for very important numerous discussions
and comments on the calculation of Feynman's integrals. We thank
Manfried Faber for reading the manuscript and discussions, and Ren\'e
Sedmik for discussions.  The work of A. N. Ivanov was supported by the
Austrian ``Fonds zur F\"orderung der Wissenschaftlichen Forschung''
(FWF) under contracts P31702-N27 and P26636-N20, and ``Deutsche
F\"orderungsgemeinschaft'' (DFG) AB 128/5-2. The work of
R. H\"ollwieser was supported by the Deutsche Forschungsgemeinschaft
in the SFB/TR 55. The work of M. Wellenzohn was supported by the MA
23.

\newpage

\section{\bf Supplemental material}
\label{sec:appendix}

\section*{Appendix A: Analytical expressions for the Feynman diagrams 
in Fig.\,\ref{fig:fig1} - Fig.\,\ref{fig:fig4} }
\renewcommand{\theequation}{A-\arabic{equation}}
\setcounter{equation}{0}

Using the Lagrangian Eq.(44) of the effective quantum field theory of
strong low-energy and electroweak interactions (see
Ref.\,\cite{Ivanov2019a}), we obtain the following analytical
expressions for the Feynman diagrams in Fig.\,\ref{fig:fig1} -
Fig.\,\ref{fig:fig4}:
\begin{eqnarray*}
&&M(n \to p e^- \bar{\nu}_e)^{(\pi^0)}_{\rm Fig.\,\ref{fig:fig1}a} = +
  2 e^2 g^2_{\pi N} M^2_W G_V\nonumber\\ &&\times \int
  \frac{d^4k}{(2\pi)^4i}\int
  \frac{d^4p}{(2\pi)^4i}\,\Big[\bar{u}_e(\vec{k}_e, \sigma_e)
    \gamma^{\mu}(1 - \gamma^5) v_{\bar{\nu}}(\vec{k}_{\bar{\nu}}, +
    \frac{1}{2})\Big]\,\Big[\bar{u}_p(\vec{k}_p,\sigma_p)\gamma^5
    \frac{1}{m_N - \hat{k}_n - \hat{k} - i0}\,\gamma^5 u_n(\vec{k}_n,
    \sigma_n)\Big]\nonumber\\ &&\times \frac{(2 k + p - q)^{\alpha_2}
    (p + 2 k - 2 q)^{\beta_2}}{[m^2_{\pi} - k^2 - i0][m^2_{\pi} - (k -
      q)^2 - i0][m^2_{\pi} - (k + p - q)^2 - i0]}\big[(p -
    q)^{\nu}\eta^{\beta_1\alpha_1} - (p - 2q)^{\beta_1} \eta^{\alpha_1
      \nu} - q^{\alpha_1} \eta^{\nu \beta_1}\big] \nonumber\\ &&\times
  \,D^{(\gamma)}_{\beta_1 \beta_2}(p)\, D^{(W)}_{\alpha_1 \alpha_2}(p
  - q)\, D^{(W)}_{\mu\nu}(- q),\nonumber\\ &&M(n \to p e^-
  \bar{\nu}_e)^{(\sigma)}_{\rm Fig.\,\ref{fig:fig1}a} = - 2 e^2
  g^2_{\pi N} M^2_W G_V\nonumber\\ &&\times \int
  \frac{d^4k}{(2\pi)^4i}\int
  \frac{d^4p}{(2\pi)^4i}\,\Big[\bar{u}_e(\vec{k}_e, \sigma_e)
    \gamma^{\mu}(1 - \gamma^5) v_{\bar{\nu}}(\vec{k}_{\bar{\nu}}, +
    \frac{1}{2})\Big]\,\Big[\bar{u}_p(\vec{k}_p,\sigma_p)\gamma^5
    \frac{1}{m_N - \hat{k}_n - \hat{k} - i0}\, u_n(\vec{k}_n,
    \sigma_n)\Big]\nonumber\\ &&\times \frac{(2 k + p - q)^{\alpha_2}
    (k + 2 p - 2 q)^{\beta_2}}{[m^2_{\sigma} - k^2 - i0][m^2_{\pi} -
      (k - q)^2 - i0][m^2_{\pi} - (k + p - q)^2 - i0]}\big[(p -
    q)^{\nu}\eta^{\beta_1\alpha_1} - (p - 2q)^{\beta_1} \eta^{\alpha_1
      \nu} - q^{\alpha_1} \eta^{\nu \beta_1}\big] \nonumber\\ &&\times
  \,D^{(\gamma)}_{\beta_1 \beta_2}(p)\, D^{(W)}_{\alpha_1 \alpha_2}(p
  - q)\, D^{(W)}_{\mu\nu}(- q),\nonumber\\ &&M(n \to p e^-
  \bar{\nu}_e)^{(\pi^0)}_{\rm Fig.\,\ref{fig:fig1}b} = - 2 e^2
  g^2_{\pi N} M^2_W G_V\nonumber\\ &&\times \int
  \frac{d^4k}{(2\pi)^4i}\int
  \frac{d^4p}{(2\pi)^4i}\,\Big[\bar{u}_e(\vec{k}_e, \sigma_e)
    \gamma^{\mu}(1 - \gamma^5) v_{\bar{\nu}}(\vec{k}_{\bar{\nu}}, +
    \frac{1}{2})\Big]\,\Big[\bar{u}_p(\vec{k}_p,\sigma_p)\gamma^{\beta_1}
    \frac{1}{m_N - \hat{k}_p + \hat{p} - i0} \gamma^5 \frac{1}{m_N -
      \hat{k}_n - \hat{k} - i0} \nonumber\\ &&\times\,\gamma^5
    u_n(\vec{k}_n, \sigma_n)\Big]\,\frac{(2 k - q)^{\nu} (p + 2 k - 2
    q)^{\beta_2}}{[m^2_{\pi} - k^2 - i0][m^2_{\pi} - (k - q)^2 -
      i0][m^2_{\pi} - (k + p - q)^2 - i0]}\,D^{(\gamma)}_{\beta_1
    \beta_2}(p)\, D^{(W)}_{\mu\nu}(- q),\nonumber\\ &&M(n \to p e^-
  \bar{\nu}_e)^{(\sigma)}_{\rm Fig.\,\ref{fig:fig1}b} = + 2 e^2
  g^2_{\pi N} M^2_W G_V\nonumber\\ &&\times \int
  \frac{d^4k}{(2\pi)^4i}\int
  \frac{d^4p}{(2\pi)^4i}\,\Big[\bar{u}_e(\vec{k}_e, \sigma_e)
    \gamma^{\mu}(1 - \gamma^5) v_{\bar{\nu}}(\vec{k}_{\bar{\nu}}, +
    \frac{1}{2})\Big]\,\Big[\bar{u}_p(\vec{k}_p,\sigma_p)\gamma^{\beta_1}
    \frac{1}{m_N - \hat{k}_p + \hat{p} - i0} \gamma^5 \frac{1}{m_N -
      \hat{k}_n - \hat{k} - i0} \nonumber\\ &&\times\, u_n(\vec{k}_n,
    \sigma_n)\Big]\,\frac{(2 k - q)^{\nu} (p + 2 k - 2
    q)^{\beta_2}}{[m^2_{\sigma} - k^2 - i0][m^2_{\pi} - (k - q)^2 -
      i0][m^2_{\pi} - (k + p - q)^2 - i0]}\,D^{(\gamma)}_{\beta_1
    \beta_2}(p)\, D^{(W)}_{\mu\nu}(- q),\nonumber\\ &&M(n \to p e^-
  \bar{\nu}_e)^{(\pi^0)}_{\rm Fig.\,\ref{fig:fig1}c} = + 2 e^2
  g^2_{\pi N} M^2_W G_V\nonumber\\ &&\times \int
  \frac{d^4k}{(2\pi)^4i}\int
  \frac{d^4p}{(2\pi)^4i}\,\Big[\bar{u}_e(\vec{k}_e,
    \sigma_e)\gamma^{\beta_1}\,\frac{1}{m_e - \hat{p} - \hat{k}_e -
      i0}\, \gamma^{\mu}(1 - \gamma^5)
    v_{\bar{\nu}}(\vec{k}_{\bar{\nu}}, +
    \frac{1}{2})\Big]\,\Big[\bar{u}_p(\vec{k}_p,\sigma_p)\gamma^5
    \frac{1}{m_N - \hat{k}_n - \hat{k} - i0}\nonumber\\ &&\times\,
    \gamma^5\, u_n(\vec{k}_n, \sigma_n)\Big]\,\frac{(2 k + p -
    q)^{\nu} (p + 2 k - 2 q)^{\beta_2}}{[m^2_{\pi} - k^2 -
      i0][m^2_{\pi} - (k - q)^2 - i0][m^2_{\pi} - (k + p - q)^2 -
      i0]}\,D^{(\gamma)}_{\beta_1\beta_2}(p)\,D^{(W)}_{\mu\nu}(p - q),
  \nonumber\\ &&M(n \to p e^- \bar{\nu}_e)^{(\sigma)}_{\rm
    Fig.\,\ref{fig:fig1}c} = - 2 e^2 g^2_{\pi N} M^2_W
  G_V\nonumber\\ &&\times \int \frac{d^4k}{(2\pi)^4i}\int
  \frac{d^4p}{(2\pi)^4i}\,\Big[\bar{u}_e(\vec{k}_e,
    \sigma_e)\gamma^{\beta_1}\,\frac{1}{m_e - \hat{p} - \hat{k}_e -
      i0}\, \gamma^{\mu}(1 - \gamma^5)
    v_{\bar{\nu}}(\vec{k}_{\bar{\nu}}, +
    \frac{1}{2})\Big]\Big[\bar{u}_p(\vec{k}_p,\sigma_p)\gamma^5
    \frac{1}{m_N - \hat{k}_n - \hat{k} - i0}\nonumber\\ &&\times\,
    u_n(\vec{k}_n, \sigma_n)\Big]\, \frac{(2 k + p - q)^{\nu} (p + 2 k
    - 2 q)^{\beta_2}}{[m^2_{\sigma} - k^2 - i0][m^2_{\pi} - (k - q)^2
      - i0][m^2_{\pi} - (k + p - q)^2 -
      i0]}\,D^{(\gamma)}_{\beta_1\beta_2}(p) \,D^{(W)}_{\mu\nu}(p -
  q),\nonumber\\ &&M(n \to p e^- \bar{\nu}_e)^{(\pi^0)}_{\rm
    Fig.\,\ref{fig:fig1}d} = - 2 e^2 g^2_{\pi N} M^2_W
  G_V\nonumber\\ &&\times \int \frac{d^4k}{(2\pi)^4i}\int
  \frac{d^4p}{(2\pi)^4i}\,\Big[\bar{u}_e(\vec{k}_e, \sigma_e)
    \gamma^{\mu}(1 - \gamma^5) v_{\bar{\nu}}(\vec{k}_{\bar{\nu}}, +
    \frac{1}{2})\Big]\,\Big[\bar{u}_p(\vec{k}_p,\sigma_p)\gamma^5
    \frac{1}{m_N - \hat{k}_n - \hat{k} - i0}\,\gamma^5\,
    u_n(\vec{k}_n, \sigma_n)\Big]\nonumber\\ &&\times \, \frac{(2 k -
    p - q)^{\alpha_2} (2 k - p)^{\beta_2}}{[m^2_{\pi} - k^2 -
      i0][m^2_{\pi} - (k - q)^2 - i0][m^2_{\pi} - (k - p)^2 -
      i0]}\,\big[(p - q)^{\nu}\eta^{\beta_1\alpha_1} - (p - 2
    q)^{\beta_1}\eta^{\nu \alpha_1} -
    q^{\alpha_1}\eta^{\beta_1\nu}\big]\nonumber\\ &&\times
  \,D^{(\gamma)}_{\beta_1\beta_2}(p) \,D^{(W)}_{\alpha_1\alpha_2}(p -
  q)\,D^{(W)}_{\mu\nu}(- q),\nonumber\\ &&M(n \to p e^-
  \bar{\nu}_e)^{(\sigma)}_{\rm Fig.\,\ref{fig:fig1}d} = - 2 e^2
  g^2_{\pi N} M^2_W G_V\nonumber\\ &&\times \int
  \frac{d^4k}{(2\pi)^4i}\int
  \frac{d^4p}{(2\pi)^4i}\,\Big[\bar{u}_e(\vec{k}_e, \sigma_e)
    \gamma^{\mu}(1 - \gamma^5) v_{\bar{\nu}}(\vec{k}_{\bar{\nu}}, +
    \frac{1}{2})\Big]\,\Big[\bar{u}_p(\vec{k}_p,\sigma_p) \frac{1}{m_N
      - \hat{k}_n - \hat{k} - i0}\,\gamma^5\, u_n(\vec{k}_n,
    \sigma_n)\Big]\nonumber\\ &&\times \, \frac{(2 k - p -
    q)^{\alpha_2} (2k - p)^{\beta_2}}{[m^2_{\pi} - k^2 -
      i0][m^2_{\sigma} - (k - q)^2 - i0][m^2_{\pi} - (k - p)^2 -
      i0]}\,\big[(p - q)^{\nu}\eta^{\beta_1\alpha_1} - (p - 2
    q)^{\beta_1}\eta^{\nu\alpha_1} -
    q^{\alpha_1}\eta^{\beta_1\nu}\big]\nonumber\\ &&\times
  \,D^{(\gamma)}_{\beta_1\beta_2}(p) \,D^{(W)}_{\alpha_1\alpha_2}(p -
  q)\,D^{(W)}_{\mu\nu}(- q),\nonumber\\
\end{eqnarray*}
\begin{eqnarray*}
  &&M(n \to p e^- \bar{\nu}_e)^{(\pi^0)}_{\rm Fig.\,\ref{fig:fig1}e} =
  + 2 e^2 g^2_{\pi N} M^2_W G_V\nonumber\\ &&\times \int
  \frac{d^4k}{(2\pi)^4i}\int
  \frac{d^4p}{(2\pi)^4i}\,\Big[\bar{u}_e(\vec{k}_e, \sigma_e)
    \gamma^{\mu}(1 - \gamma^5) v_{\bar{\nu}}(\vec{k}_{\bar{\nu}}, +
    \frac{1}{2})\Big]\,\Big[\bar{u}_p(\vec{k}_p,\sigma_p)\gamma^5
    \frac{1}{m_N - \hat{k}_n -\hat{k} - \hat{p} - i0} \gamma^{\beta_1}
    \frac{1}{m_N - \hat{k}_n - \hat{k} - i0}
    \nonumber\\ &&\times\,\gamma^5 u_n(\vec{k}_n,
    \sigma_n)\Big]\,\frac{(2 k + 2 p - q)^{\nu} (2 k + p
    )^{\beta_2}}{[m^2_{\pi} - k^2 - i0][m^2_{\pi} - (k + p)^2 -
      i0][m^2_{\pi} - (k + p - q)^2 - i0]}\,D^{(\gamma)}_{\beta_1
    \beta_2}(p)\, D^{(W)}_{\mu\nu}(- q),\nonumber\\ &&M(n \to p e^-
  \bar{\nu}_e)^{(\sigma)}_{\rm Fig.\,\ref{fig:fig1}e} = + 2 e^2
  g^2_{\pi N} M^2_W G_V\nonumber\\ &&\times \int
  \frac{d^4k}{(2\pi)^4i}\int
  \frac{d^4p}{(2\pi)^4i}\,\Big[\bar{u}_e(\vec{k}_e, \sigma_e)
    \gamma^{\mu}(1 - \gamma^5) v_{\bar{\nu}}(\vec{k}_{\bar{\nu}}, +
    \frac{1}{2})\Big]\,\Big[\bar{u}_p(\vec{k}_p,\sigma_p)\,
    \frac{1}{m_N - \hat{k}_n -\hat{k} - \hat{p} - i0} \gamma^{\beta_1}
    \frac{1}{m_N - \hat{k}_n - \hat{k} - i0}
    \nonumber\\ &&\times\,\gamma^5 u_n(\vec{k}_n,
    \sigma_n)\Big]\,\frac{(2 k + 2 p - q)^{\nu} (2 k + p
    )^{\beta_2}}{[m^2_{\pi} - k^2 - i0][m^2_{\pi} - (k + p)^2 -
      i0][m^2_{\sigma} - (k + p - q)^2 - i0]}\,D^{(\gamma)}_{\beta_1
    \beta_2}(p)\, D^{(W)}_{\mu\nu}(- q),\nonumber\\ &&M(n \to p e^-
  \bar{\nu}_e)^{(\pi^0)}_{\rm Fig.\,\ref{fig:fig1}f} = + 2 e^2
  g^2_{\pi N} M^2_W G_V\nonumber\\ &&\times \int
  \frac{d^4k}{(2\pi)^4i}\int
  \frac{d^4p}{(2\pi)^4i}\,\Big[\bar{u}_e(\vec{k}_e, \sigma_e)
    \gamma^{\mu}(1 - \gamma^5) v_{\bar{\nu}}(\vec{k}_{\bar{\nu}}, +
    \frac{1}{2})\Big]\,\Big[\bar{u}_p(\vec{k}_p,\sigma_p)\gamma^{\beta_1}
    \frac{1}{m_N - \hat{k}_p + \hat{p} - i0} \gamma^5 \frac{1}{m_N -
      \hat{k}_n - \hat{k} - i0} \nonumber\\ &&\times\,\gamma^5
    u_n(\vec{k}_n, \sigma_n)\Big]\,\frac{(2 k + 2 p - q)^{\nu} (2 k +
    p )^{\beta_2}}{[m^2_{\pi} - k^2 - i0][m^2_{\pi} - (k + p)^2 -
      i0][m^2_{\pi} - (k + p - q)^2 - i0]}\,D^{(\gamma)}_{\beta_1
    \beta_2}(p)\, D^{(W)}_{\mu\nu}(- q),\nonumber\\ &&M(n \to p e^-
  \bar{\nu}_e)^{(\sigma)}_{\rm Fig.\,\ref{fig:fig1}f} = + 2 e^2
  g^2_{\pi N} M^2_W G_V\nonumber\\ &&\times \int
  \frac{d^4k}{(2\pi)^4i}\int
  \frac{d^4p}{(2\pi)^4i}\,\Big[\bar{u}_e(\vec{k}_e, \sigma_e)
    \gamma^{\mu}(1 - \gamma^5) v_{\bar{\nu}}(\vec{k}_{\bar{\nu}}, +
    \frac{1}{2})\Big]\,\Big[\bar{u}_p(\vec{k}_p,\sigma_p)\gamma^{\beta_1}
    \frac{1}{m_N - \hat{k}_p + \hat{p} - i0}\,\frac{1}{m_N - \hat{k}_n
      - \hat{k} - i0} \nonumber\\ &&\times\,\gamma^5 u_n(\vec{k}_n,
    \sigma_n)\Big]\,\frac{(2 k + 2 p - q)^{\nu} (2 k + p
    )^{\beta_2}}{[m^2_{\pi} - k^2 - i0][m^2_{\pi} - (k + p)^2 -
      i0][m^2_{\sigma} - (k + p - q)^2 - i0]}\,D^{(\gamma)}_{\beta_1
    \beta_2}(p)\, D^{(W)}_{\mu\nu}(- q),\nonumber\\ &&M(n \to p e^-
  \bar{\nu}_e)^{(\pi^0)}_{\rm Fig.\,\ref{fig:fig1}g} = - 2 e^2
  g^2_{\pi N} M^2_W G_V\nonumber\\ &&\times \int
  \frac{d^4k}{(2\pi)^4i}\int
  \frac{d^4p}{(2\pi)^4i}\,\Big[\bar{u}_e(\vec{k}_e,
    \sigma_e)\gamma^{\beta_1} \,\frac{1}{m_e - \hat{p} - \hat{k}_e -
      i0}\,\gamma^{\mu}(1 - \gamma^5)
    v_{\bar{\nu}}(\vec{k}_{\bar{\nu}}, +
    \frac{1}{2})\Big]\,\Big[\bar{u}_p(\vec{k}_p,\sigma_p) \gamma^5
    \frac{1}{m_N - \hat{k}_n - \hat{k} - i0}
    \nonumber\\ &&\times\,\gamma^5 u_n(\vec{k}_n,
    \sigma_n)\Big]\,\frac{(2 k - p - q)^{\nu} (2 k - p
    )^{\beta_2}}{[m^2_{\pi} - k^2 - i0][m^2_{\pi} - (k - p)^2 -
      i0][m^2_{\pi} - (k - q)^2 - i0]}\,D^{(\gamma)}_{\beta_1
    \beta_2}(p)\, D^{(W)}_{\mu\nu}(p - q),\nonumber\\ &&M(n \to p e^-
  \bar{\nu}_e)^{(\sigma)}_{\rm Fig.\,\ref{fig:fig1}g} = - 2 e^2
  g^2_{\pi N} M^2_W G_V\nonumber\\ &&\times \int
  \frac{d^4k}{(2\pi)^4i}\int
  \frac{d^4p}{(2\pi)^4i}\,\Big[\bar{u}_e(\vec{k}_e,
    \sigma_e)\gamma^{\beta_1}\,\frac{1}{m_e - \hat{p} - \hat{k}_e -
      i0}\, \gamma^{\mu}(1 - \gamma^5)
    v_{\bar{\nu}}(\vec{k}_{\bar{\nu}}, +
    \frac{1}{2})\Big]\,\Big[\bar{u}_p(\vec{k}_p,\sigma_p)\,
    \frac{1}{m_N - \hat{k}_n - \hat{k} - i0}
    \nonumber\\ &&\times\,\gamma^5 u_n(\vec{k}_n,
    \sigma_n)\Big]\,\frac{(2 k - p - q)^{\nu} (2 k - p
    )^{\beta_2}}{[m^2_{\pi} - k^2 - i0][m^2_{\pi} - (k - p)^2 -
      i0][m^2_{\sigma} - (k - q)^2 - i0]}\,D^{(\gamma)}_{\beta_1
    \beta_2}(p)\, D^{(W)}_{\mu\nu}(p - q),\nonumber\\ &&M(n \to p e^-
  \bar{\nu}_e)^{(\pi^0)}_{\rm Fig.\,\ref{fig:fig1}h} = - 2 e^2
  g^2_{\pi N} M^2_W G_V\nonumber\\ &&\times \int
  \frac{d^4k}{(2\pi)^4i}\int
  \frac{d^4p}{(2\pi)^4i}\,\Big[\bar{u}_e(\vec{k}_e,
    \sigma_e)\gamma^{\beta_1}\,\frac{1}{m_e - \hat{p} - \hat{k}_e -
      i0}\, \gamma^{\mu}(1 - \gamma^5)
    v_{\bar{\nu}}(\vec{k}_{\bar{\nu}}, +
    \frac{1}{2})\Big]\,\Big[\bar{u}_p(\vec{k}_p,\sigma_p) \gamma^5
    \frac{1}{m_N - \hat{k}_n - \hat{k} - \hat{p} -
      i0}\nonumber\\ &&\times \gamma^{\beta_2}\frac{1}{m_N - \hat{k}_n
      - \hat{k} - i0}\,\gamma^5 u_n(\vec{k}_n,
    \sigma_n)\Big]\,\frac{(2 k + p - q)^{\nu}}{[m^2_{\pi} - k^2 -
      i0][m^2_{\pi} - (k + p -q)^2 - i0]}\,D^{(\gamma)}_{\beta_1
    \beta_2}(p)\, D^{(W)}_{\mu\nu}(p - q),\nonumber\\ &&M(n \to p e^-
  \bar{\nu}_e)^{(\sigma)}_{\rm Fig.\,\ref{fig:fig1}h} = - 2 e^2
  g^2_{\pi N} M^2_W G_V\nonumber\\ &&\times \int
  \frac{d^4k}{(2\pi)^4i}\int
  \frac{d^4p}{(2\pi)^4i}\,\Big[\bar{u}_e(\vec{k}_e,
    \sigma_e)\gamma^{\beta_1}\,\frac{1}{m_e - \hat{p} - \hat{k}_e -
      i0}\, \gamma^{\mu}(1 - \gamma^5)
    v_{\bar{\nu}}(\vec{k}_{\bar{\nu}}, +
    \frac{1}{2})\Big]\,\Big[\bar{u}_p(\vec{k}_p,\sigma_p)\,
    \frac{1}{m_N - \hat{k}_n - \hat{k} - \hat{p} -
      i0}\nonumber\\ &&\times \gamma^{\beta_2}\frac{1}{m_N - \hat{k}_n
      - \hat{k} - i0}\,\gamma^5 u_n(\vec{k}_n,
    \sigma_n)\Big]\,\frac{(2 k + p - q)^{\nu}}{[m^2_{\pi} - k^2 -
      i0][m^2_{\sigma} - (k + p -q)^2 - i0]}\,D^{(\gamma)}_{\beta_1
    \beta_2}(p)\, D^{(W)}_{\mu\nu}(p - q)\nonumber\\ &&M(n \to p e^-
  \bar{\nu}_e)^{(\pi^0)}_{\rm Fig.\,\ref{fig:fig1}i} = - 2 e^2
  g^2_{\pi N} M^2_W G_V\nonumber\\ &&\times \int
  \frac{d^4k}{(2\pi)^4i}\int
  \frac{d^4p}{(2\pi)^4i}\,\Big[\bar{u}_e(\vec{k}_e, \sigma_e)
    \gamma^{\mu}(1 - \gamma^5) v_{\bar{\nu}}(\vec{k}_{\bar{\nu}}, +
    \frac{1}{2})\Big]\,\Big[\bar{u}_p(\vec{k}_p,\sigma_p)\gamma^5
    \frac{1}{m_N - \hat{k}_n -\hat{k} - \hat{p} - i0} \gamma^{\beta_1}
    \frac{1}{m_N - \hat{k}_n - \hat{k} - i0}
    \nonumber\\ &&\times\,\gamma^5 u_n(\vec{k}_n,
    \sigma_n)\Big]\,\frac{(2 k + p - q)^{\alpha_2}}{[m^2_{\pi} - k^2 -
      i0][m^2_{\pi} - (k + p - q)^2 - i0]}\,\big[(p -
    q)^{\nu}\,\eta^{\beta_2 \alpha_1} - (p - 2q)^{\beta_2}
    \eta^{\alpha_1 \nu} - q^{\alpha_1} \eta^{\nu \beta_2}\big]\,
  D^{(\gamma)}_{\beta_1 \beta_2}(p)\nonumber\\ &&\times\,
  D^{(W)}_{\alpha_1\alpha_2}(p - q) D^{(W)}_{\mu\nu}(- q),\nonumber\\ 
\end{eqnarray*}
\begin{eqnarray*}
&&M(n \to p e^- \bar{\nu}_e)^{(\sigma)}_{\rm Fig.\,\ref{fig:fig1}i} =
  - 2 e^2 g^2_{\pi N} M^2_W G_V\nonumber\\ &&\times \int
  \frac{d^4k}{(2\pi)^4i}\int
  \frac{d^4p}{(2\pi)^4i}\,\Big[\bar{u}_e(\vec{k}_e, \sigma_e)
    \gamma^{\mu}(1 - \gamma^5) v_{\bar{\nu}}(\vec{k}_{\bar{\nu}}, +
    \frac{1}{2})\Big]\,\Big[\bar{u}_p(\vec{k}_p,\sigma_p)\,
    \frac{1}{m_N - \hat{k}_n - \hat{k} - \hat{p} - i0}
    \gamma^{\beta_1} \frac{1}{m_N - \hat{k}_n - \hat{k} - i0}
    \nonumber\\ &&\times\,\gamma^5 u_n(\vec{k}_n,
    \sigma_n)\Big]\,\frac{(2 k + p - q)^{\alpha_2}}{[m^2_{\pi} - k^2 -
      i0][m^2_{\sigma} - (k + p - q)^2 - i0]}\,\big[(p -
    q)^{\nu}\,\eta^{\beta_2 \alpha_1} - (p - 2q)^{\beta_2}
    \eta^{\alpha_1 \nu} - q^{\alpha_1} \eta^{\nu \beta_2}\big]\,
  D^{(\gamma)}_{\beta_1 \beta_2}(p)\nonumber\\ &&\times\,
  D^{(W)}_{\alpha_1\alpha_2}(p - q) D^{(W)}_{\mu\nu}(- q),
  \nonumber\\ &&M(n \to p e^- \bar{\nu}_e)^{(\pi^0)}_{\rm
    Fig.\,\ref{fig:fig1}j} = + 2 e^2 g^2_{\pi N} M^2_W
  G_V\nonumber\\ &&\times \int \frac{d^4k}{(2\pi)^4i}\int
  \frac{d^4p}{(2\pi)^4i}\,\Big[\bar{u}_e(\vec{k}_e, \sigma_e)
    \gamma^{\mu}(1 - \gamma^5) v_{\bar{\nu}}(\vec{k}_{\bar{\nu}}, +
    \frac{1}{2})\Big]\,\Big[\bar{u}_p(\vec{k}_p,\sigma_p)\gamma^{\beta_1}
    \frac{1}{m_N - \hat{k}_p + \hat{p} - i0}\, \gamma^5 \frac{1}{m_N -
      \hat{k}_n - \hat{k} + \hat{p} -
      i0}\nonumber\\ &&\times\,\gamma^{\beta_2}\, \frac{1}{m_N -
      \hat{k}_n - \hat{k} - i0}\, \gamma^5 u_n(\vec{k}_n,
    \sigma_n)\Big]\,\frac{(2 k - q)^{\nu} }{[m^2_{\pi} - k^2 -
      i0][m^2_{\pi} - (k - q)^2 - i0]}\,D^{(\gamma)}_{\beta_1
    \beta_2}(p)\, D^{(W)}_{\mu\nu}(- q),\nonumber\\ &&M(n \to p e^-
  \bar{\nu}_e)^{(\sigma)}_{\rm Fig.\,\ref{fig:fig1}j} = + 2 e^2
  g^2_{\pi N} M^2_W G_V\nonumber\\ &&\times \int
  \frac{d^4k}{(2\pi)^4i}\int
  \frac{d^4p}{(2\pi)^4i}\,\Big[\bar{u}_e(\vec{k}_e, \sigma_e)
    \gamma^{\mu}(1 - \gamma^5) v_{\bar{\nu}}(\vec{k}_{\bar{\nu}}, +
    \frac{1}{2})\Big]\,\Big[\bar{u}_p(\vec{k}_p,\sigma_p)\gamma^{\beta_1}
    \frac{1}{m_N - \hat{k}_p + \hat{p} - i0}\, \frac{1}{m_N -
      \hat{k}_n - \hat{k} + \hat{p} - i0}\nonumber\\ &&\times\,
    \gamma^{\beta_2}\, \frac{1}{m_N - \hat{k}_n - \hat{k} -
      i0}\,\gamma^5 u_n(\vec{k}_n, \sigma_n)\Big]\,\frac{(2 k -
    q)^{\nu}}{[m^2_{\pi} - k^2 - i0][m^2_{\sigma} - (k - q)^2 -
      i0]}\,D^{(\gamma)}_{\beta_1 \beta_2}(p)\, D^{(W)}_{\mu\nu}(-
  q),\nonumber\\ &&M(n \to p e^- \bar{\nu}_e)^{(\pi^0)}_{\rm
    Fig.\,\ref{fig:fig1}k} = - e^2 g^2_{\pi N} M^2_W
  G_V\nonumber\\ &&\times \int \frac{d^4k}{(2\pi)^4i}\int
  \frac{d^4p}{(2\pi)^4i}\,\Big[\bar{u}_e(\vec{k}_e, \sigma_e)
    \gamma^{\mu}(1 - \gamma^5) v_{\bar{\nu}}(\vec{k}_{\bar{\nu}}, +
    \frac{1}{2})\Big]\,\Big[\bar{u}_p(\vec{k}_p,\sigma_p)\,\gamma^5\,
    \frac{1}{m_N - \hat{k}_p -\hat{k} - i0}\, \gamma^{\beta_1}
    \frac{1}{m_N - \hat{k}_p - \hat{k} + \hat{p} - i0}
    \nonumber\\ &&\times\,\gamma^{\alpha_2}(1 -
    \gamma^5)\,\frac{1}{m_N - \hat{k}_n -\hat{k} - i0}\,\gamma^5
    u_n(\vec{k}_n, \sigma_n)\Big]\,\frac{1}{m^2_{\pi} - k^2 -
    i0}\,\big[(p - q)^{\nu}\,\eta^{\beta_2 \alpha_1} - (p -
    2q)^{\beta_2} \eta^{\alpha_1 \nu} - q^{\alpha_1} \eta^{\nu
      \beta_2}\big]\nonumber\\ &&\times \, D^{(\gamma)}_{\beta_1
    \beta_2}(p)\, D^{(W)}_{\alpha_1\alpha_2}(p - q) D^{(W)}_{\mu\nu}(-
  q), \nonumber\\ &&M(n \to p e^- \bar{\nu}_e)^{(\sigma)}_{\rm
    Fig.\,\ref{fig:fig1}k} = - e^2 g^2_{\pi N} M^2_W
  G_V\nonumber\\ &&\times \int \frac{d^4k}{(2\pi)^4i}\int
  \frac{d^4p}{(2\pi)^4i}\,\Big[\bar{u}_e(\vec{k}_e, \sigma_e)
    \gamma^{\mu}(1 - \gamma^5) v_{\bar{\nu}}(\vec{k}_{\bar{\nu}}, +
    \frac{1}{2})\Big]\,\Big[\bar{u}_p(\vec{k}_p,\sigma_p)\,\gamma^5\,
    \frac{1}{m_N - \hat{k}_p -\hat{k} - i0}\, \gamma^{\beta_1}
    \frac{1}{m_N - \hat{k}_p - \hat{k} + \hat{p} - i0}
    \nonumber\\ &&\times\,\gamma^{\alpha_2}(1 -
    \gamma^5)\,\frac{1}{m_N - \hat{k}_n -\hat{k} - i0}\,\gamma^5
    u_n(\vec{k}_n, \sigma_n)\Big]\,\frac{1}{m^2_{\sigma} - k^2 -
    i0}\,\big[(p - q)^{\nu}\,\eta^{\beta_2 \alpha_1} - (p -
    2q)^{\beta_2} \eta^{\alpha_1 \nu} - q^{\alpha_1} \eta^{\nu
      \beta_2}\big]\nonumber\\ &&\times \,D^{(\gamma)}_{\beta_1
    \beta_2}(p)\, D^{(W)}_{\alpha_1\alpha_2}(p - q) D^{(W)}_{\mu\nu}(-
  q), \nonumber\\ &&M(n \to p e^- \bar{\nu}_e)^{(\pi^0)}_{\rm
    Fig.\,\ref{fig:fig1}l} = - e^2 g^2_{\pi N} M^2_W
  G_V\nonumber\\ &&\times \int \frac{d^4k}{(2\pi)^4i}\int
  \frac{d^4p}{(2\pi)^4i}\,\Big[\bar{u}_e(\vec{k}_e, \sigma_e)
    \gamma^{\mu}(1 - \gamma^5) v_{\bar{\nu}}(\vec{k}_{\bar{\nu}}, +
    \frac{1}{2})\Big]\,\Big[\bar{u}_p(\vec{k}_p,\sigma_p)\,\gamma^{\beta_1}\,
    \frac{1}{m_N - \hat{k}_p + \hat{p} - i0}\nonumber\\ &&\times \,
    \gamma^5 \frac{1}{m_N - \hat{k}_p - \hat{k} + \hat{p} -
      i0}\,\gamma^{\beta_2}\,\frac{1}{m_N - \hat{k}_p - \hat{k} -
      i0}\,\gamma^{\nu} (1 - \gamma^5)\,\frac{1}{m_N - \hat{k}_n -
      \hat{k} - i0}\,\gamma^5 u_n(\vec{k}_n, \sigma_n)\Big]
  \nonumber\\ &&\times\,\frac{1}{m^2_{\pi} - k^2 -
    i0}\,D^{(\gamma)}_{\beta_1 \beta_2}(p)\, D^{(W)}_{\mu\nu}(- q),
  \nonumber\\ &&M(n \to p e^- \bar{\nu}_e)^{(\sigma)}_{\rm
    Fig.\,\ref{fig:fig1}l} = - e^2 g^2_{\pi N} M^2_W
  G_V\nonumber\\ &&\times \int \frac{d^4k}{(2\pi)^4i}\int
  \frac{d^4p}{(2\pi)^4i}\,\Big[\bar{u}_e(\vec{k}_e, \sigma_e)
    \gamma^{\mu}(1 - \gamma^5) v_{\bar{\nu}}(\vec{k}_{\bar{\nu}}, +
    \frac{1}{2})\Big]\,\Big[\bar{u}_p(\vec{k}_p,\sigma_p)\,\gamma^{\beta_1}\,
    \frac{1}{m_N - \hat{k}_p + \hat{p} - i0} \nonumber\\ &&\times\,
    \frac{1}{m_N - \hat{k}_p - \hat{k} + \hat{p} -
      i0}\,\gamma^{\beta_2}\,\frac{1}{m_N - \hat{k}_p - \hat{k} -
      i0}\,\gamma^{\nu} (1 - \gamma^5)\,\frac{1}{m_N - \hat{k}_n -
      \hat{k} - i0}\, u_n(\vec{k}_n,
    \sigma_n)\Big]\nonumber\\ &&\times\,\frac{1}{m^2_{\sigma} - k^2 -
    i0}\,D^{(\gamma)}_{\beta_1 \beta_2}(p)\, D^{(W)}_{\mu\nu}(- q),
  \nonumber\\ &&M(n \to p e^- \bar{\nu}_e)^{(\pi^0)}_{\rm
    Fig.\,\ref{fig:fig1}m} = - e^2 g^2_{\pi N} M^2_W
  G_V\nonumber\\ &&\times \int \frac{d^4k}{(2\pi)^4i}\int
  \frac{d^4p}{(2\pi)^4i}\,\Big[\bar{u}_e(\vec{k}_e,
    \sigma_e)\,\gamma^{\beta_1}\,\frac{1}{m_e - \hat{p} - \hat{k}_e -
      i0}\,\gamma^{\mu}(1 -
    \gamma^5)\,v_{\bar{\nu}}(\vec{k}_{\bar{\nu}}, +
    \frac{1}{2})\Big]\,\Big[\bar{u}_p(\vec{k}_p,\sigma_p)\, \gamma^5\,
    \frac{1}{m_N - \hat{k}_p - \hat{k} - i0}
    \nonumber\\ &&\times\,\gamma^{\beta_2}\, \frac{1}{m_N - \hat{k}_p
      - \hat{k} + \hat{p} - i0}\,\gamma^{\nu} (1 -
    \gamma^5)\,\frac{1}{m_N - \hat{k}_n - \hat{k} - i0}\,\gamma^5
    u_n(\vec{k}_n, \sigma_n)\Big] \, \frac{1}{m^2_{\pi} - k^2 -
    i0}\,D^{(\gamma)}_{\beta_1\beta_2}(p)\, D^{(W)}_{\mu\nu}(p - q),
  \nonumber\\
\end{eqnarray*}
\begin{eqnarray*}
&&M(n \to p e^- \bar{\nu}_e)^{(\sigma)}_{\rm Fig.\,\ref{fig:fig1}m} =
  - e^2 g^2_{\pi N} M^2_W G_V\nonumber\\ &&\times \int
  \frac{d^4k}{(2\pi)^4i}\int
  \frac{d^4p}{(2\pi)^4i}\,\Big[\bar{u}_e(\vec{k}_e,
    \sigma_e)\,\gamma^{\beta_1}\,\frac{1}{m_e - \hat{p} - \hat{k}_e -
      i0}\,\gamma^{\mu}(1 -
    \gamma^5)\,v_{\bar{\nu}}(\vec{k}_{\bar{\nu}}, +
    \frac{1}{2})\Big]\,\Big[\bar{u}_p(\vec{k}_p,\sigma_p)\,
    \frac{1}{m_N - \hat{k}_p - \hat{k} - i0}
    \nonumber\\ &&\times\,\gamma^{\beta_2} \,\gamma^{\nu} (1 -
    \gamma^5)\,\frac{1}{m_N - \hat{k}_n - \hat{k} - i0}\,
    u_n(\vec{k}_n, \sigma_n)\Big] \, \frac{1}{m^2_{\sigma} - k^2 -
    i0}\,D^{(\gamma)}_{\beta_1\beta_2}(p)\, D^{(W)}_{\mu\nu}(p -
  q),\nonumber\\&&M(n \to p e^- \bar{\nu}_e)^{(\pi^0)}_{\rm
    Fig.\,\ref{fig:fig1}n} = - 2 e^2 g^2_{\pi N} M^2_W
  G_V\nonumber\\ &&\times \int \frac{d^4k}{(2\pi)^4i}\int
  \frac{d^4p}{(2\pi)^4i}\,\Big[\bar{u}_e(\vec{k}_e, \sigma_e)
    \gamma^{\mu}(1 - \gamma^5) v_{\bar{\nu}}(\vec{k}_{\bar{\nu}}, +
    \frac{1}{2})\Big]\,\Big[\bar{u}_p(\vec{k}_p,\sigma_p)\,\gamma^{\beta_1}
    \frac{1}{m_N - \hat{k}_p + \hat{p} - i0}\,\gamma^5\, \frac{1}{m_N
      - \hat{k}_n - \hat{k} - i0}\nonumber\\ &&\times\,\gamma^5\,
    u_n(\vec{k}_n, \sigma_n)\Big] \, \frac{(2 k + p -
    q)^{\alpha_2}}{[m^2_{\pi} - k^2 - i0][m^2_{\pi} - (k + p - q)^2 -
      i0]}\, \big[(p - q)^{\nu}\eta^{\beta_2 \alpha_1} - (p - 2
    q)^{\beta_2} \eta^{\alpha_1\nu} - q^{\alpha_1}
    \eta^{\nu\beta_2}\big]\nonumber\\ &&\times\,
  D^{(\gamma)}_{\beta_1\beta_2}(p)\, D^{(W)}_{\alpha_1 \alpha_2}(p -
  q) \, D^{(W)}_{\mu\nu}(- q),\nonumber\\&&M(n \to p e^-
  \bar{\nu}_e)^{(\sigma)}_{\rm Fig.\,\ref{fig:fig1}n} = - 2 e^2
  g^2_{\pi N} M^2_W G_V\nonumber\\ &&\times \int
  \frac{d^4k}{(2\pi)^4i}\int
  \frac{d^4p}{(2\pi)^4i}\,\Big[\bar{u}_e(\vec{k}_e, \sigma_e)
    \gamma^{\mu}(1 - \gamma^5) v_{\bar{\nu}}(\vec{k}_{\bar{\nu}}, +
    \frac{1}{2})\Big]\,\Big[\bar{u}_p(\vec{k}_p,\sigma_p)\,
    \gamma^{\beta_1} \frac{1}{m_N - \hat{k}_p + \hat{p} -
      i0}\,\gamma^5\, \frac{1}{m_N - \hat{k}_n - \hat{k} -
      i0}\nonumber\\ &&\times\, u_n(\vec{k}_n, \sigma_n)\Big] \,
  \frac{(2 k + p - q)^{\alpha_2}}{[m^2_{\sigma} - k^2 - i0][m^2_{\pi}
      - (k + p - q)^2 - i0]}\, \big[(p - q)^{\nu}\eta^{\beta_2
      \alpha_1} - (p - 2 q)^{\beta_2} \eta^{\alpha_1\nu} -
    q^{\alpha_1} \eta^{\nu\beta_2}\big]\nonumber\\ &&\times\,
  D^{(\gamma)}_{\beta_1\beta_2}(p)\, D^{(W)}_{\alpha_1 \alpha_2}(p -
  q) \, D^{(W)}_{\mu\nu}(- q),\nonumber\\&&M(n \to p e^-
  \bar{\nu}_e)^{(\pi^0)}_{\rm Fig.\,\ref{fig:fig1}o} = - 2 e^2
  g^2_{\pi N} M^2_W G_V\nonumber\\ &&\times \int
  \frac{d^4k}{(2\pi)^4i}\int
  \frac{d^4p}{(2\pi)^4i}\,\Big[\bar{u}_e(\vec{k}_e,
    \sigma_e)\,\gamma^{\beta_1}\,\frac{1}{m_e - \hat{k}_e - \hat{p} -
      i0}\, \gamma^{\mu}(1 - \gamma^5)
    v_{\bar{\nu}}(\vec{k}_{\bar{\nu}}, +
    \frac{1}{2})\Big]\,\Big[\bar{u}_p(\vec{k}_p,\sigma_p)\,\gamma^{\beta_2}
    \frac{1}{m_N - \hat{k}_p + \hat{p} - i0}\nonumber\\ &&\times
    \,\gamma^5\, \frac{1}{m_N - \hat{k}_n - \hat{k} - i0}\,\gamma^5\,
    u_n(\vec{k}_n, \sigma_n)\Big] \, \frac{(2 k + p -
    q)^{\nu}}{[m^2_{\pi} - k^2 - i0][m^2_{\pi} - (k + p - q)^2 -
      i0]}\,D^{(\gamma)}_{\beta_1\beta_2}(p)\, D^{(W)}_{\mu\nu}(p -
  q),\nonumber\\&&M(n \to p e^- \bar{\nu}_e)^{(\sigma)}_{\rm
    Fig.\,\ref{fig:fig1}o} = - 2 e^2 g^2_{\pi N} M^2_W
  G_V\nonumber\\ &&\times \int \frac{d^4k}{(2\pi)^4i}\int
  \frac{d^4p}{(2\pi)^4i}\,\Big[\bar{u}_e(\vec{k}_e,
    \sigma_e)\,\gamma^{\beta_1}\,\frac{1}{m_e - \hat{k}_e - \hat{p} -
      i0}\, \gamma^{\mu}(1 - \gamma^5)
    v_{\bar{\nu}}(\vec{k}_{\bar{\nu}}, +
    \frac{1}{2})\Big]\,\Big[\bar{u}_p(\vec{k}_p,\sigma_p)\,\gamma^{\beta_2}
    \frac{1}{m_N - \hat{k}_p + \hat{p} - i0}\nonumber\\ &&\times
    \,\gamma^5\, \frac{1}{m_N - \hat{k}_n - \hat{k} - i0}\,
    u_n(\vec{k}_n, \sigma_n)\Big] \, \frac{(2 k + p -
    q)^{\nu}}{[m^2_{\sigma} - k^2 - i0][m^2_{\pi} - (k + p - q)^2 -
      i0]}\,D^{(\gamma)}_{\beta_1\beta_2}(p)\, D^{(W)}_{\mu\nu}(p -
  q),\nonumber\\&&M(n \to p e^- \bar{\nu}_e)^{(\pi^0)}_{\rm
    Fig.\,\ref{fig:fig1}p} = - 2 e^2 g^2_{\pi N} M^2_W
  G_V\nonumber\\ &&\times \int \frac{d^4k}{(2\pi)^4i}\int
  \frac{d^4p}{(2\pi)^4i}\,\Big[\bar{u}_e(\vec{k}_e, \sigma_e)
    \gamma^{\mu}(1 - \gamma^5) v_{\bar{\nu}}(\vec{k}_{\bar{\nu}}, +
    \frac{1}{2})\Big]\,\Big[\bar{u}_p(\vec{k}_p,\sigma_p)\,\gamma^{\beta_1}
    \frac{1}{m_N - \hat{k}_p + \hat{p} - i0}\,\gamma^5\, \frac{1}{m_N
      - \hat{k}_n - \hat{k} - i0}\nonumber\\ &&\times\,\gamma^5\,
    u_n(\vec{k}_n, \sigma_n)\Big] \, \frac{(2 k + p -
    q)^{\alpha_2}}{[m^2_{\pi} - k^2 - i0][m^2_{\pi} - (k + p - q)^2 -
      i0]}\, \big[(p - q)^{\nu}\eta^{\beta_2 \alpha_1} - (p - 2
    q)^{\beta_2} \eta^{\alpha_1\nu} - q^{\alpha_1}
    \eta^{\nu\beta_2}\big]\nonumber\\ &&\times\,
  D^{(\gamma)}_{\beta_1\beta_2}(p)\, D^{(W)}_{\alpha_1 \alpha_2}(p -
  q) \, D^{(W)}_{\mu\nu}(- q),\nonumber\\&&M(n \to p e^-
  \bar{\nu}_e)^{(\sigma)}_{\rm Fig.\,\ref{fig:fig1}p} = - 2 e^2
  g^2_{\pi N} M^2_W G_V\nonumber\\ &&\times \int
  \frac{d^4k}{(2\pi)^4i}\int
  \frac{d^4p}{(2\pi)^4i}\,\Big[\bar{u}_e(\vec{k}_e, \sigma_e)
    \gamma^{\mu}(1 - \gamma^5) v_{\bar{\nu}}(\vec{k}_{\bar{\nu}}, +
    \frac{1}{2})\Big]\,\Big[\bar{u}_p(\vec{k}_p,\sigma_p)\,
    \gamma^{\beta_1} \frac{1}{m_N - \hat{k}_p + \hat{p} -
      i0}\,\gamma^5\, \frac{1}{m_N - \hat{k}_n - \hat{k} -
      i0}\nonumber\\ &&\times\, u_n(\vec{k}_n, \sigma_n)\Big] \,
  \frac{(2 k + p - q)^{\alpha_2}}{[m^2_{\pi} - k^2 - i0][m^2_{\sigma}
      - (k + p - q)^2 - i0]}\, \big[(p - q)^{\nu}\eta^{\beta_2
      \alpha_1} - (p - 2 q)^{\beta_2} \eta^{\alpha_1\nu} -
    q^{\alpha_1} \eta^{\nu\beta_2}\big]\nonumber\\ &&\times\,
  D^{(\gamma)}_{\beta_1\beta_2}(p)\, D^{(W)}_{\alpha_1 \alpha_2}(p -
  q) \, D^{(W)}_{\mu\nu}(- q),\nonumber\\&&M(n \to p e^-
  \bar{\nu}_e)^{(\pi^0)}_{\rm Fig.\,\ref{fig:fig1}q} = - 2 e^2
  g^2_{\pi N} M^2_W G_V\nonumber\\ &&\times \int
  \frac{d^4k}{(2\pi)^4i}\int
  \frac{d^4p}{(2\pi)^4i}\,\Big[\bar{u}_e(\vec{k}_e,
    \sigma_e)\,\gamma^{\beta_1}\,\frac{1}{m_e - \hat{k}_e - \hat{p} -
      i0}\, \gamma^{\mu}(1 - \gamma^5)
    v_{\bar{\nu}}(\vec{k}_{\bar{\nu}}, +
    \frac{1}{2})\Big]\,\Big[\bar{u}_p(\vec{k}_p,\sigma_p)\,\gamma^{\beta_2}
    \frac{1}{m_N - \hat{k}_p + \hat{p} - i0}\nonumber\\ &&\times
    \,\gamma^5\, \frac{1}{m_N - \hat{k}_n - \hat{k} - i0}\,\gamma^5\,
    u_n(\vec{k}_n, \sigma_n)\Big] \, \frac{(2 k + p -
    q)^{\nu}}{[m^2_{\pi} - k^2 - i0][m^2_{\pi} - (k + p - q)^2 -
      i0]}\,D^{(\gamma)}_{\beta_1\beta_2}(p)\, D^{(W)}_{\mu \nu}(p -
  q),\nonumber\\&&M(n \to p e^- \bar{\nu}_e)^{(\sigma)}_{\rm
    Fig.\,\ref{fig:fig1}q} = - 2 e^2 g^2_{\pi N} M^2_W
  G_V\nonumber\\ &&\times \int \frac{d^4k}{(2\pi)^4i}\int
  \frac{d^4p}{(2\pi)^4i}\,\Big[\bar{u}_e(\vec{k}_e,
    \sigma_e)\,\gamma^{\beta_1}\,\frac{1}{m_e - \hat{k}_e - \hat{p} -
      i0}\, \gamma^{\mu}(1 - \gamma^5)
    v_{\bar{\nu}}(\vec{k}_{\bar{\nu}}, +
    \frac{1}{2})\Big]\,\Big[\bar{u}_p(\vec{k}_p,\sigma_p)\,\gamma^{\beta_2}
    \frac{1}{m_N - \hat{k}_p + \hat{p} - i0}\nonumber\\ &&\times
    \,\gamma^5\, \frac{1}{m_N - \hat{k}_n - \hat{k} - i0}\,
    u_n(\vec{k}_n, \sigma_n)\Big] \, \frac{(2 k + p -
    q)^{\alpha_2}}{[m^2_{\pi} - k^2 - i0][m^2_{\sigma} - (k + p - q)^2
      - i0]}\,D^{(\gamma)}_{\beta_1\beta_2}(p)\, D^{(W)}_{\alpha_1
    \alpha_2}(p - q),\nonumber\\
\end{eqnarray*}
\begin{eqnarray}\label{eq:A.1}
&&M(n \to p e^- \bar{\nu}_e)^{(\pi^0)}_{\rm Fig.\,\ref{fig:fig1}r} = -
  e^2 g^2_{\pi N} M^2_W G_V\nonumber\\ &&\times \int
  \frac{d^4k}{(2\pi)^4i}\int
  \frac{d^4p}{(2\pi)^4i}\,\Big[\bar{u}_e(\vec{k}_e, \sigma_e)
    \gamma^{\mu}(1 - \gamma^5) v_{\bar{\nu}}(\vec{k}_{\bar{\nu}}, +
    \frac{1}{2})\Big]\,\Big[\bar{u}_p(\vec{k}_p,\sigma_p)\,\gamma^{\beta_1}\,
    \frac{1}{m_N - \hat{k}_p + \hat{p} - i0}\, \gamma^5 \frac{1}{m_N -
      \hat{k}_p - \hat{k} + \hat{p} - i0}
    \nonumber\\ &&\times\,\gamma^{\alpha_2}(1 -
    \gamma^5)\,\frac{1}{m_N - \hat{k}_n -\hat{k} - i0}\,\gamma^5
    u_n(\vec{k}_n, \sigma_n)\Big]\,\frac{1}{m^2_{\pi} - k^2 -
    i0}\,\big[(p - q)^{\nu}\,\eta^{\beta_2 \alpha_1} - (p -
    2q)^{\beta_2} \eta^{\alpha_1 \nu} - q^{\alpha_1} \eta^{\nu
      \beta_2}\big]\nonumber\\ &&\times \, D^{(\gamma)}_{\beta_1
    \beta_2}(p)\, D^{(W)}_{\alpha_1\alpha_2}(p - q) D^{(W)}_{\mu\nu}(-
  q), \nonumber\\ &&M(n \to p e^- \bar{\nu}_e)^{(\sigma)}_{\rm
    Fig.\,\ref{fig:fig1}r} = - e^2 g^2_{\pi N} M^2_W
  G_V\nonumber\\ &&\times \int \frac{d^4k}{(2\pi)^4i}\int
  \frac{d^4p}{(2\pi)^4i}\,\Big[\bar{u}_e(\vec{k}_e, \sigma_e)
    \gamma^{\mu}(1 - \gamma^5) v_{\bar{\nu}}(\vec{k}_{\bar{\nu}}, +
    \frac{1}{2})\Big]\,\Big[\bar{u}_p(\vec{k}_p,\sigma_p)\,\gamma^{\beta_1}\,
    \frac{1}{m_N - \hat{k}_p + \hat{p} - i0}\, \gamma^{\beta_1}
    \frac{1}{m_N - \hat{k}_p - \hat{k} + \hat{p} - i0}
    \nonumber\\ &&\times\,\gamma^{\alpha_2}(1 -
    \gamma^5)\,\frac{1}{m_N - \hat{k}_n -\hat{k} - i0}\,
    u_n(\vec{k}_n, \sigma_n)\Big]\,\frac{1}{m^2_{\sigma} - k^2 -
    i0}\,\big[(p - q)^{\nu}\,\eta^{\beta_2 \alpha_1} - (p -
    2q)^{\beta_2} \eta^{\alpha_1 \nu} - q^{\alpha_1} \eta^{\nu
      \beta_2}\big]\nonumber\\ &&\times \,D^{(\gamma)}_{\beta_1
    \beta_2}(p)\, D^{(W)}_{\alpha_1\alpha_2}(p - q) D^{(W)}_{\mu\nu}(-
  q), \nonumber\\ &&M(n \to p e^- \bar{\nu}_e)^{(\pi^0)}_{\rm
    Fig.\,\ref{fig:fig1}s} = - e^2 g^2_{\pi N} M^2_W
  G_V\nonumber\\ &&\times \int \frac{d^4k}{(2\pi)^4i}\int
  \frac{d^4p}{(2\pi)^4i}\,\Big[\bar{u}_e(\vec{k}_e,
    \sigma_e)\,\gamma^{\beta_1}\,\frac{1}{m_e - \hat{k}_e - \hat{p} -
      i0}\,\gamma^{\mu}(1 -
    \gamma^5)\,v_{\bar{\nu}}(\vec{k}_{\bar{\nu}}, +
    \frac{1}{2})\Big]\,\Big[\bar{u}_p(\vec{k}_p,\sigma_p)\,
    \gamma^{\beta_1}\, \frac{1}{m_N - \hat{k}_p + \hat{p} - i0}
    \nonumber\\ &&\times\,\gamma^5\, \frac{1}{m_N - \hat{k}_p -
      \hat{k} + \hat{p} - i0}\,\gamma^{\nu} (1 -
    \gamma^5)\,\frac{1}{m_N - \hat{k}_n - \hat{k} - i0}\,\gamma^5
    u_n(\vec{k}_n, \sigma_n)\Big] \, \frac{1}{m^2_{\pi} - k^2 -
    i0}\,D^{(\gamma)}_{\beta_1\beta_2}(p)\, D^{(W)}_{\mu\nu}(p - q),
  \nonumber\\ &&M(n \to p e^- \bar{\nu}_e)^{(\sigma)}_{\rm
    Fig.\,\ref{fig:fig1}s} = - e^2 g^2_{\pi N} M^2_W
  G_V\nonumber\\ &&\times \int \frac{d^4k}{(2\pi)^4i}\int
  \frac{d^4p}{(2\pi)^4i}\,\Big[\bar{u}_e(\vec{k}_e,
    \sigma_e)\,\gamma^{\beta_1}\,\frac{1}{m_e - \hat{k}_e - \hat{p} -
      i0}\,\gamma^{\mu}(1 -
    \gamma^5)\,v_{\bar{\nu}}(\vec{k}_{\bar{\nu}}, +
    \frac{1}{2})\Big]\,\Big[\bar{u}_p(\vec{k}_p,\sigma_p)\,
    \gamma^{\beta_1}\, \frac{1}{m_N - \hat{k}_p + \hat{p} - i0}
    \nonumber\\ &&\times\,\gamma^5\, \frac{1}{m_N - \hat{k}_p -
      \hat{k} + \hat{p} - i0}\,\gamma^{\nu} (1 -
    \gamma^5)\,\frac{1}{m_N - \hat{k}_n - \hat{k} - i0}\,\gamma^5
    u_n(\vec{k}_n, \sigma_n)\Big] \, \frac{1}{m^2_{\sigma} - k^2 -
    i0}\,D^{(\gamma)}_{\beta_1\beta_2}(p)\, D^{(W)}_{\mu\nu}(p - q),
\end{eqnarray}
where $G_V = g^2/8M^2_W$ is the Fermi weak coupling constant, $M_W$ is
the mass of the electroweak $W^-$-boson \cite{PDG2020},
$D^{(W)}_{\alpha \beta}(Q)$ and $D^{(\gamma)}_{\alpha\beta}(p)$ are
the electroweak $W^-$-boson and photon propagators, respectively,
equal to \cite{Itzykson1980}
\begin{eqnarray}\label{eq:A.2}
D^{(W)}_{\alpha \beta}(Q) &=& \frac{1}{M^2_W - Q^2 - i0}\,\Big( -
\eta_{\alpha\beta} + \frac{Q_{\alpha}Q_{\beta}}{M^2_W}\Big)\quad,\quad
D^{(\gamma)}_{\alpha\beta}(p) = \frac{1}{p^2 +
  i0}\Big(\eta_{\alpha\beta} - \xi\,\frac{p_{\alpha}p_{\beta}}{p^2 +
  i0}\Big),\nonumber\\ D^{(W)-1}_{\alpha \beta}(Q) &=& -
Q_{\alpha}Q_{\beta} - (M^2_W - Q^2)\,\eta_{\alpha\beta}\quad,\quad
D^{(W)-1}_{\alpha \rho}(Q)D^{(W)\rho}{}_{\beta}(Q) = \eta_{\alpha
  \beta},
\end{eqnarray}
where $\xi$ is a gauge parameter describing a contribution of a
longitudinal polarization state of a virtual photon
\cite{Itzykson1980}.  The analytical expressions for the Feynman
diagrams in Fig.\,\ref{fig:fig2} are given by
\begin{eqnarray*}
&& M(n \to p e^- \bar{\nu}_e)^{(\pi^0)}_{\rm Fig.\,\ref{fig:fig2}a} =
  + 2 e^2 g^2_{\pi N} M^2_W G_V\nonumber\\ &&\times \int
  \frac{d^4k}{(2\pi)^4i}\int
  \frac{d^4p}{(2\pi)^4i}\,\Big[\bar{u}_e(\vec{k}_e, \sigma_e)
    \gamma^{\mu}(1 - \gamma^5) v_{\bar{\nu}}(\vec{k}_{\bar{\nu}}, +
    \frac{1}{2})\Big]\,\,\Big[\bar{u}_p(\vec{k}_p,\sigma_p)\,\gamma^5\,
    \frac{1}{m_N - \hat{k}_n - \hat{k} - i0}\,\gamma^5 u_n(\vec{k}_n,
    \sigma_n)\Big]\nonumber\\ &&\times \frac{(2 k - q)^{\nu}(p + 2k -
    2 q)^{\beta_1}(p + 2k - 2 q)^{\beta_2}}{[m^2_{\pi} - k^2 -
      i0][m^2_{\pi} - (k - q)^2 - i0]^2[m^2_{\pi} - (k + p - q)^2 -
      i0]}\,D^{(\gamma)}_{\beta_1 \beta_2}(p)\, D^{(W)}_{\mu\nu}(-
  q),\nonumber\\ && M(n \to p e^- \bar{\nu}_e)^{(\sigma)}_{\rm
    Fig.\,\ref{fig:fig2}a} = - 2 e^2 g^2_{\pi N} M^2_W
  G_V\nonumber\\ &&\times \int \frac{d^4k}{(2\pi)^4i}\int
  \frac{d^4p}{(2\pi)^4i}\,\Big[\bar{u}_e(\vec{k}_e, \sigma_e)
    \gamma^{\mu}(1 - \gamma^5) v_{\bar{\nu}}(\vec{k}_{\bar{\nu}}, +
    \frac{1}{2})\Big]\,\,\Big[\bar{u}_p(\vec{k}_p,\sigma_p)\,
    \frac{1}{m_N - \hat{k}_n - \hat{k} - i0}\, u_n(\vec{k}_n,
    \sigma_n)\Big]\nonumber\\ &&\times \frac{(2 k - q)^{\nu}(p + 2k -
    2 q)^{\beta_1}(p + 2k - 2 q)^{\beta_2}}{[m^2_{\sigma} - k^2 -
      i0][m^2_{\pi} - (k - q)^2 - i0]^2[m^2_{\pi} - (k + p - q)^2 -
      i0]}\,D^{(\gamma)}_{\beta_1 \beta_2}(p)\, D^{(W)}_{\mu\nu}(-
  q),\nonumber\\
  && M(n \to p e^-
  \bar{\nu}_e)^{(\pi^0)}_{\rm Fig.\,\ref{fig:fig2}b} = + 2 e^2
  g^2_{\pi N} M^2_W G_V\nonumber\\ &&\times \int
  \frac{d^4k}{(2\pi)^4i}\int
  \frac{d^4p}{(2\pi)^4i}\,\Big[\bar{u}_e(\vec{k}_e, \sigma_e)
    \gamma^{\mu}(1 - \gamma^5) v_{\bar{\nu}}(\vec{k}_{\bar{\nu}}, +
    \frac{1}{2})\Big]\,\Big[\bar{u}_p(\vec{k}_p,\sigma_p)\,
    \gamma^5\,\frac{1}{m_N - \hat{k}_n - \hat{k} - i0}\,\gamma^5
    u_n(\vec{k}_n, \sigma_n)\Big]\nonumber\\
\end{eqnarray*}
\begin{eqnarray}\label{eq:A.3}
  &&\times\, \frac{(2 k -
    q)^{\nu}(2 k - p)^{\beta_1}(2 k - p)^{\beta_2}}{[m^2_{\pi} - k^2 -
      i0]^2[m^2_{\pi} - (k - p)^2 - i0][m^2_{\pi} - (k - q)^2 - i0]}\,
  D^{(\gamma)}_{\beta_1 \beta_2}(p)\, D^{(W)}_{\mu\nu}(-
  q),\nonumber\\
 && M(n \to p e^- \bar{\nu}_e)^{(\sigma)}_{\rm
    Fig.\,\ref{fig:fig2}b} = + 2 e^2 g^2_{\pi N} M^2_W
  G_V\nonumber\\ &&\times \int \frac{d^4k}{(2\pi)^4i}\int
  \frac{d^4p}{(2\pi)^4i}\,\Big[\bar{u}_e(\vec{k}_e, \sigma_e)
    \gamma^{\mu}(1 - \gamma^5) v_{\bar{\nu}}(\vec{k}_{\bar{\nu}}, +
    \frac{1}{2})\Big]\,\Big[\bar{u}_p(\vec{k}_p,\sigma_p)\,
    \frac{1}{m_N - \hat{k}_n - \hat{k} - i0}\,\gamma^5 u_n(\vec{k}_n,
    \sigma_n)\Big]\nonumber\\ &&\times\, \frac{(2 k - q)^{\nu}(2 k -
    p)^{\beta_1}(2 k - p)^{\beta_2}}{[m^2_{\pi} - k^2 -
      i0]^2[m^2_{\pi} - (k - p)^2 - i0][m^2_{\sigma} - (k - q)^2 -
      i0]}\, D^{(\gamma)}_{\beta_1 \beta_2}(p)\, D^{(W)}_{\mu\nu}(-
  q),\nonumber\\
  && M(n \to p e^- \bar{\nu}_e)^{(\pi^0)}_{\rm
    Fig.\,\ref{fig:fig2}c} = + 2 e^2 g^2_{\pi N} M^2_W
  G_V\nonumber\\ &&\times \int \frac{d^4k}{(2\pi)^4i}\int
  \frac{d^4p}{(2\pi)^4i}\,\Big[\bar{u}_e(\vec{k}_e, \sigma_e)
    \gamma^{\mu}(1 - \gamma^5) v_{\bar{\nu}}(\vec{k}_{\bar{\nu}}, +
    \frac{1}{2})\Big]\,\Big[\bar{u}_p(\vec{k}_p,\sigma_p)\,
    \gamma^5\,\frac{1}{m_N - \hat{k}_n - \hat{k} -
      i0}\,\gamma^{\beta_1} \frac{1}{m_N - \hat{k}_n - \hat{k} +
      \hat{p} - i0}\nonumber\\ &&\times
    \,\gamma^{\beta_2}\,\frac{1}{m_N - \hat{k}_n - \hat{k} -
      i0}\,\gamma^5 u_n(\vec{k}_n, \sigma_n)\Big]\,\frac{(2 k -
    q)^{\nu}}{[m^2_{\pi} - k^2 - i0][m^2_{\pi} - (k - q)^2 -
      i0]}\,D^{(\gamma)}_{\beta_1 \beta_2}(p)\, D^{(W)}_{\mu\nu}(-
  q),\nonumber\\
  && M(n \to p e^- \bar{\nu}_e)^{(\sigma)}_{\rm
    Fig.\,\ref{fig:fig2}c} = + 2 e^2 g^2_{\pi N} M^2_W
  G_V\nonumber\\ &&\times \int \frac{d^4k}{(2\pi)^4i}\int
  \frac{d^4p}{(2\pi)^4i}\,\Big[\bar{u}_e(\vec{k}_e, \sigma_e)
    \gamma^{\mu}(1 - \gamma^5) v_{\bar{\nu}}(\vec{k}_{\bar{\nu}}, +
    \frac{1}{2})\Big]\,\Big[\bar{u}_p(\vec{k}_p,\sigma_p)\,
    \frac{1}{m_N - \hat{k}_n - \hat{k} - i0}\,\gamma^{\beta_1}
    \frac{1}{m_N - \hat{k}_n - \hat{k} + \hat{p} -
      i0}\nonumber\\ &&\times \,\gamma^{\beta_2}\,\frac{1}{m_N -
      \hat{k}_n - \hat{k} - i0}\,\gamma^5 u_n(\vec{k}_n,
    \sigma_n)\Big]\,\frac{(2 k - q)^{\nu}}{[m^2_{\pi} - k^2 -
      i0][m^2_{\sigma} - (k - q)^2 - i0]}\,D^{(\gamma)}_{\beta_1
    \beta_2}(p)\, D^{(W)}_{\mu\nu}(- q),\nonumber\\
 && M(n \to p e^- \bar{\nu}_e)^{(\pi^0)}_{\rm Fig.\,\ref{fig:fig2}d}
  = + e^2 g^2_{\pi N} M^2_W G_V\nonumber\\ &&\times \int
  \frac{d^4k}{(2\pi)^4i}\int
  \frac{d^4p}{(2\pi)^4i}\,\Big[\bar{u}_e(\vec{k}_e, \sigma_e)
    \gamma^{\mu}(1 - \gamma^5) v_{\bar{\nu}}(\vec{k}_{\bar{\nu}}, +
    \frac{1}{2})\Big]\,\Big[\bar{u}_p(\vec{k}_p,\sigma_p)\,
    \gamma^5\,\frac{1}{m_N - \hat{k}_p - \hat{k} -
      i0}\,\gamma^{\beta_1} \,\frac{1}{m_N - \hat{k}_p - \hat{k} +
      \hat{p} - i0}\nonumber\\ &&\times \,\gamma^{\beta_2}
    \,\frac{1}{m_N - \hat{k}_p - \hat{k} - i0}\,\gamma^{\nu} (1 -
    \gamma^5)\,\frac{1}{m_N - \hat{k}_n - \hat{k} - i0} \,\gamma^5\,
    u_n(\vec{k}_n, \sigma_n)\Big]\, \frac{1}{m^2_{\pi} - k^2 - i0}\,
  D^{(\gamma)}_{\beta_1\beta_2} (p) \,D^{(W)}_{\mu \nu}( -
  q),\nonumber\\ && M(n \to p e^- \bar{\nu}_e)^{(\sigma)}_{\rm
    Fig.\,\ref{fig:fig2}d} = + e^2 g^2_{\pi N} M^2_W
  G_V\nonumber\\ &&\times \int \frac{d^4k}{(2\pi)^4i}\int
  \frac{d^4p}{(2\pi)^4i}\,\Big[\bar{u}_e(\vec{k}_e, \sigma_e)
    \gamma^{\mu}(1 - \gamma^5) v_{\bar{\nu}}(\vec{k}_{\bar{\nu}}, +
    \frac{1}{2})\Big]\,\Big[\bar{u}_p(\vec{k}_p,\sigma_p)\,
    \gamma^5\,\frac{1}{m_N - \hat{k}_p - \hat{k} -
      i0}\,\gamma^{\beta_1} \,\frac{1}{m_N - \hat{k}_p - \hat{k} +
      \hat{p} - i0}\nonumber\\ &&\times \,\gamma^{\beta_2}
    \,\frac{1}{m_N - \hat{k}_p - \hat{k} - i0}\,\gamma^{\nu} (1 -
    \gamma^5)\,\frac{1}{m_N - \hat{k}_n - \hat{k} - i0} \,
    u_n(\vec{k}_n, \sigma_n)\Big]\, \frac{1}{m^2_{\sigma} - k^2 -
    i0}\, D^{(\gamma)}_{\beta_1\beta_2} (p) \,D^{(W)}_{\mu \nu}( - q).
\end{eqnarray}
For the Feynman diagrams in Fig.\,\ref{fig:fig3} we obtain the
following analytical expressions
\begin{eqnarray*}
&& M(n \to p e^- \bar{\nu}_e)^{(\pi^0)}_{\rm Fig.\,\ref{fig:fig3}a} =
  + 2 e^2 g^2_{\pi N} M^2_W G_V\nonumber\\ &&\times \int
  \frac{d^4k}{(2\pi)^4i}\int
  \frac{d^4p}{(2\pi)^4i}\,\Big[\bar{u}_e(\vec{k}_e, \sigma_e)
    \gamma^{\mu}(1 - \gamma^5) v_{\bar{\nu}}(\vec{k}_{\bar{\nu}}, +
    \frac{1}{2})\Big]\,\Big[\bar{u}_p(\vec{k}_p,\sigma_p)\,
    \gamma^5\,\frac{1}{m_N - \hat{k}_n - \hat{k} - i0}\,\gamma^5
    u_n(\vec{k}_n, \sigma_n)\Big]\nonumber\\ &&\times \,\frac{(p + 2k
    - 2 q)^{\beta_1}}{[m^2_{\pi} - k^2 - i0][m^2_{\pi} - (k - q)^2 -
      i0][m^2_{\pi} - (k + p - q)^2 - i0]}\,
  D^{(\gamma)}_{\beta_1\beta_2}(p)\, D^{(W)}_{\mu}{}^{\beta_2}(-
  q),\nonumber\\&& M(n \to p e^- \bar{\nu}_e)^{(\sigma)}_{\rm
    Fig.\,\ref{fig:fig3}a} = + 2 e^2 g^2_{\pi N} M^2_W
  G_V\nonumber\\ &&\times \int \frac{d^4k}{(2\pi)^4i}\int
  \frac{d^4p}{(2\pi)^4i}\,\Big[\bar{u}_e(\vec{k}_e, \sigma_e)
    \gamma^{\mu}(1 - \gamma^5) v_{\bar{\nu}}(\vec{k}_{\bar{\nu}}, +
    \frac{1}{2})\Big]\,\Big[\bar{u}_p(\vec{k}_p,\sigma_p)\,
    \gamma^5\,\frac{1}{m_N - \hat{k}_n - \hat{k} - i0}\,\gamma^5
    u_n(\vec{k}_n, \sigma_n)\Big]\nonumber\\ &&\times \,\frac{(p + 2k
    - 2 q)^{\beta_1}}{[m^2_{\sigma} - k^2 - i0][m^2_{\pi} - (k - q)^2
      - i0][m^2_{\pi} - (k + p - q)^2 - i0]}\,
  D^{(\gamma)}_{\beta_1\beta_2}(p)\, D^{(W)}_{\mu}{}^{\beta_2}(-
  q),\nonumber\\&& M(n \to p e^- \bar{\nu}_e)^{(\pi^0)}_{\rm
    Fig.\,\ref{fig:fig3}b} = - 2 e^2 g^2_{\pi N} M^2_W
  G_V\nonumber\\ &&\times \int \frac{d^4k}{(2\pi)^4i}\int
  \frac{d^4p}{(2\pi)^4i}\,\Big[\bar{u}_e(\vec{k}_e, \sigma_e)
    \gamma^{\mu}(1 - \gamma^5) v_{\bar{\nu}}(\vec{k}_{\bar{\nu}}, +
    \frac{1}{2})\Big]\,\Big[\bar{u}_p(\vec{k}_p,\sigma_p)\,
    \gamma^{\beta_1}\,\frac{1}{m_N - \hat{k}_p + \hat{p} -
      i0}\,\gamma^5\,\frac{1}{m_N - \hat{k}_n - \hat{k} -
      i0}\nonumber\\ &&\times \,\gamma^5 u_n(\vec{k}_n, \sigma_n)\Big]
  \,\frac{1}{[m^2_{\pi} - k^2 - i0][m^2_{\pi} - (k + p - q)^2 - i0]}\,
  D^{(\gamma)}_{\beta_1\beta_2}(p)\, D^{(W)}_{\mu}{}^{\beta_2}(-
  q),\nonumber\\
\end{eqnarray*}
\begin{eqnarray*}
  && M(n \to p e^- \bar{\nu}_e)^{(\sigma)}_{\rm
    Fig.\,\ref{fig:fig3}b} = - 2 e^2 g^2_{\pi N} M^2_W
  G_V\nonumber\\ &&\times \int \frac{d^4k}{(2\pi)^4i}\int
  \frac{d^4p}{(2\pi)^4i}\,\Big[\bar{u}_e(\vec{k}_e, \sigma_e)
    \gamma^{\mu}(1 - \gamma^5) v_{\bar{\nu}}(\vec{k}_{\bar{\nu}}, +
    \frac{1}{2})\Big]\,\Big[\bar{u}_p(\vec{k}_p,\sigma_p)\,
    \gamma^{\beta_1}\,\frac{1}{m_N - \hat{k}_p + \hat{p} -
      i0}\,\gamma^5\,\frac{1}{m_N - \hat{k}_n - \hat{k} -
      i0}\nonumber\\ &&\times \, u_n(\vec{k}_n, \sigma_n)\Big]
  \,\frac{1}{[m^2_{\sigma} - k^2 - i0][m^2_{\pi} - (k + p - q)^2 -
      i0]}\, D^{(\gamma)}_{\beta_1\beta_2}(p)\,
  D^{(W)}_{\mu}{}^{\beta_2}(- q),\nonumber\\
  && M(n \to p e^-
  \bar{\nu}_e)^{(\pi^0)}_{\rm Fig.\,\ref{fig:fig3}c} = + 2 e^2
  g^2_{\pi N} M^2_W G_V\nonumber\\ &&\times \int
  \frac{d^4k}{(2\pi)^4i}\int
  \frac{d^4p}{(2\pi)^4i}\,\Big[\bar{u}_e(\vec{k}_e, \sigma_e)
    \gamma^{\mu}(1 - \gamma^5) v_{\bar{\nu}}(\vec{k}_{\bar{\nu}}, +
    \frac{1}{2})\Big]\,\Big[\bar{u}_p(\vec{k}_p,\sigma_p)\,
    \gamma^5\,\frac{1}{m_N - \hat{k}_n - \hat{k} - i0}\,\gamma^5
    u_n(\vec{k}_n, \sigma_n)\Big]\nonumber\\ &&\times
  \,\frac{1}{[m^2_{\pi} - k^2 - i0][m^2_{\pi} - (k - q)^2 -
      i0]}\,\big[(p - q)^{\nu}\eta^{\beta_1\alpha_1} -
    p^{\beta_1}\eta^{\alpha_1 \nu} +
    q^{\alpha_1}\eta^{\nu\beta_1}\big]\,
  D^{(\gamma)}_{\beta_1\beta_2}(p)\, D^{(W)}_{\alpha_1}{}^{\beta_2}(p
  - q)\, D^{(W)}_{\mu\nu}(- q),\nonumber\\
  && M(n \to p e^-
  \bar{\nu}_e)^{(\sigma)}_{\rm Fig.\,\ref{fig:fig3}c} = + 2 e^2
  g^2_{\pi N} M^2_W G_V\nonumber\\ &&\times \int
  \frac{d^4k}{(2\pi)^4i}\int
  \frac{d^4p}{(2\pi)^4i}\,\Big[\bar{u}_e(\vec{k}_e, \sigma_e)
    \gamma^{\mu}(1 - \gamma^5) v_{\bar{\nu}}(\vec{k}_{\bar{\nu}}, +
    \frac{1}{2})\Big]\,\Big[\bar{u}_p(\vec{k}_p,\sigma_p)\,
    \gamma^5\,\frac{1}{m_N - \hat{k}_n - \hat{k} - i0}\,
    u_n(\vec{k}_n, \sigma_n)\Big]\nonumber\\ &&\times
  \,\frac{1}{[m^2_{\sigma} - k^2 - i0][m^2_{\pi} - (k - q)^2 -
      i0]}\,\big[(p - q)^{\nu}\eta^{\beta_1\alpha_1} -
    p^{\beta_1}\eta^{\alpha_1 \nu} +
    q^{\alpha_1}\eta^{\nu\beta_1}\big]\,
  D^{(\gamma)}_{\beta_1\beta_2}(p)\, D^{(W)}_{\alpha_1}{}^{\beta_2}(p
  - q)\, D^{(W)}_{\mu\nu}(- q),\nonumber\\
  && M(n \to p e^-
  \bar{\nu}_e)^{(\pi^0)}_{\rm Fig.\,\ref{fig:fig3}d} =+ 2 e^2 g^2_{\pi
    N} M^2_W G_V\nonumber\\ &&\times \int \frac{d^4k}{(2\pi)^4i}\int
  \frac{d^4p}{(2\pi)^4i}\,\Big[\bar{u}_e(\vec{k}_e, \sigma_e)
    \gamma^{\beta_1}\,\frac{1}{m_e - \hat{k}_e - \hat{p} -
      i0}\,\gamma^{\mu}(1 - \gamma^5)
    v_{\bar{\nu}}(\vec{k}_{\bar{\nu}}, +
    \frac{1}{2})\Big]\,\,\Big[\bar{u}_p(\vec{k}_p,\sigma_p)\,
    \gamma^5\,\frac{1}{m_N - \hat{k}_n - \hat{k} -
      i0}\nonumber\\ &&\times \,\gamma^5 u_n(\vec{k}_n, \sigma_n)\Big]
  \,\frac{1}{[m^2_{\pi} - k^2 - i0][m^2_{\pi} - (k - q)^2 -
      i0]}\,D^{(\gamma)}_{\beta_1\beta_2}(p)
  \,D^{(W)}_{\alpha_1}{}^{\beta_2}(p - q),\nonumber\\
  && M(n \to p e^- \bar{\nu}_e)^{(\sigma)}_{\rm Fig.\,\ref{fig:fig3}d}
  =- 2 e^2 g^2_{\pi N} M^2_W G_V\nonumber\\ &&\times \int
  \frac{d^4k}{(2\pi)^4i}\int
  \frac{d^4p}{(2\pi)^4i}\,\Big[\bar{u}_e(\vec{k}_e, \sigma_e)
    \gamma^{\beta_1}\,\frac{1}{m_e - \hat{k}_e - \hat{p} -
      i0}\,\gamma^{\mu}(1 - \gamma^5)
    v_{\bar{\nu}}(\vec{k}_{\bar{\nu}}, +
    \frac{1}{2})\Big]\,\,\Big[\bar{u}_p(\vec{k}_p,\sigma_p)\,
    \gamma^5\,\frac{1}{m_N - \hat{k}_n - \hat{k} -
      i0}\nonumber\\ &&\times \,u_n(\vec{k}_n, \sigma_n)\Big]
  \,\frac{1}{[m^2_{\sigma} - k^2 - i0][m^2_{\pi} - (k - q)^2 -
      i0]}\,D^{(\gamma)}_{\beta_1\beta_2}(p)
  \,D^{(W)}_{\alpha_1}{}^{\beta_2}(p - q),\nonumber\\&& M(n \to p e^-
  \bar{\nu}_e)^{(\pi^0)}_{\rm Fig.\,\ref{fig:fig3}e} =+ 2 e^2 g^2_{\pi
    N} M^2_W G_V\nonumber\\ &&\times \int \frac{d^4k}{(2\pi)^4i}\int
  \frac{d^4p}{(2\pi)^4i}\,\Big[\bar{u}_e(\vec{k}_e, \sigma_e)
    \,\gamma^{\mu}(1 - \gamma^5) v_{\bar{\nu}}(\vec{k}_{\bar{\nu}}, +
    \frac{1}{2})\Big]\,\Big[\bar{u}_p(\vec{k}_p,\sigma_p)\,
    \gamma^5\,\frac{1}{m_N - \hat{k}_n - \hat{k} - i0}\,\gamma^5\,
    u_n(\vec{k}_n, \sigma_n)\Big]\nonumber\\ &&\times \,\frac{(2 k -
    p)^{\beta_1}}{[m^2_{\pi} - k^2 - i0][m^2_{\pi} - (k - p)^2 -
      i0][m^2_{\pi} - (k - q)^2 -
      i0]}\,D^{(\gamma)}_{\beta_1\beta_2}(p)
  \,D^{(W)}_{\mu}{}^{\beta_2}(- q),\nonumber\\&& M(n \to p e^-
  \bar{\nu}_e)^{(\sigma)}_{\rm Fig.\,\ref{fig:fig3}e} =+ 2 e^2
  g^2_{\pi N} M^2_W G_V\nonumber\\ &&\times \int
  \frac{d^4k}{(2\pi)^4i}\int
  \frac{d^4p}{(2\pi)^4i}\,\Big[\bar{u}_e(\vec{k}_e, \sigma_e)
    \,\gamma^{\mu}(1 - \gamma^5) v_{\bar{\nu}}(\vec{k}_{\bar{\nu}}, +
    \frac{1}{2})\Big]\,\Big[\bar{u}_p(\vec{k}_p,\sigma_p)\,
    \gamma^5\,\frac{1}{m_N - \hat{k}_n - \hat{k} - i0}\,u_n(\vec{k}_n,
    \sigma_n)\Big]\nonumber\\ &&\times \,\frac{(2 k -
    p)^{\beta_1}}{[m^2_{\pi} - k^2 - i0][m^2_{\pi} - (k - p)^2 -
      i0][m^2_{\sigma} - (k - q)^2 -
      i0]}\,D^{(\gamma)}_{\beta_1\beta_2}(p)
  \,D^{(W)}_{\mu}{}^{\beta_2}(- q),\nonumber\\&& M(n \to p e^-
  \bar{\nu}_e)^{(\pi^0)}_{\rm Fig.\,\ref{fig:fig3}f} = + 2 e^2 g^2_{\pi
    N} M^2_W G_V\nonumber\\ &&\times \int \frac{d^4k}{(2\pi)^4i}\int
  \frac{d^4p}{(2\pi)^4i}\,\Big[\bar{u}_e(\vec{k}_e, \sigma_e)
    \,\gamma^{\mu}(1 - \gamma^5) v_{\bar{\nu}}(\vec{k}_{\bar{\nu}}, +
    \frac{1}{2})\Big]\,\Big[\bar{u}_p(\vec{k}_p,\sigma_p)\,
    \gamma^5\,\frac{1}{m_N - \hat{k}_n - \hat{k} - \hat{p} -
      i0}\,\gamma^{\beta_1}\nonumber\\ &&\times\, \frac{1}{m_N -
      \hat{k}_n - \hat{k} - i0}\,\gamma^5 u_n(\vec{k}_n,
    \sigma_n)\Big]\,\frac{1}{[m^2_{\pi} - k^2 - i0][m^2_{\pi} - (k + p
      - q)^2 - i0]}\,D^{(\gamma)}_{\beta_1\beta_2}(p)
  \,D^{(W)}_{\mu}{}^{\beta_2}(- q),\nonumber\\&& M(n \to p e^-
  \bar{\nu}_e)^{(\sigma)}_{\rm Fig.\,\ref{fig:fig3}f} = + 2 e^2
  g^2_{\pi N} M^2_W G_V\nonumber\\ &&\times \int
  \frac{d^4k}{(2\pi)^4i}\int
  \frac{d^4p}{(2\pi)^4i}\,\Big[\bar{u}_e(\vec{k}_e, \sigma_e)
    \,\gamma^{\mu}(1 - \gamma^5) v_{\bar{\nu}}(\vec{k}_{\bar{\nu}}, +
    \frac{1}{2})\Big]\,\Big[\bar{u}_p(\vec{k}_p,\sigma_p)\,
    \gamma^5\,\frac{1}{m_N - \hat{k}_n - \hat{k} - \hat{p} -
      i0}\,\gamma^{\beta_1}\nonumber\\ &&\times\, \frac{1}{m_N -
      \hat{k}_n - \hat{k} - i0}\,\gamma^5 u_n(\vec{k}_n,
    \sigma_n)\Big]\,\frac{1}{[m^2_{\pi} - k^2 - i0][m^2_{\sigma} - (k
      + p - q)^2 - i0]}\,D^{(\gamma)}_{\beta_1\beta_2}(p)
  \,D^{(W)}_{\mu}{}^{\beta_2}(- q),\nonumber\\&& M(n \to p e^-
  \bar{\nu}_e)^{(\pi^0)}_{\rm Fig.\,\ref{fig:fig3}g} = + 2 e^2 g^2_{\pi
    N} M^2_W G_V\nonumber\\ &&\times \int \frac{d^4k}{(2\pi)^4i}\int
  \frac{d^4p}{(2\pi)^4i}\,\Big[\bar{u}_e(\vec{k}_e, \sigma_e)
    \,\gamma^{\mu}(1 - \gamma^5) v_{\bar{\nu}}(\vec{k}_{\bar{\nu}}, +
    \frac{1}{2})\Big]\,\Big[\bar{u}_p(\vec{k}_p,\sigma_p)\,\gamma^{\beta_1}
    \frac{1}{m_N - \hat{k}_p + \hat{p} - i0}\, \gamma^5\,\frac{1}{m_N
      - \hat{k}_n - \hat{k} - i0}\nonumber\\ &&\times \,\gamma^5
    u_n(\vec{k}_n, \sigma_n)\Big]\,\frac{1}{[m^2_{\pi} - k^2 -
      i0][m^2_{\pi} - (k + p - q)^2 -
      i0]}\,D^{(\gamma)}_{\beta_1\beta_2}(p)
  \,D^{(W)}_{\mu}{}^{\beta_2}(- q),\nonumber\\
\end{eqnarray*}
\begin{eqnarray}\label{eq:A.4}
  && M(n \to p e^-
  \bar{\nu}_e)^{(\sigma)}_{\rm Fig.\,\ref{fig:fig3}g} = + 2 e^2
  g^2_{\pi N} M^2_W G_V\nonumber\\ &&\times \int
  \frac{d^4k}{(2\pi)^4i}\int
  \frac{d^4p}{(2\pi)^4i}\,\Big[\bar{u}_e(\vec{k}_e, \sigma_e)
    \,\gamma^{\mu}(1 - \gamma^5) v_{\bar{\nu}}(\vec{k}_{\bar{\nu}}, +
    \frac{1}{2})\Big]\,\Big[\bar{u}_p(\vec{k}_p,\sigma_p)\,\gamma^{\beta_1}
    \frac{1}{m_N - \hat{k}_p + \hat{p} - i0}\, \gamma^5\,\frac{1}{m_N
      - \hat{k}_n - \hat{k} - i0}\nonumber\\ &&\times \,\gamma^5
    u_n(\vec{k}_n, \sigma_n)\Big]\,\frac{1}{[m^2_{\pi} - k^2 -
      i0][m^2_{\sigma} - (k + p - q)^2 -
      i0]}\,D^{(\gamma)}_{\beta_1\beta_2}(p)
  \,D^{(W)}_{\mu}{}^{\beta_2}(- q),\nonumber\\&& M(n \to p e^-
  \bar{\nu}_e)^{(\pi^0)}_{\rm Fig.\,\ref{fig:fig3}h} = - 2 e^2
  g^2_{\pi N} M^2_W G_V\nonumber\\ &&\times \int
  \frac{d^4k}{(2\pi)^4i}\int
  \frac{d^4p}{(2\pi)^4i}\,\Big[\bar{u}_e(\vec{k}_e, \sigma_e)
    \,\gamma^{\mu}(1 - \gamma^5) v_{\bar{\nu}}(\vec{k}_{\bar{\nu}}, +
    \frac{1}{2})\Big]\,\Big[\bar{u}_p(\vec{k}_p,\sigma_p)\,
    \gamma^5\,\frac{1}{m_N - \hat{k}_n - \hat{k} -
      i0}\nonumber\\ &&\times \,\gamma^5 u_n(\vec{k}_n,
    \sigma_n)\Big]\,\frac{1}{[m^2_{\pi} - k^2 - i0][m^2_{\pi} - (k -
      q)^2 - i0]}\,\big[(p - q)^{\nu} \eta^{\beta_1 \alpha_1} -
    p^{\beta_1} \eta^{\alpha_1 \nu} + q^{\alpha_1}
    \eta^{\nu\beta_1}\big]\nonumber\\ &&\times
  \,D^{(\gamma)}_{\beta_1\beta_2}(p)
  \,D^{(W)}_{\alpha_1}{}^{\beta_2}(p - q)\, D^{(W)}_{\mu\nu}(-
  q),\nonumber\\&& M(n \to p e^- \bar{\nu}_e)^{(\sigma)}_{\rm
    Fig.\,\ref{fig:fig3}h} = - 2 e^2 g^2_{\pi N} M^2_W
  G_V\nonumber\\ &&\times \int \frac{d^4k}{(2\pi)^4i}\int
  \frac{d^4p}{(2\pi)^4i}\,\Big[\bar{u}_e(\vec{k}_e, \sigma_e)
    \,\gamma^{\mu}(1 - \gamma^5) v_{\bar{\nu}}(\vec{k}_{\bar{\nu}}, +
    \frac{1}{2})\Big]\,\Big[\bar{u}_p(\vec{k}_p,\sigma_p)\,
    \gamma^5\,\frac{1}{m_N - \hat{k}_n - \hat{k} -
      i0}\nonumber\\ &&\times \,\gamma^5 u_n(\vec{k}_n,
    \sigma_n)\Big]\,\frac{1}{[m^2_{\pi} - k^2 - i0][m^2_{\sigma} - (k
      - q)^2 - i0]}\,\big[(p - q)^{\nu} \eta^{\beta_1 \alpha_1} -
    p^{\beta_1} \eta^{\alpha_1 \nu} + q^{\alpha_1}
    \eta^{\nu\beta_1}\big]\nonumber\\ &&\times
  \,D^{(\gamma)}_{\beta_1\beta_2}(p)
  \,D^{(W)}_{\alpha_1}{}^{\beta_2}(p - q)\, D^{(W)}_{\mu\nu}(-
  q),\nonumber\\
 && M(n \to p e^- \bar{\nu}_e)^{(\pi^0)}_{\rm Fig.\,\ref{fig:fig3}i} =
  - 2 e^2 g^2_{\pi N} M^2_W G_V\nonumber\\ &&\times \int
  \frac{d^4k}{(2\pi)^4i}\int \frac{d^4p}{(2\pi)^4i}\,
  \Big[\bar{u}_e(\vec{k}_e, \sigma_e)\, \gamma^{\beta_1} \frac{1}{m_e
      - \hat{k}_e - \hat{p} - i0} \,\gamma^{\mu}(1 - \gamma^5)
    v_{\bar{\nu}}(\vec{k}_{\bar{\nu}}, +
    \frac{1}{2})\Big]\,\Big[\bar{u}_p(\vec{k}_p,\sigma_p)\,
    \gamma^5\,\frac{1}{m_N - \hat{k}_n - \hat{k} -
      i0}\nonumber\\ &&\times \,\gamma^5 u_n(\vec{k}_n,
    \sigma_n)\Big]\,\frac{1}{[m^2_{\pi} - k^2 - i0][m^2_{\pi} - (k -
      q)^2 - i0]}\, D^{(\gamma)}_{\beta_1\beta_2}(p)
  \,D^{(W)}_{\mu}{}^{\beta_2}(p - q),\nonumber\\&& M(n \to p e^-
  \bar{\nu}_e)^{(\sigma)}_{\rm Fig.\,\ref{fig:fig3}i} = - 2 e^2
  g^2_{\pi N} M^2_W G_V\nonumber\\ &&\times \int
  \frac{d^4k}{(2\pi)^4i}\int \frac{d^4p}{(2\pi)^4i}\,
  \Big[\bar{u}_e(\vec{k}_e, \sigma_e)\, \gamma^{\beta_1} \frac{1}{m_e
      - \hat{k}_e - \hat{p} - i0} \,\gamma^{\mu}(1 - \gamma^5)
    v_{\bar{\nu}}(\vec{k}_{\bar{\nu}}, +
    \frac{1}{2})\Big]\,\Big[\bar{u}_p(\vec{k}_p,\sigma_p)\,
    \gamma^5\,\frac{1}{m_N - \hat{k}_n - \hat{k} -
      i0}\nonumber\\ &&\times \,\gamma^5 u_n(\vec{k}_n,
    \sigma_n)\Big]\,\frac{1}{[m^2_{\pi} - k^2 - i0][m^2_{\sigma} - (k -
      q)^2 - i0]}\, D^{(\gamma)}_{\beta_1\beta_2}(p)
  \,D^{(W)}_{\mu}{}^{\beta_2}(p - q).
\end{eqnarray}
The analytical expressions for the Feynman diagrams in
Fig.\,\ref{fig:fig4} are equal to
\begin{eqnarray*}
 && M(n \to p e^- \bar{\nu}_e)_{\rm Fig.\,\ref{fig:fig4}a} = - 2 e^2
  g^2_{\pi N} M^2_W G_V \nonumber\\ &&\times \int
  \frac{d^4k}{(2\pi)^4i}\int
  \frac{d^4p}{(2\pi)^4i}\,\Big[\bar{u}_e(\vec{k}_e, \sigma_e)\,
    \gamma^{\mu}(1 - \gamma^5) v_{\bar{\nu}}(\vec{k}_{\bar{\nu}}, +
    \frac{1}{2})\Big]\,\Big[\bar{u}_p(\vec{k}_p,\sigma_p)\,
    \gamma^{\beta_1}\,\frac{1}{m_N - \hat{k}_p - \hat{p} -
      i0}\,\gamma^{\nu}(1 - \gamma^5) \nonumber\\ &&\times \,
    \frac{1}{m_N - \hat{k}_n - \hat{p} - i0}\,\gamma^5\, \frac{1}{m_N
      - \hat{k}_n - \hat{k} - i0}\,\gamma^5 u_n(\vec{k}_n,
    \sigma_n)\Big]\,\frac{(2 k - p)^{\beta_2}}{[m^2_{\pi} - k^2 -
      i0][m^2_{\pi} - (k - p)^2 - i0]} \nonumber\\ &&\times\,
  D^{(\gamma)}_{\beta_1\beta_2}(p)\, D^{(W)}_{\mu\nu}(-
  q),\nonumber\\ && M(n \to p e^- \bar{\nu}_e)_{\rm
    Fig.\,\ref{fig:fig4}b} = + 2 e^2 g^2_{\pi N} M^2_W G_V
  \nonumber\\ &&\times \int \frac{d^4k}{(2\pi)^4i}\int
  \frac{d^4p}{(2\pi)^4i}\,\Big[\bar{u}_e(\vec{k}_e, \sigma_e)\,
    \gamma^{\mu}(1 - \gamma^5) v_{\bar{\nu}}(\vec{k}_{\bar{\nu}}, +
    \frac{1}{2})\Big]\,\Big[\bar{u}_p(\vec{k}_p,\sigma_p)\,
    \,\gamma^{\alpha_2}(1 - \gamma^5)\, \frac{1}{m_N - \hat{k}_n -
      \hat{p} - i0}\, \gamma^5 \nonumber\\ &&\times \, \frac{1}{m_N -
      \hat{k}_n - \hat{k} - i0}\,\gamma^5 u_n(\vec{k}_n,
    \sigma_n)\Big] \frac{(2 k - p)^{\beta_1}}{[m^2_{\pi} - k^2 -
      i0][m^2_{\pi} - (k - p)^2 - i0]} \big[(p -
    q)^{\nu}\eta^{\beta_2\alpha_1} - (p - 2 q)^{\beta_2}
    \eta^{\alpha_2\nu} -
    q^{\alpha_2}\eta^{\nu\beta_2}\big]\nonumber\\ &&\times\,
  D^{(\gamma)}_{\beta_1\beta_2}(p)\, D^{(W)}_{\alpha_1\alpha_2}(p -
  q)\, D^{(W)}_{\mu\nu}( - q),\nonumber\\
  && M(n \to p e^-
  \bar{\nu}_e)_{\rm Fig.\,\ref{fig:fig4}c} = + 2 e^2 g^2_{\pi N} M^2_W
  G_V \nonumber\\ &&\times \int \frac{d^4k}{(2\pi)^4i}\int
  \frac{d^4p}{(2\pi)^4i}\,\Big[\bar{u}_e(\vec{k}_e,
    \sigma_e)\,\gamma^{\beta_1}\,\frac{1}{m_e - \hat{k}_e - \hat{p} -
      i0} \gamma^{\mu}(1 - \gamma^5)
    v_{\bar{\nu}}(\vec{k}_{\bar{\nu}}, +
    \frac{1}{2})\Big]\,\Big[\bar{u}_p(\vec{k}_p,\sigma_p)\,
    \gamma^{\nu}(1 - \gamma^5) \nonumber\\
\end{eqnarray*}
\begin{eqnarray}\label{eq:A.5}
 &&\times \, \frac{1}{m_N -
      \hat{k}_n - \hat{p} - i0}\,\gamma^5\, \frac{1}{m_N - \hat{k}_n -
      \hat{k} - i0}\,\gamma^5 u_n(\vec{k}_n, \sigma_n)\Big]\,\frac{(2
    k - p)^{\beta_2}}{[m^2_{\pi} - k^2 - i0][m^2_{\pi} - (k - p)^2 -
      i0]}\nonumber\\ &&\times\, D^{(\gamma)}_{\beta_1\beta_2}(p)\,
  D^{(W)}_{\mu\nu}(p - q)\nonumber\\
  && M(n \to p e^-
  \bar{\nu}_e)_{\rm Fig.\,\ref{fig:fig4}d} = - 2 e^2 g^2_{\pi N} M^2_W
  G_V \nonumber\\ &&\times \int \frac{d^4k}{(2\pi)^4i}\int
  \frac{d^4p}{(2\pi)^4i}\,\Big[\bar{u}_e(\vec{k}_e, \sigma_e)\,
    \gamma^{\mu}(1 - \gamma^5) v_{\bar{\nu}}(\vec{k}_{\bar{\nu}}, +
    \frac{1}{2})\Big]\,\Big[\bar{u}_p(\vec{k}_p,\sigma_p)\,\gamma^{\beta_1}\frac{1}{m_N
      - \hat{k}_p - \hat{p} - i0}\, \gamma^{\nu}(1 - \gamma^5)
    \nonumber\\ &&\times \, \frac{1}{m_N - \hat{k}_n - \hat{p} -
      i0}\,\gamma^5\,\frac{1}{m_N - \hat{k}_n - \hat{k} - \hat{p} -
      i0}\,\gamma^{\beta_2} \frac{1}{m_N - \hat{k}_n - \hat{k} -
      i0}\,\gamma^5 u_n(\vec{k}_n, \sigma_n)\Big]\,\frac{1}{m^2_{\pi}
    - k^2 - i0}\nonumber\\ &&\times\,
  D^{(\gamma)}_{\beta_1\beta_2}(p)\, D^{(W)}_{\mu\nu}(-
  q),\nonumber\\ && M(n \to p e^- \bar{\nu}_e)_{\rm
    Fig.\,\ref{fig:fig4}e} = + 2 e^2 g^2_{\pi N} M^2_W G_V
  \nonumber\\ &&\times \int \frac{d^4k}{(2\pi)^4i}\int
  \frac{d^4p}{(2\pi)^4i}\,\Big[\bar{u}_e(\vec{k}_e, \sigma_e)\,
    \gamma^{\mu}(1 - \gamma^5) v_{\bar{\nu}}(\vec{k}_{\bar{\nu}}, +
    \frac{1}{2})\Big]\,\Big[\bar{u}_p(\vec{k}_p,\sigma_p)\,
    \gamma^{\alpha_2}(1 - \gamma^5)\,\frac{1}{m_N - \hat{k}_n -
      \hat{p} - i0}\, \gamma^5 \nonumber\\ &&\times \, \frac{1}{m_N -
      \hat{k}_n - \hat{k}- \hat{p} - i0}\,\gamma^{\beta_2}
    \,\frac{1}{m_N - \hat{k}_n - \hat{k} - i0}\,\gamma^5
    u_n(\vec{k}_n, \sigma_n)\Big]\,\big[(p -
    q)^{\nu}\eta^{\beta_2\alpha_1} - (p - 2 q)^{\beta_2}\eta^{\alpha_1
      \nu} - q^{\alpha_1}\eta^{\nu\beta_2}\big]\nonumber\\ &&\times\,
  \frac{1}{m^2_{\pi} - k^2 - i0}\, D^{(\gamma)}_{\beta_1\beta_2}(p)\,
  D^{(W)}_{\alpha_1\alpha_2}(p - q)\, D^{(W)}_{\mu\nu}(-
  q),\nonumber\\ && M(n \to p e^- \bar{\nu}_e)_{\rm
    Fig.\,\ref{fig:fig4}f} = + 2 e^2 g^2_{\pi N} M^2_W G_V
  \nonumber\\ &&\times \int \frac{d^4k}{(2\pi)^4i}\int
  \frac{d^4p}{(2\pi)^4i}\,\Big[\bar{u}_e(\vec{k}_e,
    \sigma_e)\,\gamma^{\beta_1}\,\frac{1}{m_e - \hat{k}_e - \hat{p} -
      i0}\, \gamma^{\mu}(1 - \gamma^5)
    v_{\bar{\nu}}(\vec{k}_{\bar{\nu}}, +
    \frac{1}{2})\Big]\,\Big[\bar{u}_p(\vec{k}_p,\sigma_p)\,
    \gamma^{\nu}(1 - \gamma^5)\nonumber\\ &&\times \,\frac{1}{m_N -
      \hat{k}_n - \hat{p} - i0}\, \gamma^5 \, \frac{1}{m_N - \hat{k}_n
      - \hat{k}- \hat{p} - i0}\,\gamma^{\beta_2} \,\frac{1}{m_N -
      \hat{k}_n - \hat{k} - i0}\,\gamma^5 u_n(\vec{k}_n,
    \sigma_n)\Big]\nonumber\\ &&\times\, \frac{1}{m^2_{\pi} - k^2 -
    i0}\, D^{(\gamma)}_{\beta_1\beta_2}(p)\,D^{(W)}_{\mu\nu}(p - q).
\end{eqnarray}
Now we may proceed to the analysis of gauge invariance of the Feynman
diagrams in Fig.\,\ref{fig:fig1} - Fig.\,\ref{fig:fig4}. Since in the
limit of the infinite mass of the $\sigma$-meson $m_{\sigma} \to
\infty$, the Feynman diagrams in Fig.\,\ref{fig:fig1} -
Fig.\,\ref{fig:fig4} with $\sigma$-meson exchanges vanish, we analyze
gauge invariance of the Feynman diagrams with pion-exchanges only.

\section*{Appendix B: Gauge invariance of  the Feynman diagrams 
in Fig.\,\ref{fig:fig1} - Fig.\,\ref{fig:fig4} }
\renewcommand{\theequation}{B-\arabic{equation}}
\setcounter{equation}{0}

In this Appendix we analyze gauge invariance or independence of the
Feynman diagrams in Fig.\,\ref{fig:fig1} - Fig.\,\ref{fig:fig4} of a
gauge parameter $\xi$ of the photon propagator
$D^{(\gamma)}_{\alpha\beta}(p)$ defined in Eq.(\ref{eq:A.2}). For this
aim we follow \cite{Ivanov2019a} and calculate by using dimensional
regularization the partial derivative with respect to a gauge
parameter $\xi$ of the analytical expression of every Feynman diagram
in Fig.\,\ref{fig:fig1} - Fig.\,\ref{fig:fig4},
i.e. $\partial/\partial \xi M(n \to p e^- \bar{\nu}_e)_{\rm Fig.j}$
for $j = 1,2,3,4$, and sum up the contributions.

\subsection*{\bf Gauge-dependent contributions of the Feynman diagrams in
  Fig.\,\ref{fig:fig1}}

The analytical expressions for the gauge-dependent contributions of
the Feynman diagrams in Fig.\,\ref{fig:fig1} are
\begin{eqnarray*}
\hspace{-0.35in}&&\frac{\partial}{\partial \xi}M(n \to p e^-
\bar{\nu}_e)^{(\pi^0)}_{\rm Fig.\,\ref{fig:fig1}a} = - 2 e^2 g^2_{\pi
  N} M^2_W G_V\nonumber\\\hspace{-0.35in} &&\times \int
\frac{d^4k}{(2\pi)^4i}\int \frac{d^4p}{(2\pi)^4i}\,\frac{1}{(p^2 +
  i0)^2}\, \Big[\bar{u}_e \gamma^{\mu}(1 - \gamma^5)
  v_{\bar{\nu}})\Big]\,\Big[\bar{u}_p\,\gamma^5 \, \frac{1}{m_N -
    \hat{k}_n - \hat{k} - i0}\,\gamma^5
  u_n\Big]\nonumber\\ \hspace{-0.35in}&&\times \, \Big(\frac{ (2 k + p
  - q)^{\alpha_2} }{[m^2_{\pi} - k^2 - i0][m^2_{\pi} - (k + p - q)^2 -
    i0]} - \frac{ (2 k + p - q)^{\alpha_2} }{[m^2_{\pi} - k^2 -
    i0][m^2_{\pi} - (k - q)^2 -
    i0]}\Big)\nonumber\\ \hspace{-0.35in}&&\times\,
\big[D^{(W)-1}_{\alpha_1\nu}(- q) - D^{(W)-1}_{\alpha_1\nu}(p -
  q)\big]\, D^{(W)\alpha_1}{}_{ \alpha_2}(p - q)\,
D^{(W)}_{\mu}{}^{\nu}(- q) = \nonumber\\
\end{eqnarray*}
\begin{eqnarray}\label{eq:B.1}
\hspace{-0.35in} &&= - 2 e^2
g^2_{\pi N} M^2_W G_V \nonumber\\ \hspace{-0.35in}&&\times \int
\frac{d^4k}{(2\pi)^4i}\int \frac{d^4p}{(2\pi)^4i}\,\frac{1}{(p^2 +
  i0)^2}\,\Big[\bar{u}_e\, \gamma^{\mu}(1 - \gamma^5)
  v_{\bar{\nu}}\Big]\, \Big[\bar{u}_p\,\gamma^5 \frac{1}{m_N -
    \hat{k}_n - \hat{k} - i0}\,\gamma^5
  u_n\Big]\nonumber\\
\hspace{-0.35in}&&\times\, \Big(\frac{(2 k + p -
  q)^{\nu}}{[m^2_{\pi} - k^2 - i0][m^2_{\pi} - (k + p - q)^2 - i0]} -
\frac{(2 k + p - q)^{\nu}}{[m^2_{\pi} - k^2 - i0][m^2_{\pi} - (k -
    q)^2 - i0]}\Big)\nonumber\\ \hspace{-0.35in}&&\times\,
\big[D^{(W)}_{\mu \nu }(p - q) - D^{(W)}_{\mu\nu}(- q)\big]
\end{eqnarray}
and 
\begin{eqnarray}\label{eq:B.2}
\hspace{-0.35in}&&\frac{\partial}{\partial \xi} M(n \to p e^-
\bar{\nu}_e)^{(\pi^0)}_{\rm Fig.\,\ref{fig:fig1}b} = + 2 e^2 g^2_{\pi
  N} M^2_W G_V \nonumber\\\hspace{-0.35in}&&\times \int
\frac{d^4k}{(2\pi)^4i}\int \frac{d^4p}{(2\pi)^4i}\,\frac{1}{(p^2 +
  i0)^2}\, \Big[\bar{u}_e\, \gamma^{\mu}(1 - \gamma^5)
  v_{\bar{\nu}}\Big]\,\Big[\bar{u}_p\,\hat{p}\, \frac{1}{m_N -
    \hat{k}_p + \hat{p} - i0} \gamma^5 \frac{1}{m_N - \hat{k}_n -
    \hat{k} - i0} \,\gamma^5 u_n\Big]\nonumber\\ \hspace{-0.35in}&&
\,\Big(\frac{(2 k - q)^{\nu}}{[m^2_{\pi} - k^2 - i0][m^2_{\pi} - (k +
    p - q)^2 - i0]} - \frac{(2 k - q)^{\nu}}{[m^2_{\pi} - k^2 -
    i0][m^2_{\pi} - (k - q)^2 - i0]}\Big)\, D^{(W)}_{\mu\nu}(- q) =
\nonumber\\\hspace{-0.35in}&&= + 2 e^2 g^2_{\pi N} M^2_W G_V
\nonumber\\\hspace{-0.35in}&&\times \int \frac{d^4k}{(2\pi)^4i}\int
\frac{d^4p}{(2\pi)^4i}\,\frac{1}{(p^2 + i0)^2}\, \Big[\bar{u}_e\,
  \gamma^{\mu}(1 - \gamma^5) v_{\bar{\nu}}\Big]\,\Big[\bar{u}_p
  \,\gamma^5 \frac{1}{m_N - \hat{k}_n - \hat{k} - i0} \,\gamma^5
  u_n\Big]\nonumber\\ \hspace{-0.35in}&& \,\frac{(2 k -
  q)^{\nu}}{[m^2_{\pi} - k^2 - i0][m^2_{\pi} - (k + p - q)^2 - i0]} \,
D^{(W)}_{\mu\nu}(- q)
\end{eqnarray}
and
\begin{eqnarray}\label{eq:B.3}
  \hspace{-0.35in}&&\frac{\partial}{\partial \xi}M(n \to p e^-
  \bar{\nu}_e)^{(\pi^0)}_{\rm Fig.\,\ref{fig:fig1}c} = - 2 e^2
  g^2_{\pi N} M^2_W G_V \nonumber\\
  \hspace{-0.35in}&& \times \int
  \frac{d^4k}{(2\pi)^4i}\int \frac{d^4p}{(2\pi)^4i}\,\frac{1}{(p^2 +
    i0)^2}\,\Big[\bar{u}_e\hat{p}\,\frac{1}{m_e - \hat{p} - \hat{k}_e
      - i0}\, \gamma^{\mu}(1 - \gamma^5)
    v_{\bar{\nu}}\Big]\,\Big[\bar{u}_p\,\gamma^5 \frac{1}{m_N -
      \hat{k}_n - \hat{k} - i0}\, \gamma^5\,
    u_n\Big]\nonumber\\\hspace{-0.35in}
  \hspace{-0.35in}&&\times\, \Big(\frac{(2 k + p -
    q)^{\nu}}{[m^2_{\pi} - k^2 - i0][m^2_{\pi} - (k + p - q)^2 - i0]}
  - \frac{(2 k + p - q)^{\nu}}{[m^2_{\pi} - k^2 - i0][m^2_{\pi} - (k -
      q)^2 - i0]}\Big)\,D^{(W)}_{\mu\nu}(p - q) = \nonumber\\
  \hspace{-0.35in}&&= + 2 e^2 g^2_{\pi N} M^2_W G_V
  \nonumber\\\hspace{-0.35in}
  \hspace{-0.35in}&&\times \int
  \frac{d^4k}{(2\pi)^4i}\int \frac{d^4p}{(2\pi)^4i}\,\frac{1}{(p^2 +
    i0)^2}\,\Big[\bar{u}_e\, \gamma^{\mu}(1 - \gamma^5)
    v_{\bar{\nu}}\Big]\,\Big[\bar{u}_p\,\gamma^5 \frac{1}{m_N -
      \hat{k}_n - \hat{k} - i0}\, \gamma^5\,
    u_n\Big]\nonumber\\ \hspace{-0.35in}&&\times\, \Big(\frac{(2 k + p
    - q)^{\nu}}{[m^2_{\pi} - k^2 - i0][m^2_{\pi} - (k + p - q)^2 -
      i0]} - \frac{2 k + p - q)^{\nu}}{[m^2_{\pi} - k^2 -
      i0][m^2_{\pi} - (k - q)^2 - i0]}\Big)\,D^{(W)}_{\mu\nu}(p - q)
\end{eqnarray}
and 
\begin{eqnarray}\label{eq:B.4}
  \hspace{-0.35in}&&\frac{\partial}{\partial \xi}M(n \to p e^-
  \bar{\nu}_e)^{(\pi^0)}_{\rm Fig.\,\ref{fig:fig1}d} = + 2 e^2
  g^2_{\pi N} M^2_W G_V \nonumber\\ \hspace{-0.35in}&&\times \int
  \frac{d^4k}{(2\pi)^4i}\int \frac{d^4p}{(2\pi)^4i}\,\frac{1}{(p^2 +
    i0)^2}\,\Big[\bar{u}_e\, \gamma^{\mu}(1 - \gamma^5)
    v_{\bar{\nu}}\Big]\,\Big[\bar{u}_p\,\gamma^5 \frac{1}{m_N -
      \hat{k}_n - \hat{k} - i0}\,\gamma^5\,
    u_n\Big]\nonumber\\ \hspace{-0.35in}&&\times \, \Big(\frac{(2 k -
    p - q)^{\alpha_2}}{[m^2_{\pi} - k^2 - i0][m^2_{\pi} - (k - q)^2 -
      i0]} - \frac{(2 k - p - q)^{\alpha_2}}{[m^2_{\pi} - (k - q)^2 -
      i0][m^2_{\pi} - (k - p)^2 - i0]}\Big)
  \nonumber\\ \hspace{-0.35in}&&\times\,
  \big[D^{(W)-1}_{\alpha_1\nu}(- q) - D^{(W)-1}_{\alpha_1\nu}(p -
    q)\big]\, D^{(W)\alpha_1}{}_{ \alpha_2}(p - q)\,
  D^{(W)}_{\mu}{}^{\nu}(- q) = \nonumber\\ \hspace{-0.35in}&& = + 2
  e^2 g^2_{\pi N} M^2_W G_V \nonumber\\ \hspace{-0.35in}&&\times \int
  \frac{d^4k}{(2\pi)^4i}\int \frac{d^4p}{(2\pi)^4i}\,\frac{1}{(p^2 +
    i0)^2}\,\Big[\bar{u}_e\, \gamma^{\mu}(1 - \gamma^5)
    v_{\bar{\nu}}\Big]\,\Big[\bar{u}_p\,\gamma^5 \frac{1}{m_N -
      \hat{k}_n - \hat{k} - i0}\,\gamma^5\,
    u_n\Big]\nonumber\\ \hspace{-0.35in}&&\times \, \Big(\frac{(2 k -
    p - q)^{\nu}}{[m^2_{\pi} - k^2 - i0][m^2_{\pi} - (k - q)^2 - i0]}
  - \frac{(2 k - p - q)^{\nu}}{[m^2_{\pi} - (k - q)^2 - i0][m^2_{\pi}
      - (k - p)^2 - i0]}\Big)\nonumber\\ \hspace{-0.35in}&&\times
  \,\big[D^{(W)}_{\mu\nu}(p - q) - D^{(W)}_{\mu\nu}(- q)\big]
\end{eqnarray}
and 
\begin{eqnarray}\label{eq:B.5}
\hspace{-0.35in}&&\frac{\partial}{\partial \xi}M(n \to p e^-
\bar{\nu}_e)^{(\pi^0)}_{\rm Fig.\,\ref{fig:fig1}e} = - 2 e^2 g^2_{\pi
  N} M^2_W G_V \nonumber\\ \hspace{-0.35in}&& \times \int
\frac{d^4k}{(2\pi)^4i}\int \frac{d^4p}{(2\pi)^4i}\,\frac{1}{(p^2 +
  i0)^2}\,\Big[\bar{u}_e\, \gamma^{\mu}(1 - \gamma^5)
  v_{\bar{\nu}}\Big]\,\Big\{\Big[\bar{u}_p\, \gamma^5 \frac{1}{m_N -
    \hat{k}_n -\hat{k} - \hat{p} - i0}\,\gamma^5
  u_n\Big]\nonumber\\ \hspace{-0.35in}&& - \Big[\bar{u}_p\,\gamma^5
  \frac{1}{m_N - \hat{k}_n - \hat{k} - i0}\,\gamma^5
  u_n\Big]\Big\}\,\Big(\frac{(2 k + 2 p - q)^{\nu}}{[m^2_{\pi} - (k +
    p)^2 - i0][m^2_{\pi} - (k + p - q)^2 -
    i0]}\nonumber\\ \hspace{-0.35in}&&- \frac{(2 k + 2 p - q)^{\nu}
}{[m^2_{\pi} - k^2 - i0][m^2_{\pi} - (k + p - q)^2 - i0]}\Big) \,
D^{(W)}_{\mu\nu}(- q)
\end{eqnarray}
and 
\begin{eqnarray}\label{eq:B.6}
 \hspace{-0.35in}&&\frac{\partial}{\partial \xi} M(n \to p e^-
 \bar{\nu}_e)^{(\pi^0)}_{\rm Fig.\,\ref{fig:fig1}f} = - 2 e^2 g^2_{\pi
   N} M^2_W G_V\nonumber\\ \hspace{-0.35in}&&\times \int \frac{d^4k}{(2\pi)^4i}\int
 \frac{d^4p}{(2\pi)^4i}\,\frac{1}{(p^2 +
   i0)^2}\,\Big[\bar{u}_e\,\gamma^{\mu}(1 - \gamma^5)
   v_{\bar{\nu}}\Big]\,\Big[\bar{u}_p \,\hat{p}\, \frac{1}{m_N -
     \hat{k}_p + \hat{p} - i0} \gamma^5 \frac{1}{m_N - \hat{k}_n -
     \hat{k} - i0}\,\gamma^5
   u_n\Big]\nonumber\\ \hspace{-0.35in}&&\times\,\Big(\frac{(2 k + 2 p -
   q)^{\nu}}{[m^2_{\pi} - (k + p)^2 - i0][m^2_{\pi} - (k + p - q)^2 -
     i0]} - \frac{(2 k + 2 p - q)^{\nu} }{[m^2_{\pi} - k^2 -
     i0][m^2_{\pi} - (k + p - q)^2 - i0]}\Big) \, D^{(W)}_{\mu\nu}(-
 q) = \nonumber\\ \hspace{-0.35in}&& = - 2 e^2 g^2_{\pi N} M^2_W
 G_V\nonumber\\ \hspace{-0.35in}&&\times \int
 \frac{d^4k}{(2\pi)^4i}\int \frac{d^4p}{(2\pi)^4i}\,\frac{1}{(p^2 +
   i0)^2}\,\Big[\bar{u}_e\,\gamma^{\mu}(1 - \gamma^5)
   v_{\bar{\nu}}\Big]\,\Big[\bar{u}_p \, \gamma^5 \frac{1}{m_N -
     \hat{k}_n - \hat{k} - i0}\,\gamma^5
   u_n\Big]\nonumber\\ \hspace{-0.35in}&&\times\,\Big(\frac{(2 k + 2 p
   - q)^{\nu}}{[m^2_{\pi} - (k + p)^2 - i0][m^2_{\pi} - (k + p - q)^2
     - i0]} - \frac{(2 k + 2 p - q)^{\nu} }{[m^2_{\pi} - k^2 -
     i0][m^2_{\pi} - (k + p - q)^2 - i0]}\Big) \, D^{(W)}_{\mu\nu}(-
 q)
\end{eqnarray}
and 
\begin{eqnarray}\label{eq:B.7}
 \hspace{-0.35in}&&\frac{\partial}{\partial \xi} M(n \to p e^-
 \bar{\nu}_e)^{(\pi^0)}_{\rm Fig.\,\ref{fig:fig1}g} = + 2 e^2 g^2_{\pi
   N} M^2_W G_V \nonumber\\ \hspace{-0.35in}&&\times \int
 \frac{d^4k}{(2\pi)^4i}\int \frac{d^4p}{(2\pi)^4i}\,\frac{1}{(p^2 +
   i0)^2}\,\Big[\bar{u}_e \,\hat{p} \,\frac{1}{m_e - \hat{p} -
     \hat{k}_e - i0}\,\gamma^{\mu}(1 - \gamma^5)
   v_{\bar{\nu}}\Big]\,\Big[\bar{u}_p\, \gamma^5 \frac{1}{m_N -
     \hat{k}_n - \hat{k} - i0}\,\gamma^5
   u_n\Big]\nonumber\\ \hspace{-0.35in}&&\times \,\Big(\frac{(2 k - p
   - q)^{\nu} }{[m^2_{\pi} - k^2 - i0][m^2_{\pi} - (k - q)^2 - i0]}-
 \frac{(2 k - p - q)^{\nu}}{[m^2_{\pi} - (k - p)^2 - i0][m^2_{\pi} - (k
   - q)^2 - i0]}\Big)\, D^{(W)}_{\mu\nu}(p - q)=
   \nonumber\\ \hspace{-0.35in}&& = - 2 e^2 g^2_{\pi N} M^2_W G_V
   \nonumber\\ \hspace{-0.35in}&&\times \int
   \frac{d^4k}{(2\pi)^4i}\int \frac{d^4p}{(2\pi)^4i}\,\frac{1}{(p^2 +
     i0)^2}\,\Big[\bar{u}_e \,\gamma^{\mu}(1 - \gamma^5)
     v_{\bar{\nu}}\Big]\,\Big[\bar{u}_p\, \gamma^5 \frac{1}{m_N -
       \hat{k}_n - \hat{k} - i0}\,\gamma^5
     u_n\Big]\nonumber\\ \hspace{-0.35in}&&\times \,\Big(\frac{(2 k -
     p - q)^{\nu} }{[m^2_{\pi} - k^2 - i0][m^2_{\pi} - (k - q)^2 -
       i0]}- \frac{(2 k - p - q)^{\nu}}{[m^2_{\pi} - (k - p)^2 -
     i0][m^2_{\pi} - (k - q)^2 - i0]}\Big)\, D^{(W)}_{\mu\nu}(p - q)
\end{eqnarray}
and 
\begin{eqnarray}\label{eq:B.8}
 \hspace{-0.35in}&&\frac{\partial}{\partial \xi} M(n \to p e^-
 \bar{\nu}_e)^{(\pi^0)}_{\rm Fig.\,\ref{fig:fig1}h} = + 2 e^2 g^2_{\pi
   N} M^2_W G_V \nonumber\\ \hspace{-0.35in}&&\times \int
 \frac{d^4k}{(2\pi)^4i}\int \frac{d^4p}{(2\pi)^4i}\,\frac{1}{(p^2 +
   i0)^2}\,\Big[\bar{u}_e\,\hat{p}\,\frac{1}{m_e - \hat{p} - \hat{k}_e
     - i0}\, \gamma^{\mu}(1 - \gamma^5)
   v_{\bar{\nu}}\Big]\,\Big[\bar{u}_p\, \gamma^5 \frac{1}{m_N -
     \hat{k}_n - \hat{k} - \hat{p} - i0}\nonumber\\ &&\times
   \hat{p}\,\frac{1}{m_N - \hat{k}_n - \hat{k} - i0}\,\gamma^5
   u_n\Big]\,\frac{(2 k + p - q)^{\nu}}{[m^2_{\pi} - k^2 -
     i0][m^2_{\pi} - (k + p -q)^2 - i0]}\, D^{(W)}_{\mu\nu}(p -
 q)=\nonumber\\ \hspace{-0.35in}&&= - 2 e^2 g^2_{\pi N} M^2_W G_V
 \nonumber\\ \hspace{-0.35in}&&\times \int \frac{d^4k}{(2\pi)^4i}\int
 \frac{d^4p}{(2\pi)^4i}\,\frac{1}{(p^2 + i0)^2}\,\Big[\bar{u}_e\,
   \gamma^{\mu}(1 - \gamma^5)
   v_{\bar{\nu}}\Big]\,\Big\{\Big[\bar{u}_p\, \gamma^5 \frac{1}{m_N -
     \hat{k}_n - \hat{k} - \hat{p} - i0}\,\gamma^5
   u_n\Big]\nonumber\\ \hspace{-0.35in}&& - \Big[\bar{u}_p\,
   \gamma^5\,\frac{1}{m_N - \hat{k}_n - \hat{k} - i0}\,\gamma^5
   u_n\Big]\Big\}\,\frac{(2 k + p - q)^{\nu}}{[m^2_{\pi} - k^2 -
     i0][m^2_{\pi} - (k + p -q)^2 - i0]}\, D^{(W)}_{\mu\nu}(p - q)
\end{eqnarray}
and 
\begin{eqnarray}\label{eq:B.9}
 \hspace{-0.35in}&&\frac{\partial}{\partial \xi} M(n \to p e^-
 \bar{\nu}_e)^{(\pi^0)}_{\rm Fig.\,\ref{fig:fig1}i} = + 2 e^2 g^2_{\pi
   N} M^2_W G_V \nonumber\\ \hspace{-0.35in}&&\times \int
 \frac{d^4k}{(2\pi)^4i}\int \frac{d^4p}{(2\pi)^4i}\,\frac{1}{(p^2 +
   i0)^2}\,\Big[\bar{u}_e\, \gamma^{\mu}(1 - \gamma^5)
   v_{\bar{\nu}}\Big]\,\Big[\bar{u}_p\, \gamma^5 \frac{1}{m_N -
     \hat{k}_n -\hat{k} - \hat{p} - i0} \,\hat{p}\, \frac{1}{m_N -
     \hat{k}_n - \hat{k} - i0} \nonumber\\\hspace{-0.35in}
   &&\times\,\gamma^5 u_n\Big]\,\frac{(2 k + p -
   q)^{\alpha_2}}{[m^2_{\pi} - k^2 - i0][m^2_{\pi} - (k + p - q)^2 -
     i0]}\,\big[D^{(W)-1}_{\nu\alpha_1}(- q) -
   D^{(W)-1}_{\nu\alpha_1}(p - q)\big]\,
 D^{(W)\alpha_1}{}_{\alpha_2}(p - q) D^{(W)}_{\mu}{}^{\nu}(- q)=
 \nonumber\\ \hspace{-0.35in}&& = + 2 e^2 g^2_{\pi N} M^2_W G_V
 \nonumber\\ \hspace{-0.35in}&&\times \int \frac{d^4k}{(2\pi)^4i}\int
 \frac{d^4p}{(2\pi)^4i}\,\frac{1}{(p^2 + i0)^2}\,\Big[\bar{u}_e\,
   \gamma^{\mu}(1 - \gamma^5)
   v_{\bar{\nu}}\Big]\,\Big\{\Big[\bar{u}_p\, \gamma^5 \frac{1}{m_N -
     \hat{k}_n -\hat{k} - \hat{p} - i0}\,\gamma^5 u_n\Big]
 \nonumber\\\hspace{-0.35in} && - \Big[\bar{u}_p\, \gamma^5 \,
   \frac{1}{m_N - \hat{k}_n - \hat{k} - i0} \,\gamma^5
   u_n\Big]\Big\}\,\frac{(2 k + p - q)^{\nu}}{[m^2_{\pi} - k^2 -
     i0][m^2_{\pi} - (k + p - q)^2 - i0]}\,\big[D^{(W)}_{\mu\nu}(p -
   q) - D^{(W)}_{\mu \nu}(- q)\big]
\end{eqnarray}
and
\begin{eqnarray}\label{eq:B.10}
 \hspace{-0.35in}&&\frac{\partial}{\partial \xi} M(n \to p e^-
 \bar{\nu}_e)^{(\pi^0)}_{\rm Fig.\,\ref{fig:fig1}j} = + 2 e^2 g^2_{\pi
   N} M^2_W G_V\nonumber\\ \hspace{-0.35in}&&\times \int
 \frac{d^4k}{(2\pi)^4i}\int \frac{d^4p}{(2\pi)^4i}\,\frac{1}{(p^2 +
   i0)^2}\,\Big[\bar{u}_e\, \gamma^{\mu}(1 - \gamma^5)
   v_{\bar{\nu}}\Big]\,\Big[\bar{u}_p \, \gamma^5 \frac{1}{m_N -
     \hat{k}_n - \hat{k} + \hat{p} - i0}\, \gamma^5
   u_n\Big]\nonumber\\ \hspace{-0.35in}&&\,\frac{(2 k - q)^{\nu}
 }{[m^2_{\pi} - k^2 - i0][m^2_{\pi} - (k - q)^2 - i0]} \,
 D^{(W)}_{\mu\nu}(- q)
\end{eqnarray}
and
\begin{eqnarray}\label{eq:B.11}
 \hspace{-0.35in}&&\frac{\partial}{\partial \xi} M(n \to p e^-
 \bar{\nu}_e)^{(\pi^0)}_{\rm Fig.\,\ref{fig:fig1}k} = +  e^2 g^2_{\pi
   N} M^2_W G_V \nonumber\\ \hspace{-0.35in}&&\times \int
 \frac{d^4k}{(2\pi)^4i}\int \frac{d^4p}{(2\pi)^4i}\,\frac{1}{(p^2 +
   i0)^2}\,\Big[\bar{u}_e\, \gamma^{\mu}(1 - \gamma^5)
   v_{\bar{\nu}}\Big]\,\Big[\bar{u}_p\,\gamma^5\, \frac{1}{m_N -
     \hat{k}_p -\hat{k} - i0}\, \hat{p}\, \frac{1}{m_N - \hat{k}_p -
     \hat{k} + \hat{p} - i0}
   \nonumber\\ \hspace{-0.35in}
  \hspace{-0.35in}&&\times\,\gamma^{\alpha_2}(1 -
  \gamma^5)\,\frac{1}{m_N - \hat{k}_n -\hat{k} - i0}\,\gamma^5
  u_n\Big]\,\frac{1}{m^2_{\pi} - k^2 - i0}\,\big[D^{(W)-1}_{\nu
     \alpha_1}(- q) - D^{(W)-1}_{\nu \alpha_1}(p - q)\big]
 D^{(W)\alpha_1}{}_{\alpha_2}(p - q) D^{(W)}_{\mu}{}^{\nu}(- q)
 =\nonumber\\
\hspace{-0.35in}&&= + e^2 g^2_{\pi N} M^2_W G_V
 \nonumber\\ \hspace{-0.35in}&&\times \int \frac{d^4k}{(2\pi)^4i}\int
 \frac{d^4p}{(2\pi)^4i}\,\frac{1}{(p^2 + i0)^2}\,\Big[\bar{u}_e\,
   \gamma^{\mu}(1 - \gamma^5)
   v_{\bar{\nu}}\Big]\,\Big\{\Big[\bar{u}_p\,\gamma^5\, \frac{1}{m_N -
     \hat{k}_p -\hat{k} - i0}\,\gamma^{\nu}(1 -
   \gamma^5)\,\frac{1}{m_N - \hat{k}_n -\hat{k} - i0}\,\gamma^5
   u_n\Big] \nonumber\\ \hspace{-0.35in}&& -
 \Big[\bar{u}_p\,\gamma^5\,\frac{1}{m_N - \hat{k}_p - \hat{k} +
     \hat{p} - i0}\,\gamma^{\nu}(1 - \gamma^5)\,\frac{1}{m_N -
     \hat{k}_n -\hat{k} - i0}\,\gamma^5
   u_n\Big]\Big\}\,\frac{1}{m^2_{\pi} - k^2 -
   i0}\,\big[D^{(W)}_{\mu\nu}(p - q) - D^{(W)}_{\mu\nu}(- q)\big]
\end{eqnarray}
and 
\begin{eqnarray}\label{eq:B.12}
 \hspace{-0.35in}&&\frac{\partial}{\partial \xi} M(n \to p e^-
 \bar{\nu}_e)^{(\pi^0)}_{\rm Fig.\,\ref{fig:fig1}l} = - e^2 g^2_{\pi
   N} M^2_W G_V \nonumber\\ \hspace{-0.35in}&&\times \int
 \frac{d^4k}{(2\pi)^4i}\int \frac{d^4p}{(2\pi)^4i}\,\frac{1}{(p^2 +
   i0)^2}\,\Big[\bar{u}_e\, \gamma^{\mu}(1 - \gamma^5)
   v_{\bar{\nu}}\Big]\,\Big[\bar{u}_p\,\hat{p}\, \frac{1}{m_N -
     \hat{k}_p + \hat{p} - i0} \, \gamma^5 \frac{1}{m_N - \hat{k}_p -
     \hat{k} + \hat{p} -
     i0}\nonumber\\ \hspace{-0.35in}&&\times\,\hat{p}\,\frac{1}{m_N -
     \hat{k}_p - \hat{k} - i0}\,\gamma^{\nu} (1 -
   \gamma^5)\,\frac{1}{m_N - \hat{k}_n - \hat{k} - i0}\,\gamma^5
   u_n\Big]\,\frac{1}{m^2_{\pi} - k^2 - i0}\, D^{(W)}_{\mu\nu}(-
 q)=\nonumber\\ \hspace{-0.35in}&& = - e^2 g^2_{\pi N} M^2_W G_V
 \nonumber\\ \hspace{-0.35in}&&\times \int \frac{d^4k}{(2\pi)^4i}\int
 \frac{d^4p}{(2\pi)^4i}\,\frac{1}{(p^2 + i0)^2}\,\Big[\bar{u}_e\,
   \gamma^{\mu}(1 - \gamma^5)
   v_{\bar{\nu}}\Big]\,\Big\{\Big[\bar{u}_p\,\gamma^5 \,\frac{1}{m_N -
     \hat{k}_p - \hat{k} - i0}\,\gamma^{\nu} (1 -
   \gamma^5)\,\frac{1}{m_N - \hat{k}_n - \hat{k} - i0}
   \nonumber\\ \hspace{-0.35in}&&\times\,\gamma^5 u_n(\vec{k}_n,
   \sigma_n)\Big] - \Big[\bar{u}_p\,\gamma^5 \,\frac{1}{m_N -
     \hat{k}_p - \hat{k} + \hat{p} - i0}\,\gamma^{\nu} (1 -
   \gamma^5)\,\frac{1}{m_N - \hat{k}_n - \hat{k} - i0}\,\gamma^5
   u_n\Big]\Big\}\nonumber\\ \hspace{-0.35in}&&\times\,
 \frac{1}{m^2_{\pi} - k^2 - i0}\, D^{(W)}_{\mu\nu}(- q)
\end{eqnarray}
and 
\begin{eqnarray}\label{eq:B.13}
 \hspace{-0.35in}&&\frac{\partial}{\partial \xi} M(n \to p e^-
 \bar{\nu}_e)^{(\pi^0)}_{\rm Fig.\,\ref{fig:fig1}m} = + e^2 g^2_{\pi
   N} M^2_W G_V \nonumber\\ \hspace{-0.35in}&&\times \int
 \frac{d^4k}{(2\pi)^4i}\int \frac{d^4p}{(2\pi)^4i}\,\frac{1}{(p^2 +
   i0)^2}\,\Big[\bar{u}_e \,\hat{p}\,\frac{1}{m_e - \hat{p} -
     \hat{k}_e - i0}\,\gamma^{\mu}(1 -
   \gamma^5)\,v_{\bar{\nu}}\Big]\,\Big[\bar{u}_p\, \gamma^5\,
   \frac{1}{m_N - \hat{k}_p - \hat{k} - i0}
   \nonumber\\ \hspace{-0.35in}&&\times \,\hat{p}\, \frac{1}{m_N -
     \hat{k}_p - \hat{k} + \hat{p} - i0}\,\gamma^{\nu} (1 -
   \gamma^5)\,\frac{1}{m_N - \hat{k}_n - \hat{k} - i0}\,\gamma^5
   u_n\Big] \, \frac{1}{m^2_{\pi} - k^2 - i0}\, D^{(W)}_{\mu\nu}(p -
 q)= \nonumber\\ \hspace{-0.35in}&& = - e^2 g^2_{\pi N} M^2_W G_V
 \nonumber\\ \hspace{-0.35in}&&\times \int \frac{d^4k}{(2\pi)^4i}\int
 \frac{d^4p}{(2\pi)^4i}\,\frac{1}{(p^2 + i0)^2}\,\Big[\bar{u}_e
   \,\gamma^{\mu}(1 -
   \gamma^5)\,v_{\bar{\nu}}\Big]\,\Big\{\Big[\bar{u}_p\, \gamma^5\,
   \frac{1}{m_N - \hat{k}_p - \hat{k} - i0}\,\gamma^{\nu} (1 -
   \gamma^5)\,\frac{1}{m_N - \hat{k}_n - \hat{k} - i0}\,\gamma^5
   u_n\Big] \nonumber\\ \hspace{-0.35in}&& - \Big[\bar{u}_p\,
   \gamma^5\, \frac{1}{m_N - \hat{k}_p - \hat{k} + \hat{p} -
     i0}\,\gamma^{\nu} (1 - \gamma^5)\,\frac{1}{m_N - \hat{k}_n -
     \hat{k} - i0}\,\gamma^5 u_n\Big]\Big\} \, \frac{1}{m^2_{\pi} -
   k^2 - i0}\, D^{(W)}_{\mu\nu}(p - q)
\end{eqnarray}
and
\begin{eqnarray}\label{eq:B.14}
 \hspace{-0.35in}&&\frac{\partial}{\partial \xi} M(n \to p e^-
 \bar{\nu}_e)^{(\pi^0)}_{\rm Fig.\,\ref{fig:fig1}n} = + 2 e^2 g^2_{\pi
   N} M^2_W G_V \nonumber\\ \hspace{-0.35in}&&\times \int
 \frac{d^4k}{(2\pi)^4i}\int \frac{d^4p}{(2\pi)^4i}\,\frac{1}{(p^2 +
   i0)^2}\,\Big[\bar{u}_e\, \gamma^{\mu}(1 - \gamma^5)
   v_{\bar{\nu}}\Big]\,\Big[\bar{u}_p \,\gamma^5\, \frac{1}{m_N -
     \hat{k}_n - \hat{k} - i0}\,\gamma^5\, u_n
   \Big]\nonumber\\ \hspace{-0.35in}&&\times \, \frac{(2 k + p -
   q)^{\nu}}{[m^2_{\pi} - k^2 - i0][m^2_{\pi} - (k + p - q)^2 -
     i0]}\,\big[D^{(W)}_{\mu\nu}(p - q) - D^{(W)}_{\mu\nu}(- q)\big]
\end{eqnarray}
and 
\begin{eqnarray}\label{eq:B.15}
 \hspace{-0.35in}&&\frac{\partial}{\partial \xi} M(n \to p e^-
 \bar{\nu}_e)^{(\pi^0)}_{\rm Fig.\,\ref{fig:fig1}o} = - 2 e^2 g^2_{\pi
   N} M^2_W G_V \nonumber\\ \hspace{-0.35in}&&\times \int
 \frac{d^4k}{(2\pi)^4i}\int \frac{d^4p}{(2\pi)^4i}\,\frac{1}{(p^2 +
   i0)^2}\,\Big[\bar{u}_e\, \gamma^{\mu}(1 - \gamma^5)
   v_{\bar{\nu}}\Big]\,\Big[\bar{u}_p \,\gamma^5\, \frac{1}{m_N -
     \hat{k}_n - \hat{k} - i0}\,\gamma^5\, u_n
   \Big]\nonumber\\ \hspace{-0.35in}&&\times \, \frac{(2 k + p -
   q)^{\nu}}{[m^2_{\pi} - k^2 - i0][m^2_{\pi} - (k + p - q)^2 -
     i0]}\,D^{(W)}_{\mu\nu}(p - q)
\end{eqnarray}
and 
\begin{eqnarray}\label{eq:B.16}
 \hspace{-0.35in}&&\frac{\partial}{\partial \xi} M(n \to p e^-
 \bar{\nu}_e)^{(\pi^0)}_{\rm Fig.\,\ref{fig:fig1}p} = + 2 e^2 g^2_{\pi
   N} M^2_W G_V \nonumber\\ \hspace{-0.35in}&&\times \int
 \frac{d^4k}{(2\pi)^4i}\int \frac{d^4p}{(2\pi)^4i}\,\frac{1}{(p^2 +
   i0)^2}\,\Big[\bar{u}_e\, \gamma^{\mu}(1 - \gamma^5) v_{\bar{\nu}}
   \Big]\,\Big[\bar{u}_p \,\gamma^5\, \frac{1}{m_N - \hat{k}_n -
     \hat{k} - i0}\,\gamma^5\, u_n
   \Big]\nonumber\\ \hspace{-0.35in}&&\times \, \frac{(2 k + p -
   q)^{\nu}}{[m^2_{\pi} - k^2 - i0][m^2_{\pi} - (k + p - q)^2 -
     i0]}\,\big[D^{(W)}_{\mu\nu}(p - q) - D^{(W)}_{\mu\nu}(- q)\big]
\end{eqnarray}
and 
\begin{eqnarray}\label{eq:B.17}
 \hspace{-0.35in}&&\frac{\partial}{\partial \xi} M(n \to p e^-
 \bar{\nu}_e)^{(\pi^0)}_{\rm Fig.\,\ref{fig:fig1}q} = - 2 e^2 g^2_{\pi
   N} M^2_W G_V \nonumber\\ \hspace{-0.35in}&&\times \int
 \frac{d^4k}{(2\pi)^4i}\int \frac{d^4p}{(2\pi)^4i}\,\frac{1}{(p^2 +
   i0)^2}\,\Big[\bar{u}_e\, \gamma^{\mu}(1 - \gamma^5) v_{\bar{\nu}}
   \Big]\,\Big[\bar{u}_p \,\gamma^5\, \frac{1}{m_N - \hat{k}_n -
     \hat{k} - i0}\,\gamma^5\, u_n
   \Big]\nonumber\\ \hspace{-0.35in}&&\times \, \frac{(2 k + p -
   q)^{\nu}}{[m^2_{\pi} - k^2 - i0][m^2_{\pi} - (k + p - q)^2 -
     i0]}\,D^{(W)}_{\mu\nu}(p - q)
\end{eqnarray}
and
\begin{eqnarray}\label{eq:B.18}
\hspace{-0.35in}&&\frac{\partial}{\partial \xi} M(n \to p e^-
\bar{\nu}_e)^{(\pi^0)}_{\rm Fig.\,\ref{fig:fig1}r} = + e^2 g^2_{\pi N}
M^2_W G_V\nonumber\\ &&\times \int \frac{d^4k}{(2\pi)^4i}\int
\frac{d^4p}{(2\pi)^4i}\,\frac{1}{(p^2 + i0)^2}\,\Big[\bar{u}_e\,
  \gamma^{\mu}(1 - \gamma^5) v_{\bar{\nu}} \Big]\,\Big[\bar{u}_p
  \,\gamma^5 \frac{1}{m_N - \hat{k}_p - \hat{k} + \hat{p} - i0}
  \nonumber\\ &&\times\,\gamma^{\nu}(1 - \gamma^5)\,\frac{1}{m_N -
    \hat{k}_n -\hat{k} - i0}\,\gamma^5 u_n \Big]\,\frac{1}{m^2_{\pi} -
  k^2 - i0}\,\big[D^{(W)}_{\mu\nu}(p - q) - D^{(W)}_{\mu\nu}(- q)\big]
\end{eqnarray}
and
\begin{eqnarray}\label{eq:B.19}
\hspace{-0.35in}&&\frac{\partial}{\partial \xi} M(n \to p e^-
\bar{\nu}_e)^{(\pi^0)}_{\rm Fig.\,\ref{fig:fig1}s} = - e^2 g^2_{\pi N}
M^2_W G_V\nonumber\\ &&\times \int \frac{d^4k}{(2\pi)^4i}\int
\frac{d^4p}{(2\pi)^4i}\,\frac{1}{(p^2 + i0)^2}\,\Big[\bar{u}_e \,
  \gamma^{\mu}(1 - \gamma^5) v_{\bar{\nu}} \Big]\,\Big[\bar{u}_p
  \,\gamma^5 \frac{1}{m_N - \hat{k}_p - \hat{k} + \hat{p} - i0}
  \nonumber\\ &&\times\,\gamma^{\nu}(1 - \gamma^5)\,\frac{1}{m_N -
    \hat{k}_n -\hat{k} - i0}\,\gamma^5 u_n \Big]\,\frac{1}{m^2_{\pi} -
  k^2 - i0}\,D^{(W)}_{\mu\nu}(p - q)
\end{eqnarray}
For the derivation of Eqs.(\ref{eq:B.1}) - (\ref{eq:B.19}) we have
used the orthogonality relation for the electroweak $W^-$-boson
propagators (see Eq.(\ref{eq:A.2})) and Dirac equations for free
fermions. We have also omitted some terms, which vanish within the
framework of dimensional regularization \cite{Hooft1972} -
\cite{Capper1973} (see also \cite{Ivanov2019a}).

\subsection*{\bf Gauge-dependent contributions of the Feynman diagrams
  in Fig.\,\ref{fig:fig2}}

The gauge-dependent contributions of the Feynman diagrams in
Fig.\,\ref{fig:fig2} are equal to
\begin{eqnarray}\label{eq:B.20}
\hspace{-0.35in}&& \frac{\partial}{\partial \xi} M(n \to p e^-
\bar{\nu}_e)^{(\pi^0)}_{\rm Fig.\,\ref{fig:fig2}a} = - 2 e^2 g^2_{\pi
  N} M^2_W G_V\nonumber\\ &&\times \int \frac{d^4k}{(2\pi)^4i}\int
\frac{d^4p}{(2\pi)^4i}\,\frac{1}{(p^2 + i0)^2}\,
\Big[\bar{u}_e\,\gamma^{\mu}(1 - \gamma^5)
  v_{\bar{\nu}}\Big]\,\,\Big[\bar{u}_p \,\gamma^5\, \frac{1}{m_N -
    \hat{k}_n - \hat{k} - i0}\,\gamma^5
  u_n\Big]\nonumber\\ \hspace{-0.35in}&&\times \frac{(2 k -
  q)^{\nu}}{[m^2_{\pi} - k^2 - i0][m^2_{\pi} - (k + p - q)^2 - i0]}\,
D^{(W)}_{\mu\nu}(- q)
\end{eqnarray}
and 
\begin{eqnarray}\label{eq:B.21}
\hspace{-0.35in}&& \frac{\partial}{\partial \xi} M(n \to p e^-
\bar{\nu}_e)^{(\pi^0)}_{\rm Fig.\,\ref{fig:fig2}b} = - 2 e^2 g^2_{\pi
  N} M^2_W G_V\nonumber\\ \hspace{-0.35in}&&\times \int
\frac{d^4k}{(2\pi)^4i}\int \frac{d^4p}{(2\pi)^4i}\,\frac{1}{(p^2 +
  i0)^2}\,\Big[\bar{u}_e\, \gamma^{\mu}(1 - \gamma^5)
  v_{\bar{\nu}}\Big]\,\Big[\bar{u}_p\, \gamma^5\,\frac{1}{m_N -
    \hat{k}_n - \hat{k} - i0}\,\gamma^5 u_n\Big]\nonumber\\ &&\times\,
\frac{(2 k - q)^{\nu}}{[m^2_{\pi} - (k - p)^2 - i0][m^2_{\pi} - (k -
    q)^2 - i0]} \, D^{(W)}_{\mu\nu}(- q)
\end{eqnarray}
and 
\begin{eqnarray}\label{eq:B.22}
\hspace{-0.35in}&& \frac{\partial}{\partial \xi} M(n \to p e^-
\bar{\nu}_e)^{(\pi^0)}_{\rm Fig.\,\ref{fig:fig2}c} = - 2 e^2 g^2_{\pi
  N} M^2_W G_V\nonumber\\ \hspace{-0.35in}&&\times \int
\frac{d^4k}{(2\pi)^4i}\int \frac{d^4p}{(2\pi)^4i}\,\frac{1}{(p^2 +
  i0)^2}\,\Big[\bar{u}_e\, \gamma^{\mu}(1 - \gamma^5)
  v_{\bar{\nu}}\Big]\,\Big[\bar{u}_p(\vec{k}_p,\sigma_p)\, \gamma^5\,
  \frac{1}{m_N - \hat{k}_n - \hat{k} + \hat{p} - i0}\,\gamma^5
  u_n\Big]\nonumber\\ \hspace{-0.35in}&&\times \,\frac{(2 k -
  q)^{\nu}}{[m^2_{\pi} - k^2 - i0][m^2_{\pi} - (k - q)^2 - i0]}\,
D^{(W)}_{\mu\nu}(- q)  
\end{eqnarray}
and 
\begin{eqnarray}\label{eq:B.23}
\hspace{-0.35in}&& \frac{\partial}{\partial \xi} M(n \to p e^-
\bar{\nu}_e)^{(\pi^0)}_{\rm Fig.\,\ref{fig:fig2}d} = - e^2 g^2_{\pi N}
M^2_W G_V\nonumber\\ \hspace{-0.35in}&&\times \int
\frac{d^4k}{(2\pi)^4i}\int \frac{d^4p}{(2\pi)^4i}\,\frac{1}{(p^2 +
  i0)^2}\,\Big[\bar{u}_e\, \gamma^{\mu}(1 - \gamma^5)
  v_{\bar{\nu}}\Big]\,\Big[\bar{u}_p\, \gamma^5\,\frac{1}{m_N -
    \hat{k}_p - \hat{k} + \hat{p} - i0}\,\gamma^{\nu} (1 -
  \gamma^5)\,\frac{1}{m_N - \hat{k}_n - \hat{k} - i0}\,\gamma^5
  u_n\Big]\nonumber\\ \hspace{-0.35in}&&\times \, \frac{1}{m^2_{\pi} -
  k^2 - i0} \,D^{(W)}_{\mu \nu}( - q).
\end{eqnarray}
For the derivation of the r.h.s. of Eqs.(\ref{eq:B.20}) -
(\ref{eq:B.23}) we have deleted some terms, which vanish within the
dimensional regularization \cite{Ivanov2019a}.

\subsection*{\bf Gauge-dependent contributions of the Feynman diagrams in
  Fig.\,\ref{fig:fig3}}

The Feynman diagrams in Fig.\,\ref{fig:fig3} possess the following
analytical expressions for gauge-dependent parts
\begin{eqnarray}\label{eq:B.24}
\hspace{-0.35in}&& \frac{\partial}{\partial \xi} M(n \to p e^-
\bar{\nu}_e)^{(\pi^0)}_{\rm Fig.\,\ref{fig:fig3}a} = - 2 e^2 g^2_{\pi
  N} M^2_W G_V\nonumber\\ \hspace{-0.35in}&&\times \int
\frac{d^4k}{(2\pi)^4i}\int \frac{d^4p}{(2\pi)^4i}\,\frac{1}{(p^2 +
  i0)^2}\,\Big[\bar{u}_e\, \gamma^{\mu}(1 - \gamma^5)
  v_{\bar{\nu}}\Big]\,\Big[\bar{u}_p\, \gamma^5\,\frac{1}{m_N -
    \hat{k}_n - \hat{k} - i0}\,\gamma^5 u_n\Big]\nonumber\\ &&\times
\,\frac{p^{\nu}}{[m^2_{\pi} - k^2 - i0][m^2_{\pi} - (k + p - q)^2 -
    i0]}\, D^{(W)}_{\mu\nu}(- q)
\end{eqnarray}
and 
\begin{eqnarray}\label{eq:B.25}
\hspace{-0.35in}&& \frac{\partial}{\partial \xi} M(n \to p e^-
\bar{\nu}_e)^{(\pi^0)}_{\rm Fig.\,\ref{fig:fig3}b} = + 2 e^2 g^2_{\pi
  N} M^2_W G_V\nonumber\\ \hspace{-0.35in}&&\times \int
\frac{d^4k}{(2\pi)^4i}\int \frac{d^4p}{(2\pi)^4i}\,\frac{1}{(p^2 +
  i0)^2}\,\Big[\bar{u}_e\, \gamma^{\mu}(1 - \gamma^5)
  v_{\bar{\nu}}\Big]\,\Big[\bar{u}_p\,\gamma^5\,\frac{1}{m_N -
    \hat{k}_n - \hat{k} - i0} \,\gamma^5
  u_n\Big]\nonumber\\ \hspace{-0.35in}&&\times
\,\frac{p^{\nu}}{[m^2_{\pi} - k^2 - i0][m^2_{\pi} - (k + p - q)^2 -
    i0]}\, D^{(W)}_{\mu\nu}(- q)
\end{eqnarray}
and 
\begin{eqnarray}\label{eq:B.26}
\hspace{-0.35in}&& \frac{\partial}{\partial \xi} M(n \to p e^-
\bar{\nu}_e)^{(\pi^0)}_{\rm Fig.\,\ref{fig:fig3}c} = - 2 e^2 g^2_{\pi
  N} M^2_W G_V\nonumber\\ \hspace{-0.35in}&&\times \int
\frac{d^4k}{(2\pi)^4i}\int \frac{d^4p}{(2\pi)^4i}\,\frac{1}{(p^2 +
  i0)^2}\,\Big[\bar{u}_e\, \gamma^{\mu}(1 - \gamma^5)
  v_{\bar{\nu}}\Big]\,\Big[\bar{u}_p\, \gamma^5\,\frac{1}{m_N -
    \hat{k}_n - \hat{k} - i0}\,\gamma^5 u_n\Big]\nonumber\\ &&\times
\,\frac{p^{\nu}}{[m^2_{\pi} - k^2 - i0][m^2_{\pi} - (k - q)^2 -
    i0]}\,D^{(W)}_{\mu\nu}(p - q)
\end{eqnarray}
and 
\begin{eqnarray}\label{eq:B.27}
\hspace{-0.35in}&& \frac{\partial}{\partial \xi} M(n \to p e^-
\bar{\nu}_e)^{(\pi^0)}_{\rm Fig.\,\ref{fig:fig3}d} = + 2 e^2 g^2_{\pi
  N} M^2_W G_V\nonumber\\ \hspace{-0.35in}&&\times \int
\frac{d^4k}{(2\pi)^4i}\int \frac{d^4p}{(2\pi)^4i}\,\frac{1}{(p^2 +
  i0)^2}\,\Big[\bar{u}_e\,\gamma^{\mu}(1 - \gamma^5)
  v_{\bar{\nu}}\Big]\,\,\Big[\bar{u}_p\, \gamma^5\,\frac{1}{m_N -
    \hat{k}_n - \hat{k} - i0} \,\gamma^5 u_n\Big] \nonumber\\ &&\times
\,\frac{p^{\nu}}{[m^2_{\pi} - k^2 - i0][m^2_{\pi} - (k - q)^2 -
    i0]}\,D^{(W)}_{\mu\nu}(p - q)
\end{eqnarray}
and 
\begin{eqnarray}\label{eq:B.28}
\hspace{-0.35in}&& \frac{\partial}{\partial \xi} M(n \to p e^-
\bar{\nu}_e)^{(\pi^0)}_{\rm Fig.\,\ref{fig:fig3}e} = + 2 e^2 g^2_{\pi
  N} M^2_W G_V\nonumber\\ \hspace{-0.35in}&&\times \int
\frac{d^4k}{(2\pi)^4i}\int \frac{d^4p}{(2\pi)^4i}\,\frac{1}{(p^2 +
  i0)^2}\,\Big[\bar{u}_e \,\gamma^{\mu}(1 - \gamma^5)
  v_{\bar{\nu}}\Big]\,\Big[\bar{u}_p\, \gamma^5\,\frac{1}{m_N -
    \hat{k}_n - \hat{k} - i0}\,\gamma^5\, u_n\Big]\nonumber\\ &&\times
\,\frac{p^{\nu}}{[m^2_{\pi} - (k - p)^2 - i0][m^2_{\pi} - (k - q)^2 -
    i0]} \,D^{(W)}_{\mu\nu}(- q)
\end{eqnarray}
and 
\begin{eqnarray}\label{eq:B.29}
\hspace{-0.35in}&& \frac{\partial}{\partial \xi} M(n \to p e^-
\bar{\nu}_e)^{(\pi^0)}_{\rm Fig.\,\ref{fig:fig3}f} = - 2 e^2 g^2_{\pi
  N} M^2_W G_V\nonumber\\ \hspace{-0.35in}&&\times \int
\frac{d^4k}{(2\pi)^4i}\int \frac{d^4p}{(2\pi)^4i}\,\frac{1}{(p^2 +
  i0)^2}\,\Big[\bar{u}_e \,\gamma^{\mu}(1 - \gamma^5)
  v_{\bar{\nu}}\Big]\,\Big\{\Big[\bar{u}_p\, \gamma^5\,\frac{1}{m_N -
    \hat{k}_n - \hat{k} - \hat{p} - i0}\,\gamma^5
  u_n\Big]\nonumber\\ \hspace{-0.35in}&& - \Big[\bar{u}_p\, \gamma^5\,
  \frac{1}{m_N - \hat{k}_n - \hat{k} - i0}\,\gamma^5
  u_n\Big]\Big\}\,\frac{p^{\nu}}{[m^2_{\pi} - k^2 - i0][m^2_{\pi} - (k
    + p - q)^2 - i0]} \,D^{(W)}_{\mu\nu}(- q)
\end{eqnarray}
and 
\begin{eqnarray}\label{eq:B.30}
\hspace{-0.35in}&& \frac{\partial}{\partial \xi} M(n \to p e^-
\bar{\nu}_e)^{(\pi^0)}_{\rm Fig.\,\ref{fig:fig3}g} = - 2 e^2 g^2_{\pi
  N} M^2_W G_V\nonumber\\ \hspace{-0.35in}&&\times \int
\frac{d^4k}{(2\pi)^4i}\int \frac{d^4p}{(2\pi)^4i}\,\frac{1}{(p^2 +
  i0)^2}\,\Big[\bar{u}_e \,\gamma^{\mu}(1 - \gamma^5)
  v_{\bar{\nu}}\Big]\,\Big[\bar{u}_p\, \gamma^5\,\frac{1}{m_N -
    \hat{k}_n - \hat{k} - i0}\,\gamma^5
  u_n\Big]\nonumber\\ \hspace{-0.35in}&&\times
\,\frac{p^{\nu}}{[m^2_{\pi} - k^2 - i0][m^2_{\pi} - (k + p - q)^2 -
    i0]} \,D^{(W)}_{\mu\nu}(- q)
\end{eqnarray}
and 
\begin{eqnarray}\label{eq:B.31}
\hspace{-0.35in}&& \frac{\partial}{\partial \xi} M(n \to p e^-
\bar{\nu}_e)^{(\pi^0)}_{\rm Fig.\,\ref{fig:fig3}h} = + 2 e^2 g^2_{\pi
  N} M^2_W G_V\nonumber\\ \hspace{-0.35in}&&\times \int
\frac{d^4k}{(2\pi)^4i}\int \frac{d^4p}{(2\pi)^4i}\,\frac{1}{(p^2 +
  i0)^2}\,\Big[\bar{u}_e\,\gamma^{\mu}(1 - \gamma^5)
  v_{\bar{\nu}}\Big]\,\,\Big[\bar{u}_p\, \gamma^5\,\frac{1}{m_N -
    \hat{k}_n - \hat{k} - i0} \,\gamma^5 u_n\Big] \nonumber\\ &&\times
\,\frac{p^{\nu}}{[m^2_{\pi} - k^2 - i0][m^2_{\pi} - (k - q)^2 -
    i0]}\,D^{(W)}_{\mu\nu}(p - q)
\end{eqnarray}
and 
\begin{eqnarray}\label{eq:B.32}
\hspace{-0.35in}&& \frac{\partial}{\partial \xi} M(n \to p e^-
\bar{\nu}_e)^{(\pi^0)}_{\rm Fig.\,\ref{fig:fig3}i} = - 2 e^2 g^2_{\pi
  N} M^2_W G_V\nonumber\\ \hspace{-0.35in}&&\times \int
\frac{d^4k}{(2\pi)^4i}\int \frac{d^4p}{(2\pi)^4i}\,\frac{1}{(p^2 +
  i0)^2}\,\Big[\bar{u}_e\,\gamma^{\mu}(1 - \gamma^5)
  v_{\bar{\nu}}\Big]\,\,\Big[\bar{u}_p\, \gamma^5\,\frac{1}{m_N -
    \hat{k}_n - \hat{k} - i0} \,\gamma^5 u_n\Big] \nonumber\\ &&\times
\,\frac{p^{\nu}}{[m^2_{\pi} - k^2 - i0][m^2_{\pi} - (k - q)^2 -
    i0]}\,D^{(W)}_{\mu\nu}(p - q).
\end{eqnarray}
For the derivation of the r.h.s. of Eqs.(\ref{eq:B.24}) -
(\ref{eq:B.32}) we have made necessary shifts of virtual momenta and
omitted some terms, which vanish within the dimensional regularization
\cite{Hooft1972} - \cite{Capper1973} (see also \cite{Ivanov2019a}).

We would like to notice that the contributions of the Feynman diagrams
in Fig.\,\ref{fig:fig3} such as Fig.\,\ref{fig:fig3}b and
Fig.\,\ref{fig:fig3}g, Fig.\,\ref{fig:fig3}d and
Fig.\,\ref{fig:fig3}i, and Fig.\,\ref{fig:fig3}c and
Fig.\,\ref{fig:fig3}h cancel each other pairwise (see Appendix A). As
a result, a non-trivial contribution comes from the Feynman diagrams
in Fig.\,\ref{fig:fig3}a, Fig.\,\ref{fig:fig3}e and
Fig.\,\ref{fig:fig3}f only. As can be seen, the sum of these Feynman
diagrams is gauge invariant. Such a gauge invariance of these Feynman
diagrams is a consequence of Ward-Takahashi-like identity
\cite{Itzykson1980}.

\subsection*{\bf Gauge-dependent contributions of the Feynman diagrams
  in Fig.\,\ref{fig:fig4}}

The analytical expressions for the gauge-dependent parts of the
Feynman diagrams in Fig.\,\ref{fig:fig4} are given by
\begin{eqnarray}\label{eq:B.33}
\hspace{-0.35in}&& \frac{\partial}{\partial \xi} M(n \to p e^-
\bar{\nu}_e)_{\rm Fig.\,\ref{fig:fig4}a} = - 2 e^2 g^2_{\pi
  N} M^2_W G_V\nonumber\\ \hspace{-0.35in}&&\times \int
\frac{d^4k}{(2\pi)^4i}\int \frac{d^4p}{(2\pi)^4i}\,\frac{1}{(p^2 +
  i0)^2}\,\Big[\bar{u}_e\,\gamma^{\mu}(1 - \gamma^5)
  v_{\bar{\nu}}\Big]\,\,\Big[\bar{u}_p\,\gamma^{\nu}(1 -
  \gamma^5)\,\frac{1}{m_N - \hat{k}_n - \hat{p} - i0}\,\gamma^5
  \frac{1}{m_N - \hat{k}_n - \hat{k} - i0} \,\gamma^5 u_n\Big]
\nonumber\\ &&\times \,\Big[\frac{1}{m^2_{\pi} - k^2 - i0}-
\frac{1}{m^2_{\pi} - (k - p)^2 - i0}\Big]\,D^{(W)}_{\mu\nu}(- q)
\end{eqnarray}
and 
\begin{eqnarray}\label{eq:B.34}
\hspace{-0.35in}&& \frac{\partial}{\partial \xi} M(n \to p e^-
\bar{\nu}_e)_{\rm Fig.\,\ref{fig:fig4}b} = - 2 e^2 g^2_{\pi
  N} M^2_W G_V\nonumber\\ \hspace{-0.35in}&&\times \int
\frac{d^4k}{(2\pi)^4i}\int \frac{d^4p}{(2\pi)^4i}\,\frac{1}{(p^2 +
  i0)^2}\,\Big[\bar{u}_e\,\gamma^{\mu}(1 - \gamma^5)
  v_{\bar{\nu}}\Big]\,\,\Big[\bar{u}_p\,\gamma^{\nu}(1 -
  \gamma^5)\,\frac{1}{m_N - \hat{k}_n - \hat{p} - i0}\,\gamma^5
  \frac{1}{m_N - \hat{k}_n - \hat{k} - i0} \,\gamma^5 u_n\Big]
\nonumber\\ &&\times \,\Big[\frac{1}{m^2_{\pi} - k^2 - i0}-
\frac{1}{m^2_{\pi} - (k - p)^2 - i0}\Big]\,\big[D^{(W)}_{\mu\nu}(p - q) -
  D^{(W)}_{\mu\nu}(- q)\big]
\end{eqnarray}
and
\begin{eqnarray}\label{eq:B.35}
\hspace{-0.35in}&& \frac{\partial}{\partial \xi} M(n \to p e^-
\bar{\nu}_e)_{\rm Fig.\,\ref{fig:fig4}c} = + 2 e^2 g^2_{\pi
  N} M^2_W G_V\nonumber\\ \hspace{-0.35in}&&\times \int
\frac{d^4k}{(2\pi)^4i}\int \frac{d^4p}{(2\pi)^4i}\,\frac{1}{(p^2 +
  i0)^2}\,\Big[\bar{u}_e\,\gamma^{\mu}(1 - \gamma^5)
  v_{\bar{\nu}}\Big]\,\,\Big[\bar{u}_p\,\gamma^{\nu}(1 -
  \gamma^5)\,\frac{1}{m_N - \hat{k}_n - \hat{p} - i0}\,\gamma^5
  \frac{1}{m_N - \hat{k}_n - \hat{k} - i0} \,\gamma^5 u_n\Big]
\nonumber\\ &&\times \,\Big[\frac{1}{m^2_{\pi} - k^2 - i0}-
\frac{1}{m^2_{\pi} - (k - p)^2 - i0}\Big]\,D^{(W)}_{\mu\nu}(p - q)
\end{eqnarray}
and
\begin{eqnarray}\label{eq:B.36}
\hspace{-0.35in}&& \frac{\partial}{\partial \xi} M(n \to p e^-
\bar{\nu}_e)^{(\pi^0)}_{\rm Fig.\,\ref{fig:fig4}d} = - 2 e^2 g^2_{\pi
  N} M^2_W G_V\nonumber\\ \hspace{-0.35in}&&\times \int
\frac{d^4k}{(2\pi)^4i}\int \frac{d^4p}{(2\pi)^4i}\,\frac{1}{(p^2 +
  i0)^2}\,\Big[\bar{u}_e\,\gamma^{\mu}(1 - \gamma^5)
  v_{\bar{\nu}}\Big]\,\,\Big[\bar{u}_p\,\gamma^{\nu}(1 -
  \gamma^5)\,\frac{1}{m_N - \hat{k}_n - \hat{p} -
    i0}\,\gamma^5\nonumber\\ \hspace{-0.35in}&&\times \Big(
  \frac{1}{m_N - \hat{k}_n - \hat{k} - \hat{p} - i0} - \frac{1}{m_N -
    \hat{k}_n - \hat{k} - i0}\Big) \,\gamma^5 u_n\Big]\,
\frac{1}{m^2_{\pi} - k^2 - i0}\,D^{(W)}_{\mu\nu}(- q)
\end{eqnarray}
and
\begin{eqnarray}\label{eq:B.37}
\hspace{-0.35in}&& \frac{\partial}{\partial \xi} M(n \to p e^-
\bar{\nu}_e)^{(\pi^0)}_{\rm Fig.\,\ref{fig:fig4}e} = - 2 e^2 g^2_{\pi
  N} M^2_W G_V\nonumber\\ \hspace{-0.35in}&&\times \int
\frac{d^4k}{(2\pi)^4i}\int \frac{d^4p}{(2\pi)^4i}\,\frac{1}{(p^2 +
  i0)^2}\,\Big[\bar{u}_e\,\gamma^{\mu}(1 - \gamma^5)
  v_{\bar{\nu}}\Big]\,\,\Big[\bar{u}_p\,\gamma^{\nu}(1 -
  \gamma^5)\,\frac{1}{m_N - \hat{k}_n - \hat{p} -
    i0}\,\gamma^5\nonumber\\ \hspace{-0.35in}&&\times \Big(
  \frac{1}{m_N - \hat{k}_n - \hat{k} - \hat{p} - i0} - \frac{1}{m_N -
    \hat{k}_n - \hat{k} - i0}\Big) \,\gamma^5 u_n\Big]\,
\frac{1}{m^2_{\pi} - k^2 - i0}\,\big[D^{(W)}_{\mu\nu}(p - q) -
  D^{(W)}_{\mu\nu}(- q)\big]
\end{eqnarray}
and
\begin{eqnarray}\label{eq:B.38}
\hspace{-0.35in}&& \frac{\partial}{\partial \xi} M(n \to p e^-
\bar{\nu}_e)^{(\pi^0)}_{\rm Fig.\,\ref{fig:fig4}f} = + 2 e^2 g^2_{\pi
  N} M^2_W G_V\nonumber\\ \hspace{-0.35in}&&\times \int
\frac{d^4k}{(2\pi)^4i}\int \frac{d^4p}{(2\pi)^4i}\,\frac{1}{(p^2 +
  i0)^2}\,\Big[\bar{u}_e\,\gamma^{\mu}(1 - \gamma^5)
  v_{\bar{\nu}}\Big]\,\,\Big[\bar{u}_p\,\gamma^{\nu}(1 -
  \gamma^5)\,\frac{1}{m_N - \hat{k}_n - \hat{p} -
    i0}\,\gamma^5\nonumber\\ \hspace{-0.35in}&&\times \Big(
  \frac{1}{m_N - \hat{k}_n - \hat{k} - \hat{p} - i0} - \frac{1}{m_N -
    \hat{k}_n - \hat{k} - i0}\Big) \,\gamma^5 u_n\Big]\,
\frac{1}{m^2_{\pi} - k^2 - i0}\,D^{(W)}_{\mu\nu}(p - q).
\end{eqnarray}
Now we may analyze the contributions of gauge-dependent terms of the
Feynman diagrams in Fig.\,\ref{fig:fig1} - Fig.\,\ref{fig:fig4},
describing the {\it inner} radiative corrections to the neutron beta
decay, induced by the hadronic structure of the neutron.

\subsection*{\bf Contributions of gauge-dependent terms of the Feynman
  diagrams in Fig.\,\ref{fig:fig1} - Fig.\,\ref{fig:fig4}}

Summing up the contributions of gauge dependent terms of the Feynman
diagrams in Fig.\,\ref{fig:fig1} - Fig.\,\ref{fig:fig4} we get
\begin{eqnarray}\label{eq:B.39}
\hspace{-0.15in}&& \frac{\partial}{\partial \xi} \sum^s_{J_1 = a} M(n
\to p e^- \bar{\nu}_e)_{\rm Fig.\,\ref{fig:fig1}J_1} = 2 e^2
g^2_{\pi N} M^2_W G_V \nonumber\\ \hspace{-0.15in}&&\times\int
\frac{d^4k}{(2\pi)^4i}\int \frac{d^4p}{(2\pi)^4i}\,\frac{1}{(p^2 +
  i0)^2}\,\Big[\bar{u}_e\,\gamma^{\mu}(1 - \gamma^5)
  v_{\bar{\nu}}\Big]\, \Big[\bar{u}_p \gamma^5 \frac{1}{m_N -
    \hat{k}_n - \hat{k} - i0} \gamma^5
  u_n\Big]\nonumber\\ \hspace{-0.15in}&&\times\, \Big\{\frac{(2k -
  q)^{\nu}}{[m^2_{\pi} - k^2 - i0][m^2_{\pi} - (k + p - q)^2 - i0]} +
\frac{(2k - q)^{\nu}}{[m^2_{\pi} - (k - p)^2 - i0][m^2_{\pi} - (k -
    q)^2 - i0]} \nonumber\\ \hspace{-0.15in}&& + \frac{(2 k + 2 p -
  q)^{\nu}}{[m^2_{\pi} - (k + p)^2 - i0][m^2_{\pi} - (k + p - q)^2 -
    i0]}\Big\}\,D^{(W)}_{\mu\nu}(- q) + e^2 g^2_{\pi N} M^2_W G_V\int
\frac{d^4k}{(2\pi)^4i}\int \frac{d^4p}{(2\pi)^4i}\,\frac{1}{(p^2 +
  i0)^2} \nonumber\\ \hspace{-0.15in}&&\times \,\Big[\bar{u}_e\,
  \gamma^{\mu}(1 - \gamma^5)
  v_{\bar{\nu}}\Big]\,\Big[\bar{u}_p\,\gamma^5 \frac{1}{m_N -
    \hat{k}_p - \hat{k} + \hat{p} - i0}\,\gamma^{\nu}(1 -
  \gamma^5)\,\frac{1}{m_N - \hat{k}_n -\hat{k} - i0}\,\gamma^5
  u_n\Big]\nonumber\\ &&\times \,\frac{1}{m^2_{\pi} - k^2 -
  i0}\,D^{(W)}_{\mu\nu}(- q)
\end{eqnarray}
and
\begin{eqnarray}\label{eq:B.40}
\hspace{-0.15in}&& \frac{\partial}{\partial \xi} \sum^d_{J_2 = a} M(n
\to p e^- \bar{\nu}_e)_{\rm Fig.\,\ref{fig:fig2}J_2} = 2
e^2 g^2_{\pi N} M^2_W G_V\nonumber\\ \hspace{-0.15in}&&\times \int
\frac{d^4k}{(2\pi)^4i}\int \frac{d^4p}{(2\pi)^4i}\,\frac{1}{(p^2 +
  i0)^2}\,\Big[\bar{u}_e\,\gamma^{\mu}(1 - \gamma^5)
  v_{\bar{\nu}}\Big]\, \Big[\bar{u}_p \gamma^5 \frac{1}{m_N -
    \hat{k}_n - \hat{k} - i0} \gamma^5
  u_n\Big]\nonumber\\ \hspace{-0.15in}&&\times \Big\{ - \frac{(2k -
  q)^{\nu}}{[m^2_{\pi} - k^2 - i0][m^2_{\pi} - (k + p - q)^2 - i0]} -
\frac{(2k + 2 p - q)^{\nu}}{[m^2_{\pi} - (k + p)^2 - i0][m^2_{\pi} -
    (k + p - q)^2 - i0]}\nonumber\\ \hspace{-0.15in}&& - \frac{(2k -
  q)^{\nu}}{[m^2_{\pi} - (k - p)^2 - i0][m^2_{\pi} - (k - q)^2 - i0]}
\Big\}\,D^{(W)}_{\mu\nu}(- q) - e^2 g^2_{\pi N} M^2_W G_V \int
\frac{d^4k}{(2\pi)^4i}\int \frac{d^4p}{(2\pi)^4i}\,\frac{1}{(p^2 +
  i0)^2}\nonumber\\ &&\times\,\Big[\bar{u}_e \, \gamma^{\mu}(1 -
  \gamma^5) v_{\bar{\nu}}\Big]\,\Big[\bar{u}_p \,\gamma^5 \frac{1}{m_N
    - \hat{k}_p - \hat{k} + \hat{p} - i0}\,\gamma^{\nu}(1 -
  \gamma^5)\,\frac{1}{m_N - \hat{k}_n -\hat{k} - i0}\,\gamma^5 u_n
  \Big]\nonumber\\ \hspace{-0.15in}&&\times \,\frac{1}{m^2_{\pi} - k^2
  - i0}\,D^{(W)}_{\mu\nu}(- q)
\end{eqnarray} 
and
\begin{eqnarray}\label{eq:B.41}
\hspace{-0.15in}&& \frac{\partial}{\partial \xi} \sum^i_{J_3 = a} M(n
\to p e^- \bar{\nu}_e)_{\rm Fig.\,\ref{fig:fig3}J_3} = 0.
\end{eqnarray}
and
\begin{eqnarray}\label{eq:B.42}
\hspace{-0.15in}&& \frac{\partial}{\partial \xi} \sum^f_{J_4 = a} M(n
\to p e^- \bar{\nu}_e)_{\rm Fig.\,\ref{fig:fig4}J_4} = 0.
\end{eqnarray}
One may see that the gauge dependent terms in the Feynman diagrams in
Fig.\,\ref{fig:fig1} are cancelled by the contributions of the gauge
dependent terms of the Feynman diagrams in Fig.\,\ref{fig:fig2}.
Summing up Eqs.(\ref{eq:B.39}) - (\ref{eq:B.42}) we get
\begin{eqnarray}\label{eq:B.43}
\hspace{-0.15in}&& \frac{\partial}{\partial \xi}\Big(\sum^s_{J_1 = a}
M(n \to p e^- \bar{\nu}_e)_{\rm Fig.\,\ref{fig:fig1}J_1 } +
\sum^d_{J_2 = a} M(n \to p e^- \bar{\nu}_e)_{\rm
  Fig.\,\ref{fig:fig2}J_2}\nonumber\\
\hspace{-0.15in}&&+ \sum^i_{J_3 = a} M(n \to p e^- \bar{\nu}_e)_{\rm
  Fig.\,\ref{fig:fig3}J_3} + \sum^f_{J_4 = a} M(n \to p e^-
\bar{\nu}_e)_{\rm Fig.\,\ref{fig:fig4}J_4}\Big) = 0.
\end{eqnarray}
This testifies gauge invariance of the Feynman diagrams in
Fig.\,\ref{fig:fig1} - Fig.\,\ref{fig:fig4} with $\pi$-meson
exchanges, defining contributions of the hadronic structure of the
neutron to the radiative corrections of order $O(\alpha/\pi)$ with an
arbitrary dependence on the nucleon mass $m_N$, calculated in the
effective quantum theory L$\sigma$M$\&$SET of strong and electroweak
low-energy interactions \cite{Ivanov2019a}.

Below following the standard procedure for the calculation of the
Feynman diagrams \cite{Feynman1950} - \cite{Smirnov2012} we proceed to
the analytical calculation of the Feynman diagrams in
Fig.\,\ref{fig:fig1} - Fig.\,\ref{fig:fig4}.

\newpage

\section*{Appendix C: Analytical calculation of the Feynman diagrams 
in Fig.\,\ref{fig:fig1} and Fig.\,\ref{fig:fig2} }
\renewcommand{\theequation}{C-\arabic{equation}}
\setcounter{equation}{0}

\subsection*{C1. Analytical calculation of the Feynman diagram in
  Fig.\,\ref{fig:fig1} to the amplitude of the neutron beta decay}
\renewcommand{\theequation}{C1-\arabic{equation}}
\setcounter{equation}{0}

For the calculation of the Feynman diagrams in Fig.\,\ref{fig:fig1} we
use the standard procedure of the calculation of the Feynman diagrams
\cite{Feynman1950} - \cite{Smirnov2012} and the dimensional
regularization \cite{Hooft1972}-\cite{Capper1973} (see also the
Supplemental Material of Ref. \cite{Ivanov2019a}), i.e. the
integration over virtual momenta in the $n$-dimensional momentum
space. Because of gauge invariance of the Feynman diagrams in
Fig.\,\ref{fig:fig1} - Fig.\,\ref{fig:fig4} we perform the calculation
in the Feynman gauge for the photon propagator and in the limit
$m_{\sigma} \to \infty$ of the infinite mass of the $\sigma$-meson
\cite{Weinberg1967a} (see also \cite{Ivanov2019a}) that allows to keep
the contributions of the Feynman diagrams with the $\pi$-meson
exchanges only. Then, we calculate the Feynman diagrams in
Fig.\,\ref{fig:fig1} in the leading logarithmic approximation (LLA)
\cite{Czarnecki2002, Bissegger2007}. In our work such an approximation
is justified by the finding \cite{Bissegger2007} that in the LLA the
linear $\sigma$-model without a nucleon is equivalent to ChPT by
Gasser-Leutwyler \cite{Gasser1984}. Then, the LLA allows i) to deal
with the contributions of the Feynman diagrams in
Fig.\,\ref{fig:fig1}, which preserve the chiral $SU(2) \times SU(2)$
symmetry of the L$\sigma$M \cite{Bijnens1996} and ii) to obtain the
contributions of the Feynman diagrams in Fig.\,\ref{fig:fig1} to the
amplitude of the neutron beta decay with the standard $V - A$
structure in agreement with Sirlin's analysis of the influence of
strong low-energy interactions on the inner radiative corrections of
order $O(\alpha/\pi)$, carried out within the current algebra approach
\cite{Sirlin1967, Sirlin1978}.

\subsection*{C1a. Analytical calculation of the Feynman diagram in
  Fig.\,\ref{fig:fig1}a}
\renewcommand{\theequation}{C1a-\arabic{equation}}
\setcounter{equation}{0}

The analytical expression of the Feynman diagram in
Fig.\,\ref{fig:fig1}a is given by (see Eq.(\ref{eq:A.1}))
\begin{eqnarray}\label{eq:C1a.1}
\hspace{-0.30in}&&M(n \to p e^- \bar{\nu}_e)^{(\pi^0)}_{\rm
  Fig.\,\ref{fig:fig1}a} = + 2 e^2 g^2_{\pi N} M^2_W G_V \int
\frac{d^4k}{(2\pi)^4i}\int \frac{d^4p}{(2\pi)^4i}\,\Big[\bar{u}_e
  \gamma^{\mu}(1 - \gamma^5) v_{\bar{\nu}}\Big]\,\Big[\bar{u}_p
  \gamma^5 \frac{1}{m_N - \hat{k}_n - \hat{k} - i0}\,\gamma^5
  u_n\Big] \nonumber\\
\hspace{-0.30in}&&\times \frac{(2 k + p -
  q)^{\alpha_2} (p + 2 k - 2 q)^{\beta_2}}{[m^2_{\pi} - k^2 -
    i0][m^2_{\pi} - (k - q)^2 - i0][m^2_{\pi} - (k + p - q)^2 -
    i0]}\big[(p - q)^{\nu}\eta^{\beta_1\alpha_1} - (p - 2q)^{\beta_1}
  \eta^{\alpha_1 \nu} - q^{\alpha_1} \eta^{\nu \beta_1}\big]
\nonumber\\
\hspace{-0.30in} &&\times \,D^{(\gamma)}_{\beta_1 \beta_2}(p)\,
D^{(W)}_{\alpha_1 \alpha_2}(p - q)\, D^{(W)}_{\mu\nu}(- q).
\end{eqnarray}
The propagator $D^{(W)}_{\alpha_1\alpha_2}(p - q)$ of the electroweak
$W^-$-boson we take in the physical gauge in its standard form
\cite{Weinberg1971} (see also \cite{Ivanov2019a})
\begin{eqnarray}\label{eq:C1a.2}
D^{(W)}_{\alpha_1\alpha_2}(p - q) = \frac{1}{M^2_W - (p - q)^2 -
  i0}\Big(- \eta_{\alpha_1\alpha_2} + \frac{(p - q)_{\alpha_1}(p -
  q)_{\alpha_2}}{M^2_W}\Big).
\end{eqnarray}
Plugging Eq.(\ref{eq:C1a.2}) into Eq.(\ref{eq:C1a.1}) and taking the
photon propagator $D^{(\gamma)}_{\beta_1\beta_2}(p)$ in the Feynman
gauge $D^{(\gamma)}_{\beta_1\beta_2}(p) = \eta_{\beta_1\beta_2}/(p^2 +
i0)$ we transcribe Eq.(\ref{eq:C1a.1}) into the form
\begin{eqnarray}\label{eq:C1a.3}
\hspace{-0.30in}&&M(n \to p e^- \bar{\nu}_e)^{(\pi^0)}_{\rm
  Fig.\,\ref{fig:fig1}a} = - 2 e^2 g^2_{\pi N} G_V \int
\frac{d^4k}{(2\pi)^4i}\int \frac{d^4p}{(2\pi)^4i}\,\Big[\bar{u}_e
  \gamma^{\mu}(1 - \gamma^5) v_{\bar{\nu}}\Big]\,\Big[\bar{u}_p
  \hat{k}u_n\Big] \frac{1}{m^2_N - (k_n + k)^2 - i0}\nonumber\\
  \hspace{-0.30in}&&\times \frac{(4 k^2 + 2 k\cdot p) p_{\mu} - (2 p^2
    + 4 k\cdot p) k_{\mu}}{[m^2_{\pi} - k^2 - i0][m^2_{\pi} - (k -
      q)^2 - i0][m^2_{\pi} - (k + p - q)^2 - i0][p^2 + i0][M^2_W - (p
      - q)^2 - i0]},
\end{eqnarray}
where we have kept only the leading contributions and used the Dirac
equation for a free neutron. Then, we have neglected the terms of
order $O(k_n\cdot q/M^2_W) \sim 10^{-7}$ and higher and made a
replacement $M^2_W D^{(W)}_{\mu \nu}(- q) = - \eta_{\mu\nu}$ that is
valid up to the contributions of order $m_em_N/M^2_W \sim 10^{-7}$
relative to unity. In such a LLA we may keep in
$D^{(W)}_{\alpha_1\alpha_2}(p - q)$ only the term proportional to
$\eta_{\alpha_1\alpha_2}$. For the estimate we have used the electron
mass $m_e = 0.5110\,{\rm MeV}$, the nucleon mass $m_N = (m_n + m_p)/2
= 938.9188,{\rm MeV}$ calculated at $m_n = 939.5654\,{\rm MeV}$ and
$m_p = 938.2721\, {\rm MeV}$, where $m_n$ and $m_p$ are the neutron
and proton masses, respectively, and the electroweak $W^-$-boson mass
$M_W = 80.379\,{\rm GeV}$ \cite{PDG2020}.

Following the standard procedure of the calculation of the Feynman
integrals \cite{Feynman1950} - \cite{Smirnov2012}, we merge the
denominators of the integrand by using the Feynman parametrization
\cite{Feynman1950}
\begin{eqnarray*}
\hspace{-0.30in}&&M(n \to p e^- \bar{\nu}_e)^{(\pi^0)}_{\rm
  Fig.\,\ref{fig:fig1}a} = 2 e^2 g^2_{\pi N} G_V \int
\frac{d^4k}{(2\pi)^4i}\int \frac{d^4p}{(2\pi)^4i} 5! \int^1_0
dx_1\int^1_0 dx_2 \int^1_0 dx_3 \int^1_0 dx_4 \int^1_0 dx_5 \int^1_0
dx_6 \nonumber\\
 \end{eqnarray*}
\begin{eqnarray}\label{eq:C1a.4}
  \hspace{-0.30in}&&\times \,\delta(1 - x_1 - x_2 - x_3 - x_4 - x_5 -
  x_6) \,\Big[\bar{u}_e \gamma^{\mu}(1 - \gamma^5) v_{\bar{\nu}}\Big]
  \Big[\bar{u}_p \hat{k}u_n\Big] \big(4 k^2 + 2
  k\cdot p) p_{\mu} - (2 p^2 + 4 k\cdot p) k_{\mu}\big)\nonumber\\
  \hspace{-0.30in}&&\times \, \Big(x_1 (m^2_{\pi} - k^2) + x_2
  (m^2_{\pi} - (k - q)^2) + x_3 (m^2_{\pi} - (k + p - q)^2) + x_4
  (m^2_N - (k_n + k)^2) + x_5 (- p^2)  \nonumber\\
\hspace{-0.30in}&& + x_6 (M^2_W - (p - q)^2) - i0\Big)^{-6} ,
\end{eqnarray}
where $x_j$ for $j = 1,2,\ldots,6$ are the Feynman parameters
\cite{Feynman1950}.  For the regularization of the Feynman integral we
use the dimensional regularization
\cite{Hooft1972}-\cite{Capper1973}. In other words we integrate over
virtual momenta in the $n$-dimensional momentum space. This gives for
the right-hand-side (r.h.s) of Eq.(\ref{eq:C1a.4}) the following
expression
\begin{eqnarray}\label{eq:C1a.5}
\hspace{-0.30in}&&M(n \to p e^- \bar{\nu}_e)^{(\pi^0)}_{\rm
  Fig.\,\ref{fig:fig1}a} = 2 e^2 g^2_{\pi N} G_V \int
\frac{d^nk}{(2\pi)^ni}\int \frac{d^np}{(2\pi)^ni}\,\mu^{2(4-n)}\, 5!
\int^1_0 dx_1\int^1_0 dx_2 \int^1_0 dx_3 \int^1_0 dx_4 \int^1_0 dx_5
\int^1_0 dx_6 \nonumber\\
  \hspace{-0.30in}&&\times \,\delta(1 - x_1 - x_2 - x_3 - x_4 - x_5 -
  x_6) \,\Big[\bar{u}_e \gamma^{\mu}(1 - \gamma^5) v_{\bar{\nu}}\Big]
  \Big[\bar{u}_p \hat{k}u_n\Big] \big(4 k^2 + 2
k\cdot p) p_{\mu} - (2 p^2 + 4 k\cdot p) k_{\mu}\big)\nonumber\\
  \hspace{-0.30in}&&\times \, \Big(x_1 (m^2_{\pi} - k^2) + x_2
  (m^2_{\pi} - (k - q)^2) + x_3 (m^2_{\pi} - (k + p - q)^2) + x_4
  (m^2_N - (k_n + k)^2) + x_5 (- p^2)  \nonumber\\
\hspace{-0.30in}&& + x_6 (M^2_W - (p - q)^2) - i0\Big)^{-6},
\end{eqnarray}
where $\mu$ is an energy scale
\cite{Hooft1972}-\cite{Capper1973}. Below we set $\mu = m_N$, since
the nucleon mass $m_N$ is a typical scale of our analysis. For
diagonalization of the denominator we make a change of virtual momenta
$k \to k + b_1$ and $p \to p + b_2$. We determine vectors $b_1$ and
$b_2$ in terms of external momenta of interacting particles by
removing non-diagonal terms. After diagonalization of the denominator
and integration over the virtual momenta, we arrive at the expression
\begin{eqnarray}\label{eq:C1a.6}
\hspace{-0.30in}&&M(n \to p e^- \bar{\nu}_e)^{(\pi^0)}_{\rm
  Fig.\,\ref{fig:fig1}a} = 2 e^2 g^2_{\pi N} G_V \int^1_0\!
dx_1\int^1_0 \! dx_2 \int^1_0 \!dx_3 \int^1_0\! dx_4 \int^1_0\! dx_5
\int^1_0\!  dx_6 \frac{\delta(1 - x_1 - x_2 - x_3 - x_4 - x_5 -
  x_6)}{2^{2n} \pi^n U^{n/2}}\nonumber\\
  \hspace{-0.30in}&&\times \bigg\{ - \frac{n -
    1}{2}\,\frac{\displaystyle \Gamma\Big(2 -
    \frac{n}{2}\Big)\Gamma\Big(2 - \frac{n - 1}{2}\Big)}{2^{n - 3}
    \displaystyle
    \Gamma\Big(\frac{1}{2}\Big)}\Big(\frac{Q}{m^2_N}\Big)^{-4 + n}
  \Big[\bar{u}_e \gamma^{\mu}(1 - \gamma^5) v_{\bar{\nu}}\Big]
  \Big[\bar{u}_p \gamma_{\mu} u_n\Big] - \frac{n}{2} \,\Gamma(5 - n)
  Q^{-5 + n} m^{2(4-n)}_N \nonumber\\
  \hspace{-0.30in}&& \times \Big\{ - \frac{2}{n}\, (b^2_2 + 2 b_1
  \cdot b_2) \Big[\bar{u}_e \gamma^{\mu} (1 - \gamma^5)
    v_{\bar{\nu}}\Big] \Big[\bar{u}_p \gamma_{\mu} u_n\Big] + 4\,
  \frac{n + 1}{n} \Big[\bar{u}_e \hat{b}_2 (1 - \gamma^5)
    v_{\bar{\nu}}\Big] \Big[\bar{u}_p \hat{b}_1 u_n\Big] - 2\, \frac{n
    - 1}{n}\Big[\bar{u}_e \hat{b}_1 (1 - \gamma^5)
    v_{\bar{\nu}}\Big]\nonumber\\
  \hspace{-0.30in}&& \times \Big[\bar{u}_p \hat{b}_1 u_n\Big] +
  \frac{2}{n}\, \Big[\bar{u}_e \hat{b}_2 (1 - \gamma^5)
    v_{\bar{\nu}}\Big] \Big[\bar{u}_p \hat{b}_2 u_n\Big] -
  \frac{4}{n}\, \Big[\bar{u}_e \hat{b}_1 (1 - \gamma^5)
    v_{\bar{\nu}}\Big] \Big[\bar{u}_p \hat{b}_2 u_n\Big]\Big\}
  \bigg\},
\end{eqnarray}
where we have denoted
\begin{eqnarray}\label{eq:C1a.7}
  \hspace{-0.30in}U &=& (x_1 + x_2 + x_4)(x_3 + x_5 + x_6) + (x_5 +
  x_6) x_3,\nonumber\\
  \hspace{-0.30in} b_1 &=& \frac{(x_x + x_5 + x_6) a_1 - x_3 a_2}{U}=
  - \frac{(x_3 + x_5 + x_6)x_4}{U}\,k_n = \frac{Y}{U}\,k_n, \nonumber\\
  \hspace{-0.30in} b_2 &=&
  \frac{(x_1 + x_2 + x_3 + x_4) a_2 - x_3 a_1}{U} = \frac{x_3 x_4}{U}
  \, k_n = \frac{\bar{Y}}{U}\, k_n,\nonumber\\
  \hspace{-0.30in}Q &=& x_6 M^2_W + \frac{(x_3 + x_5 + x_6) a^2_1 - 2
    x_3 a_1 \cdot a_2 + (x_1 + x_2 + x_3 + x_4) a^2_2}{U}
\end{eqnarray}
with $a_1 = - x_4 k_n$ and $a_2 = 0$, where we have neglected the
contributions proportional to $q$ and $q^2$. For the derivation of the
Lorentz structure of the matrix element Eq.(\ref{eq:C1a.6}) we have
used the algebra of the Dirac $\gamma$-matrices in the $n$-dimensional
space-time $\gamma_{\mu} \gamma_{\nu} + \gamma_{\nu} \gamma_{\mu} =
2\eta_{\mu\nu}$, $\gamma^{\lambda} \gamma_{\lambda} = n$,
$\gamma^{\lambda}\hat{a}\gamma_{\lambda} = - (n - 2) \hat{a}$,
$\gamma^{\lambda}\hat{a}\hat{b} \gamma_{\lambda} = 4\, a\cdot b + (n -
4)\, \hat{a} \hat{b}$ and $\gamma^{\lambda} \hat{a} \hat{b} \hat{c}
\gamma_{\lambda} = - 2 \hat{c} \hat{b} \hat{a} - (n - 4) \hat{a}
\hat{b} \hat{c}$, where $\hat{a} = \gamma^{\alpha} a_{\alpha}$ and so
on \cite{Itzykson1980}.  Taking the limit $n \to 4$ in
Eq.(\ref{eq:C1a.6}), keeping the divergent contributions in the form
proportional to $\Gamma(2 - n/2)$ and using the Dirac equation for a
free neutron, we arrive at the expression
\begin{eqnarray}\label{eq:C1a.8}
\hspace{-0.21in}&&M(n \to p e^- \bar{\nu}_e)^{(\pi^0)}_{\rm
  Fig.\,\ref{fig:fig1}a} = 2 e^2 g^2_{\pi N} G_V \int^1_0\!
dx_1\int^1_0 \! dx_2 \int^1_0 \!dx_3 \int^1_0\! dx_4 \int^1_0\! dx_5
\int^1_0\!  dx_6 \frac{\delta(1 - x_1 - x_2 - x_3 - x_4 - x_5 -
  x_6)}{256 \pi^4 U^2}\nonumber\\
  \hspace{-0.21in}&&\times \Big\{\Big( - \frac{3}{4}\, \Gamma\Big(2 -
  \frac{n}{2}\Big) + f^{(V)}_{1a}(x_1, \ldots, x_6) +
  g^{(W)}_{1a}(x_1,\ldots, x_6)\,\frac{m^2_N}{M^2_W}\Big)
  \Big[\bar{u}_e \gamma^{\mu}(1 - \gamma^5) v_{\bar{\nu}}\Big]
  \Big[\bar{u}_p \gamma_{\mu} u_n\Big] + f^{(W)}_{1a}(x_1,\ldots,
  x_6)\,\frac{m^2_N}{M^2_W} \nonumber\\
  \hspace{-0.30in}&& \times \Big[\bar{u}_e \frac{\hat{k}_n}{m_N}(1 -
    \gamma^5) v_{\bar{\nu}}\Big] \Big[\bar{u}_p u_n\Big]\Big\}.
\end{eqnarray}
where we have denoted
\begin{eqnarray}\label{eq:C1a.9}
  \hspace{-0.30in}f^{(V)}_{1a}(x_1, \ldots, x_6) &=& \frac{3}{2}
  \,{\ell n}\frac{M^2_W}{m^2_N} + \frac{3}{2}\,{\ell
    n}\bar{Q},\nonumber\\
  \hspace{-0.30in}g^{(W)}_{1a}(x_1, \ldots, x_6) &=&
  \,\frac{1}{\bar{Q}}\Big[\frac{\bar{Y}^2}{U^2} +
    2\,\frac{Y\bar{Y}}{U^2}\Big],\nonumber\\
  \hspace{-0.30in}f^{(W)}_{1a}(x_1, \ldots, x_6) &=&
  \frac{1}{\bar{Q}}\,\Big[3\, \frac{Y^2}{U^2} - 8\,
    \frac{Y\bar{Y}}{U^2} - \frac{\bar{Y}^2}{U^2}\Big].
\end{eqnarray}
where we have denoted
\begin{eqnarray}\label{eq:C1a.10}
  \hspace{-0.30in}\frac{Q}{m^2_N} &=& \frac{M^2_W}{m^2_N}\,
  \bar{Q} \quad, \quad \bar{Q} = \frac{1}{U}\Big(x_6 U + (x_3 + x_5 +
  x_6) x^2_4 \frac{m^2_N}{M^2_W}\Big),\nonumber\\
  \hspace{-0.30in} Y&=& - (x_3 + x_5 + x_6)x_4 \;,\; \bar{Y} =
  x_3 x_4.
\end{eqnarray}
After the integration over the Feynman parameters, the contribution of
the Feynman diagram in Fig.\,\ref{fig:fig1}a to the amplitude of the
neutron beta decay is equal to
\begin{eqnarray}\label{eq:C1a.11}
\hspace{-0.30in}M(n \to p e^- \bar{\nu}_e)^{(\pi^0)}_{\rm
  Fig.\,\ref{fig:fig1}a} &=& \frac{\alpha}{2\pi}\frac{g^2_{\pi
    N}}{16\pi^2} G_V \Big\{\Big(A_{1a}\, \Gamma\Big(2 -
\frac{n}{2}\Big) + F^{(V)}_{1a} + G^{(W)}_{1a}\frac{m^2_N}{M^2_W}\Big)
\Big[\bar{u}_e \gamma^{\mu}(1 - \gamma^5) v_{\bar{\nu}}\Big]
\Big[\bar{u}_p \gamma_{\mu} u_n\Big] \nonumber\\
  \hspace{-0.30in}&+& F^{(W)}_{1a}\frac{m^2_N}{M^2_W}\Big[\bar{u}_e
    \frac{\hat{k}_n}{m_N}(1 - \gamma^5) v_{\bar{\nu}}\Big]
  \Big[\bar{u}_p u_n\Big]\Big\},
\end{eqnarray}
where we have used $e^2 = 4\pi \alpha$ \cite{PDG2020}. The structure
constants are equal to $A_{1a} = - 1/8$, $F^{(V)}_{1a} = 1.5303$,
$G^{(W)}_{1a} = - 1.9009$ and $F^{(W)}_{1a} = 19.4989$, respectively. The
Lorentz structure of Eq.(\ref{eq:C1a.11}) is calculated at the neglect
of the contributions of order $O(k_n\cdot q/M^2_W) \sim 10^{-7}$ and
$O(m^2_{\pi}/M^2_W) \sim 10^{-6}$, respectively. We would like to
emphasize that all contributions to the matrix element
Eq.(\ref{eq:C1a.11}) are $G$-even \cite{Lee1956a} and induced by the
contributions of the first class currents \cite{Weinberg1958} (see
also \cite{Ivanov2018}).

\subsection*{C1b. Analytical calculation of the Feynman diagram in
  Fig.\,\ref{fig:fig1}b}
\renewcommand{\theequation}{C1b-\arabic{equation}}
\setcounter{equation}{0}

In the Feynman gauge for the photon propagator the analytical
expression of the Feynman diagram in Fig.\,\ref{fig:fig1}b is given by
(see Eq.(\ref{eq:A.1}))
\begin{eqnarray}\label{eq:C1b.1}
\hspace{-0.3in}&&M(n \to p e^- \bar{\nu}_e)^{(\pi^0)}_{\rm
  Fig.\,\ref{fig:fig1}b} = - 2 e^2 g^2_{\pi N} M^2_W G_V \int
\frac{d^4k}{(2\pi)^4i}\int \frac{d^4p}{(2\pi)^4i}\Big[\bar{u}_e
  \gamma^{\mu}(1 - \gamma^5) v_{\bar{\nu}}\Big] \Big[\bar{u}_p
  \gamma^{\beta} \frac{1}{m_N - \hat{k}_p + \hat{p} - i0} \gamma^5
  \nonumber\\\hspace{-0.3in} &&\times \frac{1}{m_N - \hat{k}_n -
    \hat{k} - i0}\gamma^5 u_n\Big]\frac{(2 k - q)^{\nu} (p + 2 k - 2
  q)_{\beta}}{[m^2_{\pi} - k^2 - i0][m^2_{\pi} - (k - q)^2 -
    i0][m^2_{\pi} - (k + p - q)^2 - i0]}\,\frac{1}{p^2 +
  i0}\, D^{(W)}_{\mu\nu}(- q).~~~
\end{eqnarray}
At the first step of the calculation we rewrite the r.h.s. of
Eq.(\ref{eq:C1b.1}) as follows
\begin{eqnarray}\label{eq:C1b.2}
\hspace{-0.3in}&&M(n \to p e^- \bar{\nu}_e)^{(\pi^0)}_{\rm
  Fig.\,\ref{fig:fig1}b} =\nonumber\\
\hspace{-0.3in}&&= + 2 e^2 g^2_{\pi N} G_V \int
\frac{d^4k}{(2\pi)^4i}\int \frac{d^4p}{(2\pi)^4i}\Big[\bar{u}_e
  \gamma^{\mu}(1 - \gamma^5) v_{\bar{\nu}}\Big] \Big[\bar{u}_p \frac{2
    k^{\beta}_p - \gamma^{\beta}\hat{p}}{m^2_N - (k_p - p)^2 - i0}
  \frac{\hat{k}}{m^2_N - (k_n + k)^2 - i0} u_n\Big]\nonumber\\
\hspace{-0.3in}&&\times \frac{(2 k - q)^{\nu} (p + 2 k - 2
  q)_{\beta}}{[m^2_{\pi} - k^2 - i0][m^2_{\pi} - (k - q)^2 -
    i0][m^2_{\pi} - (k + p - q)^2 - i0]}\,\frac{1}{p^2 + i0}\, M^2_W
D^{(W)}_{\mu\nu}(- q).
\end{eqnarray}
where we have used the Dirac equations for a free neutron and a free
proton. Then, making the replacement $M^2_W D^{(W)}_{\mu\nu}(- q) \to
- \eta_{\mu\nu}$, which is valid at the neglect of the terms of order
$O(m_N m_e/M^2_W) \sim 10^{-7}$, we proceed to the calculation of the
integral
\begin{eqnarray}\label{eq:C1b.3}
\hspace{-0.21in}{\cal F}(k_n, k_e, q)_{1b} &=& - \int
\frac{d^4p}{(2\pi)^4i} \int \frac{d^4k}{(2\pi)^4i}\,\frac{1}{m^2_N -
  (k_p - p)^2 - i0} \frac{1}{m^2_N - (k_n + k)^2 - i0}\nonumber\\
\hspace{-0.21in}&&\times \,\frac{2 \hat{k} (1 - \gamma^5) \otimes (-
  p^2 \hat{k} + 2 k^2 \hat{p} - 4(k\cdot p) \hat{k}) }{[m^2_{\pi} -
    k^2 - i0][m^2_{\pi} - (k - q)^2 - i0][m^2_{\pi} - (k + p - q)^2 -
    i0]}\,\frac{1}{p^2 + i0},
\end{eqnarray}
where we have kept only the divergent contributions that corresponds
to the LLA.  Following the standard procedure of the calculation of
the Feynman integrals \cite{Feynman1950} - \cite{Smirnov2012}, we
transcribe the r.h.s. of Eq.(\ref{eq:C1b.3}) into the form
\begin{eqnarray}\label{eq:C1b.4}
\hspace{-0.21in}&&{\cal F}(k_n, k_e, q)_{1b} = 5!\int
\frac{d^4p}{(2\pi)^4i}\int \frac{d^4k}{(2\pi)^4i} \int^1_0\!\!\!dx_1
\int^1_0\!\!\! dx_2 \int^1_0\!\!\! dx_3 \int^1_0\!\!\! dx_4
\int^1_0\!\!\! dx_5 \int^1_0\!\!\!  dx_6\delta(1 - x_1 - x_2 - x_3 -
x_4 - x_5 - x_6)\nonumber\\
\hspace{-0.21in}&&\times \Big(2 \hat{k}(1 - \gamma^5) \otimes \big(-
p^2 \hat{k} + 2 k^2 \hat{p} - 4 (k\cdot p) \hat{k}\big)\Big)\,
\Big(x_1(m^2_{\pi} - k^2) + x_2(m^2_{\pi} - (k - q)^2) + x_3
(m^2_{\pi} - (k + p - q)^2)\nonumber\\
\hspace{-0.21in}&& + x_4(m^2_N - (k_n + k)^2) + x_5(m^2_N - (k_p -
p)^2) + x_6(- p^2) - i0\Big)^{-6}\,,
\end{eqnarray}
where $x_j$ for $j = 1,2,\ldots,6$ are the Feynman parameters
\cite{Feynman1950}.  Then, we proceed to the $n$-dimensional momentum
space \cite{Smirnov2004, Smirnov2006, Smirnov2012} and rewrite
Eq.(\ref{eq:C1b.4}) in the more convenient form
\begin{eqnarray}\label{eq:C1b.5}
\hspace{-0.3in}&&{\cal F}(k_n, k_e, q)_{1b} = 5!  \int^1_0dx_1
\int^1_0dx_2 \int^1_0dx_3 \int^1_0dx_4 \int^1_0dx_5 \int^1_0
dx_6\delta(1 - x_1 - x_2 - x_3 - x_4 - x_5 - x_6) \nonumber\\
\hspace{-0.3in}&& \times \int \frac{d^np}{(2\pi)^ni}\int
\frac{d^nk}{(2\pi)^ni}\,m^{2(4 - n)}_N \Big(2 \hat{k} (1 - \gamma^5)
\otimes \big(- p^2 \hat{k} + 2 k^2 \hat{p} - 4 (k\cdot p)
\hat{k}\big)\Big)\, \Big((x_1 + x_2 + x_3)m^2_{\pi} - (x_2 + x_3)q^2
\nonumber\\
\hspace{-0.3in}&& - (x_1 + x_2 + x_3 + x_4) k^2 - 2 x_3 k\cdot p -
(x_3 + x_5 + x_6) p^2 + 2 k\cdot a_1 + 2 p\cdot a_2 - i0\Big)^{-6},
\end{eqnarray}
where $a_1 = (x_2 + x_3) q - x_4 k_n$ and $a_2 = x_3q + x_5 k_p = (x_3
+ x_5) q + x_5 k_n$ as $k_p = k_n + q$. Following the standard
procedure of the diagonalization of the quadratic forms for the
calculation of Feynman integrals, expounded in subsection C1a, making
a Wick rotation \cite{Itzykson1980} and integrating over virtual
momenta in the $n$-dimensional momentum space \cite{Smirnov2004} -
\cite{Smirnov2012}, we arrive at the expression
\begin{eqnarray}\label{eq:C1b.6}
\hspace{-0.35in}&&{\cal F}(k_n, k_e, q)_{1b} = \int^1_0dx_1
\int^1_0dx_2 \int^1_0dx_3 \int^1_0dx_4 \int^1_0dx_5 \int^1_0
dx_6\,\frac{\delta(1 - x_1 - x_2 - x_3 - x_4 - x_5 - x_6)}{2^{2n}
  \pi^n U^{n/2}} \nonumber\\
\hspace{-0.35in}&&\times \bigg\{- \frac{n}{2}\,\frac{\displaystyle
  \Gamma\Big(2 - \frac{n}{2}\Big) \Gamma\Big(2 - \frac{n - 1}{2}\Big)
}{2^{n - 3}\displaystyle
  \Gamma\Big(\frac{1}{2}\Big)}\,\Big(\frac{Q}{m^2_N}\Big)^{-4 + n} \,
\gamma^{\mu} (1 - \gamma^5) \otimes \gamma_{\mu} + \ldots\bigg\},
\end{eqnarray}
where we have denoted
\begin{eqnarray}\label{eq:C1b.7}
  \hspace{-0.3in}U &=& (x_1 + x_2 + x_4)(x_3 + x_5 + x_6) + x_3(x_5 +
  x_6),\nonumber\\
  \hspace{-0.3in}b_1 &=& \frac{(x_3 + x_5 + x_6)\, a_1 - x_3 a_2}{U} =
  \nonumber\\
  \hspace{-0.3in} &=& \frac{(x_2 (x_3 + x_5) + (x_2 + x_3)x_6)\, q -
    (x_3 x_5 + (x_3 + x_5 + x_6)x_4)\, k_n}{U} = \frac{X q + Y
    k_n}{U},\nonumber\\
   \hspace{-0.3in}b_2 &=& \frac{(x_1 + x_2 + x_3 + x_4)\, a_2 - x_3
     a_1}{U} =  \nonumber\\
  \hspace{-0.3in}&=& \frac{((x_1 + x_4)x_3 + (x_1 + x_2 + x_3 + x_4)
    x_5)\, q + (x_3 x_4 + (x_1 + x_2 + x_3 + x_4) x_5) \, k_n }{U} =
  \frac{\bar{X} q + \bar{Y} k_n}{U},\nonumber\\
   \hspace{-0.3in}Q &=&(x_1 + x_2 + x_3) m^2_{\pi} - (x_2 + x_3) q^2 +
   \frac{(x_3 + x_5 + x_6) a^2_1 - 2 x_3 a_1\cdot a_2 + (x_1 + x_2 +
     x_3 + x_4) a^2_2}{U}.
\end{eqnarray}
The 4-vectors $b_1$ and $b_2$ appear as shifts of virtual momenta $k
\to k + b_1$ and $p \to p + b_2$, which are necessary for a
diagonalization of the denominator in Eq.(\ref{eq:C1b.5}). Taking the
limit $n \to 4$ and keeping divergent contributions proportional to
$\Gamma(2 - n/2)$, we arrive at the expression
\begin{eqnarray}\label{eq:C1b.8}
\hspace{-0.3in}&&{\cal F}(k_n, k_e, q)_{1b} = \int^1_0dx_1
\int^1_0dx_2 \int^1_0dx_3 \int^1_0dx_4 \int^1_0dx_5\,\frac{\delta(1 -
  x_1 - x_2 - x_3 - x_4 - x_5)}{256 \pi^4 U^2} \nonumber\\
\hspace{-0.35in}&&\times\Big\{\Big(- \Gamma\Big(2 - \frac{n}{2}\Big) +
2 \,{\ell n}\frac{Q}{m^2_N}\Big) \, \gamma^{\mu} (1 - \gamma^5)
\otimes\gamma_{\mu} + \ldots\Big\}.
\end{eqnarray}
Using the Dirac equations for free fermions we transcribe
Eq.(\ref{eq:C1b.8}) into the form
\begin{eqnarray*}
\hspace{-0.5in}&&{\cal F}(k_n, k_e, q)_{1b} = \int^1_0dx_1
\int^1_0dx_2 \int^1_0dx_3 \int^1_0dx_4 \int^1_0dx_5 \int^1_0
dx_6\,\frac{\delta(1 - x_1 - x_2 - x_3 - x_4 - x_5 - x_6)}{256 \pi^4
  U^2} \nonumber\\
\end{eqnarray*}
\begin{eqnarray}\label{eq:C1b.9}
\hspace{-0.5in}&&\times \Big\{\Big( - \Gamma\Big(2 - \frac{n}{2}\Big)
+ f^{(V)}_{1b}(x_1, \ldots,x_6) + g^{(V)}_{1b}(x_1,
\ldots,x_6)\,\frac{k_n\cdot q}{m^2_N}\Big) \gamma^{\mu} (1 - \gamma^5)
\otimes\gamma_{\mu}\Big\},
\end{eqnarray}
where we have denoted
\begin{eqnarray}\label{eq:C1b.10}
\hspace{-0.3in}f^{(V)}_{1b}(x_1, \ldots,x_6) = 2 {\ell
  n}\frac{\bar{Q}}{m^2_N} \quad, \quad g^{(V)}_{1b}(x_1, \ldots,x_6) =
4\, \frac{V}{U}\, \frac{m^2_N}{\bar{Q}}.
\end{eqnarray}
For the calculation of the integrals over the Feynman parameters we use
\begin{eqnarray}\label{eq:C1b.11}
\hspace{-0.3in}\frac{\bar{Q}}{m^2_N} &=& \frac{1}{U}\Big((x_1 + x_2 +
x_3)U\,\frac{m^2_{\pi}}{m^2_N} + (x_3 + x_5 + x_6)x^2_4 + 2 x_3 x_4x_5
+ (x_1 + x_2 + x_3 + x_4)x^2_5\Big),\nonumber\\
\hspace{-0.3in} V &=& (x_1 + x_4)(x_3 + x_5)x_5 + (x_2 + x_3) x^2_5 -
(x_3 + x_5)x_2 x_4 - (x_2 + x_3)x_4 x_6
\end{eqnarray}
for $m_{\pi} = 139.5706\,{\rm MeV}$ and $m_N = (m_n + m_p)/2$ with
$m_n = 939.5654\,{\rm MeV}$ and $m_p = 938.2721\,{\rm MeV}$
\cite{PDG2020}, where $\bar{Q} = Q\big|_{q = 0}$.  A deviation of $Q$
from $\bar{Q}$ is taken into account in the linear approximation in
powers of $k_n \cdot q/m^2_N$. The function $g^{(V)}_{1b}(x_1,
\ldots,x_6)$ is defined by the expansion of ${\ell n}Q/m^2_N$ in
powers of $k_n \cdot q/m^2_N$ (or $q_0/m_N$).

After the integration over the Feynman parameters, the contribution of
the Feynman diagram in Fig.\,\ref{fig:fig1}b to the amplitude of the
neutron beta decay is equal to
\begin{eqnarray}\label{eq:C1b.12}
\hspace{-0.3in}M(n \to p e^- \bar{\nu}_e)^{(\pi^0)}_{\rm
  Fig.\,\ref{fig:fig1}b} = \frac{\alpha}{2\pi}\frac{g^2_{\pi
    N}}{16\pi^2}\,G_V \Big\{\Big(A_{1b}\Gamma\Big(2 - \frac{n}{2}\Big)
+ F^{(V)}_{1b} + G^{(V)}_{1b}\,\frac{k_n\cdot q}{m^2_N}\Big)
\Big[\bar{u}_e \gamma^{\mu}(1 - \gamma^5) v_{\bar{\nu}}\Big]\,
\Big[\bar{u}_p\gamma_{\mu}u_n\Big] \Big\},
\end{eqnarray}
where the structure constants are equal to $A_{1b} = - 1/6$,
$F^{(V)}_{1b} = - 0.6752$ and $G^{(V)}_{1b} = - 0.0234$. The Lorentz
structure of Eq.(\ref{eq:C1b.12}) is obtained at the neglect the
contributions of order $O(E^2_0/m^2_N)\sim 10^{-6}$. We would like to
emphasize that all contributions to the matrix element
Eq.(\ref{eq:C1b.12}) are induced by the contributions of the hadronic
{\it first class} currents \cite{Weinberg1958}, which are $G$-even
\cite{Lee1956a} (see also \cite{Ivanov2018}).

\subsection*{C1c. Analytical calculation of the Feynman diagram in
  Fig.\,\ref{fig:fig1}c}
\renewcommand{\theequation}{C1c-\arabic{equation}}
\setcounter{equation}{0}

In the Feynman gauge for the photon propagator the analytical
expression of the Feynman diagram in Fig.\,\ref{fig:fig1}c is given by
(see Eq.(\ref{eq:A.1}))
\begin{eqnarray}\label{eq:C1c.1}
\hspace{-0.3in}&&M(n \to p e^- \bar{\nu}_e)^{(\pi^0)}_{\rm
  Fig.\,\ref{fig:fig1}c} = + 2 e^2 g^2_{\pi N} M^2_W
G_V\nonumber\\ &&\times \int \frac{d^4k}{(2\pi)^4i}\int
\frac{d^4p}{(2\pi)^4i}\,\Big[\bar{u}_e \gamma^{\beta}\,\frac{1}{m_e -
    \hat{p} - \hat{k}_e - i0}\, \gamma^{\mu}(1 - \gamma^5)
  v_{\bar{\nu}}\Big]\,\Big[\bar{u}_p\gamma^5 \frac{1}{m_N - \hat{k}_n
    - \hat{k} - i0}\, \gamma^5\,
  u_n\Big]\nonumber\\ \hspace{-0.3in}&&\times\,\frac{(2 k + p -
  q)^{\nu} (p + 2 k - 2 q)_{\beta}}{[m^2_{\pi} - k^2 - i0][m^2_{\pi} -
    (k - q)^2 - i0][m^2_{\pi} - (k + p - q)^2 - i0]}\,\frac{1}{p^2 +
  i0} \,D^{(W)}_{\mu\nu}(p - q).
\end{eqnarray}
Using the Dirac equations for a free electron and a free neutron, we
transcribe Eq.(\ref{eq:C1c.1}) into the form
\begin{eqnarray}\label{eq:C1c.2}
\hspace{-0.3in}&&M(n \to p e^- \bar{\nu}_e)^{(\pi^0)}_{\rm
  Fig.\,\ref{fig:fig1}c} = - 2 e^2 g^2_{\pi N} M^2_W
G_V\nonumber\\ &&\times \int \frac{d^4k}{(2\pi)^4i}\int
\frac{d^4p}{(2\pi)^4i}\,\Big[\bar{u}_e\,\frac{2k^{\beta}_e +
    \gamma^{\beta}\hat{p}}{m^2_e - (p + k_e)^2 - i0}\, \gamma^{\mu}(1
  - \gamma^5) v_{\bar{\nu}}\Big]\,\Big[\bar{u}_p
  \frac{\hat{k}}{m^2_N - (k_n + k)^2 - i0}\, u_n\Big]\nonumber\\
\hspace{-0.3in}&&\times\, \frac{(2 k + p - q)^{\nu} (p + 2 k - 2
  q)_{\beta}}{[m^2_{\pi} - k^2 - i0][m^2_{\pi} - (k - q)^2 -
    i0][m^2_{\pi} - (k + p - q)^2 - i0]}\,\frac{1}{p^2 + i0}
\,D^{(W)}_{\mu\nu}(p - q).
\end{eqnarray}
Then, we take the propagator of the electroweak $W^-$-boson
$D^{(W)}_{\mu\nu}(p - q)$ in the form
\begin{eqnarray*}
  D^{(W)}_{\mu\nu}(p - q) = - \frac{1}{M^2_W}\Big(\eta_{\mu\nu} +
  \frac{(p - q)^2 \eta_{\mu\nu} - (p - q)_{\mu} (p - q)_{\nu}}{M^2_W -
    (p - q)^2 - i0}\Big)
\end{eqnarray*}
and represent the matrix element $M(n \to p e^-
\bar{\nu}_e)^{(\pi^0)}_{\rm Fig.\,\ref{fig:fig1}c}$ as the
superposition of two terms
\begin{eqnarray*}
\hspace{-0.3in}&&M(n \to p e^- \bar{\nu}_e)^{(\pi^0)}_{\rm
  Fig.\,\ref{fig:fig1}c}  =\nonumber\\
\hspace{-0.3in}&& = 2 e^2 g^2_{\pi N} G_V\int
\frac{d^4k}{(2\pi)^4i}\int
\frac{d^4p}{(2\pi)^4i}\,\Big[\bar{u}_e\,\frac{
    \gamma^{\beta}\hat{p}}{m^2_e - (p + k_e)^2 - i0}\, \gamma^{\mu}(1
  - \gamma^5) v_{\bar{\nu}}\Big]\,\Big[\bar{u}_p \frac{\hat{k}}{m^2_N
    - (k_n + k)^2 - i0}\, u_n\Big] \nonumber\\
\end{eqnarray*}
\begin{eqnarray}\label{eq:C1c.3}
\hspace{-0.3in}&&\times\, \frac{(2 k + p)_{\mu} (p + 2
  k)_{\beta}}{[m^2_{\pi} - k^2 - i0][m^2_{\pi} - (k - q)^2 -
    i0][m^2_{\pi} - (k + p - q)^2 - i0]}\,\frac{1}{p^2 +
  i0}\nonumber\\
\hspace{-0.3in}&&+ 4 e^2 g^2_{\pi N} G_V \int
\frac{d^4k}{(2\pi)^4i}\int
\frac{d^4p}{(2\pi)^4i}\,\Big[\bar{u}_e\,\frac{
    \gamma^{\beta}\hat{p}}{m^2_e - (p + k_e)^2 - i0}\, \gamma^{\mu}(1
  - \gamma^5) v_{\bar{\nu}}\Big]\,\Big[\bar{u}_p \frac{\hat{k}}{m^2_N
    - (k_n + k)^2 - i0}\, u_n\Big] \nonumber\\
\hspace{-0.3in}&&\times\, \frac{ (p + 2 k)_{\beta}}{[m^2_{\pi} - k^2 -
    i0][m^2_{\pi} - (k - q)^2 - i0][m^2_{\pi} - (k + p - q)^2 -
    i0]}\,\frac{1}{p^2 + i0}\,\frac{k_{\mu}p^2 - (k\cdot
  p)\,p_{\mu}}{M^2_W - (p - q)^2 - i0},
\end{eqnarray}
where we have kept only leading divergent contributions that is
equivalent to the use of the LLA.  Then, it is convenient to rewrite
the r.h.s. of Eq.(\ref{eq:C1c.3}) as follows
\begin{eqnarray*}
M(n \to p e^- \bar{\nu}_e)^{(\pi^0)}_{\rm Fig.\,\ref{fig:fig1}c} = - 2
e^2 g^2_{\pi N}G_V \Big[\bar{u}_e \Big({\cal F}^{(1)}(k_n, k_e,
  q)^{\mu}_{1c} + 2 {\cal F}^{(2)}(k_n, k_e, q)^{\mu}_{1c}\Big)
  v_{\bar{\nu}}\Big]\,\Big[\bar{u}_p\gamma_{\mu} u_n\Big].
\end{eqnarray*}
 This reduces the calculation of the Feynman diagram in
 Fig.\,\ref{fig:fig1}c to the calculation of the integrals ${\cal
   F}^{(1)}(k_n, k_e, q)_{1c} = {\cal F}^{(1)}(k_n, k_e,
 q)^{\mu}_{1c}\otimes \gamma_{\mu}$ and ${\cal F}^{(2)}(k_n, k_e,
 q)_{1c} = {\cal F}^{(2)}(k_n, k_e, q)^{\mu}_{1c}\otimes
 \gamma_{\mu}$, where the integrals ${\cal F}^{(1)}(k_n, q)_{1c}$ and
 ${\cal F}^{(2)}(k_n, k_e, q)_{1c}$ are given by
\begin{eqnarray}\label{eq:C1c.4}
\hspace{-0.21in}&&{\cal F}^{(1)}(k_n, k_e, q)_{1c} = - \int
\frac{d^4k}{(2\pi)^4i}\int \frac{d^4p}{(2\pi)^4i}\,\frac{1}{m^2_e - (p
  + k_e)^2 - i0}\,\frac{1}{m^2_N - (k_n + k)^2 -
  i0}\,\frac{1}{m^2_{\pi} - k^2 - i0} \,\frac{1}{m^2_{\pi} - (k - q)^2
  - i0} \nonumber\\
\hspace{-0.21in}&& \times \, \frac{1}{m^2_{\pi} - (k + p - q)^2 -
  i0}\,\frac{1}{p^2 + i0}\Big(\big(4 p^2 \hat{k} + p^2 \hat{p} - 4 k^2
\hat{p} + 8 (k \cdot p) \hat{k}\big)(1 - \gamma^5) \otimes
\hat{k}\Big)
\end{eqnarray}
and 
\begin{eqnarray}\label{eq:C1c.5}
\hspace{-0.21in}&&{\cal F}^{(2)}(k_n, k_e, q)_{1c} = - \int
\frac{d^4k}{(2\pi)^4i}\int \frac{d^4p}{(2\pi)^4i}\, \frac{1}{m^2_e -
  (p + k_e)^2 - i0}\,\frac{1}{m^2_N - (k_n + k)^2 -
  i0}\,\frac{1}{m^2_{\pi} - k^2 - i0} \,\frac{1}{m^2_{\pi} - (k - q)^2
  - i0} \nonumber\\
\hspace{-0.21in}&& \times \, \frac{1}{m^2_{\pi} - (k + p - q)^2 - i0}
\, \frac{1}{p^2 + i0}\,\frac{1}{M^2_W - (p - q)^2 -
  i0}\Big(p^2\big(p^2 \hat{k} - 2 k^2 \hat{p} + 2 (k\cdot p) \hat{k} -
(k\cdot p) \hat{p}\big) \,(1 - \gamma^5) \otimes \hat{k}\Big).
\end{eqnarray}
The calculation of these integrals is similar to the calculation of
the integral ${\cal F}(k_n,q)_{1b}$ in the subsection C1b.

\subsubsection*{\bf Calculation of the integral
  ${\cal F}^{(1)}(k_n, k_e, q)_{1c}$}

Skipping standard intermediate calculations \cite{Smirnov2004,
  Smirnov2006, Smirnov2012}, we define the integral ${\cal
  F}^{(1)}(k_n, k_e, q)_{1c}$ in terms of the integrals over the
Feynman parameters. We get
\begin{eqnarray}\label{eq:C1c.6}
\hspace{-0.3in}&&{\cal F}^{(1)}(k_n, k_e, q)_{1c} = \int^1_0dx_1
\int^1_0dx_2 \int^1_0dx_3 \int^1_0dx_4 \int^1_0dx_5 \int^1_0
dx_6\,\frac{\delta(1 - x_1 - x_2 - x_3 - x_4 - x_5 - x_6)}{2^{2n}
  \pi^n U^{n/2}}\nonumber\\
\hspace{-0.3in}&&\times \Bigg\{n\,\frac{\displaystyle \Gamma\Big(2 -
  \frac{n}{2}\Big)\Gamma\Big(2 - \frac{n - 1}{2}\Big)}{\displaystyle
  2^{n - 3}\Gamma\Big(\frac{1}{2}\Big)}\,\Big(\frac{Q}{m^2_N}\Big)^{-4
  + n}\,\gamma^{\mu} (1 - \gamma^5) \otimes \gamma_{\mu} +
\ldots\Bigg\},
\end{eqnarray}
where we have denoted
\begin{eqnarray}\label{eq:C1c.7}
  \hspace{-0.3in}U &=& (x_1 + x_2 + x_4)(x_3 + x_5 + x_6) + (x_5 +
  x_6) x_3,\nonumber\\
  \hspace{-0.3in}b_1 &=& \frac{(x_3 + x_5 + x_6)a_1 - x_3 a_2}{U} =
  \frac{(x_2 x_3 + (x_2 + x_3)(x_5 + x_6))q - (x_3 + x_5 + x_6)x_4 k_n
    + x_3 x_5 k_e}{U} =  \nonumber\\
  \hspace{-0.3in} &=&\frac{X q + Y k_n + Z k_e}{U},\nonumber\\
   \hspace{-0.3in}b_2 &=& \frac{(x_1 + x_2 + x_3 + x_4)a_2 - x_3
     a_1}{U} = \frac{(x_1 + x_4)x_3 q + x_3 x_4 k_n - (x_1 + x_2 + x_3
     + x_4) x_5 k_e}{U} =  \nonumber\\
  \hspace{-0.3in} &=&\frac{\bar{X} q + \bar{Y} k_n + \bar{Z}
    k_e}{U},\nonumber\\
   \hspace{-0.3in}Q &=&(x_1 + x_2 + x_3) m^2_{\pi} -(x_2 + x_3) q^2 +
   \frac{(x_3 + x_5 + x_6) a^2_1 - 2 x_3 a_1\cdot a_2 + (x_1 + x_2 +
     x_3 + x_4) a^2_2}{U}
\end{eqnarray}
with $a_1 = (x_2 + x_3)q - x_4 k_n$ and $a_2 = x_3q - x_5 k_e$.
Taking the limit $n \to 4$ and keeping the divergent part proportional
to $\Gamma(2 - n/2)$, we transcribe the integral Eq.(\ref{eq:C1c.6})
into the form
\begin{eqnarray}\label{eq:C1c.8}
\hspace{-0.3in}&&{\cal F}^{(1)}(k_n, k_e, q)_{1c} = \int^1_0dx_1
\int^1_0dx_2 \int^1_0dx_3 \int^1_0dx_4 \int^1_0dx_5 \int^1_0
dx_6\,\frac{\delta(1 - x_1 - x_2 - x_3 - x_4 - x_5 - x_6)}{256 \pi^4
  U^2}\nonumber\\
\hspace{-0.3in}&&\times \Big\{\Big(2\, \Gamma(2 - \frac{n}{2}\Big) -
  4\,{\ell n}\frac{Q}{m^2_N} \Big)\,\gamma^{\mu}(1 -
\gamma^5) \otimes \gamma_{\mu} + \ldots\Big\}.
\end{eqnarray}
Taking into account that in the matrix element $M(n \to p e^-
\bar{\nu}_e)^{(\pi^0)}_{\rm Fig.\,\ref{fig:fig1}c}$ the integral
${\cal F}^{(1)}(k_n, k_e, q)_{1c}$ sandwiches between the Dirac wave
functions of a free neutron and free decay fermions, we may define its
irreducible Lorentz structure. Having neglected the contributions of
order $O(E^2_0/m^2_N) \sim 10^{-6}$, we arrive at the expression
\begin{eqnarray}\label{eq:C1c.9}
\hspace{-0.3in}&&{\cal F}^{(1)}(k_n, k_e, q)_{1c} = \int^1_0dx_1
\int^1_0dx_2 \int^1_0dx_3 \int^1_0dx_4 \int^1_0dx_5 \int^1_0
dx_6\,\frac{\delta(1 - x_1 - x_2 - x_3 - x_4 - x_5 - x_6)}{256 \pi^4
  U^2} \nonumber\\
\hspace{-0.3in}&&\times \Big\{\Big( 2\, \Gamma\Big(2 -
\frac{n}{2}\Big) + f^{(V)}_{1c^{(1)}}(x_1,\ldots,x_6) +
g^{(V)}_{1c^{(1)}}(x_1,\ldots,x_6)\,\frac{k_n\cdot q}{m^2_N}+
h^{(V)}_{1c^{(1)}}(x_1,\ldots,x_6) \,\frac{k_n\cdot k_e}{m^2_N}\Big)
\,\gamma^{\mu}(1 - \gamma^5) \otimes \gamma_{\mu} + \ldots \Big\},
\nonumber\\
\hspace{-0.3in}&&
\end{eqnarray}
where we have denoted
\begin{eqnarray}\label{eq:C1c.10} 
\hspace{-0.3in}f^{(V)}_{1c^{(1)}}(x_1,\ldots,x_6) = - 4\,{\ell
  n}\frac{\bar{Q}}{m^2_N}\quad,\quad
g^{(V)}_{1c^{(1)}}(x_1,\ldots,x_6) = - 8\,\frac{V}{U}\,\frac{
  m^2_N}{\bar{Q}}\quad,\quad h^{(V)}_{1c^{(1)}}(x_1,\ldots,x_6) = -
8\,\frac{\bar{V}}{U}\,\frac{ m^2_N}{\bar{Q}}.
\end{eqnarray}
For the calculation of the integrals over the Feynman parameters we use
\begin{eqnarray}\label{eq:C1c.11}
\hspace{-0.3in}\frac{\bar{Q}}{m^2_N} &=& \frac{1}{U}\Big((x_1 + x_2 +
x_3)\,U \frac{m^2_{\pi}}{m^2_N} + (x_3 + x_5 +
x_6)x^2_4\Big),\nonumber\\
\hspace{-0.3in}V &=& - (x_5 + x_6)(x_2 + x_3)x_4 - x_2 x_3 x_4+
\;,\;\bar{V} = - x_3x_4x_5
\end{eqnarray}
for $m_{\pi} = 139.5706\,{\rm MeV}$ and $m_N = (m_n + m_p)/2 =
938.9188\, {\rm MeV}$ with $m_n = 939.5654\,{\rm MeV}$ and $m_p =
938.2721\,{\rm MeV}$ \cite{PDG2020}, where $\bar{Q} = Q\big|_{q = k_e
  = 0}$.  A deviation of $Q$ from $\bar{Q}$ is taken into account in
the linear approximation in powers of $k_n \cdot q/m^2_N$ (or
$q_0/m_N$) and $k_n \cdot k_e/m^2_N$ (or $E_e/m_N$), respectively. The
functions $g^{(V)}_{1c^{(1)}}(x_1,\ldots,x_6)$ and
$h^{(V)}_{1c^{(1)}}(x_1,\ldots,x_6)$ appear as an expansion of ${\ell
  n}Q/m^2_N$ in powers of of $k_n \cdot q/m^2_N$ (or $q_0/m_N$) and
$k_n \cdot k_e/m^2_N$ (or $E_e/m_N$), respectively. After the
integration over the Feynman parameters, the contribution of the
integral ${\cal F}^{(1)}(k_n, q)_{1c}$ to the matrix element $M(n \to
p e^-\bar{\nu}_e)^{(\pi^0)}_{\rm Fig.\,\ref{fig:fig1}c}$ takes the
form
\begin{eqnarray}\label{eq:C1c.12}
\hspace{-0.3in}&&M(n \to p e^- \bar{\nu}_e)^{(\pi^0)}_{\rm
  Fig.\,\ref{fig:fig1}c^{(1)}} = - \frac{\alpha}{2\pi}\frac{g^2_{\pi
    N}}{16\pi^2}\,G_V \Big\{\Big(A_{1c^{(1)}}\Gamma\Big(2 -
\frac{n}{2}\Big) + F^{(V)}_{1c^{(1)}} + G^{(V)}_{1c^{(1)}}\,\frac{k_n\cdot
  q}{m^2_N} + H^{(V)}_{1c^{(1)}}\,\frac{k_n\cdot
  k_e}{m^2_N}\Big)\Big[\bar{u}_e \gamma^{\mu}(1 - \gamma^5)
  v_{\bar{\nu}}\Big] \nonumber\\
\hspace{-0.3in}&& \times \Big[\bar{u}_p\gamma_{\mu}u_n\Big]\Big\},
\end{eqnarray}
where the structure constants are equal to $A_{1c^{(1)}} = 1/3$,
$F^{(V)}_{1c^{(1)}} = 1.9119$, $G^{(V)}_{1c^{(1)}} = 1.3393$,
$H^{(V)}_{1c^{(1)}} = 0.1791$. The Lorentz structure of the matrix
element Eq.(\ref{eq:C1c.12}) is calculated at the neglect of the
contributions of order $O(E^2_0/m^2_N)\sim 10^{-6}$ to the integral
${\cal F}^{(1)}(k_n, k_e, q)_{1c}$.

\subsubsection*{\bf Calculation of the integral
  ${\cal F}^{(2)}(k_n, k_e, q)_{1c}$}

Skipping standard intermediate calculations, we define the integral
${\cal F}^{(2)}(k_n, k_e, q)_{1c}$ in terms of the integrals over the
Feynman parameters. We get
\begin{eqnarray*}
\hspace{-0.3in}&&{\cal F}^{(2)}(k_n, k_e, q)_{1c} = \int^1_0 \!\!
dx_1 \int^1_0 \!\! dx_2 \int^1_0 \!\! dx_3 \int^1_0 \!\! dx_4 \int^1_0
\!\! dx_5 \int^1_0 \!\! dx_6 \int^1_0 \!\! dx_7\,\frac{\delta(1 - x_1
  - x_2 - x_3 - x_4 - x_5 - x_6 - x_7)}{2^{2n} \pi^n
  U^{n/2}} \nonumber\\
\hspace{-0.3in}&&\times \Bigg\{ - \frac{(n - 1)(n +
  2)}{8}\,\frac{\displaystyle \Gamma\Big(2 -
  \frac{n}{2}\Big)\Gamma\Big(2 - \frac{n - 1}{2}\Big)}{\displaystyle
  2^{n - 3}\Gamma\Big(\frac{1}{2}\Big)}\,\Big(\frac{Q}{m^2_N}\Big)^{-4
  + n}\,\gamma^{\mu} (1 - \gamma^5) \otimes \gamma_{\mu} +
\frac{n^2}{4} \Gamma(5 - n)\,Q^{-5+n} m^{2(4-n)}_N \nonumber\\
\hspace{-0.3in}&& \times \Big[\Big(2\, \frac{n + 2}{n^2}\,b^2_2 -
  \frac{1}{n^2}\, b^2_1 + 2\,\frac{n + 2}{n^2}\,b_1\cdot b_2
  \Big)\gamma^{\mu}(1 - \gamma^5) \otimes \gamma_{\mu} - 2\, \frac{n^2
    - 3 n + 2}{n^2}\,\hat{b}_2 (1 - \gamma^5) \otimes \hat{b}_1 + 2\,
  \frac{n + 2}{n^2}\, \hat{b}_1 (1 - \gamma^5) \otimes \hat{b}_2
  \nonumber\\
\end{eqnarray*}
\begin{eqnarray}\label{eq:C1c.13}
\hspace{-0.3in}&& - \frac{n + 2}{n^2}\, \hat{b}_1 (1 - \gamma^5)
\otimes \hat{b}_1 - \frac{n + 4}{n^2}\, \hat{b}_2 (1 - \gamma^5)
\otimes \hat{b}_2 \Big] \Bigg\},
\end{eqnarray}
where we have denoted
\begin{eqnarray}\label{eq:C1c.14}
  \hspace{-0.3in}U &=& (x_1 + x_2 + x_4)(x_3 + x_5 + x_6 + x_7) +
  x_3(x_5 + x_6 + x_7),\nonumber\\
  \hspace{-0.3in}b_1 &=& \frac{(x_3 + x_5 + x_6 + x_7)a_1 - x_3
    a_2}{U} = \frac{- (x_3 + x_5 + x_6 + x_7)x_4 k_n }{U} =
  \frac{Y}{U}\,k_n ,\nonumber\\
   \hspace{-0.3in}b_2 &=& \frac{(x_1 + x_2 + x_3 + x_4)a_2 - x_3
     a_1}{U} = \frac{x_3 x_4 k_n}{U} =
   \frac{\bar{Y}}{U}\,k_n,\nonumber\\
   \hspace{-0.3in}Q &=&x_7 M^2_W + \frac{(x_3 + x_5 + x_6 + x_7) a^2_1
     - 2 x_3 a_1\cdot a_2 + (x_1 + x_2 + x_3 + x_4) a^2_2}{U}
\end{eqnarray}
with $a_1 = - x_4 k_n$ and $a_2 = 0$, where we have neglected the
contributions of the terms proportional to $q$ and $k_e$, appearing in
${\cal F}^{(2)}(k_n, k_e, q)_{1c}$ in the form of an expansion in
powers of $k_n\cdot q/M^2_W$ and $k_n\cdot k_e/M^2_W$ of order
$10^{-7}$, and the terms of order $m^2_{\pi}/M^2_W \sim
10^{-6}$. Taking the limit $n \to 4$ and keeping i) a divergent
contribution proportional to $\Gamma(2 - n/2)$ and ii) the
contributions of order $O(m^2_N/M^2_W)$ in the large electroweak
$W^-$-boson mass $M_W$ expansion, we transcribe Eq.(\ref{eq:C1c.13})
into the form
\begin{eqnarray}\label{eq:C1c.15}
\hspace{-0.3in}&&{\cal F}^{(2)}(k_n, k_e, q)_{1c} = \int^1_0dx_1
\int^1_0dx_2 \int^1_0dx_3 \int^1_0dx_4 \int^1_0dx_5 \int^1_0 dx_6
\int^1_0 dx_7\,\frac{\delta(1 - x_1 - x_2 - x_3 - x_4 - x_5 - x_6 -
  x_7)}{256\pi^4 U^2}\nonumber\\
\hspace{-0.3in}&&\times \Big\{\Big( - \frac{9}{8}\,\Gamma\Big(2 -
\frac{n}{2}\Big) + f^{(V)}_{1c^{(2)}}(x_1,\ldots,x_7) +
g^{(W)}_{1c^{(2)}}(x_1, \ldots, x_7)\,
\frac{m^2_N}{M^2_W}\Big)\,\gamma^{\mu} (1 - \gamma^5) \otimes
\gamma_{\mu} + f^{(W)}_{1c^{(2)}}(x_1, \ldots, x_7)\,
\frac{m^2_N}{M^2_W} \nonumber\\
\hspace{-0.3in}&& \times \,\frac{\hat{k}_n}{m_N}(1 - \gamma^5) \otimes
1 \Big\}.
\end{eqnarray}
The functions of the Feynman parameters are defined by
\begin{eqnarray}\label{eq:C1c.16}
\hspace{-0.3in}f^{(V)}_{1c^{(2)}}(x_1,\ldots,x_7) &=& 
\frac{9}{4}\,{\ell n}\frac{M^2_W}{m^2_N} + \frac{9}{4}\,{\ell
  n}\bar{Q}, \nonumber\\
\hspace{-0.3in} g^{(W)}_{1c^{(2)}}(x_1,\ldots,x_7) &=&
\frac{1}{\bar{Q}}\Big[- \frac{1}{4}\,\frac{Y^2}{U^2} + 3\,
  \frac{Y\bar{Y}}{U^2} + 3\, \frac{\bar{Y}^2}{U^2}\Big], \nonumber\\
\hspace{-0.3in} f^{(W)}_{1c^{(2)}}(x_1,\ldots,x_7) &=&
\frac{1}{\bar{Q}}\Big[- \frac{3}{2}\,\frac{Y^2}{U^2} +
  \frac{Y\bar{Y}}{U^2} - 2\, \frac{\bar{Y}^2}{U^2} \Big].
\end{eqnarray}
For the integration over the Feynman parameters we use
\begin{eqnarray}\label{eq:C1c.17}
\hspace{-0.3in}\bar{Q} &=& \frac{1}{U}\Big(x_7 U + (x_3 + x_5 + x_6 +
x_7)x^2_4\,\frac{m^2_N}{M^2_W}\Big),\nonumber\\
\hspace{-0.3in}Y&=&- (x_3 + x_5 + x_6 + x_7) x_4\;,\; \bar{Y} = x_3
x_4.
\end{eqnarray}
where $M_W = 80.379\,{\rm GeV}$ is the electroweak $W^-$-boson mass
\cite{PDG2020}. After the integration over the Feynman parameters, the
contribution of the integral ${\cal F}^{(1)}(k_n, q)$ to the matrix
element $M(n \to p e^-\bar{\nu}_e)^{(\pi^0)}_{\rm
  Fig.\,\ref{fig:fig1}c}$ takes the form
\begin{eqnarray}\label{eq:C1c.18}
\hspace{-0.3in}M(n \to p e^- \bar{\nu}_e)^{(\pi^0)}_{\rm
  Fig.\,\ref{fig:fig1}c^{(2)}} &=& - \frac{\alpha}{2\pi}\frac{g^2_{\pi
    N}}{16\pi^2}\,G_V \, 2\,\Big\{\Big(A_{1c^{(2)}}\Gamma\Big(2 -
\frac{n}{2}\Big) + F^{(V)}_{1c^{(2)}} + G^{(W)}_{1c^{(2)}}\,
\frac{m^2_N}{M^2_W}\Big) \Big[\bar{u}_e \gamma^{\mu}(1 - \gamma^5)
  v_{\bar{\nu}}\Big]\Big[\bar{u}_p \gamma_{\mu} u_n\Big]\nonumber\\
\hspace{-0.30in}&+&
F^{(W)}_{1c^{(2)}}\,\frac{m^2_N}{M^2_W}\Big[\bar{u}_e
  \frac{\hat{k}_n}{m_N}(1 - \gamma^5) v_{\bar{\nu}}\Big]\Big[\bar{u}_p
  u_n\Big]\Big\},
\end{eqnarray}
where the structure constants are equal to $A_{1c^{(2)}} = - 0.0254$,
$F^{(V)}_{1c^{(2)}} = 0.3257$, $G^{(W)}_{1c^{(2)}} = - 0.1180$
and $F^{(W)}_{1c^{(2)}} = - 0.3520$, respectively. The Lorentz
structure of the matrix element Eq.(\ref{eq:C1c.18}) is obtained at
the neglect of the contributions of the terms proportional to
$k_n\cdot q/M^2_W$ and $k_n\cdot k_e/M^2_W$, respectively, which are
of order $10^{-7}$ or even smaller.

Thus, the contribution of the Feynman diagram in Fig.\,\ref{fig:fig1}c
to the amplitude of the neutron beta decay is equal to
\begin{eqnarray}\label{eq:C1c.19}
\hspace{-0.3in}&&M(n \to p e^- \bar{\nu}_e)^{(\pi^0)}_{\rm
  Fig.\,\ref{fig:fig1}c} = - \frac{\alpha}{2\pi}\frac{g^2_{\pi
    N}}{16\pi^2}\,G_V \Big\{\Big(A_{1c}\Gamma\Big(2 - \frac{n}{2}\Big)
+ F^{(V)}_{1c} + G^{(V)}_{1c}\,\frac{k_n\cdot q}{m^2_N} +
H^{(V)}_{1c}\,\frac{k_n\cdot k_e}{m^2_N} + G^{(W)}_{1c}\,
\frac{m^2_N}{M^2_W}\Big) \nonumber\\
\hspace{-0.3in}&&\times \Big[\bar{u}_e \gamma^{\mu}(1 - \gamma^5)
  v_{\bar{\nu}}\Big] \Big[\bar{u}_p\gamma_{\mu}u_n\Big] +
F^{(W)}_{1c^{(2)}}\,\frac{m^2_N}{M^2_W} \,
\Big[\bar{u}_e\frac{\hat{k}_n}{m_n} (1 - \gamma^5) v_{\bar{\nu}}\Big]
\Big[\bar{u}_p u_n\Big]\Big\},
\end{eqnarray}
where the structure constants are equal to $A_{1c} = A_{1c^{(1)}} +
2\,A_{1c^{(2)}} = 0.2825$, $F^{(V)}_{1c} = F^{(V)}_{1c^{(1)}} + 2\,
F^{(V)}_{1c^{(2)}} = 2.5633$, $G^{(V)}_{1c} = G^{(V)}_{1c^{(1)}} =
1.3393$, $H^{(V)}_{1c} = H^{(V)}_{1c^{(1)}} = 0.1791$, $G^{(W)}_{1c} =
2\, G^{(W)}_{1c^{(2)}} = - 0.2360$ and $F^{(W)}_{1c} = 2\,
F^{(W)}_{1c^{(2)}} = - 0.7040$.  The Lorentz structure of
Eq.(\ref{eq:C1c.19}) is calculated at the neglect the contributions of
order $O(E^2_0/m^2_N)\sim 10^{-6}$ and even smaller to the integrals
${\cal F}^{(1)}(k_n, k_e, q)_{1c}$ and ${\cal F}^{(2)}(k_n, k_e,
q)_{1c}$, respectively.

\subsection*{C1d. Analytical calculation of the Feynman diagram in
  Fig.\,\ref{fig:fig1}d}
\renewcommand{\theequation}
            {C1d-\arabic{equation}} \setcounter{equation}{0}

In the Feynman gauge for the photon propagator the analytical
expression for the Feynman diagram in Fig.\,\ref{fig:fig1}d is given
by (see Eq.(\ref{eq:A.1}))
\begin{eqnarray}\label{eq:C1d.1}
\hspace{-0.3in}&&M(n \to p e^- \bar{\nu}_e)^{(\pi^0)}_{\rm
  Fig.\,\ref{fig:fig1}d} = - 2 e^2 g^2_{\pi N} M^2_W G_V \int
\frac{d^4k}{(2\pi)^4i}\int \frac{d^4p}{(2\pi)^4i}\,\Big[\bar{u}_e
  \gamma^{\mu}(1 - \gamma^5) v_{\bar{\nu}}\Big]\,\Big[\bar{u}_p
  \gamma^5 \frac{1}{m_N - \hat{k}_n - \hat{k} - i0}\,\gamma^5\,
  u_n\Big]\nonumber\\ &&\times \, \frac{(2 k - p - q)^{\alpha_2} (2 k
  - p)_{\beta}}{[m^2_{\pi} - k^2 - i0][m^2_{\pi} - (k - q)^2 -
    i0][m^2_{\pi} - (k - p)^2 - i0]}\,\big[(p -
  q)^{\nu}\eta^{\beta\alpha_1} - (p - 2 q)^{\beta}\eta^{\nu \alpha_1}
  - q^{\alpha_1}\eta^{\beta \nu}\big]\, \frac{1}{p^2 +
  i0}\nonumber\\ &&\times \,D^{(W)}_{\alpha_1\alpha_2}(p -
q)\,D^{(W)}_{\mu\nu}(- q).
\end{eqnarray}
The calculation of Eq.(\ref{eq:C1d.1}) is similar to the calculation
of Eq.(\ref{eq:C1a.1}). Skipping standard intermediate
calculations, we adduce the final result only. We get
\begin{eqnarray}\label{eq:C1d.2}
\hspace{-0.30in}&&M(n \to p e^- \bar{\nu}_e)^{(\pi^0)}_{\rm
  Fig.\,\ref{fig:fig1}d} = - \frac{\alpha}{2\pi} \frac{g^2_{\pi
    N}}{16\pi^2}G_V \int^1_0 \!\! dx_1 \int^1_0 \!\! dx_2 \int^1_0
\!\! dx_3 \int^1_0\!\!  dx_4 \int^1_0 \!\! dx_5 \int^1_0 \!\!dx_6
\frac{\delta(1 - x_1 - x_2 - x_3 - x_4 - x_5 - x_6 )}{U^2}\nonumber\\
\hspace{-0.30in}&&\times \Big\{\Big( \frac{3}{4}\,\Gamma\Big(2 -
\frac{n}{2}\Big) + f^{(V)}_{1d}(x_1, \ldots, x_6) + g^{(W)}_{1d}(x_1,
\ldots, x_6)\, \frac{m^2_N}{M^2_W}\Big) \Big[\bar{u}_e \gamma^{\mu} (1
  - \gamma^5) v_{\bar{\nu}}\Big] \Big[\bar{u}_p \gamma_{\mu} u_n\Big]
+ f^{(W)}_{1d}(x_1, \ldots, x_6)\, \frac{m^2_N}{M^2_W} \nonumber\\
\hspace{-0.30in}&&\times \, \Big[\bar{u}_e \frac{\hat{k}_n}{m_N} (1 -
  \gamma^5) v_{\bar{\nu}}\Big]\Big[\bar{u}_pu_n\Big]\Big\},
\end{eqnarray}
where we have denoted
\begin{eqnarray}\label{eq:C1d.3}
\hspace{-0.3in} f^{(V)}_{1d}(x_1, \ldots, x_6) &=& -
\frac{3}{2}\,{\ell n}\frac{M^2_W}{m^2_N} - \frac{3}{2}\,{\ell n}
\bar{Q}, \nonumber\\
\hspace{-0.3in} g^{(W)}_{1d}(x_1, \ldots, x_6) &=&
\frac{1}{\bar{Q}}\Big[2\, \frac{Y\bar{Y}}{U^2} -
  \frac{\bar{Y}^2}{U^2}\Big] \nonumber\\
\hspace{-0.3in} f^{(W)}_{1d}(x_1, \ldots, x_6)
&=&\frac{1}{\bar{Q}}\Big[- 3\, \frac{Y^2}{U^2} - 8\,
  \frac{Y\bar{Y}}{U^2} + \frac{\bar{Y}^2}{U^2}\Big].
\end{eqnarray}
For the calculation of the integrals over the Feynman parameters we
use
\begin{eqnarray}\label{eq:C1d.4}
\hspace{-0.3in} \bar{Q} &=& \frac{1}{U}\Big(x_6 U + (x_3 + x_5 + x_6)
x^2_4\, \frac{m^2_N}{M^2_W}\Big),\nonumber\\
\hspace{-0.3in} U &=& (x_1 + x_2 + x_4)(x_3 + x_5 + x_6) + (x_5 + x_6)
x_3, \nonumber\\
\hspace{-0.3in} Y &=& - (x_3 + x_5 + x_6)x_4\;,\; \bar{Y} = - x_3 x_4.
\end{eqnarray}
After the integration over the Feynman parameters, the contribution of
the Feynman diagram in Fig.\,\ref{fig:fig1}d to the amplitude of the
neutron beta decay is equal to
\begin{eqnarray}\label{eq:C1d.5}
\hspace{-0.3in}M(n \to p e^- \bar{\nu}_e)^{(\pi^0)}_{\rm
  Fig.\,\ref{fig:fig1}d} &=& - \frac{\alpha}{2\pi} \frac{g^2_{\pi
    N}}{16\pi^2}G_V \Big\{\Big( A_{1d}\,\Gamma\Big(2 -
\frac{n}{2}\Big) + F^{(V)}_{1d} + G^{(W)}_{1d}\,
\frac{m^2_N}{M^2_W}\Big)\, \Big[\bar{u}_e \gamma^{\mu} (1 - \gamma^5)
  v_{\bar{\nu}}\Big] \Big[\bar{u}_p \gamma_{\mu} u_n\Big]\nonumber\\
\hspace{-0.3in}&+& F^{(W)}_{1d}\, \frac{m^2_N}{M^2_W}\, \Big[\bar{u}_e
  \frac{\hat{k}_n}{m_N} (1 - \gamma^5)
  v_{\bar{\nu}}\Big]\Big[\bar{u}_pu_n\Big]\Big\}.
\end{eqnarray}
The structure constants are equal to $A_{1d} = 1/8$, $F^{(V)}_{1d} = -
1.5303$, $G^{(W)}_{1d} = 1.9009$ and $F^{(W)}_{1d} = - 19.4989$,
respectively. The Lorentz structure of the matrix element
Eq.(\ref{eq:C1d.5}) is calculated at the neglect of the contributions
of order $O(k_n\cdot q/M^2_W) \sim 10^{-7}$ and $O(E^2_0/m^2_N) \sim
O(m^2_{\pi}/M^2_W) \sim 10^{-6}$, respectively.

\subsection*{C1e. Analytical calculation of the Feynman diagram in
  Fig.\,\ref{fig:fig1}e}
\renewcommand{\theequation}{C1e-\arabic{equation}}
\setcounter{equation}{0}

In the Feynman gauge for the photon propagator the analytical
expression for the Feynman diagram in Fig.\,\ref{fig:fig1}e is given
by (see Eq.(\ref{eq:A.1}))
\begin{eqnarray}\label{eq:C1e.1}
 &&M(n \to p e^- \bar{\nu}_e)^{(\pi^0)}_{\rm Fig.\,\ref{fig:fig1}e} =
  + 2 e^2 g^2_{\pi N} M^2_W G_V \nonumber\\ &&\times \int
  \frac{d^4k}{(2\pi)^4i}\int \frac{d^4p}{(2\pi)^4i}\,\Big[\bar{u}_e
    \gamma^{\mu}(1 - \gamma^5) v_{\bar{\nu}}\Big]\,\Big[\bar{u}_p
    \gamma^5 \frac{1}{m_N - \hat{k}_n -\hat{k} - \hat{p} - i0}
    \gamma^{\beta} \frac{1}{m_N - \hat{k}_n - \hat{k} - i0}\,\gamma^5
    u_n \Big] \nonumber\\ &&\times\,\frac{(2 k + 2 p - q)^{\nu} (2 k +
    p )_{\beta}}{[m^2_{\pi} - k^2 - i0][m^2_{\pi} - (k + p)^2 -
      i0][m^2_{\pi} - (k + p - q)^2 - i0]}\,\frac{1}{p^2 + i0}\,
  D^{(W)}_{\mu\nu}(- q).
\end{eqnarray}
Using the Dirac equations for a free neutron and a free proton, we
transcribe the integrand in Eq.(\ref{eq:C1e.1}) into the form
\begin{eqnarray}\label{eq:C1e.2}
 M(n \to p e^- \bar{\nu}_e)^{(\pi^0)}_{\rm Fig.\,\ref{fig:fig1}e} = +
 2 e^2 g^2_{\pi N} G_V\,\Big[\bar{u}_e \gamma_{\mu}(1 - \gamma^5)
   v_{\bar{\nu}}\Big]\,\Big[\bar{u}_p\, {\cal F}(k_n, k_e,
   q)^{\mu}_{1e} u_n\Big],
\end{eqnarray}
where we have made the replacement $M^2_W D^{(W)}_{\mu\nu}(- q) \to -
\eta_{\mu\nu}$, which is valid at the neglect of the contributions of
order $O(m_Nm_e/M^2_W) \sim 10^{-7}$, and introduced the integral
${\cal F}(k_n, k_e, q)_{1e} = \gamma_{\mu}(1 - \gamma^5) \otimes {\cal
  F}(k_n, k_e, q)^{\mu}_{1e}$ defined by
\begin{eqnarray}\label{eq:C1e.3}
 &&{\cal F}(k_n, k_e, q)_{1e} = \int \frac{d^4k}{(2\pi)^4i}\int
  \frac{d^4p}{(2\pi)^4i}\,\frac{1}{m^2_N - (k_n + k + p)^2 -
    i0}\frac{1}{m^2_N - (k_n + k)^2 - i0}\,\frac{1}{m^2_{\pi} - k^2 -
    i0} \nonumber\\
\hspace{-0.30in}&&\times \,\frac{1}{m^2_{\pi} - (k + p)^2 - i0}\,
\frac{1}{m^2_{\pi} - (k + p - q)^2 - i0}\,\frac{1}{p^2 + i0}\, \Big((2
\hat{k} + 2 \hat{p}) (1 - \gamma^5) \otimes \big(2 k^2
\hat{k} + p^2 \hat{k} + k^2 \hat{p}  \nonumber\\
\hspace{-0.30in}&& + 2 (k\cdot p) \hat{k}\big)\Big),
\end{eqnarray}
where we have kept only the leading divergent contributions that is
equivalent to the LLA.  In terms of the Feynman parameters the
integral ${\cal F}(k_n, k_e, q)_{1e}$ reads
\begin{eqnarray}\label{eq:C1e.4}
\hspace{-0.21in}&&{\cal F}(k_n, k_e, q)_{1e} = - 5!\int
\frac{d^4p}{(2\pi)^4i}\int \frac{d^4k}{(2\pi)^4i} \int^1_0\!\!\!dx_1
\int^1_0\!\!\! dx_2 \int^1_0\!\!\! dx_3 \int^1_0\!\!\! dx_4
\int^1_0\!\!\! dx_5 \int^1_0\!\!\!  dx_6\delta(1 - x_1 - x_2 - x_3 -
x_4 - x_5 - x_6)\nonumber\\
\hspace{-0.21in}&&\times (2 \hat{k} + 2\hat{p})(1 - \gamma^5) \otimes
\big(2 k^2 \hat{k} + p^2 \hat{k} + k^2 \hat{p} + 2 (k\cdot p)
\hat{k}\big) \Big(x_1(m^2_{\pi} - k^2) + x_2(m^2_{\pi} - (k + p)^2) +
x_3 (m^2_{\pi} - (k + p - q)^2) \nonumber\\
\hspace{-0.21in}&& + x_4(m^2_N - (k_n + k)^2) + x_5(m^2_N - (k_n + k +
p)^2) + x_6(- p^2) - i0\Big)^{-6}\,,
\end{eqnarray}
where $x_j$ for $j = 1,2,\ldots,6$ are the Feynman parameters
\cite{Feynman1950}.

Skipping standard intermediate calculations (see, for example,
subsections C1a and C1b), we define the integral ${\cal F}(k_n,
q)_{1e}$ in terms of the integrals over the Feynman parameters. We get
\begin{eqnarray}\label{eq:C1e.5}
\hspace{-0.21in}&&{\cal F}(k_n, k_e, q)_{1e} = \int^1_0\!\!\!dx_1
\int^1_0\!\!\! dx_2 \int^1_0\!\!\! dx_3 \int^1_0\!\!\! dx_4
\int^1_0\!\!\! dx_5 \int^1_0\!\!\!  dx_6 \frac{\delta(1 - x_1 - x_2 -
  x_3 - x_4 - x_5 - x_6)}{2^{2n}\pi^n U^{n/2}} \Big\{ - (2 n + 3)
\nonumber\\
\hspace{-0.21in}&& \times \,\frac{\displaystyle \Gamma\Big(2 -
  \frac{n}{2}\Big)\Gamma\Big(2 - \frac{n - 1}{2}\Big)}{\displaystyle
  2^{n - 3} \Gamma\Big(\frac{1}{2}\Big)}\,\Big(\frac{Q}{m^2_N}\Big)^{-
  4 + n}\gamma^{\mu}(1 - \gamma^5) \otimes \gamma_{\mu} + \ldots
\Big\},
\end{eqnarray}
where we have integrated over the virtual momenta in the
$n$-dimensional momentum space and denoted
\begin{eqnarray*}
  \hspace{-0.3in}U &=& (x_1 + x_4 + x_6)(x_2 + x_3 + x_5) + (x_1 +
  x_4)x_6,\nonumber\\
  \hspace{-0.3in}b_1 &=& \frac{(x_2 + x_3 + x_5 + x_6)a_1 - (x_2 + x_3
    + x_5) a_2}{U}
  = \nonumber\\
  \hspace{-0.3in} &=& \frac{x_3 x_6 q - (x_5 x_6 + (x_2 + x_3 + x_5 +
    x_6) x_4) k_n}{U} = \frac{X q + Y k_n}{U},\nonumber\\
   \hspace{-0.3in}b_2 &=& \frac{(x_1 + x_2 + x_3 + x_4 + x_5)a_2 -
     (x_2 + x_3 + x_5) a_1}{U} = \nonumber\\
  \hspace{-0.3in}&=& \frac{(x_1 + x_4)x_3 q + (- x_1 x_5 + (x_2 + x_3)
    x_4) k_n}{U} = \frac{\bar{X} q + \bar{Y} k_n}{U},\nonumber\\
\end{eqnarray*}
 \begin{eqnarray}\label{eq:C1e.6}    
   \hspace{-0.3in}Q &=& (x_1 + x_2 + x_3) m^2_{\pi} - x_3 q^2 +
   \frac{(x_2 + x_3 + x_5 + x_6) a^2_1 - 2 (x_2 + x_3 + x_5) a_1\cdot
     a_2 + (x_1 + x_2 + x_3 + x_4 + x_5) a^2_2}{U}\nonumber\\
   \hspace{-0.3in}&&
\end{eqnarray}
with $a_1 = x_3 q - (x_4 + x_5) k_n$ and $a_2 = x_3 q - x_5
k_n$. Taking the limit $n \to 4$ and keeping a divergent contribution
proportional to $\Gamma(2 - n/2)$, we transcribe the integral ${\cal
  F}(k_n, k_e, q)_{1e}$ into the form
\begin{eqnarray}\label{eq:C1e.7}
\hspace{-0.3in}&&{\cal F}(k_n, k_e, q)_{1e} = \int^1_0\!\!\!dx_1
\int^1_0\!\!\! dx_2 \int^1_0\!\!\! dx_3 \int^1_0\!\!\! dx_4
\int^1_0\!\!\! dx_5 \int^1_0\!\!\!  dx_6 \frac{\delta(1 - x_1 - x_2 -
  x_3 - x_4 - x_5 - x_6)}{256\pi^4 U^2} \Big\{ \Big(- \frac{11}{2}\,
  \Gamma\Big(2 - \frac{n}{2}\Big) \nonumber\\
\hspace{-0.3in}&& + 11\,{\ell n}\frac{Q}{m^2_N}\Big)\, \gamma^{\mu}(1
- \gamma^5) \otimes \gamma_{\mu} + \ldots\Big\}.
\end{eqnarray}
Having neglected the contributions of order $O(E^2_0/m^2_N) \sim
10^{-6}$, we arrive at the expression
\begin{eqnarray}\label{eq:C1e.8}
\hspace{-0.3in}&&{\cal F}(k_n, k_e, q)_{1e} = \int^1_0dx_1
\int^1_0dx_2 \int^1_0dx_3 \int^1_0dx_4 \int^1_0dx_5 \int^1_0
dx_6\,\frac{\delta(1 - x_1 - x_2 - x_3 - x_4 - x_5 - x_6)}{256 \pi^4
  U^2}\nonumber\\
\hspace{-0.3in}&&\times \Big\{- \frac{11}{2}\, \Gamma\Big(2 -
\frac{n}{2}\Big)\,\gamma^{\mu}(1 - \gamma^5) \otimes \gamma_{\mu} +
\Big(f^{(V)}_{1e}(x_1,\ldots,x_6) +
g^{(V)}_{1e}(x_1,\ldots,x_6)\,\frac{k_n\cdot q}{m^2_N}\Big)
\,\gamma^{\mu}(1 - \gamma^5) \otimes \gamma_{\mu} + \ldots\Big\},
\nonumber\\
\hspace{-0.21in}&&
\end{eqnarray}
where we have denoted
\begin{eqnarray}\label{eq:C1e.9}
\hspace{-0.3in}f^{(V)}_{1e}(x_1,\ldots,x_6) = 11\,{\ell
  n}\frac{\bar{Q}}{m^2_N}\quad,\quad g^{(V)}_{1e}(x_1,\ldots,x_6) =
22\,\frac{V}{U}\,\frac{m^2_N}{\bar{Q}}.
\end{eqnarray}
For the calculation of the integrals over the Feynman parameters
we use
\begin{eqnarray}\label{eq:C1e.10}
\hspace{-0.3in} \frac{\bar{Q}}{m^2_N} &=& \frac{1}{U}\Big((x_1 + x_2 +
x_3) U \frac{m^2_{\pi}}{m^2_N} + x_6 (x_4 + x_5)^2 + (x_2 + x_3 + x_5)
x^2_4 + (x_1 + x_4) x^2_5\Big),\nonumber\\
\hspace{-0.3in} V &=& - (x_1 + x_4) x_3 x_5 - (x_4 + x_5) x_3 x_6,
\end{eqnarray}
where $\bar{Q} = Q\big|_{q = 0}$. A deviation of $Q$ from $\bar{Q}$ is
taken into account in the linear approximation in powers of $k_n \cdot
q/m^2_N$ (or $q_0/m_N)$. The structure function
$g^{(V)}_{1e}(x_1,\ldots,x_6)$ we obtain expanding the logarithmic
function $11\,{\ell n}Q/m^2_N$ in powers of $k_n \cdot q/m^2_N$ (or
$q_0/m_N$).

Thus, the contribution of the Feynman diagram in Fig.\,\ref{fig:fig1}e
to the amplitude of the neutron beta decay is equal to
\begin{eqnarray}\label{eq:C1e.11}
\hspace{-0.3in}M(n \to p e^- \bar{\nu}_e)^{(\pi^0)}_{\rm
  Fig.\,\ref{fig:fig1}e} = \frac{\alpha}{2\pi}\frac{g^2_{\pi
    N}}{16\pi^2}\,G_V \Big\{\Big(A_{1e}\Gamma\Big(2 - \frac{n}{2}\Big)
+ F^{(V)}_{1e} + G^{(V)}_{1e}\,\frac{k_n\cdot
  q}{m^2_N}\Big)\Big[\bar{u}_e \gamma^{\mu}(1 - \gamma^5)
  v_{\bar{\nu}}\Big]\Big[\bar{u}_p\gamma_{\mu}u_n\Big] \Big\},
\end{eqnarray}
where the structure constants are equal to $A_{1e} = - 11/12$,
$F^{(V)}_{1e} = - 3.6949$ and $G^{(V)}_{1e} = - 1.7594$.  The Lorentz
structure of Eq.(\ref{eq:C1e.11}) is calculated at the neglect the
contributions of order $O(E^2_0/m^2_N)\sim 10^{-6}$.

\subsection*{C1f. Analytical calculation of the Feynman diagram in
  Fig.\,\ref{fig:fig1}f}
\renewcommand{\theequation}{C1f-\arabic{equation}}
\setcounter{equation}{0}

In the Feynman gauge for the photon propagator the analytical
expression for the Feynman diagram in Fig.\,\ref{fig:fig1}f is given
by (see Eq.(\ref{eq:A.1}))
\begin{eqnarray}\label{eq:C1f.1}
\hspace{-0.3in}&&M(n \to p e^- \bar{\nu}_e)^{(\pi^0)}_{\rm
  Fig.\,\ref{fig:fig1}f} = + 2 e^2 g^2_{\pi N} M^2_W
G_V\nonumber\\ &&\times \int \frac{d^4k}{(2\pi)^4i}\int
\frac{d^4p}{(2\pi)^4i}\,\Big[\bar{u}_e \gamma^{\mu}(1 - \gamma^5)
  v_{\bar{\nu}} \Big]\,\Big[\bar{u}_p \gamma^{\beta} \frac{1}{m_N -
    \hat{k}_p + \hat{p} - i0} \gamma^5 \frac{1}{m_N - \hat{k}_n -
    \hat{k} - i0} \,\gamma^5 u_n\Big] \nonumber\\
\hspace{-0.3in}&& \times \,\frac{(2 k + 2 p - q)^{\nu} (2 k +
    p )_{\beta}}{[m^2_{\pi} - k^2 - i0][m^2_{\pi} - (k + p)^2 -
      i0][m^2_{\pi} - (k + p - q)^2 - i0]}\, \frac{1}{p^2 + i 0} \,
  D^{(W)}_{\mu\nu}(- q).
\end{eqnarray}
Using the Dirac equations for a free neutron and a free proton, we
define the r.h.s. of Eq.(\ref{eq:C1f.1}) as follows
\begin{eqnarray}\label{eq:C1f.2}
 M(n \to p e^- \bar{\nu}_e)^{(\pi^0)}_{\rm Fig.\,\ref{fig:fig1}f} = +
 2 e^2 g^2_{\pi N} G_V\,\Big[\bar{u}_e \gamma_{\mu}(1 - \gamma^5)
   v_{\bar{\nu}}\Big]\,\Big[\bar{u}_p\, {\cal F}(k_n, k_e, q)^{\mu}_{1f}
   u_n\Big],
\end{eqnarray}
where we have made the replacement $M^2_W D^{(W)}_{\mu\nu}(- q) \to -
\eta_{\mu\nu}$, which is valid at the neglect of the contributions of
order $O(m_Nm_e/M^2_W) \sim 10^{-7}$, and introduced the integral
${\cal F}(k_n, k_e, q)_{1f} = \gamma_{\mu}(1 - \gamma^5) \otimes {\cal
  F}(k_n, k_e, q)^{\mu}_{1f}$ defined by
\begin{eqnarray}\label{eq:C1f.3}
\hspace{-0.3in} &&{\cal F}(k_n, k_e, q)_{1f} = - \int
\frac{d^4k}{(2\pi)^4i}\int \frac{d^4p}{(2\pi)^4i}\,\frac{1}{m^2_N -
  (k_p - p)^2 - i0}\frac{1}{m^2_N - (k_n + k)^2 -
  i0}\frac{1}{m^2_{\pi} - k^2 - i0}\frac{1}{m^2_{\pi} - (k + p)^2 -
  i0} \nonumber\\
\hspace{-0.3in}&& \times \, \frac{1}{m^2_{\pi} - (k + p - q)^2 -
  i0}\frac{1}{p^2 + i0}\,\Big((2 \hat{k} + 2 \hat{p}) (1 - \gamma^5)
\otimes \big( p^2 \hat{k}- 2 k^2 \hat{p} + 4 (k\cdot
p)\hat{k}\big)\Big),
\end{eqnarray}
where we have kept only leading divergent contributions.  In terms of
the integrals over the Feynman parameters the integral ${\cal
  F}(k_n,q)_{1f}$ reads
\begin{eqnarray}\label{eq:C1f.4}
\hspace{-0.21in}&&{\cal F}(k_n, k_e, q)_{1f} =  5!\int
\frac{d^4p}{(2\pi)^4i}\int \frac{d^4k}{(2\pi)^4i} \int^1_0\!\!\!dx_1
\int^1_0\!\!\! dx_2 \int^1_0\!\!\! dx_3 \int^1_0\!\!\! dx_4
\int^1_0\!\!\! dx_5 \int^1_0\!\!\!  dx_6\delta(1 - x_1 - x_2 - x_3 -
x_4 - x_5 - x_6) \nonumber\\
\hspace{-0.21in}&& \times \Big((2 \hat{k} + 2\hat{p})(1 - \gamma^5)
\otimes \big(p^2 \hat{k} - 2 k^2 \hat{p} + 4 (k\cdot
p)\hat{k}\big)\Big) \Big(x_1(m^2_{\pi} - k^2) + x_2(m^2_{\pi} - (k +
p)^2) + x_3 (m^2_{\pi} - (k + p - q)^2) \nonumber\\
\hspace{-0.21in}&& + x_4(m^2_N - (k_n + k)^2) + x_5(m^2_N - (k_p -
p)^2) + x_6(- p^2) - i0\Big)^{-6},
\end{eqnarray}
where $x_j$ for $j = 1,2,\ldots,6$ are the Feynman parameters
\cite{Feynman1950}.  Skipping standard intermediate calculations (see
subsection C1b), we define the integral ${\cal F}(k_n, k_e, q)_{1f}$ in
terms of the integrals over the Feynman parameters. We get
\begin{eqnarray}\label{eq:C1f.5}
\hspace{-0.21in}&&{\cal F}(k_n, k_e, q)_{1f} = \int^1_0\!\!\!dx_1
\int^1_0\!\!\! dx_2 \int^1_0\!\!\! dx_3 \int^1_0\!\!\! dx_4
\int^1_0\!\!\! dx_5 \int^1_0\!\!\!  dx_6 \frac{\delta(1 - x_1 - x_2 -
  x_3 - x_4 - x_5 - x_6)}{2^{2n}\pi^n U^{n/2}} \nonumber\\
\hspace{-0.21in}&& \times \Bigg\{ \Big(2 -
\frac{n}{2}\Big)\frac{\displaystyle \Gamma\Big(2 -
  \frac{n}{2}\Big)\Gamma\Big(2 - \frac{n - 1}{2}\Big)}{\displaystyle
  2^{n - 3} \Gamma\Big(\frac{1}{2}\Big)}\,\Big(\frac{Q}{m^2_N}\Big)^{-
  4 + n}\gamma^{\mu}(1 - \gamma^5) \otimes \gamma_{\mu} + \ldots
\Bigg\},
\end{eqnarray}
where we have integrated over the virtual momenta in the
$n$-dimensional momentum space and denoted
\begin{eqnarray}\label{eq:C1f.6}
  \hspace{-0.3in}U &=& (x_1 + x_2 + x_3 + x_4)(x_5 + x_6) + (x_1 +
  x_4) (x_2 + x_3),\nonumber\\
  \hspace{-0.3in}b_1 &=& \frac{(x_2 + x_3 + x_5 + x_6) a_1 - (x_2 +
    x_3) a_2}{U} = \nonumber\\
  \hspace{-0.3in} &=& \frac{(x_3 x_6 - x_2 x_5) q - ((x_2 + x_3)(x_4 +
    x_5) + (x_5 + x_6) x_4) k_n}{U} = \frac{X q + Y k_n}{U}
  ,\nonumber\\
   \hspace{-0.3in}b_2 &=& \frac{(x_1 + x_2 + x_3 + x_4)a_2 - (x_2 +
     x_3) a_1}{U} = \nonumber\\
  \hspace{-0.3in}&=& \frac{((x_1 + x_4)x_3 + (x_1 + x_2 + x_3 + x_4)
    x_5) q + ( (x_2 + x_3) x_4 + (x_1 + x_2 + x_3 + x_4) x_5) k_n}{U}
  = \frac{\bar{X} q + \bar{Y} k_n}{U},\nonumber\\
   \hspace{-0.3in}Q &=& (x_1 + x_2 + x_3) m^2_{\pi} 
   - x_3  q^2 \nonumber\\
  \hspace{-0.3in}&+& \frac{(x_2 + x_3 + x_5 + x_6) a^2_1 - 2 (x_2 +
    x_3) a_1\cdot a_2 + (x_1 + x_2 + x_3 + x_4) a^2_2}{U}
\end{eqnarray}
with $a_1 = x_3 q - x_4 k_n$ and $a_2 = (x_3 + x_5) q + x_5
k_n$. Taking the limit $n \to 4$ we transcribe the integral ${\cal
  F}(k_n, k_e, q)_{1f}$ into the form
\begin{eqnarray}\label{eq:C1f.7}
\hspace{-0.3in}&&{\cal F}(k_n, k_e, q)_{1f} = \int^1_0dx_1
\int^1_0dx_2 \int^1_0dx_3 \int^1_0dx_4 \int^1_0dx_5 \int^1_0
dx_6\,\frac{\delta(1 - x_1 - x_2 - x_3 - x_4 - x_5 - x_6)}{256 \pi^4
  U^2}\nonumber\\
\hspace{-0.3in}&& \times \,f^{(V)}_{1f}(x_1,\ldots,x_6)\, \gamma^{\mu}
(1 - \gamma^5) \otimes \gamma_{\mu},
\end{eqnarray}
where we have denoted
\begin{eqnarray}\label{eq:C1f.8}
\hspace{-0.3in}f^{(V)}_{1f}(x_1,\ldots,x_6) = \frac{1}{2}.
\end{eqnarray}
Thus, the contribution of the Feynman diagram in Fig.\,\ref{fig:fig1}f
to the amplitude of the neutron beta decay is equal to
\begin{eqnarray}\label{eq:C1f.9}
\hspace{-0.3in}&&M(n \to p e^- \bar{\nu}_e)^{(\pi^0)}_{\rm
  Fig.\,\ref{fig:fig1}f} = \frac{\alpha}{2\pi}\frac{g^2_{\pi
    N}}{16\pi^2}\,G_V F^{(V)}_{1f}\Big[\bar{u}_e \gamma^{\mu}(1 -
  \gamma^5) v_{\bar{\nu}}\Big]\Big[\bar{u}_p \gamma_{\mu} u_n\Big],
\end{eqnarray}
where the structure constant is equal to $F^{(V)}_{1f} = 0.0573$.  The
Lorentz structure of Eq.(\ref{eq:C1f.9}) is calculated at the neglect
of the contributions of order $O(E^2_0/m^2_N)\sim 10^{-6}$.

\subsection*{C1g. Analytical calculation of the Feynman diagram in
  Fig.\,\ref{fig:fig1}g}
\renewcommand{\theequation}{C1g-\arabic{equation}}
\setcounter{equation}{0}

In the Feynman gauge for the photon propagator the analytical
expression for the Feynman diagram in Fig.\,\ref{fig:fig1}g is given
by (see Eq.(\ref{eq:A.1}))
\begin{eqnarray}\label{eq:C1g.1}
&&M(n \to p e^- \bar{\nu}_e)^{(\pi^0)}_{\rm Fig.\,\ref{fig:fig1}g} = -
  2 e^2 g^2_{\pi N} M^2_W G_V\nonumber\\ &&\times \int
  \frac{d^4k}{(2\pi)^4i}\int \frac{d^4p}{(2\pi)^4i}\,\Big[\bar{u}_e
    \gamma^{\beta} \,\frac{1}{m_e - \hat{p} - \hat{k}_e -
      i0}\,\gamma^{\mu}(1 - \gamma^5)
    v_{\bar{\nu}}\Big]\,\Big[\bar{u}_p \gamma^5 \frac{1}{m_N -
      \hat{k}_n - \hat{k} - i0}\,\gamma^5 u_n \Big]
  \nonumber\\ &&\times\,\frac{(2 k - p - q)^{\nu} (2 k - p
    )_{\beta}}{[m^2_{\pi} - k^2 - i0][m^2_{\pi} - (k - p)^2 -
      i0][m^2_{\pi} - (k - q)^2 - i0]}\, \frac{1}{p^2 + i0} \,
  D^{(W)}_{\mu\nu}(p - q).
\end{eqnarray}
Taking the propagator of the electroweak $W^-$-boson in the form
Eq.(\ref{eq:C1a.2})
\begin{eqnarray*}
 D^{(W)}_{\mu\nu}(p - q) = - \frac{1}{M^2_W}\Big(\eta_{\mu\nu} +
 \frac{(p - q)^2 \eta_{\mu\nu} - (p - q)_{\mu}(p - q)_{\nu}}{M^2_W -
   (p - q)^2 - i0}\Big)
\end{eqnarray*}
and using the Dirac equations for free fermions we transcribe the
matrix element Eq.(\ref{eq:C1g.1}) as follows
\begin{eqnarray}\label{eq:C1g.2}
M(n \to p e^- \bar{\nu}_e)^{(\pi^0)}_{\rm Fig.\,\ref{fig:fig1}g} = - 2
e^2 g^2_{\pi N}G_V \Big[\bar{u}_e \Big({\cal F}^{(1)}(k_n, k_e,
  q)^{\mu}_{1g} + 2 {\cal F}^{(2)}(k_n, k_e, q)^{\mu}_{1g}\Big)
  v_{\bar{\nu}}\Big]\,\Big[\bar{u}_p\gamma_{\mu} u_n\Big].
\end{eqnarray}
Thus, the calculation of the Feynman diagram in Fig.\,\ref{fig:fig1}g
reduces to the calculation of the integrals ${\cal F}^{(1)}(k_n, k_e,
q)_{1g} =\\ = {\cal F}^{(1)}(k_n, k_e, q)^{\mu}_{1g}\otimes \gamma_{\mu}$
and ${\cal F}^{(2)}(k_n, k_e, q)_{1g} = {\cal F}^{(2)}(k_n, k_e,
q)^{\mu}_{1g}\otimes \gamma_{\mu}$, where the integrals ${\cal
  F}^{(1)}(k_n, q)_{1g}$ and ${\cal F}^{(2)}(k_n, k_e, q)_{1g}$ are
given by
\begin{eqnarray}\label{eq:C1g.3}
  \hspace{-0.3in}&&{\cal F}^{(1)}(k_n, k_e, q)_{1g} = \int
  \frac{d^4k}{(2\pi)^4i}\int \frac{d^4p}{(2\pi)^4i}\,\frac{1}{m^2_e -
    (p + k_e)^2 - i0}\,\frac{1}{m^2_N - (k_n + k)^2 -
    i0}\,\frac{1}{m^2_{\pi} - k^2 - i0}\nonumber\\
\hspace{-0.3in}&&\times \, \frac{1}{m^2_{\pi} - (k - p)^2 -
  i0}\,\frac{1}{m^2_{\pi} - (k - q)^2 - i0}\,\frac{1}{p^2 + i0}\,
\Big(\big(p^2 \hat{p} - 4 k^2 \hat{p} + 8 (p\cdot k) \hat{k} - 4 p^2
\hat{k}\big) (1 - \gamma^5) \otimes \hat{k}\Big)
\end{eqnarray}
and 
\begin{eqnarray}\label{eq:C1g.4}
\hspace{-0.3in}&&{\cal F}^{(2)}(k_n, k_e, q)_{1g} = \int
\frac{d^4k}{(2\pi)^4i}\int \frac{d^4p}{(2\pi)^4i} \frac{1}{m^2_e - (p
  + k_e)^2 - i0} \frac{1}{m^2_N - (k_n + k)^2 - i0}\nonumber\\
\hspace{-0.3in}&&\times \frac{1}{m^2_{\pi} - k^2 - i0}\,
\frac{1}{m^2_{\pi} - (k - p)^2 - i0}\,\frac{1}{m^2_{\pi} - (k - q)^2 -
  i0}\,\frac{1}{p^2 + i0}\, \frac{1}{M^2_W - (p - q)^2 -
  i0}\nonumber\\
\hspace{-0.3in}&&\times \,\Big(p^2\big(- p^2 \hat{k} - 2 k^2 \hat{p} +
2 (k\cdot p) \hat{k} + (k\cdot p) \hat{p} \big) (1 - \gamma^5) \otimes
\hat{k}\Big),
\end{eqnarray}
where we have kept only leading divergent contributions.

\subsubsection*{\bf Calculation of the integral
  ${\cal F}^{(1)}(k_n, k_e, q)_{1g}$}

Skipping standard intermediate calculations, we define the integral
${\cal F}^{(1)}(k_n, k_e, q)_{1g}$ in terms of the integrals over the
Feynman parameters. We get
\begin{eqnarray}\label{eq:C1g.5}
\hspace{-0.3in}&&{\cal F}^{(1)}(k_n, k_e, q)_{1g} = \int^1_0dx_1
\int^1_0dx_2 \int^1_0dx_3 \int^1_0dx_4 \int^1_0dx_5 \int^1_0
dx_6\,\frac{\delta(1 - x_1 - x_2 - x_3 - x_4 - x_5 - x_6)}{2^{2n}
  \pi^n U^{n/2}}\nonumber\\
\hspace{-0.3in}&&\times \Bigg\{n\,\frac{\displaystyle \Gamma\Big(2 -
  \frac{n}{2}\Big)\Gamma\Big(2 - \frac{n - 1}{2}\Big)}{\displaystyle
  2^{n - 3}\Gamma\Big(\frac{1}{2}\Big)}\,\Big(\frac{Q}{m^2_N}\Big)^{-4
  + n}\,\gamma^{\mu} (1 - \gamma^5) \otimes \gamma_{\mu} +
\ldots\Bigg\},
\end{eqnarray}
where we have denoted
\begin{eqnarray*}
  \hspace{-0.3in}U &=& (x_1 + x_3 + x_4)(x_2 + x_5 + x_6) + (x_5 +
  x_6) x_2,\nonumber\\
  \hspace{-0.3in}b_1 &=& \frac{(x_2 + x_5 + x_6)a_1 + x_2 a_2}{U} =
  \frac{(x_2 + x_5 + x_6)x_3 q - (x_2 + x_5 + x_6)x_4 k_n - x_2 x_5
    k_e}{U} = \frac{X q + Y k_n + Z k_e}{U},\nonumber\\
   \hspace{-0.3in}b_2 &=& \frac{(x_1 + x_2 + x_3 + x_4)a_2 + x_2
     a_1}{U} = \frac{x_2 x_3 q - x_2 x_4 k_n - (x_1 + x_2 + x_3 + x_4)
     x_5 k_e}{U} = \frac{\bar{X} q + \bar{Y} k_n + \bar{Z}
     k_e}{U},\nonumber\\
\end{eqnarray*}
 \begin{eqnarray}\label{eq:C1g.6}  
   \hspace{-0.3in}Q &=&(x_1 + x_2 + x_3) m^2_{\pi} - x_3 q^2 +
   \frac{(x_2 + x_5 + x_6) a^2_1 + 2 x_2 a_1\cdot a_2 + (x_1 + x_2 +
     x_3 + x_4) a^2_2}{U}
\end{eqnarray}
with $a_1 = x_3 q - x_4 k_n$ and $a_2 = - x_5 k_e$. Taking the limit
$n \to 4$ and keeping the divergent contribution proportional to
$\Gamma(2 - n/2)$, we transcribe the integral ${\cal F}^{(1)}(k_n,
q)_{1g}$ into the form
\begin{eqnarray}\label{eq:C1g.7}
\hspace{-0.3in}&&{\cal F}^{(1)}(k_n, k_e, q)_{1g} = \int^1_0dx_1
\int^1_0dx_2 \int^1_0dx_3 \int^1_0dx_4 \int^1_0dx_5 \int^1_0
dx_6\,\frac{\delta(1 - x_1 - x_2 - x_3 - x_4 - x_5 - x_6)}{256
  \pi^4 U^2}\nonumber\\
\hspace{-0.3in}&& \times \Big\{\Big(2\,\Gamma\Big(2 - \frac{n}{2}\Big)
  - 4\, {\ell n}\frac{Q}{m^2_N} \Big)\,\gamma^{\mu} (1 - \gamma^5)
\otimes \gamma_{\mu} + \ldots\Big\}.
\end{eqnarray}
Having neglected the contributions of order
$O(E^2_0/m^2_N) \sim 10^{-6}$, we arrive at the expression
\begin{eqnarray}\label{eq:C1g.8}
\hspace{-0.3in}&&{\cal F}^{(1)}(k_n, q)_{1g} = \int^1_0dx_1
\int^1_0dx_2 \int^1_0dx_3 \int^1_0dx_4 \int^1_0dx_5 \int^1_0
dx_6\,\frac{\delta(1 - x_1 - x_2 - x_3 - x_4 - x_5 - x_6)}{256 \pi^4
  U^2} \nonumber\\
 \hspace{-0.3in}&&\times \Big\{\Big(2\, \Gamma(2 - \frac{n}{2}\Big) +
 f^{(V)}_{1g^{(1)}}(x_1,\ldots,x_6) +
 g^{(V)}_{1g^{(1)}}(x_1,\ldots,x_6)\,\frac{k_n\cdot q}{m^2_N} +
 h^{(V)}_{1g^{(1)}}(x_1,\ldots,x_6)\,\frac{k_n\cdot k_e}{m^2_N}\Big)
 \,\gamma^{\mu}(1 - \gamma^5) \otimes \gamma_{\mu}\Big\},\nonumber\\
 \hspace{-0.3in}&&
\end{eqnarray}
where we have denoted
\begin{eqnarray}\label{eq:C1g.9}
\hspace{-0.3in}f^{(V)}_{1g^{(1)}}(x_1,\ldots,x_6) = - 4\,{\ell n}
\frac{\bar{Q}}{m^2_N}\quad,\quad g^{(V)}_{1g^{(1)}}(x_1,\ldots,x_6) =
- 8\, \frac{V}{U}\, \frac{m^2_N}{\bar{Q}}\quad,\quad
h^{(V)}_{1g^{(1)}}(x_1,\ldots,x_6) = - 8\, \frac{\bar{V}}{U}\,
\frac{m^2_N}{\bar{Q}}.
\end{eqnarray}
For the calculation of the integrals over the Feynman parameters we
use
\begin{eqnarray}\label{eq:C1g.10}
\hspace{-0.3in} \frac{\bar{Q}}{m^2_N} &=& \frac{1}{U}\Big((x_1 + x_2 +
x_3) U \frac{m^2_{\pi}}{m^2_N} + (x_2 + x_5 + x_6)
x^2_4\Big),\nonumber\\
\hspace{-0.3in} V &=& - (x_2 + x_5 + x_6) x_3 x_4 \;,\; \bar{V} = x_2
x_4 x_5,
\end{eqnarray}
where $\bar{Q} = Q\big|_{q = k_e = 0}$.  A deviation of $Q$ from
$\bar{Q}$ is taken into account in the linear approximation in the
form of an expansion in powers of $k_n\cdot q/m^2_N$ (or $q_0/m_N$)
and $k_n \cdot k_e/m^2_N$ (or $E_e/m_N$), respectively. The structure
functions $g^{(V)}_{1g^{(1)}}(x_1,\ldots,x_6)$ and
$h^{(V)}_{1g^{(1)}}(x_1,\ldots,x_6)$ we obtain by expanding the
logarithmic function $- 4\,{\ell n}Q/m^2_N$ in powers of $k_n\cdot
q/m^2_N$ (or $q_0/m_N$) and $k_n \cdot k_e/m^2_N$ (or $E_e/m_N$),
respectively.

The contribution of the integral ${\cal F}^{(1)}(k_n, k_e, q)$ of the
Feynman diagram in Fig.\,\ref{fig:fig1}g to the matrix element $M(n
\to p e^- \bar{\nu}_e)^{(\pi^0)}_{\rm Fig.\,\ref{fig:fig1}g}$ takes
the form
\begin{eqnarray}\label{eq:C1g.11}
\hspace{-0.3in}&&M(n \to p e^- \bar{\nu}_e)^{(\pi^0)}_{\rm
  Fig.\,\ref{fig:fig1}g^{(1)}} = - \frac{\alpha}{2\pi}\frac{g^2_{\pi
    N}}{16\pi^2}\,G_V \Big\{\Big(A_{1g^{(1)}} \Gamma\Big(2 -
\frac{n}{2}\Big) + F^{(V)}_{1g^{(1)}} +
G^{(V)}_{1g^{(1)}}\,\frac{k_n\cdot q}{m^2_N} +
H^{(V)}_{1g^{(1)}}\,\frac{k_n\cdot k_e}{m^2_N}\Big) \nonumber\\
\hspace{-0.3in}&& \times \Big[\bar{u}_e \gamma^{\mu}(1 - \gamma^5)
  v_{\bar{\nu}}\Big] \Big[\bar{u}_p \gamma_{\mu} u_n\Big] \Big\},
\end{eqnarray}
where the structure constants are equal to $A_{1g^{(1)}} = 1/3$,
$F^{(V)}_{1g^{(1)}} = 1.9119$, $G^{(V)}_{1g^{(1)}} = - 0.9811$ and
$H^{(V)}_{1g^{(1)}} = - 0.1791$.  The Lorentz structure of
Eq.(\ref{eq:C1g.11}) is calculated at the neglect of the contributions
of order $O(E^2_0/m^2_N)\sim 10^{-6}$.

\subsubsection*{\bf Calculation of the integral
  ${\cal F}^{(2)}(k_n, k_e, q)_{1g}$}

After merging of the denominators by using the Feynman
representation in terms the integrals over the Feynman parameters, we
transcribe the integral ${\cal F}^{(2)}(k_n, k_e, q)_{1g}$ into the
form
\begin{eqnarray}\label{eq:C1g.12}
\hspace{-0.3in}&&{\cal F}^{(2)}(k_n, k_e, q)_{1g} = \nonumber\\
\hspace{-0.3in}&& = - 6!\int\!\!\! \frac{d^4k}{(2\pi)^4i}\int\!\!\!
\frac{d^4p}{(2\pi)^4i}\int^1_0\!\!\! dx_1 \int^1_0\!\!\! dx_2
\int^1_0\!\!\! dx_3 \int^1_0\!\!\! dx_4 \int^1_0\!\!\! dx_5 \int^1_0
\!\!\! dx_6 \int^1_0 \!\!\! dx_7 \delta(1 - x_1 - x_2 - x_3 - x_4 -
x_5 - x_6 - x_7) \nonumber\\
\hspace{-0.3in}&& \times \Big(p^2\big(- p^2 \hat{k} - 2 k^2 \hat{p} +
2 (k\cdot p) \hat{k} + (k\cdot p) \hat{p} \big) (1 - \gamma^5) \otimes
\hat{k}\Big) \Big(x_1(m^2_{\pi} - k^2) + x_2 (m^2_{\pi} - (k - p)^2 +
x_3 (m^2_{\pi} - (k - q)^2) \nonumber\\
\hspace{-0.3in}&& + x_4 (- 2k \cdot k_n - k^2) + x_5 (- 2 p\cdot k_e -
p^2) + x_6(- p^2) + x_7 (M^2_W - (p - q)^2) - i0\Big)^{-7}.
\end{eqnarray}
Following the calculation of the Feynman diagram in
Fig.\,\ref{fig:fig1}c, we may assert that for the integral ${\cal
  F}^{(2)}(k_n, k_e, q)_{1g}$ we obtain the Lorentz structure being
similar to the Lorentz structure of the integral ${\cal F}^{(2)}(k_n,
k_e, q)_{1c}$ (see Eq.(\ref{eq:C1c.15})).  Skipping standard
intermediate calculations \cite{Smirnov2004, Smirnov2006, Smirnov2012},
we define the integral ${\cal F}^{(2)}(k_n, k_e, q)_{1g}$ in terms of
the integrals over the Feynman parameters only. We get
\begin{eqnarray}\label{eq:C1g.13}
\hspace{-0.3in}&&{\cal F}^{(2)}(k_n, k_e, q)_{1g} = \int^1_0 \!\!
dx_1 \int^1_0 \!\! dx_2 \int^1_0 \!\! dx_3 \int^1_0 \!\! dx_4 \int^1_0
\!\! dx_5 \int^1_0 \!\! dx_6 \int^1_0 \!\! dx_7\,\frac{\delta(1 - x_1
  - x_2 - x_3 - x_4 - x_5 - x_6 - x_7)}{2^{2n} \pi^n
  U^{n/2}}\nonumber\\
\hspace{-0.3in}&&\times \Bigg\{ - \frac{(n - 1)(n +
  2)}{8}\,\frac{\displaystyle \Gamma\Big(2 -
  \frac{n}{2}\Big)\Gamma\Big(2 - \frac{n - 1}{2}\Big)}{\displaystyle
  2^{n - 3}\Gamma\Big(\frac{1}{2}\Big)}\,\Big(\frac{Q}{m^2_N}\Big)^{-4
  + n}\,\gamma^{\mu} (1 - \gamma^5) \otimes \gamma_{\mu} -
\frac{n^2}{4} \Gamma(5 - n)\,Q^{-5+n} m^{2(4-n)}_N \nonumber\\
\hspace{-0.3in}&&\times\Big[\Big(- 2\, \frac{n + 2}{n^2} b^2_2 + 2\,
  \frac{n + 2}{n^2}\, b_1\cdot b_2 \Big)\gamma^{\mu}(1 - \gamma^5)
  \otimes \gamma_{\mu} - \frac{(n - 1)(n + 2)}{n^2}\, \hat{b}_1 (1 -
  \gamma^5) \otimes \hat{b}_1 - 2\, \frac{n + 1}{n}\, \hat{b}_2 (1 -
  \gamma^5) \otimes \hat{b}_1 \nonumber\\
\hspace{-0.3in}&& + 2\, \frac{n + 2}{n^2}\,\hat{b}_1 (1 - \gamma^5)
\otimes \hat{b}_2 + \frac{n + 4}{n^2}\,\hat{b}_2 (1 - \gamma^5)
\otimes \hat{b}_2\Big] \Bigg\},
\end{eqnarray} 
where we have denoted
\begin{eqnarray}\label{eq:C1g.14}
  \hspace{-0.3in}U &=& (x_1 + x_3 + x_4)(x_2 + x_5 + x_6 + x_7) +
  x_2(x_5 + x_6 + x_7),\nonumber\\
  \hspace{-0.3in}b_1 &=& \frac{(x_2 + x_5 + x_6 + x_7)a_1 + x_2
    a_2}{U} = - \frac{(x_2 + x_5 + x_6 + x_7)x_4 k_n }{U} =
  \frac{Y}{U}\,k_n ,\nonumber\\
   \hspace{-0.3in}b_2 &=& \frac{(x_1 + x_2 + x_3 + x_4)a_2 + x_2
     a_1}{U} = - \frac{x_2 x_4 k_n}{U} =
   \frac{\bar{Y}}{U}\,k_n,\nonumber\\
   \hspace{-0.3in}Q &=&x_7 M^2_W + \frac{(x_2 + x_5 + x_6 + x_7) a^2_1
     + 2 x_2 a_1\cdot a_2 + (x_1 + x_2 + x_3 + x_4) a^2_2}{U}
\end{eqnarray}
with $a_1 = - x_4 k_n$ and $a_2 = 0$, where we have neglected the
contributions of the terms proportional to $q$ and $k_e$, appearing in
${\cal F}^{(2)}(k_n, k_e, q)_{1g}$ in the form of an expansion in
powers of $k_n\cdot q/M^2_W$ and $k_n\cdot k_e/M^2_W$ of order
$10^{-7}$, and the terms of order $m^2_{\pi}/M^2_W \sim
10^{-6}$. Taking the limit $n \to 4$ and keeping the divergent
contribution proportional to $\Gamma(2 - n/2)$ and the contributions
of order $O(m^2_N/M^2_W)$ in the large electroweak $W^-$-boson mass
$M_W$ expansion, we transcribe Eq.(\ref{eq:C1g.13}) into the form
\begin{eqnarray}\label{eq:C1g.15}
\hspace{-0.3in}&&{\cal F}^{(2)}(k_n, k_e, q)_{1g} = \int^1_0dx_1
\int^1_0dx_2 \int^1_0dx_3 \int^1_0dx_4 \int^1_0dx_5 \int^1_0 dx_6
\int^1_0 dx_7 \,\frac{ \delta(1 - x_1 - x_2 - x_3 - x_4 - x_5 - x_6 -
  x_7)}{256 \pi^4 U^2} \nonumber\\
\hspace{-0.3in}&&\times \Big\{\Big( - \frac{9}{8}\, \Gamma\Big(2 -
\frac{n}{2}\Big) + f^{(V)}_{1g^{(2)}}(x_1, \ldots, x_7) +
g^{(W)}_{1g^{(2)}}(x_1, \ldots,
x_7)\,\frac{m^2_N}{M^2_W}\Big)\,\gamma^{\mu} (1 - \gamma^5) \otimes
\gamma_{\mu} + f^{(W)}_{1g^{(2)}}(x_1, \ldots,
x_7)\,\frac{m^2_N}{M^2_W} \nonumber\\
\hspace{-0.3in}&&\times \frac{\hat{k}_n}{m_N} (1 - \gamma^5) \otimes
1\Big\},
\end{eqnarray}
The functions of the Feynman parameters are defined by
\begin{eqnarray}\label{eq:C1g.16}
\hspace{-0.3in}f^{(V)}_{1g^{(2)}}(x_1,\ldots,x_7) &=& 
\frac{9}{4}\,{\ell n}\frac{M^2_W}{m^2_N} + \frac{9}{4}\,{\ell
  n}\bar{Q}, \nonumber\\
\hspace{-0.3in} g^{(W)}_{1g^{(2)}}(x_1,\ldots,x_7) &=&
\frac{1}{\bar{Q}}\Big[- 3 \, \frac{Y\bar{Y}}{U^2} +
  3\,\frac{\bar{Y}^2}{U^2}\Big], \nonumber\\
\hspace{-0.3in} f^{(W)}_{1g^{(2)}}(x_1,\ldots,x_7) &=&
\frac{1}{\bar{Q}}\Big[\frac{9}{2}\, \frac{Y^2}{U^2} + 7\,
  \frac{Y\bar{Y}}{U^2} - 2\, \frac{\bar{Y}^2}{U^2} \Big].
\end{eqnarray}
For the integration over the Feynman parameters we use
\begin{eqnarray}\label{eq:C1g.17}
\hspace{-0.3in}\bar{Q} &=& \frac{1}{U}\Big(x_7 U + (x_2 + x_5 + x_6 +
x_7)x^2_4\,\frac{m^2_N}{M^2_W}\Big),\nonumber\\
\hspace{-0.3in}Y&=&- (x_2 + x_5 + x_6 + x_7) x_4\;,\; \bar{Y} = - x_2
x_4.
\end{eqnarray}
where $m_N = 938.9188\,{\rm MeV}$ and $M_W = 80.379\,{\rm GeV}$ are
the nucleon and electroweak $W^-$-boson masses \cite{PDG2020}.

After the integration over the Feynman parameters the contribution of
the integral ${\cal F}^{(2)}(k_n, k_e, q)_{1g}$ to the matrix element
$M(n \to p e^-\bar{\nu}_e)^{(\pi^0)}_{\rm Fig.\,\ref{fig:fig1}g}$
is equal to
\begin{eqnarray}\label{eq:C1g.18}
\hspace{-0.3in}M(n \to p e^- \bar{\nu}_e)^{(\pi^0)}_{\rm
  Fig.\,\ref{fig:fig1}g^{(2)}} &=& - \frac{\alpha}{2\pi}\frac{g^2_{\pi
    N}}{16\pi^2}\,G_V \Big\{\Big(A_{1g^{(2)}}\Gamma\Big(2 -
\frac{n}{2}\Big) + F^{(V)}_{1g^{(2)}} + G^{(W)}_{1g^{(2)}}\,
\frac{m^2_N}{M^2_W}\Big) \Big[\bar{u}_e \gamma^{\mu}(1 - \gamma^5)
  v_{\bar{\nu}}\Big]\Big[\bar{u}_p \gamma_{\mu} u_n\Big]\nonumber\\
\hspace{-0.30in}&+&
F^{(W)}_{1g^{(2)}}\,\frac{m^2_N}{M^2_W}\Big[\bar{u}_e
  \frac{\hat{k}_n}{m_N}(1 - \gamma^5) v_{\bar{\nu}}\Big]\Big[\bar{u}_p
  u_n\Big]\Big\},
\end{eqnarray}
where the structure coefficients are equal to $A_{1g^{(2)}}= -
0.0384$, $F^{(V)}_{1g^{(2)}} = 0.5127$, $G^{(W)}_{1g^{(2)}} = -
0.1140$ and $F^{(W)}_{1g^{(2)}} = 1.4810$, respectively. The Lorentz
structure of the matrix element Eq.(\ref{eq:C1g.18}) is obtained at
the neglect of the contributions of the terms $k_n\cdot q/M^2_W$ and
$k_n\cdot k_e/M^2_W$, respectively, which are of order $10^{-7}$ and
the terms of order $O(m^2_{\pi}/M^2_W) \sim 10^{-6}$.

Thus, contribution of the Feynman diagram in Fig.\,\ref{fig:fig1}g to
the amplitude of the neutron beta decay  takes the form
\begin{eqnarray}\label{eq:C1g.19}
\hspace{-0.3in}&&M(n \to p e^- \bar{\nu}_e)^{(\pi^0)}_{\rm
  Fig.\,\ref{fig:fig1}g} = - \frac{\alpha}{2\pi}\frac{g^2_{\pi
    N}}{16\pi^2}\,G_V\, 2\,\Big\{\Big(A_{1g}\Gamma\Big(2 -
\frac{n}{2}\Big) + F^{(V)}_{1g} + G^{(V)}_{1g}\,\frac{k_n\cdot
  q}{m^2_N} + H^{(V)}_{1g}\,\frac{k_n\cdot k_e}{m^2_N} +
G^{(W)}_{1g}\, \frac{m^2_N}{M^2_W}\Big) \nonumber\\
\hspace{-0.3in}&&\times \Big[\bar{u}_e \gamma^{\mu}(1 - \gamma^5)
  v_{\bar{\nu}}\Big] \Big[\bar{u}_p\gamma_{\mu}u_n\Big] +
F^{(W)}_{1g}\,\frac{m^2_N}{M^2_W}\Big[\bar{u}_e
  \frac{\hat{k}_n}{m_N}(1 - \gamma^5) v_{\bar{\nu}}\Big]\Big[\bar{u}_p
  u_n\Big]\Big\},
\end{eqnarray}
where the coefficients are equal to $A_{1g} = A_{1g^{(1)}} + 2
A_{1g^{(2)}} = 0.2565$, $F^{(V)}_{1g} = F^{(V)}_{1g^{(1)}} + 2\,
F^{(V)}_{1g^{(2)}} = 2.9373$, $G^{(V)}_{1g} = G^{(V)}_{1g^{(1)}} = -
0.9811$, $H^{(V)}_{1g} = H^{(V)}_{1g^{(1)}} = - 0.1791$,
$G^{(W)}_{1g} = 2\, G^{(W)}_{1g^{(2)}} = - 0.2280$ and $F^{(W)}_{1g} = 2\,
F^{(W)}_{1g^{(2)}} = 2.9620$.  The Lorentz structure of the matrix element in
Eq.(\ref{eq:C1g.19}) is calculated at the neglect the contributions of
order $O(E^2_0/m^2_N)\sim 10^{-6}$ and even smaller of order $10^{-7}$
to the integrals ${\cal F}^{(1)}(k_n, k_e, q)_{1g}$ and ${\cal
  F}^{(2)}(k_n, k_e, q)_{1g}$, respectively.

\subsection*{C1h. Analytical calculation of the Feynman diagram in
  Fig.\,\ref{fig:fig1}h}
\renewcommand{\theequation}{C1h-\arabic{equation}}
\setcounter{equation}{0}

In the Feynman gauge for the photon propagator the analytical
expression of the Feynman diagram in Fig.\,\ref{fig:fig1}h is given
by (see Eq.(\ref{eq:A.1}))
\begin{eqnarray}\label{eq:C1h.1}
&&M(n \to p e^- \bar{\nu}_e)^{(\pi^0)}_{\rm Fig.\,\ref{fig:fig1}h} = -
  2 e^2 g^2_{\pi N} M^2_W G_V\nonumber\\ &&\times \int
  \frac{d^4k}{(2\pi)^4i}\int \frac{d^4p}{(2\pi)^4i}\Big[\bar{u}_e
    \gamma^{\beta}\,\frac{1}{m_e - \hat{p} - \hat{k}_e - i0}\,
    \gamma^{\mu}(1 - \gamma^5) v_{\bar{\nu}}\Big]\Big[\bar{u}_p
    \gamma^5 \frac{1}{m_N - \hat{k}_n - \hat{k} - \hat{p} - i0}
    \gamma_{\beta}\frac{1}{m_N - \hat{k}_n - \hat{k} - i0} \gamma^5
    u_n\Big]\nonumber\\ &&\times \,\frac{(2 k + p -
    q)^{\nu}}{[m^2_{\pi} - k^2 - i0][m^2_{\pi} - (k + p - q)^2 - i0]}\,
  \frac{1}{p^2 + i0}\, D^{(W)}_{\mu\nu}(p - q).
\end{eqnarray}
For the calculation of this diagram it is convenient to take the
propagator of the electroweak $W^-$-boson in the form
\begin{eqnarray*}
 D^{(W)}_{\mu\nu}(p - q) = - \frac{1}{M^2_W}\Big(\eta_{\mu\nu} +
 \frac{(p - q)^2 \eta_{\mu\nu} - (p - q)_{\mu}(p - q)_{\nu}}{M^2_W -
   (p - q)^2 - i0}\Big)
\end{eqnarray*}
and to transcribe Eq.(\ref{eq:C1h.1}) into the form
\begin{eqnarray}\label{eq:C1h.2}
M(n \to p e^- \bar{\nu}_e)^{(\pi^0)}_{\rm Fig.\,\ref{fig:fig1}h} = -
  2 e^2 g^2_{\pi N} G_V\Big[\bar{u}_e\Big({\cal F}^{(1)}(k_n,
    q)^{\beta\lambda\rho} + 2\, {\cal F}^{(2)}(k_n, q)^{\beta\lambda
      \rho} \Big)v_{\bar{\nu}}\Big]\Big[\bar{u}_p \gamma_{\lambda}
    \gamma_{\beta} \gamma_{\rho} u_n\Big],
\end{eqnarray}
where the integrals ${\cal F}^{(1)}(k_n ,k_e, q)^{\beta\lambda\rho}$
and ${\cal F}^{(2)}(k_n, k_e, q)^{\beta\lambda \rho}$ are related to
the integrals ${\cal F}^{(1)}(k_n, k_e, q)_{1h}$ and ${\cal
  F}^{(2)}(k_n, k_e, q)_{1h}$ as follows
\begin{eqnarray}\label{eq:C1h.3}
{\cal F}^{(1)}(k_n, k_e, q)_{1h} &=& {\cal F}^{(1)}(k_n, k_e,
  q)^{\beta\lambda\rho} \otimes \gamma_{\lambda} \gamma_{\beta}
  \gamma_{\rho},\nonumber\\
  {\cal F}^{(2)}(k_n, k_e, q)_{1h} &=& {\cal F}^{(2)}(k_n, k_e,
  q)^{\beta\lambda\rho} \otimes \gamma_{\lambda} \gamma_{\beta}
  \gamma_{\rho}
\end{eqnarray}
and are given by
\begin{eqnarray}\label{eq:C1h.4}
&&{\cal F}^{(1)}(k_n, k_e, q)_{1h} = \int\frac{d^4k}{(2\pi)^4i}\int
  \frac{d^4p}{(2\pi)^4i}\, \frac{1}{m^2_e - (p + k_e)^2 - i0}
  \frac{1}{m^2_N - (k_n + k + p)^2 - i0} \frac{1}{m^2_N - (k_n + k)^2
    - i0}\nonumber\\ &&\times \frac{1}{m^2_{\pi} - k^2 - i0}
  \frac{1}{m^2_{\pi} - (k + p - q)^2 - i0} \frac{1}{p^2 + i0}\, \Big(
  \gamma^{\beta} (2 \hat{p} \hat{k} + p^2) (1 - \gamma^5) \otimes
  (\hat{k} + \hat{p}) \gamma_{\beta} \hat{k}\Big)
\end{eqnarray}
and
\begin{eqnarray}\label{eq:C1h.5}
&&{\cal F}^{(2)}(k_n, k_e, q)_{1h} = \int\frac{d^4k}{(2\pi)^4i}\int
  \frac{d^4p}{(2\pi)^4i}\, \frac{1}{m^2_e - (p + k_e)^2 - i0}
  \frac{1}{m^2_N - (k_n + k + p)^2 - i0} \frac{1}{m^2_N - (k_n + k)^2
    - i0}\nonumber\\ &&\times \frac{1}{m^2_{\pi} - k^2 - i0}
  \frac{1}{m^2_{\pi} - (k + p - q)^2 - i0} \frac{1}{p^2 + i0}
  \frac{1}{M^2_W - (p - q)^2 - i0}\, \Big(p^2( \gamma^{\beta}
  \hat{p}\hat{k} - (k\cdot p) \gamma^{\beta}) (1 - \gamma^5) \otimes
  (\hat{k} + \hat{p}) \gamma_{\beta} \hat{k} \Big),\nonumber\\ &&
\end{eqnarray}
where we have kept only leading divergent contributions that
corresponds to the use of the LLA and used the Dirac equations for
free fermions.

\subsubsection*{\bf Calculation of the integral
  ${\cal F}^{(1)}(k_n, k_e, q)_{1h}$}

Skipping standard intermediate calculations (see subsections C1a and
C1b), we define the integral ${\cal F}^{(1)}(k_n, k_e, q)_{1h}$ in
terms of the integrals over the Feynman parameters. We get
\begin{eqnarray}\label{eq:C1h.6}
\hspace{-0.3in}&&{\cal F}^{(1)}(k_n, k_e, q)_{1h} = \int^1_0dx_1
\int^1_0dx_2 \int^1_0dx_3 \int^1_0dx_4 \int^1_0dx_5 \int^1_0
dx_6\,\frac{\delta(1 - x_1 - x_2 - x_3 - x_4 - x_5 - x_6)}{2^{2n}
  \pi^n U^{n/2}}\nonumber\\
\hspace{-0.3in}&&\times \Bigg\{\Big( \frac{n(n - 2)}{4}\,\gamma^{\mu}
(1 - \gamma^5) \otimes \gamma_{\mu} - \frac{1}{2}\,\gamma^{\beta}
\gamma^{\alpha} \gamma^{\mu} (1 - \gamma^5) \otimes \gamma_{\alpha}
\gamma_{\beta} \gamma_{\mu}\Big) \frac{\displaystyle \Gamma\Big(2 -
  \frac{n}{2}\Big)\Gamma\Big(2 - \frac{n - 1}{2}\Big)}{\displaystyle
  2^{n - 3}\Gamma\Big(\frac{1}{2}\Big)}\,\Big(\frac{Q}{m^2_N}\Big)^{-4
  + n} + \ldots\Bigg\}, \nonumber\\
\hspace{-0.3in}&&
\end{eqnarray}
where we have denoted
\begin{eqnarray}\label{eq:C1h.7}
  \hspace{-0.3in}U &=& (x_1 + x_4)(x_2 + x_3 + x_5 + x_6) + (x_2 +
  x_3)(x_5 + x_6),\nonumber\\
  \hspace{-0.3in}b_1 &=& \frac{(x_2 + x_3 + x_5 + x_6)a_1 - (x_2 +
    x_3) a_2}{U} = \nonumber\\
  \hspace{-0.3in} &=& \frac{(x_5 + x_6) x_2 q - ((x_2 + x_3 + x_5 +
    x_6)x_4 + (x_5 + x_6) x_3) k_n + (x_2 + x_3) x_5 k_e}{U} = \frac{X
    q + Y k_n + Z k_e}{U},\nonumber\\
   \hspace{-0.3in}b_2 &=& \frac{(x_1 + x_2 + x_3 + x_4)a_2 - (x_2 +
     x_3) a_1}{U} =\nonumber\\
  \hspace{-0.3in} &=& \frac{(x_1 + x_4) x_2 q + (x_2 x_4 - x_1 x_3)
    k_n - (x_1 + x_2 + x_3 + x_4) x_5 k_e}{U} = \frac{\bar{X} q +
    \bar{Y} k_n + \bar{Z} k_e}{U},\nonumber\\
   \hspace{-0.3in}Q &=&(x_1 + x_2) m^2_{\pi} - x_2 q^2 + \frac{(x_2 +
     x_3 + x_5 + x_6) a^2_1 - 2 (x_2 + x_3) a_1\cdot a_2 + (x_1 + x_2
     + x_3 + x_4) a^2_2}{U}
\end{eqnarray}
with $a_1 = x_2 q - (x_3 + x_4) k_n$ and $a_2 = x_2q - x_3 k_n - x_5
k_e$. Taking the limit $n \to 4$ and keeping a divergent contribution
proportional to $\Gamma(2 - n/2)$, we transcribe Eq.(\ref{eq:C1h.6})
into the form
\begin{eqnarray}\label{eq:C1h.8}
\hspace{-0.3in}&&{\cal F}^{(1)}(k_n, k_e, q)_{1h} = \int^1_0dx_1
\int^1_0dx_2 \int^1_0dx_3 \int^1_0dx_4 \int^1_0dx_5 \int^1_0
dx_6\,\frac{\delta(1 - x_1 - x_2 - x_3 - x_4 - x_5 - x_6)}{256 \pi^4
  U^2} \nonumber\\
\hspace{-0.3in}&& \times \Big\{\Big(\gamma^{\mu} (1 - \gamma^5)
\otimes \gamma_{\mu} - \frac{1}{4}\,\gamma^{\beta} \gamma^{\alpha}
\gamma^{\mu} (1 - \gamma^5) \otimes \gamma_{\alpha} \gamma_{\beta}
\gamma_{\mu}\Big)\Big[\Gamma\Big(2 - \frac{n}{2}\Big) - 2 {\ell
    n}\Big(\frac{Q}{m^2_N}\Big)\Big] + \ldots \Big\}.
\end{eqnarray}
Using the Dirac equations for fermions, the relation
between Dirac $\gamma$-matrices \cite{Itzykson1980}
\begin{eqnarray}\label{eq:C1h.9}
\gamma^{\mu}\gamma^{\beta}\gamma^{\alpha} = \gamma^{\mu}\eta^{\beta
  \alpha} - \gamma^{\beta} \eta^{\alpha \mu} + \gamma^{\alpha}
\eta^{\mu \beta} +
i\,\varepsilon^{\mu \beta \alpha \rho}\,\gamma_{\rho}\gamma^5,
\end{eqnarray}
where $\varepsilon^{\mu \beta \alpha \rho}$ is the Levi-Civita tensor
defined by $\varepsilon^{0123} = 1$ and $\varepsilon_{\mu \beta \alpha
  \rho}= - \varepsilon^{\mu \beta \alpha \rho}$, and neglecting the
contributions of order $O(E^2_0/m^2_N) \sim 10^{-6}$, we arrive at the
expression
\begin{eqnarray*}
\hspace{-0.3in}&&{\cal F}^{(1)}(k_n, k_e, q)_{1h} = \int^1_0dx_1
\int^1_0dx_2 \int^1_0dx_3 \int^1_0dx_4 \int^1_0dx_5 \int^1_0
dx_6\,\frac{\delta(1 - x_1 - x_2 - x_3 - x_4 - x_5 - x_6)}{256 \pi^4
  U^2}\nonumber\\
\hspace{-0.3in}&&\times \Big\{\Big(\frac{3}{2}\, \Gamma\Big(2 -
\frac{n}{2}\Big) + f^{(V)}_{1h^{(1)}}(x_1,\ldots,x_6) +
g^{(V)}_{1h^{(1)}}(x_1,\ldots,x_6)\,\frac{k_n\cdot q}{m^2_N} +
h^{(V)}_{1h^{(1)}}(x_1,\ldots,x_6)\,\frac{k_n\cdot k_e}{m^2_N}\Big)
\,\gamma^{\mu}(1 - \gamma^5) \otimes
\gamma_{\mu} \nonumber\\
\end{eqnarray*}
\begin{eqnarray}\label{eq:C1h.10}
\hspace{-0.3in}&& + \Big(- \frac{3}{2}\,\Gamma\Big(2 -
\frac{n}{2}\Big) + f^{(A)}_{1h^{(1)}}(x_1,\ldots,x_6) +
g^{(A)}_{1h^{(1)}}(x_1,\ldots,x_6)\,\frac{k_n\cdot q}{m^2_N} +
h^{(A)}_{1h^{(1)}}(x_1,\ldots,x_6)\,\frac{k_n\cdot k_e}{m^2_N}\Big)
\,\gamma^{\mu}(1 - \gamma^5) \otimes \gamma_{\mu} \gamma^5\nonumber\\
\hspace{-0.3in}&& + \ldots\Big\},
\end{eqnarray}
where we have denoted
\begin{eqnarray}\label{eq:C1h.11}
\hspace{-0.3in}f^{(V)}_{1h^{(1)}}(x_1, \ldots,x_6) = - 3\,{\ell
  n}\frac{\bar{Q}}{m^2_N}\quad,\quad g^{(V)}_{1h^{(1)}}(x_1,
\ldots,x_6) = - 6\,
  \frac{V}{U}\,\frac{m^2_N}{\bar{Q}}\quad,\quad  h^{(V)}_{1h^{(1)}}(x_1,
\ldots,x_6) = - 6\,
\frac{\bar{V}}{U}\,\frac{m^2_N}{\bar{Q}},\nonumber\\
\hspace{-0.3in}f^{(A)}_{1h^{(1)}}(x_1, \ldots,x_6) = + 3\,{\ell
  n}\frac{\bar{Q}}{m^2_N}\quad,\quad g^{(A)}_{1h^{(1)}}(x_1,
\ldots,x_6) = + 6\,
  \frac{V}{U}\,\frac{m^2_N}{\bar{Q}}\quad,\quad  h^{(A)}_{1h^{(1)}}(x_1,
\ldots,x_6) = + 6\,
\frac{\bar{V}}{U}\,\frac{m^2_N}{\bar{Q}}.
\end{eqnarray}
For the calculation of the integrals over the Feynman parameters we
use
\begin{eqnarray}\label{eq:C1h.12}
\hspace{-0.3in} \frac{\bar{Q}}{m^2_N} &=& \frac{1}{U}\Big((x_1 + x_2 )
U \frac{m^2_{\pi}}{m^2_N} + (x_5 + x_6)(x_3 + x_4)^2 + (x_2 + x_3)
x^2_4 + (x_1 + x_4) x^2_3\Big),\nonumber\\
\hspace{-0.3in} V &=& - x_2 (x_3 + x_4) (x_5 + x_6) - (x_1 + x_4) x_2
x_3 \;,\; \bar{V} = (x_1 x_3 - x_2 x_4) x_5,
\end{eqnarray}
where $\bar{Q} = Q\big|_{q = k_e = 0}$.  A deviation of $Q$ from
$\bar{Q}$ is taken into account in the linear approximation in the
form of an expansion in powers of $k_n\cdot q/m^2_N$ (or $q_0/m_N$)
and $k_n \cdot k_e/m^2_N$ (or $E_e/m_N$), respectively. The structure
functions $g^{(V)}_{1h^{(1)}}(x_1, \ldots,x_6)$ and
$h^{(V)}_{1h^{(1)}}(x_1, \ldots,x_6)$ as well as
$g^{(A)}_{1h^{(1)}}(x_1, \ldots,x_6)$ and $h^{(A)}_{1h^{(1)}}(x_1,
\ldots,x_6)$ are calculated as an expansion of the logarithmic
functions $- 3\,{\ell n}Q/m^2_N$ and $+ 3\,{\ell n}Q/m^2_N$ in powers
of $k_n\cdot q/m^2_N$ (or $q_0/m_N$) and $k_n \cdot k_e/m^2_N$ (or
$E_e/m_N$), respectively.

Having integrated over the Feynman parameters, we obtain the
contribution of the integral ${\cal F}^{(1)}(k_n, k_e, q)_{1h}$ to the
matrix element $M(n \to p e^- \bar{\nu}_e)^{(\pi^0)}_{\rm
  Fig.\,\ref{fig:fig1}h}$
\begin{eqnarray}\label{eq:C1h.13}
\hspace{-0.3in}&&M(n \to p e^- \bar{\nu}_e)^{(\pi^0)}_{\rm
  Fig.\,\ref{fig:fig1}h^{(1)}} = - \frac{\alpha}{2\pi}\frac{g^2_{\pi
    N}}{16\pi^2}\,G_V \Big\{\Big(A_{1h^{(1)}}\Gamma\Big(2 -
\frac{n}{2}\Big) + F^{(V)}_{1h^{(1)}} + G^{(V)}_{1h^{(1)}} \frac{k_n
  \cdot q}{m^2_N} + H^{(V)}_{1h^{(1)}} \frac{k_n \cdot
  k_e}{m^2_N}\Big)\nonumber\\
  \hspace{-0.3in}&&\times \Big[\bar{u}_e \gamma^{\mu}(1 - \gamma^5)
    v_{\bar{\nu}}\Big]\Big[\bar{u}_p \gamma_{\mu} u_n\Big] +
  \Big(B_{1h^{(1)}}\Gamma\Big(2 - \frac{n}{2}\Big) +
  F^{(A)}_{1h^{(1)}} + G^{(A)}_{1h^{(1)}} \frac{k_n \cdot q}{m^2_N} +
  H^{(A)}_{1h^{(1)}} \frac{k_n \cdot k_e}{m^2_N}\Big) \Big[\bar{u}_e
    \gamma^{\mu}(1 - \gamma^5) v_{\bar{\nu}}\Big] \nonumber\\
  \hspace{-0.3in}&&\times \Big[\bar{u}_p \gamma_{\mu}\gamma^5 u_n\Big]
  \Big\},
\end{eqnarray}
where the structure constants are equal to $A_{1h^{(1)}} = 0.1720$,
$B_{1h^{(1)}} = - 0.1720$, $F^{(V)}_{1h^{(1)}} = 0.6688$,
$G^{(V)}_{1h^{(1)}} = 0.3700$, $H^{(V)}_{1h^{(1)}} = 0$,
$F^{(A)}_{1h^{(1)}}= - 0.6688$, $G^{(A)}_{1h^{(1)}} = - 0.3700$ and
$H^{(A)}_{1h^{(1)}} = 0$. The Lorentz structure of
Eq.(\ref{eq:C1h.13}) is calculated at the neglect the contributions of
order $O(E^2_0/m^2_N)\sim 10^{-6}$.

\subsubsection*{\bf Calculation of the integral
  ${\cal F}^{(2)}(k_n, k_e, q)_{1h}$}

Skipping standard intermediate calculations, we define the integral
${\cal F}^{(2)}(k_n, k_e, q)_{1h}$ in terms of the integrals over the
Feynman parameters. We get
\begin{eqnarray}\label{eq:C1h.14}
\hspace{-0.3in}&&{\cal F}^{(2)}(k_n, k_e, q)_{1g} = \int^1_0 \!\!
dx_1 \int^1_0 \!\! dx_2 \int^1_0 \!\! dx_3 \int^1_0 \!\! dx_4 \int^1_0
\!\! dx_5 \int^1_0 \!\! dx_6 \int^1_0 \!\! dx_7\,\frac{\delta(1 - x_1
  - x_2 - x_3 - x_4 - x_5 - x_6 - x_7)}{2^{2n} \pi^n
  U^{n/2}} \nonumber\\
\hspace{-0.3in}&& \times \Bigg\{ \frac{(n + 2)}{8}\,\frac{\displaystyle
  \Gamma\Big(2 - \frac{n}{2}\Big)\Gamma\Big(2 - \frac{n -
    1}{2}\Big)}{\displaystyle 2^{n -
    3}\Gamma\Big(\frac{1}{2}\Big)}\,\Big(\frac{Q}{m^2_N}\Big)^{-4 +
  n}\,\Big[\gamma^{\beta}\gamma^{\alpha} \gamma^{\mu} (1 - \gamma^5)
  \otimes \gamma_{\alpha} \gamma_{\beta} \gamma_{\mu} + (n - 2)
  \,\gamma^{\mu} (1 - \gamma^5) \otimes \gamma_{\mu} \Big] \nonumber\\
  \hspace{-0.3in}&& - \frac{n^2}{4} \Gamma(5 - n)\,Q^{-5+n}
  m^{2(4-n)}_N \Big[ \Big( \frac{1}{n^2}\,b^2_2\,
    \gamma^{\beta}\gamma^{\alpha} \gamma^{\mu} (1 - \gamma^5) \otimes
    \gamma_{\alpha} \gamma_{\beta} \gamma_{\mu} + \frac{n - 2}{n^2}\,
    b^2_2\, \gamma^{\mu}(1 - \gamma^5) \otimes \gamma_{\mu} +
    \frac{n^2 - 4}{n^2}\, (b_1\cdot b_2)\nonumber\\
  \hspace{-0.3in}&& \times \, \gamma^{\mu}(1 - \gamma^5) \otimes
  \gamma_{\mu} \Big) + \gamma^{\beta} \hat{b}_2 \gamma^{\alpha} (1 -
  \gamma^5) \otimes \Big(\frac{n + 2}{n^2}\, \gamma_{\alpha}
  \gamma_{\beta} \hat{b}_1 + \frac{n + 2}{n^2}\, \hat{b}_1
  \gamma_{\beta} \gamma_{\alpha} + \frac{n + 4}{n^2}\, \hat{b}_2
  \gamma_{\beta} \gamma_{\alpha}\Big) \nonumber\\
\hspace{-0.3in}&& + \frac{n + 2}{n^2} \, \gamma^{\beta}
\gamma^{\alpha} \hat{b}_1 (1 - \gamma^5) \otimes \gamma_{\alpha}
\gamma_{\beta} \hat{b}_1 + \gamma^{\mu} (1 - \gamma^5) \otimes \Big( -
\frac{n + 2}{n^2}\, \hat{b}_1 \gamma_{\mu} \hat{b}_2 - \frac{n +
  2}{n^2}\, \hat{b}_2 \gamma_{\mu} \hat{b}_1 - \frac{n + 4}{n^2}\,
\hat{b}_2 \gamma_{\beta} \hat{b}_2\Big) \nonumber\\
\hspace{-0.3in}&& - \frac{n^2 - 4}{n^2}\, \gamma^{\mu} \hat{b}_2
\hat{b}_1 (1 - \gamma^5) \otimes \gamma_{\mu}\Big] \Bigg\},
\end{eqnarray} 
where we have denoted
\begin{eqnarray}\label{eq:C1h.15}
  \hspace{-0.3in}U &=& (x_1 + x_2 + x_3 + x_4)( x_5 + x_6 + x_7) +
  (x_1 + x_4)(x_2 + x_3),\nonumber\\
  \hspace{-0.3in}b_1 &=& \frac{(x_2 + x_3 + x_5 + x_6 + x_7)a_1 - (x_2
    + x_3) a_2}{U} = - \frac{(x_5 + x_6 + x_7)(x_3 + x_4) + (x_2 +
    x_3) x_4 }{U}\, k_n = \frac{Y}{U}\,k_n ,\nonumber\\
   \hspace{-0.3in}b_2 &=& \frac{(x_1 + x_2 + x_3 + x_4)a_2 - (x_2 +
     x_3) a_1}{U} = \frac{x_2 x_4 - x_1 x_3}{U}\, k_n =
   \frac{\bar{Y}}{U}\,k_n,\nonumber\\
   \hspace{-0.3in}Q &=&x_7 M^2_W + \frac{(x_2 + x_3 + x_5 + x_6 + x_7)
     a^2_1 - 2 (x_2 + x_3) a_1\cdot a_2 + (x_1 + x_2 + x_3 + x_4)
     a^2_2}{U}
\end{eqnarray}
with $a_1 = - (x_3 + x_4) k_n$ and $a_2 = - x_3 k_n$. In
Eq.(\ref{eq:C1h.14}) we have neglected the contributions of the terms
proportional to $q$ and $k_e$, appearing in ${\cal F}^{(2)}(k_n, k_e,
q)_{1h}$ in the form of an expansion in powers of $k_n\cdot q/M^2_W$
and $k_n\cdot k_e/M^2_W$ of order $10^{-7}$, and the terms of order
$O(m^2_{\pi}/M^2_W) \sim 10^{-6}$. Taking the limit $n \to 4$ and
keeping i) a divergent contribution proportional to $\Gamma(2 - n/2)$
and ii) the contributions of order $O(m^2_N/M^2_W)$ in the large
electroweak $W^-$-boson mass $M_W$ expansion, using the Dirac
equations for a free neutron and a free proton and the relation
between Dirac matrices Eq.(\ref{eq:C1h.9}), we transcribe
Eq.(\ref{eq:C1h.14}) into the form
\begin{eqnarray}\label{eq:C1h.16}
 \hspace{-0.3in}&&{\cal F}^{(2)}(k_n, k_e, q)_{1h} =  \int^1_0dx_1
 \int^1_0dx_2 \int^1_0dx_3 \int^1_0dx_4 \int^1_0dx_5 \int^1_0 dx_6
 \int^1_0 dx_7\,\frac{\delta(1 - x_1 - x_2 - x_3 - x_4 - x_5 - x_6 -
   x_7)}{256 \pi^4 U^2}\nonumber\\
\hspace{-0.3in}&& \times \Big\{ \Big( f^{(V)}_{1h^{(2)}}(x_1,\ldots,
x_7) + g^{(W)}_{1h^{(2)}}(x_1,\ldots, x_7)\,\frac{m^2_N}{M^2_W}\Big)\,
\gamma^{\mu}(1 - \gamma^5) \otimes \gamma_{\mu} + \Big(\frac{9}{4}\,
\Gamma\Big(2 - \frac{n}{2}\Big) + f^{(A)}_{1h^{(2)}}(x_1,\ldots,
x_7)\nonumber\\
\hspace{-0.3in}&& + h^{(W)}_{1h^{(2)}}(x_1,\ldots,
x_7)\,\frac{m^2_N}{M^2_W}\Big)\, \gamma^{\mu} (1 - \gamma^5) \otimes
\gamma_{\mu} \gamma^5 + f^{(W)}_{1h^{(2)}}(x_1,\ldots,
x_7)\,\frac{m^2_N}{M^2_W}\frac{\hat{k}_n}{m_N}(1 - \gamma^5) \otimes
1\Big\},
\end{eqnarray}
where we have denoted
\begin{eqnarray}\label{eq:C1h.17}
  \hspace{-0.3in}f^{(V)}_{1h^{(2)}}(x_1,\ldots, x_7) &=& -
  \frac{3}{4}, \nonumber\\
  \hspace{-0.3in}g^{(W)}_{1h^{(2)}}(x_1,\ldots, x_7) &=&
  \frac{1}{\bar{Q}}\Big[3\,\frac{Y^2}{U^2} + 9\,\frac{Y\bar{Y}}{U^2} -
    2\, \frac{\bar{Y}^2}{U^2}\Big], \nonumber\\
   \hspace{-0.3in}f^{(A)}_{1h^{(2)}}(x_1,\ldots, x_7) &=& -
   \frac{9}{2}{\ell n} \frac{M^2_W}{m^2_N} - \frac{9}{2}{\ell
     n} \bar{Q},\nonumber\\
   \hspace{-0.3in}h^{(W)}_{1h^{(2)}}(x_1,\ldots, x_7) &=& 
   \frac{1}{\bar{Q}}\Big[3\, \frac{Y^2}{U^2} +
     \frac{11}{2}\,\frac{\bar{Y}^2}{U^2} \Big],\nonumber\\
   \hspace{-0.3in}f^{(W)}_{1h^{(2)}}(x_1,\ldots, x_7) &=&
   \frac{1}{\bar{Q}}\Big[ - 3\,\frac{Y^2}{U^2} +
     6\,\frac{Y\bar{Y}}{U^2} + 4\, \frac{\bar{Y}^2}{U^2}\Big].
\end{eqnarray}
For the calculation of the integrals over the Feynman parameters we use
\begin{eqnarray}\label{eq:C1h.18}
   \hspace{-0.3in}\frac{Q}{m^2_N} &=& \frac{M^2_W}{m^2_N}\bar{Q} \;,\;
   \bar{Q} = \frac{1}{U}\Big(x_7 U + \big((x_5 + x_6 + x_7)(x_3 +
   x_4)^2 + (x_2 + x_3) x^2_4 + (x_1 + x_4) x^2_3\big) \,
   \frac{m^2_N}{M^2_W}\Big),\nonumber\\
   \hspace{-0.3in} Y &=& - (x_5 + x_6 + x_7)(x_3 + x_4) - (x_2 + x_3)
   x_4\;,\; \bar{Y} = x_2 x_4 - x_1 x_3.
\end{eqnarray}
Having integrated Eq.(\ref{eq:C1h.16}) over the Feynman parameters, we
arrive at the contribution of the integral ${\cal F}^{(2)}(k_n, k_e,
q)_{1h}$ to the matrix element $M(n \to p e^-
\bar{\nu}_e)^{(\pi^0)}_{\rm Fig.\,\ref{fig:fig1}h}$. It reads
\begin{eqnarray}\label{eq:C1h.19}
\hspace{-0.21in}&&M(n \to p e^- \bar{\nu}_e)^{(\pi^0)}_{\rm
  Fig.\,\ref{fig:fig1}h^{(2)}} = - \frac{\alpha}{2\pi}\frac{g^2_{\pi
    N}}{16\pi^2}\,G_V \Big\{\Big(A_{1h^{(2)}}\, \Gamma\Big(2 -
\frac{n}{2}\Big) + F^{(V)}_{1h^{(2)}} + G^{(W)}_{1h^{(2)}}
\frac{m^2_N}{M^2_W}\Big)\, \Big[\bar{u}_e \gamma^{\mu}(1 - \gamma^5)
  v_{\bar{\nu}}\Big] \Big[\bar{u}_p \gamma_{\mu} u_n\Big] \nonumber\\
\hspace{-0.21in}&& + \Big(B_{1h^{(2)}}\Gamma\Big(2 -
\frac{n}{2}\Big) + F^{(A)}_{1h^{(2)}} + H^{(W)}_{1h^{(2)}}
\frac{m^2_N}{M^2_W}\Big) \Big[\bar{u}_e \gamma^{\mu}(1 - \gamma^5)
  v_{\bar{\nu}}\Big] \Big[\bar{u}_p \gamma_{\mu}\gamma^5 u_n\Big] +
F^{(W)}_{1h^{(2)}}\, \frac{m^2_N}{M^2_W} \Big[\bar{u}_e
  \frac{\hat{k}_n}{m_n} (1 - \gamma^5) v_{\bar{\nu}}\Big]
\Big[\bar{u}_p u_n\Big]\Big\},\nonumber\\
\hspace{-0.21in}&&
\end{eqnarray}
where the structure constants are equal to $F^{(V)}_{1h^{(2)}} = -
0.0143$, $G^{(W)}_{1h^{(2)}} = 0.8780$, $B_{1h^{(2)}} = - 0.0430$,
$F^{(A)}_{1h^{(2)}}= - 0.5606$, $H^{(W)}_{1h^{(2)}} = 0.9432$, and
$F^{(W)}_{1h^{(2)}} = - 0.8021$.  The Lorentz structure of
Eq.(\ref{eq:C1h.19}) is calculated at the neglect the contributions of
order $O(k_n\cdot q/M^2_W) \sim 10^{-7}$ and $O(m^2_{\pi}/M^2_W)\sim
10^{-6}$, respectively.

Summing up Eq.(\ref{eq:C1h.13}) and Eq.(\ref{eq:C1h.19}), we obtain the
contribution of the Feynman diagram in Fig.\,\ref{fig:fig1}h to the
amplitude of the neutron beta decay. It is equal to
\begin{eqnarray}\label{eq:C1h.20}
\hspace{-0.21in}&&M(n \to p e^- \bar{\nu}_e)^{(\pi^0)}_{\rm
  Fig.\,\ref{fig:fig1}h} = - \frac{\alpha}{2\pi}\frac{g^2_{\pi
    N}}{16\pi^2}\,G_V \Big\{\Big(A_{1h}\Gamma\Big(2 - \frac{n}{2}\Big)
\Big[\bar{u}_e \gamma^{\mu}(1 - \gamma^5) v_{\bar{\nu}}\Big]
\Big[\bar{u}_p \gamma_{\mu} u_n\Big] + B_{1h}\Gamma\Big(2 -
\frac{n}{2}\Big) \Big[\bar{u}_e \gamma^{\mu}(1 - \gamma^5)
  v_{\bar{\nu}}\Big]\nonumber\\
  \hspace{-0.21in}&&\times \Big[\bar{u}_p \gamma_{\mu}\gamma^5
    u_n\Big] + \Big(F^{(V)}_{1h} + G^{(V)}_{1h} \frac{k_n \cdot
    q}{m^2_N} + H^{(V)}_{1h} \frac{k_n \cdot k_e}{m^2_N} +
  G^{(W)}_{1h} \frac{m^2_N}{M^2_W} \Big) \Big[\bar{u}_e \gamma^{\mu}(1 -
    \gamma^5) v_{\bar{\nu}}\Big]\Big[\bar{u}_p \gamma_{\mu} u_n\Big] +
  \Big(F^{(A)}_{1h} + G^{(A)}_{1h} \frac{k_n \cdot q}{m^2_N}\nonumber\\
  \hspace{-0.21in}&& + H^{(A)}_{1h} \frac{k_n \cdot k_e}{m^2_N} +
  H^{(W)}_{1h} \frac{m^2_N}{M^2_W}\Big) \Big[\bar{u}_e \gamma^{\mu}(1
    - \gamma^5) v_{\bar{\nu}}\Big] \Big[\bar{u}_p \gamma_{\mu}\gamma^5
    u_n\Big] +
  F^{(W)}_{1h} \frac{m^2_N}{M^2_W} \Big[\bar{u}_e
    \frac{\hat{k}_n}{m_n} (1 - \gamma^5) v_{\bar{\nu}}\Big]
  \Big[\bar{u}_p u_n\Big]\Big\},
\end{eqnarray}
where the structure constants are equal to $A_{1h} = 0.1720$, $B_{1h}
= B_{1h^{(1)}} + 2\,B_{1h^{(2)}} = - 0.2580$, $F^{(V)}_{1h}=
F^{(V)}_{1h^{(1)}} + 2\,F^{(V)}_{1h^{(2)}} = - 0.4524$, $G^{(V)}_{1h}
= G^{(V)}_{1h^{(1)}} =0.3700$, $H^{(V)}_{1h} = H^{(V)}_{1h^{(1)}} = -
0.1595$, $G^{(W)}_{1h} = 2\, G^{(W)}_{1h^{(2)}} = 1.7560$,
$F^{(A)}_{1h} = F^{(A)}_{1h^{(1)}} + 2\,F^{(A)}_{1h^{(2)}} = -
1.7900$, $G^{(A)}_{1h} = G^{(A)}_{1h^{(1)}} = - 0.3700$, $H^{(A)}_{1h}
= H^{(A)}_{1h^{(1)}} = 0$, $H^{(W)}_{1h} = 2\, H^{(W)}_{1h^{(2)}}
= 1.8864$ and $F^{(W)}_{1h} = 2\, F^{(W)}_{1h^{(2)}} = - 1.6042$. The
Lorentz structure of Eq.(\ref{eq:C1h.20}) is calculated at the neglect
the contributions of order $O(E^2_0/m^2_N)\sim O(m^2_{\pi}/M^2_W) \sim
10^{-6}$.

\subsection*{C1i. Analytical calculation of the Feynman diagram in
  Fig.\,\ref{fig:fig1}i}
\renewcommand{\theequation}{C1i-\arabic{equation}}
\setcounter{equation}{0}

In the Feynman gauge for the photon propagator the analytical
expression of the Feynman diagram in Fig.\,\ref{fig:fig1}i is given
by (see Eq.(\ref{eq:A.1}))
\begin{eqnarray}\label{eq:C1i.1}
 &&M(n \to p e^- \bar{\nu}_e)^{(\pi^0)}_{\rm Fig.\,\ref{fig:fig1}i} =
  - 2 e^2 g^2_{\pi N} M^2_W G_V\nonumber\\ &&\times \int
  \frac{d^4k}{(2\pi)^4i}\int \frac{d^4p}{(2\pi)^4i}\,\Big[\bar{u}_e
    \gamma^{\mu}(1 - \gamma^5) v_{\bar{\nu}}\Big]\,\Big[\bar{u}_p
    \gamma^5 \frac{1}{m_N - \hat{k}_n -\hat{k} - \hat{p} - i0}
    \gamma_{\beta} \frac{1}{m_N - \hat{k}_n - \hat{k} - i0}\,\gamma^5
    u_n\Big] \nonumber\\ &&\times\,\frac{(2 k + p -
    q)^{\alpha_2}}{[m^2_{\pi} - k^2 - i0][m^2_{\pi} - (k + p - q)^2 -
      i0]}\,\big[(p - q)^{\nu}\,\eta^{\beta \alpha_1} - (p -
    2q)^{\beta} \eta^{\alpha_1 \nu} - q^{\alpha_1} \eta^{\nu
      \beta}\big]\, \frac{1}{p^2 + i0} \nonumber\\ &&\times\,
  D^{(W)}_{\alpha_1\alpha_2}(p - q) D^{(W)}_{\mu\nu}(- q).
\end{eqnarray}
Since the calculation of this diagram is similar to the calculation of
the Feynman diagram in Fig.\,\ref{fig:fig1}a, so we adduce at once the
expression in terms of the integrals over the Feynman parameters.  It
takes the form
\begin{eqnarray}\label{eq:C1i.2}
\hspace{-0.3in} &&M(n \to p e^- \bar{\nu}_e)^{(\pi^0)}_{\rm
  Fig.\,\ref{fig:fig1}i} = - \frac{\alpha}{2\pi} \frac{g^2_{\pi
    N}}{16\pi^2} G_V \int^1_0 \!\!dx_1 \int^1_0 \!\!dx_2 \int^1_0
\!\!dx_3 \int^1_0 \!\!dx_4 \int^1_0 \!\!dx_5 \int^1_0 \!\!dx_6
\frac{\delta(1 - x_1 - x_2 - x_3 - x_4 - x_5 - x_6)}{U^2} \nonumber\\
\hspace{-0.3in}&&\Big\{ g^{(W)}_{1i}(x_1, \ldots, x_6)
\frac{m^2_N}{M^2_W}\Big[\bar{u}_e \gamma^{\mu}(1 - \gamma^5)
  v_{\bar{\nu}}\Big]\Big[\bar{u}_p \gamma_{\mu} u_n\Big] +
f^{(W)}_{1i}(x_1, \ldots, x_6) \frac{m^2_N}{M^2_W}\,\Big[\bar{u}_e
  \frac{\hat{k}_n}{m_N}(1 - \gamma^5) v_{\bar{\nu}}\Big]\Big[\bar{u}_p
  u_n\Big]\Big\},
\end{eqnarray}
where we have denoted
\begin{eqnarray}\label{eq:C1i.3}
  \hspace{-0.3in}U&=& (x_1 + x_2 + x_3 + x_4)(x_5 + x_6) + (x_1 +
  x_4)(x_2 + x_3),\nonumber\\
 \hspace{-0.3in} g^{(W)}_{1i}(x_1, \ldots, x_6)
 &=&\frac{1}{\bar{Q}}\Big[- \frac{Y^2}{U^2} + 2\, \frac{Y\bar{Y}}{U^2}
   + \frac{\bar{Y}^2}{U^2}\Big],\nonumber\\
 \hspace{-0.3in} f^{(W)}_{1i}(x_1, \ldots, x_6)
 &=&\frac{1}{\bar{Q}}\Big[ 4\,\frac{Y^2}{U^2} - 8\,
   \frac{Y\bar{Y}}{U^2} - 4\, \frac{\bar{Y}^2}{U^2} \Big].
\end{eqnarray}
For the calculation of the integrals over the Feynman parameters we
use
\begin{eqnarray}\label{eq:C1i.4}
 \hspace{-0.3in} \bar{Q} &=& \frac{1}{U}\Big(x_6 U + \big((x_5 + x_6)
 (x_3 + x_4)^2 + (x_2 + x_3) x^2_4 + (x_1 + x_4) x^2_3\big)\,
 \frac{m^2_N}{M^2_W}\Big),\nonumber\\
  \hspace{-0.3in} Y &=& - (x_5 + x_6)(x_3 +  x_4) - (x_2 + x_3)
  x_4\;,\; \bar{Y} = x_2 x_4 - x_1 x_3.
\end{eqnarray}
After the integration over the Feynman parameters, the contribution of
the Feynman diagram in Fig.\,\ref{fig:fig1}i to the amplitude of the
neutron beta decay is equal to
\begin{eqnarray}\label{eq:C1i.5}
\hspace{-0.3in}&& M(n \to p e^- \bar{\nu}_e)^{(\pi^0)}_{\rm
  Fig.\,\ref{fig:fig1}i} = \nonumber\\
\hspace{-0.3in}&&= - \frac{\alpha}{2\pi} \frac{g^2_{\pi N}}{16\pi^2}
G_V \Big\{ G^{(W)}_{1i} \frac{m^2_N}{M^2_W}\Big[\bar{u}_e
  \gamma^{\mu}(1 - \gamma^5) v_{\bar{\nu}}\Big]\Big[\bar{u}_p
  \gamma_{\mu} u_n\Big] + F^{(W)}_{1i}
\frac{m^2_N}{M^2_W}\,\Big[\bar{u}_e \frac{\hat{k}_n}{m_N}(1 -
  \gamma^5) v_{\bar{\nu}}\Big]\Big[\bar{u}_p u_n\Big]\Big\},
\end{eqnarray}
where the structure constants are equal to $G^{(W)}_{1i} = - 1.7652$
and $F^{(W)}_{1i} = 7.0609$. The Lorentz structure of
Eq.(\ref{eq:C1i.5}) is calculated at the neglect the contributions of
order $O(E^2_0/m^2_N) \sim O(m^2_{\pi}/M^2_W) \sim 10^{-6}$ and even
smaller.

\subsection*{C1j. Analytical calculation of the Feynman diagram in
  Fig.\,\ref{fig:fig1}j}
\renewcommand{\theequation}{C1j-\arabic{equation}}
\setcounter{equation}{0}

In the Feynman gauge for the photon propagator the analytical
expression of the Feynman diagram in Fig.\,\ref{fig:fig1}j is given by
(see Eq.(\ref{eq:A.1}))
\begin{eqnarray}\label{eq:C1j.1}
 &&M(n \to p e^- \bar{\nu}_e)^{(\pi^0)}_{\rm Fig.\,\ref{fig:fig1}j} =
  - 2 e^2 g^2_{\pi N} G_V \int \frac{d^4k}{(2\pi)^4i}\int
  \frac{d^4p}{(2\pi)^4i}\,\Big[\bar{u}_e \gamma^{\mu}(1 - \gamma^5)
    v_{\bar{\nu}}\Big]\nonumber\\ && \times \,\Big[\bar{u}_p
    \gamma^{\beta} \frac{1}{m_N - \hat{k}_p + \hat{p} - i0}\, \gamma^5
    \frac{1}{m_N - \hat{k}_n - \hat{k} + \hat{p} -
      i0}\,\gamma_{\beta}\, \frac{1}{m_N - \hat{k}_n - \hat{k} - i0}\,
    \gamma^5 u_n\Big] \nonumber\\ && \times \,\frac{(2 k - q)_{\mu}
  }{[m^2_{\pi} - k^2 - i0][m^2_{\pi} - (k - q)^2 - i0]}\, \frac{1}{p^2
    + i0},
\end{eqnarray}
where we have made a replacement $M^2_W D^{(W)}_{\mu\nu} (- q) \to -
\eta_{\mu\nu}/M^2_W$, which is valid up to relative corrections of
order $10^{-7}$. The calculation of the r.h.s. of Eq.(\ref{eq:C1j.1})
we reduce to the calculation of the following integral
\begin{eqnarray}\label{eq:C1j.2}
\hspace{-0.21in}&& {\cal F}(k_n,k_e, q)_{1j} = - \int
\frac{d^4k}{(2\pi)^4i}\int \frac{d^4p}{(2\pi)^4i}\frac{1}{m^2_{\pi} -
  k^2 - i0}\frac{1}{m^2_{\pi} - (k - q)^2 - i0} \frac{1}{m^2_N - (k_p
  - p)^2 - i0} \frac{1}{m^2_N - (k_n + k - p)^2 - i0}\nonumber\\
\hspace{-0.21in}&& \times \frac{1}{m^2_N - (k_n + k)^2 - i0}
\frac{1}{p^2 + i0} \Big(8 \hat{k}(1 - \gamma^5) \otimes \hat{k} (p^2 -
k\cdot p)\Big),
\end{eqnarray}
where we have kept only leading divergent contributions that is
equivalent to the use of the LLA.  In terms of the integrals over the
Feynman parameters and after the integration over the virtual momenta
in the $n$-dimensional momentum space we arrive at the following
expression
\begin{eqnarray}\label{eq:C1j.3}
\hspace{-0.30in}&& {\cal F}(k_n, k_e, q)_{1j} = \int^1_0 dx_1 \int^1_0
dx_2 \int^1_0 dx_3 \int^1_0 dx_4 \int^1_0 dx_5 \int^1_0 dx_6
\frac{\delta(1 - x_1 - x_2 - x_3 - x_4 - x_5 - x_6)}{2^{2n} \pi^n
  U^{n/2}} \nonumber\\
\hspace{-0.30in}&& \times \Bigg\{ 2 n\, \frac{\displaystyle
  \Gamma\Big(2 - \frac{n}{2}\Big)\Gamma\Big(2 - \frac{n -
    1}{2}\Big)}{\displaystyle 2^{n -
    3}\Gamma\Big(\frac{1}{2}\Big)}\,\Big(\frac{Q}{m^2_N}\Big)^{-4 + n}
\gamma^{\mu} (1 - \gamma^5) \otimes \gamma_{\mu} + \ldots \Bigg\},
\end{eqnarray}
where we have denoted
\begin{eqnarray}\label{eq:C1j.4}
\hspace{-0.30in}U&=& (x_1 + x_2 + x_5) (x_3 + x_4 + x_6) + (x_3 + x_6)
x_4, \nonumber\\
\hspace{-0.30in}b_1 &=& \frac{(x_3 + x_4 + x_6) a_1 + x_4 a_2}{U} =
\frac{((x_4 + x_6) x_2 + (x_2 + x_4) x_3) q - ((x_3 + x_4) x_5 + (x_4
  + x_5) x_6) k_n}{U} = \frac{X q + Y k_n}{U},\nonumber\\
\hspace{-0.30in}b_2 &=& \frac{(x_1 + x_2 + x_4 + x_5) a_2 + x_4
  a_1}{U} = \nonumber\\
\hspace{-0.30in}&=&\frac{((x_1 + x_2 + x_5) x_3 + (x_2 + x_3)
  x_4) q + ((x_1 + x_2 )(x_3 + x_4) + (x_4 + x_5) x_3) k_n}{U} =
\frac{\bar{X} q + \bar{Y} k_n}{U}, \nonumber\\
\hspace{-0.30in}Q &=& (x_1 + x_2) m^2_{\pi} - x_2 q^2 + \frac{(x_3 +
  x_4 + x_6) a^2_1 + 2 x_4 a_1 \cdot a_2 + (x_1 + x_2 + x_4 + x_5)
  a^2_2}{U}
\end{eqnarray}
with $a_1 = x_2 q - (x_4 + x_5) k_n$ and $a_2 = x_3 q + (x_3 + x_4)
k_n$. Taking the limit $n \to 4$ and keeping the divergent
contribution proportional to $\Gamma(2 - n/2)$, we get
\begin{eqnarray}\label{eq:C1j.5}
\hspace{-0.30in}&& {\cal F}(k_n, k_e, q)_{1j} = \int^1_0 dx_1 \int^1_0
dx_2 \int^1_0 dx_3 \int^1_0 dx_4 \int^1_0 dx_5 \int^1_0 dx_6
\frac{\delta(1 - x_1 - x_2 - x_3 - x_4 - x_5 - x_6)}{256 \pi^4 U^2}
\nonumber\\
\hspace{-0.30in}&& \times \Big\{\Big(4\, \Gamma\Big(2 -
\frac{n}{2}\Big) - 2 - 8 {\ell n}\frac{Q}{m^2_N}\Big)
\gamma^{\mu} (1 - \gamma^5) \otimes \gamma_{\mu} + \ldots \Big\}.
\end{eqnarray}
Neglecting the contributions of order $O(E^2_0/m^2_N) \sim 10^{-6}$, we
obtain the following Lorentz structure for ${\cal F}(k_n, k_e,
q)_{1j}$
\begin{eqnarray*}
\hspace{-0.30in}&& {\cal F}(k_n, k_e, q)_{1j} = \int^1_0 dx_1 \int^1_0
dx_2 \int^1_0 dx_3 \int^1_0 dx_4 \int^1_0 dx_5 \int^1_0 dx_6
\frac{\delta(1 - x_1 - x_2 - x_3 - x_4 - x_5 - x_6)}{256 \pi^4 U^2}
\nonumber\\
\end{eqnarray*}
\begin{eqnarray}\label{eq:C1j.6}
\hspace{-0.30in}&& \times \Big\{\Big(4\, \Gamma\Big(2 -
\frac{n}{2}\Big) + f^{(V)}_{1j}(x_1, \ldots, x_6) + g^{(V)}_{1j}(x_1,
\ldots, x_6)\, \frac{k_n\cdot q}{m^2_N}\Big) \gamma^{\mu} (1 -
\gamma^5) \otimes \gamma_{\mu} + \ldots \Big\}.
\end{eqnarray}
The functions of the Feynman parameters $f^{(V)}_{1j}(x_1, \ldots,
x_6)$ and $g^{(V)}_{1j}(x_1, \ldots,
x_6)$ are equal to
\begin{eqnarray}\label{eq:C1j.7}
\hspace{-0.30in}&&f^{(V)}_{1j}(x_1, \ldots, x_6) = - 8\, {\ell
  n}\frac{\bar{Q}}{m^2_N} \quad,\quad g^{(V)}_{1j}(x_1, \ldots, x_6) =
- 16\, \frac{V}{U}\, \frac{m^2_N}{\bar{Q}}.
\end{eqnarray}
For the calculation of the integrals over the Feynman parameters we
use
\begin{eqnarray}\label{eq:C1j.8}
\hspace{-0.30in}\frac{\bar{Q}}{m^2_N} &=& \frac{1}{U}\Big((x_1 +
x_2)\,U \frac{m^2_{\pi}}{m^2_N} + (x_4 + x_5)^2 x_6 + (x_1 + x_2) (x_3
+ x_4)^2 + (x_3 + x_4)(x_4 + x_5) (x_3 + x_5)\Big), \nonumber\\
\hspace{-0.30in} V &=& x_3((x_1 + x_2) (x_3 + x_4)  + (x_4 +
x_5)x_3) - x_2((x_3 + x_4) x_5 + (x_4 + x_5)x_6),
\end{eqnarray}
where $\bar{Q} = Q\big|_{q = 0}$.  The contributions of the terms
$k_n\cdot q/m^2_N$ (or $q_0/m_N$) are taken into account in the linear
approximation of the expansion of $Q$ in powers of $k_n\cdot q/m^2_N$
(or $q_0/m_N$). The structure function $g^{(V)}_{1j}(x_1, \ldots,
x_6)$ is calculated by an expansion of the logarithmic function $ -
8\, {\ell n} Q/m^2_N$ in powers of $k_n\cdot q/m^2_N$ (or $q_0/m_N$).

After the integration over the Feynman parameters, we arrive at the
contribution of the Feynman diagram in Fig.\,\ref{fig:fig1}j to the
amplitude of the neutron beta decay 
\begin{eqnarray}\label{eq:C1j.9}
\hspace{-0.30in} &&M(n \to p e^- \bar{\nu}_e)^{(\pi^0)}_{\rm
  Fig.\,\ref{fig:fig1}j} = - \frac{\alpha}{2\pi} \frac{g^2_{\pi
    N}}{16\pi^2} G_V\Big\{\Big(A_{1j}\Gamma\Big(2 - \frac{n}{2}\Big) +
 F^{(V)}_{1j} + G^{(V)}_{1j}\,\frac{k_n \cdot q}{m^2_N} \Big)
\Big[\bar{u}_e \gamma^{\mu}(1 - \gamma^5) v_{\bar{\nu}}\Big]
\Big[\bar{u}_p\gamma_{\mu} u_n\Big]\Big\}, \nonumber\\
\hspace{-0.30in}&&
\end{eqnarray}
where the structure constants are equal to $A_{1j} = 2/3$,
$F^{(V)}_{1j} = 1.9182$ and $G^{(V)}_{1j} = - 0.7666$. The Lorentz
structure of Eq.(\ref{eq:C1j.9}) is obtained at the neglect of the
contributions to the integral ${\cal F}(k_n, q)_{1j}$ of order
$O(E^2_0/M^2_N) \sim 10^{-6}$.

\subsection*{C1k. Analytical calculation of the Feynman diagram in
  Fig.\,\ref{fig:fig1}k}
\renewcommand{\theequation}{C1k-\arabic{equation}}
\setcounter{equation}{0}

In the Feynman gauge for the photon propagator the analytical
expression of the Feynman diagram in Fig.\,\ref{fig:fig1}k is given by
(see Eq.(\ref{eq:A.1}))
\begin{eqnarray}\label{eq:C1k.1}
&&M(n \to p e^- \bar{\nu}_e)^{(\pi^0)}_{\rm Fig.\,\ref{fig:fig1}k} = -
  e^2 g^2_{\pi N} M^2_W G_V \int \frac{d^4k}{(2\pi)^4i}\int
  \frac{d^4p}{(2\pi)^4i}\,\Big[\bar{u}_e \gamma^{\mu}(1 - \gamma^5)
    v_{\bar{\nu}} \Big] \nonumber\\ &&\times \,\Big[\bar{u}_p
    \,\gamma^5\, \frac{1}{m_N - \hat{k}_p -\hat{k} - i0}\,
    \gamma_{\beta} \frac{1}{m_N - \hat{k}_p - \hat{k} + \hat{p} - i0}
    \,\gamma^{\alpha_2}(1 - \gamma^5)\,\frac{1}{m_N - \hat{k}_n
      -\hat{k} - i0}\,\gamma^5 u_n\Big] \nonumber\\ &&\times
  \,\frac{1}{m^2_{\pi} - k^2 - i0}\,\big[(p - q)^{\nu}\,\eta^{\beta
      \alpha_1} - (p - 2q)^{\beta} \eta^{\alpha_1 \nu} - q^{\alpha_1}
    \eta^{\nu \beta}\big] \, \frac{1}{p^2 + i0}\,
  D^{(W)}_{\alpha_1\alpha_2}(p - q) D^{(W)}_{\mu\nu}(- q).
\end{eqnarray}
Since the calculation of Eq.(\ref{eq:C1k.1}) is similar to the
calculation of Eq.(\ref{eq:C1a.1}), we adduce at once the expression
for the matrix element Eq.(\ref{eq:C1k.1}) in terms of the integrals
over the Feynman parameters. It is given by
\begin{eqnarray}\label{eq:C1k.2}
\hspace{-0.30in}&&M(n \to p e^- \bar{\nu}_e)^{(\pi^0)}_{\rm
  Fig.\,\ref{fig:fig1}k} = - \frac{\alpha}{4\pi}\frac{g^2_{\pi
    N}}{16\pi^2}G_V \int^1_0\!\! dx_1 \int^1_0 \!\! dx_2 \int^1_0 \!\!
dx_3 \int^1_0 \!\! dx_4 \int^1_0 \!\! dx_5 \int^1_0 \!\! dx_6
\frac{\delta(1 - x_1 - x_2 - x_3 - x_4 - x_5 - x_6)}{U^2} \nonumber\\
\hspace{-0.30in}&& \times \Big\{\Big( - \frac{3}{2}\,\Gamma\Big(2 -
\frac{n}{2}\Big) + f^{(V)}_{1k}(x_1, \ldots, x_6) + g^{(W)}_{1k}(x_1,
\ldots, x_6)\, \frac{m^2_N}{M^2_W}\Big) \Big[\bar{u}_e \gamma^{\mu}(1
  - \gamma^5) v_{\bar{\nu}} \Big]\Big[\bar{u}_p\gamma_{\mu} u_n\Big] +
\Big( - \frac{3}{2}\,\Gamma\Big(2 -
\frac{n}{2}\Big) \nonumber\\
\hspace{-0.30in}&&+ f^{(A)}_{1k}(x_1, \ldots, x_6) + h^{(W)}_{1k}(x_1,
\ldots, x_6)\, \frac{m^2_N}{M^2_W}\Big) \Big[\bar{u}_e \gamma^{\mu}(1
  - \gamma^5) v_{\bar{\nu}} \Big] \Big[\bar{u}_p\gamma_{\mu} \gamma^5
  u_n\Big] + f^{(W)}_{1k} (x_1, \ldots, x_6)\, \frac{m^2_N}{M^2_W}
\nonumber\\
\hspace{-0.30in}&& \times \Big[\bar{u}_e
  \frac{\hat{k}_n}{m_N} (1 - \gamma^5) v_{\bar{\nu}} \Big]
\Big[\bar{u}_p u_n\Big]\Big\},
\end{eqnarray}
where we have denoted
\begin{eqnarray*}
\hspace{-0.30in}f^{(V)}_{1k}(x_1, \ldots, x_6) &=& 3\,{\ell
  n}\frac{M^2_W}{m^2_N} + 3\,{\ell n} \bar{Q},\nonumber\\
\hspace{-0.30in}f^{(A)}_{1k}(x_1, \ldots, x_6) &=& 3\,{\ell
  n}\frac{M^2_W}{m^2_N} + 3\,{\ell n} \bar{Q},\nonumber\\
\end{eqnarray*}
\begin{eqnarray}\label{eq:C1k.3}
\hspace{-0.30in}g^{(W)}_{1k}(x_1, \ldots, x_6) &=& \frac{1}{\bar{Q}}
\Big[ \frac{Y^2}{U^2} - 3\, \frac{Y\bar{Y}}{U^2} +
  \frac{\bar{Y}^2}{U^2} \Big],\nonumber\\
\hspace{-0.30in}h^{(W)}_{1k}(x_1, \ldots, x_6) &=& \frac{1}{\bar{Q}}
\Big[ \frac{Y^2}{U^2} - \frac{Y\bar{Y}}{U^2}
  + \frac{\bar{Y}^2}{U^2}\Big], \nonumber\\
\hspace{-0.30in}f^{(W)}_{1k}(x_1, \ldots, x_6) &=& \frac{1}{\bar{Q}}
\Big[- 2\, \frac{Y^2}{U^2} + 4\, \frac{Y\bar{Y}}{U^2} + 2\,
  \frac{\bar{Y}^2}{U^2}\Big].
\end{eqnarray}
For the calculation of the integrals over the Feynman parameters we use
\begin{eqnarray}\label{eq:C1k.4}
\hspace{-0.30in} \bar{Q} &=& \frac{1}{U}\Big(x_6 U + \big((x_5 +
x_6)(x_2 + x_3 + x_4)^2 + x_3 (x_2 + x_4)^2 + (x_1 + x_2 + x_4)
x^2_3\big)\, \frac{m^2_N}{M^2_W}\Big),\nonumber\\
\hspace{-0.30in} U &=& (x_1 + x_2 + x_4) (x_3 + x_5 + x_6) + (x_5 +
x_6) x_3,\;,\; \nonumber\\
\hspace{-0.30in} Y&=& - (x_3 + x_5 + x_6)(x_2 + x_4) - (x_5 + x_6)
x_3\;,\; \bar{Y} = x_1 x_3.
\end{eqnarray}
After the integration over the Feynman parameters, we obtain the
following contribution of the Feynman diagram in Fig\,\ref{fig:fig1}k
to the amplitude of the neutron beta decay
\begin{eqnarray}\label{eq:C1k.5}
\hspace{-0.30in}&&M(n \to p e^- \bar{\nu}_e)^{(\pi^0)}_{\rm
  Fig.\,\ref{fig:fig1}k} = - \frac{\alpha}{4\pi}\frac{g^2_{\pi
    N}}{16\pi^2}G_V \Big\{\Big(A_{1k}\,\Gamma\Big(2 - \frac{n}{2}\Big)
+ F^{(V)}_{1k} + G^{(W)}_{1k}\, \frac{m^2_N}{M^2_W}\Big)
\Big[\bar{u}_e \gamma^{\mu}(1 - \gamma^5) v_{\bar{\nu}} \Big]
\Big[\bar{u}_p\gamma_{\mu} u_n\Big] \nonumber\\
\hspace{-0.30in}&&+ \Big(B_{1k}\,\Gamma\Big(2 - \frac{n}{2}\Big) +
F^{(A)}_{1k} + H^{(W)}_{1k}\, \frac{m^2_N}{M^2_W}\Big) \Big[\bar{u}_e
  \gamma^{\mu}(1 - \gamma^5) v_{\bar{\nu}}
  \Big]\Big[\bar{u}_p\gamma_{\mu} \gamma^5 u_n\Big] + F^{(W)}_{1k} \,
\frac{m^2_N}{M^2_W} \Big[\bar{u}_e \frac{\hat{k}_n}{m_N} (1 -
  \gamma^5) v_{\bar{\nu}} \Big] \Big[\bar{u}_p u_n\Big]\Big\}.\nonumber\\
\hspace{-0.30in}&&
\end{eqnarray}
The structure constants are equal to $A_{1k} = - 1/4$, $B_{1k} = -
1/4$, $F^{(V)}_{1k} = - 3.0634$, $G^{(W)}_{1k} = 13.8057$,
$F^{(A)}_{1k} = - 3.0634$, $H^{(W)}_{1k} = 10.9898$ and $F^{(W)}_{1k}
= 20.7511$. The Lorentz structure of the matrix element
Eq.(\ref{eq:C1k.5}) is obtained at the neglect the contributions of
the terms $k_n\cdot q/M^2_W$ and $m^2_{\pi}/M^2_W$, which are of order
$10^{-7}$ and $10^{-6}$, respectively.

\subsection*{C1l. Analytical calculation of the Feynman diagram in
  Fig.\,\ref{fig:fig1}l}
\renewcommand{\theequation}{C1l-\arabic{equation}}
\setcounter{equation}{0}

In the Feynman gauge for the photon propagator the analytical
expression of the Feynman diagram in Fig.\,\ref{fig:fig1}l is given by
(see Eq.(\ref{eq:A.1}))
\begin{eqnarray}\label{eq:C1l.1}
\hspace{-0.30in}&&M(n \to p e^- \bar{\nu}_e)^{(\pi^0)}_{\rm
  Fig.\,\ref{fig:fig1}l} = + e^2 g^2_{\pi N} G_V \int
\frac{d^4k}{(2\pi)^4i}\int \frac{d^4p}{(2\pi)^4i}\,\Big[\bar{u}_e
  \gamma^{\mu}(1 - \gamma^5) v_{\bar{\nu}}\Big] \,\Big[\bar{u}_p
  \,\gamma^{\beta}\, \frac{1}{m_N - \hat{k}_p + \hat{p} - i0}\,
  \gamma^5 \nonumber\\ \hspace{-0.30in}&& \times\frac{1}{m_N -
    \hat{k}_p - \hat{k} + \hat{p} - i0}\,\gamma_{\beta}\,\frac{1}{m_N
    - \hat{k}_p - \hat{k} - i0}\,\gamma_{\mu} (1 -
  \gamma^5)\,\frac{1}{m_N - \hat{k}_n - \hat{k} - i0}\,\gamma^5
  u_n\Big]\,\frac{1}{m^2_{\pi} - k^2 - i0}\, \frac{1}{p^2 +
  i0},\nonumber\\ \hspace{-0.30in}&&
\end{eqnarray}
where we have made a replacement $M^2_W D^{(W)}_{\mu\nu}(- q) \to -
\eta_{\mu\nu}$, which is valid at the neglect of the contributions of
order $10^{-7}$. The calculation of the matrix element
Eq.(\ref{eq:C1l.1}) reduces to the calculation of the following
integral
\begin{eqnarray}\label{eq:C1l.2}
\hspace{-0.30in}&&{\cal F}(k_n, k_e, q)_{1l} = - \int
\frac{d^4k}{(2\pi)^4i}\int \frac{d^4p}{(2\pi)^4i}\,\frac{1}{m^2_N -
  (k_p - p)^2 - i0} \frac{1}{m^2_N - (k_p + k - p)^2 -
  i0}\frac{1}{m^2_N - (k_p + k)^2 - i0} \nonumber\\ \hspace{-0.30in}&&
\times \frac{1}{m^2_N - (k_n + k)^2 - i0}\frac{1}{m^2_{\pi} - k^2 -
  i0}\frac{1}{p^2 + i0}\, \Big(\big(4 p^2 - 4 p\cdot k\big)
\gamma^{\mu} (1 - \gamma^5) \otimes \hat{k}\gamma_{\mu} \hat{k} (1 +
\gamma^5)\Big),
\end{eqnarray}
where we have kept only leading divergent contributions. In terms of
the integrals over the Feynman parameters and after the integration
over virtual momenta in the $n$-dimensional momentum space the
integral ${\cal F}(k_n, q)_{1l}$ is given by
\begin{eqnarray}\label{eq:C1l.3}
\hspace{-0.30in}&&{\cal F}(k_n, k_e, q)_{1l} = \int^1_0 dx_1 \int^1_0
dx_2 \int^1_0 dx_3 \int^1_0 dx_4 \int^1_0 dx_5 \int^1_0 dx_6
\frac{\delta(1 - x_1 - x_2 - x_3 - x_4 - x_5 - x_6)}{2^{2n} \pi^n
  U^{n/2}} \nonumber\\ \hspace{-0.30in}&& \times \Bigg\{ - n(n -
2)\,\frac{\displaystyle \Gamma\Big(2 - \frac{n}{2}\Big)\Gamma\Big(2 -
  \frac{n - 1}{2}\Big)}{\displaystyle 2^{n - 3}
  \Gamma\Big(\frac{1}{2}\Big)}\Big(\frac{Q}{m^2_N}\Big)^{-4 + n}
\gamma^{\mu} (1 - \gamma^5) \otimes \gamma_{\mu} (1 + \gamma^5) +
\ldots\Bigg\},
\end{eqnarray}
where we have denoted
\begin{eqnarray}\label{eq:C1l.4}
  \hspace{-0.3in}U &=& (x_1 + x_4 + x_5)(x_2 + x_3 + x_6) + (x_2 +
  x_6) x_3,\nonumber\\
  \hspace{-0.3in}b_1 &=& - \frac{((x_2 + x_3) x_4 + (x_3 + x_4) x_6) q
    + ((x_2 + x_3)(x_4 + x_5) + (x_3 + x_4 + x_5) x_6) k_n}{U} =
  \frac{X q + Y k_n}{U},\nonumber\\
   \hspace{-0.3in}b_2 &=& + \frac{((x_1 + x_4 + x_5) x_2 + (x_1 + x_2
     + x_5)x_3) q + ((x_1 + x_4 + x_5)x_2 + (x_1 + x_2) x_3) k_n}{U} =
   \frac{\bar{X} q + \bar{Y} k_n}{U},\nonumber\\
   \hspace{-0.3in}Q &=&x_1 m^2_{\pi} + \frac{(x_2 + x_3 + x_6) a^2_1 +
     2 x_3 a_1\cdot a_2 + (x_1 + x_3 + x_4 + x_5) a^2_2}{U}
\end{eqnarray}
with $a_1 = -(x_3 + x_4)q - (x_3 + x_4 + x_5) k_n$ and $a_2 = (x_2 +
x_3) q + (x_2 + x_3) k_n$ and $k_p = k_n + q$. Taking the limit $n \to
4$ and keeping the divergent contributions proportional to $\Gamma(2 -
n/2)$, we transcribe Eq.(\ref{eq:C1l.3}) into the form
\begin{eqnarray}\label{eq:C1l.5}
\hspace{-0.30in}&&{\cal F}(k_n, k_e, q)_{1l} = \int^1_0 dx_1 \int^1_0
dx_2 \int^1_0 dx_3 \int^1_0 dx_4 \int^1_0 dx_5 \int^1_0 dx_6
\frac{\delta(1 - x_1 - x_2 - x_3 - x_4 - x_5 - x_6)}{256 \pi^4U^2}
\nonumber\\ \hspace{-0.30in}&& \times \Big\{ \Big(- 4\,\Gamma\Big(2 -
\frac{n}{2}\Big) + 8\, {\ell n}\frac{Q}{m^2_N}\Big)\,\gamma^{\mu} (1 -
\gamma^5) \otimes \gamma_{\mu} (1 + \gamma^5) + \ldots\Big\}.
\end{eqnarray}
For the integral ${\cal F}(k_n, q)_{1l}$ we obtain the following
Lorentz structure
\begin{eqnarray}\label{eq:C1l.6}
\hspace{-0.30in}&&{\cal F}(k_n, k_e, q)_{1l} = \int^1_0 dx_1 \int^1_0
dx_2 \int^1_0 dx_3 \int^1_0 dx_4 \int^1_0 dx_5 \int^1_0 dx_6
\frac{\delta(1 - x_1 - x_2 - x_3 - x_4 - x_5 - x_6)}{256 \pi^4U^2}
\nonumber\\ \hspace{-0.30in}&& \times \Big\{- 4\,\Gamma\Big(2 -
\frac{n}{2}\Big) \, \gamma^{\mu} (1 - \gamma^5) \otimes \gamma_{\mu}
(1 + \gamma^5) + \Big( f^{(V)}_{1l}(x_1, \ldots, x_6) +
g^{(V)}_{1l}(x_1, \ldots, x_6)\, \frac{k_n\cdot q}{m^2_N}\Big)
\gamma^{\mu} (1 - \gamma^5) \otimes \gamma_{\mu}
\nonumber\\ \hspace{-0.30in}&& + \Big( f^{(A)}_{1l}(x_1, \ldots, x_6)
+ g^{(A)}_{1l}(x_1, \ldots, x_6)\, \frac{k_n\cdot q}{m^2_N}\Big)
\gamma^{\mu} (1 - \gamma^5) \otimes \gamma_{\mu}\gamma^5\Big\},
\end{eqnarray}
where we have denoted
\begin{eqnarray}\label{eq:C1l.7}
\hspace{-0.30in}f^{(V)}_{1l}(x_1, \ldots, x_6) &=&  8\, {\ell
  n}\frac{\bar{Q}}{m^2_N}\quad,\quad g^{(V)}_{1l}(x_1, \ldots, x_6) =
16\,\frac{V}{U}\,\frac{m^2_N}{\bar{Q}},\nonumber\\
\hspace{-0.30in}f^{(A)}_{1l}(x_1, \ldots, x_6) &=& 8\, {\ell
  n}\frac{\bar{Q}}{m^2_N}\quad,\quad g^{(A)}_{1l}(x_1, \ldots, x_6) =
16\,\frac{V}{U}\,\frac{m^2_N}{\bar{Q}}.
\end{eqnarray}
For the calculation of the integrals over the Feynman parameters we
use
\begin{eqnarray}\label{eq:C1l.8}
\hspace{-0.3in} \frac{\bar{Q}}{m^2_N} &=& \frac{1}{U}\Big(x_1 \,U\,
\frac{m^2_{\pi}}{m^2_N} + (x_3 + x_4 + x_5)^2 x_6 + (x_2 + x_3)(x_2 +
x_4 + x_5)(x_3 + x_4 + x_5) + (x_2 + x_3)^2 x_1\Big),\nonumber\\
\hspace{-0.3in}V &=&(x_3 + x_4 + x_5)\big( (x_2 + x_3)x_4 + (x_3 +
x_4) x_6\big) + (x_2 + x_3)\big((x_1 + x_5)(x_2 + x_3) + (x_3 + x_4)
x_2\big),
\end{eqnarray}
where by $\bar{Q} = Q\big|_{q = 0}$.  A deviation of $Q$ from
$\bar{Q}$ is taken into account in the linear approximation in the
form of an expansion in powers of $k_n\cdot q/m^2_N$ (or
$q_0/m_N$). After the integration over the Feynman parameters, we
obtain the contribution of the Feynman diagram in
Fig.\,\ref{fig:fig1}l to the amplitude of the neutron beta decay
\begin{eqnarray}\label{eq:C1l.9}
\hspace{-0.30in}&&M(n \to p e^- \bar{\nu}_e)^{(\pi^0)}_{\rm
  Fig.\,\ref{fig:fig1}l} = \frac{\alpha}{4\pi} \frac{g^2_{\pi
    N}}{16\pi^2} G_V \Big\{A_{1l}\,\Gamma\Big(2 - \frac{n}{2}\Big) \,
\Big[\bar{u}_e \gamma^{\mu} (1 - \gamma^5)
  v_{\bar{\nu}}\Big]\Big[\bar{u}_p \gamma_{\mu} u_n\Big] + B_{1l}
\,\Gamma\Big(2 - \frac{n}{2}\Big) \, \Big[\bar{u}_e \gamma^{\mu} (1 -
  \gamma^5) v_{\bar{\nu}}\Big]\nonumber\\ \hspace{-0.30in}&& \times
\Big[\bar{u}_p \gamma_{\mu} \gamma^5 u_n\Big] + \Big( F^{(V)}_{1l} +
G^{(V)}_{1l} \, \frac{k_n\cdot q}{m^2_N}\Big)\, \Big[\bar{u}_e
  \gamma^{\mu} (1 - \gamma^5) v_{\bar{\nu}}\Big]\Big[\bar{u}_p
  \gamma_{\mu} u_n\Big] + \Big( F^{(A)}_{1l} + G^{(A)}_{1l}\,
\frac{k_n\cdot q}{m^2_N}\Big)\, \Big[\bar{u}_e \gamma^{\mu} (1 -
  \gamma^5) v_{\bar{\nu}}\Big] \nonumber\\ \hspace{-0.30in}&& \times
\Big[\bar{u}_p \gamma_{\mu} \gamma^5 u_n\Big]]\Big\}.
\end{eqnarray}
The structure constants are equal to $A_{1l} = - 2/3$, $B_{1l} = -
2/3$, $F^{(V)}_{1l} = - 1.0795$, $G^{(V)}_{1l} = 1.8585$, $F^{(A)}_{1l}
=  - 1.0795$ and $G^{(A)}_{1l} = 1.8585$. The Lorentz structure of the
matrix element Eq.(\ref{eq:C1l.9}) is calculated at the neglect of
the contributions of order $O(E^2_0/m^2_N) \sim 10^{-6}$.

\subsection*{C1m. Analytical calculation of the Feynman diagram in
  Fig.\,\ref{fig:fig1}m}
\renewcommand{\theequation}{C1m-\arabic{equation}}
\setcounter{equation}{0}

In the Feynman gauge for the photon propagator the analytical
expression of the Feynman diagram in Fig.\,\ref{fig:fig1}m is given by
(see Eq.(\ref{eq:A.1}))
\begin{eqnarray}\label{eq:C1m.1}
  &&M(n \to p e^- \bar{\nu}_e)^{(\pi^0)}_{\rm Fig.\,\ref{fig:fig1}m} =
  e^2 g^2_{\pi N} G_V\int \frac{d^4k}{(2\pi)^4i}\int
  \frac{d^4p}{(2\pi)^4i}\,\Big[\bar{u}_e \gamma^{\beta}\,\frac{1}{m_e
      - \hat{p} - \hat{k}_e - i0}\,\gamma^{\mu}(1 -
    \gamma^5)\,v_{\bar{\nu}}\Big]\nonumber\\ &&\times \,\Big[\bar{u}_p
    \gamma^5\, \frac{1}{m_N - \hat{k}_p - \hat{k} - i0}
    \gamma_{\beta}\, \frac{1}{m_N - \hat{k}_p - \hat{k} + \hat{p} -
      i0}\,\gamma^{\nu} (1 - \gamma^5)\,\frac{1}{m_N - \hat{k}_n -
      \hat{k} - i0}\,\gamma^5 u_n\Big] \nonumber\\ &&\times \,
  \frac{1}{m^2_{\pi} - k^2 - i0}\,\frac{1}{p^2 +
    i0}\,\Big(\eta_{\mu\nu} + \frac{(p - q)^2 \eta_{\mu\nu} - (p -
    q)_{\mu}(p - q)_{\nu}}{M^2_W - (p - q)^2 - i0}\Big),
\end{eqnarray}
where we have used the propagator of the electroweak $W^-$-boson in
the following form
\begin{eqnarray*}
D^{(W)}_{\mu\nu}(p - q) = - \frac{1}{M^2_W} \Big(\eta_{\mu\nu} +
\frac{(p - q)^2 \eta_{\mu\nu} - (p - q)_{\mu}(p - q)_{\nu}}{M^2_W - (p
  - q)^2 - i0}\Big).
\end{eqnarray*}
The calculation of the matrix element Eq.(\ref{eq:C1m.1}) reduces to
the calculation of the integrals ${\cal F}^{(1)}(k_n, k_e, q)_{1m}$
and ${\cal F}^{(2)}(k_n, k_e, q)_{1m}$ given by
\begin{eqnarray}\label{eq:C1m.2}
\hspace{-0.30in}&&{\cal F}^{(1)}(k_n, k_e, q)_{1m} = - \int
\frac{d^4k}{(2\pi)^4i}\int \frac{d^4p}{(2\pi)^4i}\,\frac{1}{m^2_N - (p
  + k_e)^2 - i0} \frac{1}{m^2_N - (k_p + k)^2 - i0} \frac{1}{m^2_N -
  (k_p + k - p)^2 - i0} \nonumber\\ \hspace{-0.30in}&& \times
\,\frac{1}{m^2_N - (k_n + k)^2 - i0} \frac{1}{m^2_{\pi} - k^2 - i0}
\frac{1}{p^2 + i0} \Big( \gamma^{\beta} \hat{p} \gamma^{\mu} (1 -
\gamma^5) \otimes \hat{k} \gamma_{\beta} \big(\hat{k} - \hat{p}\big)
\gamma_{\mu} \hat{k} (1 + \gamma^5) \Big)
\end{eqnarray}
and
\begin{eqnarray}\label{eq:C1m.3}
\hspace{-0.30in}&&{\cal F}^{(2)}(k_n, k_e, q)_{1m} = - \int
\frac{d^4k}{(2\pi)^4i}\int \frac{d^4p}{(2\pi)^4i}\,\frac{1}{m^2_N - (p
  + k_e)^2 - i0} \frac{1}{m^2_N - (k_p + k)^2 - i0} \frac{1}{m^2_N -
  (k_p + k - p)^2 - i0} \nonumber\\ \hspace{-0.30in}&& \times
\,\frac{1}{m^2_N - (k_n + k)^2 - i0} \frac{1}{m^2_{\pi} - k^2 - i0}
\frac{1}{p^2 + i0} \frac{1}{M^2_W - (p - q)^2 - i0}\Big(p^2
\gamma^{\beta} \hat{p} \gamma^{\mu} (1 - \gamma^5) \otimes \hat{k}
\gamma_{\beta} (\hat{k} - \hat{p}) \gamma_{\mu} \hat{k} (1 + \gamma^5)
\nonumber\\ \hspace{-0.30in}&& + p^2 \gamma^{\mu} (1 - \gamma^5)
\otimes \hat{k} \gamma_{\mu} \big(k^2 \hat{p} + p^2 \hat{k} - 2 (k
\cdot p) \hat{k}\big) (1 + \gamma^5)\Big),
\end{eqnarray}
where we have kept only leading divergent contributions that is
equivalent to the use of the LLA.

\subsubsection*{\bf Calculation of the integral
  ${\cal F}^{(1)}(k_n, k_e, q)_{1m}$}

In terms of the integrals over the Feynman parameters and after the 
integration over the virtual momenta in the $n$-dimensional momentum
space, we obtain for the integral ${\cal F}^{(1)}(k_n, k_e, q)_{1m}$
the following expression
\begin{eqnarray}\label{eq:C1m.4}
\hspace{-0.30in}&&{\cal F}^{(1)}(k_n, k_e, q)_{1m} = \int^1_0 dx_1
\int^1_0 dx_2 \int^1_0 dx_3 \int^1_0 dx_4 \int^1_0 dx_5 \int^1_0 dx_6
\frac{\delta(1 - x_1 - x_2 - x_3 - x_4 - x_5 - x_6)}{2^{2n} \pi^n
  U^{n/2}} \nonumber\\ \hspace{-0.30in}&& \times \Bigg\{
\frac{n^2}{4}\frac{\displaystyle \Gamma\Big(2 -
  \frac{n}{2}\Big)\Gamma\Big(2 - \frac{n - 1}{2}\Big)}{\displaystyle
  2^{n - 3}\Gamma\Big(\frac{1}{2}\Big)}\,\Big(\frac{Q}{m^2_N}\Big)^{-4
  + n}\Big[\frac{2}{n^2}\, \gamma^{\beta}\gamma^{\alpha} \gamma^{\mu}
  (1 - \gamma^5)\otimes \gamma_{\mu} \gamma_{\alpha} \gamma_{\beta}(1
  + \gamma^5) + \frac{n - 4}{n^2}\, \gamma^{\beta}\gamma^{\alpha}
  \gamma^{\mu}\nonumber\\ \hspace{-0.30in}&& \times\, (1 -
  \gamma^5)\otimes \gamma_{\beta} \gamma_{\alpha} \gamma_{\mu} (1 +
  \gamma^5)\Big] + \ldots \Bigg\},
\end{eqnarray}
where we have used the relation $\gamma^{\alpha} \hat{a} \hat{b}
\hat{c}\gamma_{\alpha} = - 2 \hat{c} \hat{b} \hat{a} - (n - 4)\,
\hat{a} \hat{b} \hat{c}$ for the Dirac $\gamma$-matrices in the
$n$-dimensional space-time and denoted
\begin{eqnarray*}
\hspace{-0.30in}U&=& (x_1 + x_2 + x_4) (x_3 + x_5 + x_6) + (x_5 + x_6)
x_3, \nonumber\\
\hspace{-0.30in}b_1 &=& \frac{(x_3 + x_5 + x_6) a_1 + x_3 a_2}{U} =
\nonumber\\
\end{eqnarray*}
\begin{eqnarray}\label{eq:C1m.5}
\hspace{-0.30in}&=& - \frac{((x_5 + x_6)(x_2 + x_3) +
  x_2x_3)q + ((x_5 + x_6)(x_2 + x_3 + x_4) + (x_2 + x_4) x_3)k_n + x_3
  x_5 k_e}{U} = \nonumber\\
\hspace{-0.30in}&=& \frac{X q + Y k_n + Z k_e}{U},\nonumber\\
\hspace{-0.30in}b_2 &=& \frac{(x_1 + x_2 + x_3 + x_4) a_2 + x_3
  a_1}{U} = \frac{(x_1 + x_4) x_3 q + x_1 x_3 k_n - (x_1 + x_2 + x_3 +
  x_4) x_5 k_e}{U} = \nonumber\\
\hspace{-0.30in}&=& \frac{\bar{X} q + \bar{Y} k_n + \bar{Z}
  k_e}{U},\nonumber\\
\hspace{-0.30in}Q &=& x_1  m^2_{\pi} +
\frac{(x_3 + x_5 + x_6) a^2_1 + 2 x_3 a_1 \cdot a_2 + (x_1 + x_2 + x_3
  + x_4) a^2_2}{U}
\end{eqnarray}
with $a_1 = -(x_2 + x_3) q - (x_2 + x_3 + x_4) k_n$ and $a_2 = x_3 q +
x_3 k_n - x_5 k_e$. Taking the limit $n \to 4$ and keeping the
divergent contribution proportional to $\Gamma(2 - n/2)$, we get
\begin{eqnarray}\label{eq:C1m.6}
\hspace{-0.30in}&&{\cal F}^{(1)}(k_n, k_e, q)_{1m} = \int^1_0 dx_1
\int^1_0 dx_2 \int^1_0 dx_3 \int^1_0 dx_4 \int^1_0 dx_5 \int^1_0 dx_6
\frac{\delta(1 - x_1 - x_2 - x_3 - x_4 - x_5 - x_6)}{256\pi^4 U^2}
\nonumber\\ \hspace{-0.30in}&& \times \Big\{\frac{5}{2}\,
  \Gamma\Big(2 - \frac{n}{2}\Big) \gamma^{\mu} (1 - \gamma^5) \otimes
  \gamma_{\mu} (1 + \gamma^5) + \Big(-1 - 5\,{\ell
    n}\frac{Q}{m^2_N}\Big)\gamma^{\mu} (1 - \gamma^5) \otimes
  \gamma_{\mu} (1 + \gamma^5) + \ldots \Big\}.
\end{eqnarray}
The integral ${\cal F}^{(1)}(k_n, k_e, q)_{1m}$ is defined by the structure
\begin{eqnarray}\label{eq:C1m.7}
\hspace{-0.30in}&&{\cal F}^{(1)}(k_n, k_e, q)_{1m} = \int^1_0 dx_1
\int^1_0 dx_2 \int^1_0 dx_3 \int^1_0 dx_4 \int^1_0 dx_5 \int^1_0 dx_6
\frac{\delta(1 - x_1 - x_2 - x_3 - x_4 - x_5 - x_6)}{256\pi^4 U^2}
\nonumber\\ \hspace{-0.30in}&& \times
\Big\{\Big(\frac{5}{2}\,\Gamma\Big(2 - \frac{n}{2}\Big) +
f^{(V)}_{1m^{(1)}}(x_1, \ldots, x_6) + g^{(V)}_{1m^{(1)}}(x_1, \ldots,
x_6)\, \frac{k_n\cdot q}{m^2_N} + h^{(V)}_{1m^{(1)}}(x_1, \ldots,
x_6)\, \frac{k_n\cdot k_e}{m^2_N}\Big)\,\gamma^{\mu} (1 - \gamma^5)
\otimes \gamma_{\mu} \nonumber\\ \hspace{-0.30in}&& +
\Big(\frac{5}{2}\, \Gamma\Big(2 - \frac{n}{2}\Big) +
f^{(A)}_{1m^{(1)}}(x_1, \ldots, x_6) + g^{(A)}_{1m^{(1)}}(x_1, \ldots,
x_6)\, \frac{k_n\cdot q}{m^2_N} + h^{(A)}_{1m^{(1)}}(x_1, \ldots,
x_6)\, \frac{k_n\cdot k_e}{m^2_N}\Big)\gamma^{\mu} (1 - \gamma^5)
\otimes \gamma_{\mu} \gamma^5 \Big\}, \nonumber\\ \hspace{-0.30in}&&
\end{eqnarray}
where we have denoted
\begin{eqnarray}\label{eq:C1m.8}
\hspace{-0.30in}f^{(V)}_{1m^{(1)}}(x_1, \ldots, x_6) &=& - 1 - 5\,
       {\ell n}\frac{\bar{Q}}{m^2_N} \quad,\quad
       g^{(V)}_{1m^{(1)}}(x_1, \ldots, x_6) = - 10\, \frac{V}{U}
       \,\frac{m^2_N}{\bar{Q}} \quad,\quad h^{(V)}_{1m^{(1)}}(x_1,
       \ldots, x_6) = - 10\, \frac{\bar{V}}{U}
       \,\frac{m^2_N}{\bar{Q}},\nonumber\\
       \hspace{-0.30in}f^{(A)}_{1m^{(1)}}(x_1, \ldots, x_6) &=& - 1 -
       5\, {\ell n}\frac{\bar{Q}}{m^2_N} \quad,\quad
       g^{(A)}_{1m^{(1)}}(x_1, \ldots, x_6) = - 10\, \frac{V}{U}
       \,\frac{m^2_N}{\bar{Q}} \quad,\quad h^{(A)}_{1m^{(1)}}(x_1,
       \ldots, x_6) = - 10\, \frac{\bar{V}}{U}
       \,\frac{m^2_N}{\bar{Q}}.\nonumber\\ \hspace{-0.30in}&&
\end{eqnarray}
For the calculation of the integrals over  the Feynman parameters we
use
\begin{eqnarray}\label{eq:C1m.9}
\hspace{-0.3in} \frac{\bar{Q}}{m^2_N} &=& \frac{1}{U}\Big(x_1 U
\frac{m^2_{\pi}}{m^2_N} + (x_5 + x_6) (x_2 + x_3 + x_4)^2 + x_3 (x_2 +
x_4)^2 + (x_1 + x_2 + x_4) x^2_3\Big), \nonumber\\
\hspace{-0.3in} V&=& (x_2 + x_3 + x_4)((x_5 + x_6)(x_2 + x_3) + x_2
x_3) + (x_1 + x_4) x^2_3\;,\; \bar{V} = - x_1 x_3 x_5,
\end{eqnarray}
where $\bar{Q} = Q\big|_{q = k_e = 0}$.  A deviation of $Q$ from
$\bar{Q}$ is taken into account in the linear approximation in the
form of an expansion in powers of $k_n\cdot q/m^2_N$ (or $q_0/m_N$)
and $k_n\cdot k_e/m^2_N$ (or $E_e/m_N$), respectively. The structure
functions $g^{(V)}_{1m^{(1)}}(x_1, \ldots, x_6)$ and
$h^{(V)}_{1m^{(1)}}(x_1, \ldots, x_6)$ as well as the structure
functions $g^{(A)}_{1m^{(1)}}(x_1, \ldots, x_6)$ and
$h^{(A)}_{1m^{(1)}}(x_1, \ldots, x_6)$ are calculated by an expansion
of the logarithmic functions $ - 5\, {\ell n} Q/m^2_N$ in powers of
$k_n\cdot q/m^2_N$ (or $q_0/m_N$) and $k_n\cdot k_e/m^2_N$ (or
$E_e/m_N$), respectively.

After the integration over the Feynman parameters, we obtain the
contribution of the integral ${\cal F}^{(1)}(k_n, k_e, q)_{1m}$ to the
matrix element $M(n \to p e^- \bar{\nu}_e)^{(\pi^0)}_{\rm
  Fig.\,\ref{fig:fig1}m}$:
\begin{eqnarray}\label{eq:C1m.10}
\hspace{-0.30in}&&M(n \to p e^- \bar{\nu}_e)^{(\pi^0)}_{\rm
  Fig.\,\ref{fig:fig1}m^{(1)}} = \frac{\alpha}{4\pi} \frac{g^2_{\pi
    N}}{16\pi^2} G_V \Big\{A_{1m^{(1)}}\,\Gamma\Big(2 -
\frac{n}{2}\Big) \, \Big[\bar{u}_e \gamma^{\mu} (1 - \gamma^5)
  v_{\bar{\nu}}\Big]\Big[\bar{u}_p \gamma_{\mu} u_n\Big] +
B_{1m^{(1)}} \,\Gamma\Big(2 - \frac{n}{2}\Big)
\nonumber\\ \hspace{-0.30in}&&\times \, \Big[\bar{u}_e \gamma^{\mu} (1
  - \gamma^5) v_{\bar{\nu}}\Big] \Big[\bar{u}_p \gamma_{\mu} \gamma^5
  u_n\Big] + \Big( F^{(V)}_{1m^{(1)}} + G^{(V)}_{1m^{(1)}} \,
\frac{k_n\cdot q}{m^2_N} + H^{(V)}_{1m^{(1)}} \, \frac{k_n\cdot
  k_e}{m^2_N} \Big)\, \Big[\bar{u}_e \gamma^{\mu} (1 - \gamma^5)
  v_{\bar{\nu}}\Big]\Big[\bar{u}_p \gamma_{\mu} u_n\Big]
\nonumber\\ \hspace{-0.30in}&& + \Big( F^{(A)}_{1m^{(1)}} +
G^{(A)}_{1m^{(1)}}\, \frac{k_n\cdot q}{m^2_N} + H^{(A)}_{1m^{(1)}}\,
\frac{k_n\cdot k_e}{m^2_N}\Big)\, \Big[\bar{u}_e \gamma^{\mu} (1 -
  \gamma^5) v_{\bar{\nu}}\Big] \Big[\bar{u}_p \gamma_{\mu} \gamma^5
  u_n\Big]\Big\}.
\end{eqnarray}
The structure constants are equal to $A_{1m^{(1)}} = 5/12$,
$B_{1m^{(1)}} = 5/12$, $F^{(V)}_{1m^{(1)}} = 0.7420$,
$G^{(V)}_{1m^{(1)}} = - 1.0198$, $H^{(V)}_{1m^{(1)}} = 0.0639$,
$F^{(A)}_{1m^{(1)}} = 0.7420$, $G^{(A)}_{1m^{(1)}} =  - 1.0198$ and
$H^{(A)}_{1m^{(1)}} = 0.0639$. The Lorentz structure of the matrix
element Eq.(\ref{eq:C1m.10}) is calculated at the neglect of the
contributions of order $O(E^2_0/m^2_N) \sim 10^{-6}$.

\subsubsection*{\bf Calculation of the integral
  ${\cal F}^{(2)}(k_n, k_e, q)_{1m}$}

Skipping intermediate calculations, which are well expounded in
previous sections, after integration over the virtual momenta in the
$n$-dimensional momentum space, we define the integral ${\cal
  F}^{(2)}(k_n, k_e, q)_{1m}$ in terms of the integrals over the
Feynman parameters. We obtain
\begin{eqnarray}\label{eq:C1m.11}
\hspace{-0.3in}&&{\cal F}^{(2)}(k_n, k_e, q)_{1m} = \int^1_0 \!\!
dx_1 \int^1_0 \!\! dx_2 \int^1_0 \!\! dx_3 \int^1_0 \!\! dx_4 \int^1_0
\!\! dx_5 \int^1_0 \!\! dx_6 \int^1_0 \!\! dx_7\,\frac{\delta(1 - x_1
  - x_2 - x_3 - x_4 - x_5 - x_6 - x_7)}{2^{2n} \pi^n
  U^{n/2}} \nonumber\\
\hspace{-0.3in}&& \times \Bigg\{ \frac{n^2 (n +
  2)}{8}\,\frac{\displaystyle \Gamma\Big(2 -
  \frac{n}{2}\Big)\Gamma\Big(2 - \frac{n - 1}{2}\Big)}{\displaystyle
  2^{n - 3}\Gamma\Big(\frac{1}{2}\Big)}\,\Big(\frac{Q}{m^2_N}\Big)^{-4
  + n}\,\Big[\frac{n - 2}{n}\, \gamma^{\mu} (1 - \gamma^5) \otimes
  \gamma_{\mu} (1 + \gamma^5) - \frac{2}{n^2}\, \gamma^{\beta}
  \gamma^{\alpha} \gamma^{\mu} (1 - \gamma^5) \nonumber\\
\hspace{-0.3in}&& \otimes  \gamma_{\mu} \gamma_{\alpha}\gamma_{\beta}
(1 + \gamma^5) - \frac{n - 4}{n^2} \, \gamma^{\beta} \gamma^{\alpha}
\gamma^{\mu} (1 - \gamma^5) \otimes \gamma_{\beta} \gamma_{\alpha}
\gamma_{\mu} (1 + \gamma^5)\Big] + \frac{n^2}{4} \Gamma(5 -
n)\,Q^{-5+n} m^{2(4-n)}_N \Big[\Big(2\, \frac{n^2 - 4}{n^2}\, (b_1
  \cdot b_2) \nonumber\\
\hspace{-0.3in}&& - \frac{(n - 2)(n + 4)}{n^2}\, b^2_2 \Big)\,
\gamma^{\mu} (1 - \gamma^5) \otimes \gamma_{\mu} (1 + \gamma^5) -
\frac{n + 2}{n^2}\, \gamma^{\beta} \gamma^{\alpha} \gamma^{\mu} (1 -
\gamma^5) \otimes \hat{b}_1 \gamma_{\beta} \gamma_{\alpha}
\gamma_{\mu} \hat{b}_1 (1 + \gamma^5) + \gamma^{\mu} (1 -
\gamma^5)\nonumber\\
\hspace{-0.3in}&& \otimes \Big( \frac{n + 2}{n^2}\, \hat{b}_1
\gamma_{\mu} \hat{b}_1 - 2\, \frac{n + 2}{n^2}\, \hat{b}_2
\gamma_{\mu} \hat{b}_1 - \frac{(n - 2)^2}{n^2}\,\hat{b}_1 \gamma_{\mu}
\hat{b}_2 \Big)(1 + \gamma^5) + \gamma^{\beta} \hat{b}_2
\gamma^{\alpha} (1 - \gamma^5) \otimes \Big(- \frac{n^2 - 4}{n^2}\,
\gamma_{\beta} \gamma_{\alpha} \hat{b}_1 - \frac{n - 2}{n}\nonumber\\
\hspace{-0.3in}&& \times \, \hat{b}_1 \gamma_{\beta} \gamma_{\alpha} -
\frac{2}{n}\, \gamma_{\alpha} \hat{b}_1 \gamma_{\beta} + 2\, \frac{n +
  2}{n^2}\,\gamma_{\alpha} \hat{b}_2 \gamma_{\beta} - \frac{n -
  4}{n}\, \gamma_{\beta} \hat{b}_1 \gamma_{\alpha} + \frac{(n + 2)(n -
  4)}{n^2}\, \gamma_{\beta} \hat{b}_2 \gamma_{\alpha}\Big) \Bigg\},
\end{eqnarray} 
where we have used the algebra of the Dirac $\gamma$-matrices in
$n$-dimensional space-time $\gamma_{\alpha}\gamma_{\mu} +
\gamma_{\mu} \gamma_{\alpha} = 2\, \eta_{\alpha \mu}$,
$\gamma^{\lambda} \gamma_{\lambda} = n$, $\gamma^{\lambda}
\gamma_{\alpha} \gamma_{\lambda} = - (n - 2)\, \gamma_{\alpha}$,
$\gamma^{\lambda} \gamma_{\alpha} \gamma_{\mu} \gamma_{\lambda} =
4\eta_{\alpha \mu} + (n - 4)\, \gamma_{\alpha} \gamma_{\mu}$ and
$\gamma^{\lambda}\gamma^{\beta} \gamma^{\alpha}
\gamma^{\mu}\gamma_{\lambda} = - 2 \gamma^{\mu} \gamma^{\alpha}
\gamma^{\beta}- (n - 4)\, \gamma^{\beta} \gamma^{\alpha} \gamma^{\mu}$
\cite{Itzykson1980} and denoted
\begin{eqnarray}\label{eq:C1m.12}
  \hspace{-0.3in}U &=& (x_1 + x_2 + x_3 + x_4)( x_5 + x_6 + x_7) +
  (x_1 + x_2 + x_4) x_3,\nonumber\\
  \hspace{-0.3in}b_1 &=& \frac{( x_3 + x_5 + x_6 + x_7)a_1 + x_3
    a_2}{U} = - \frac{(x_5 + x_6 + x_7)(x_2 + x_3 + x_4) + (x_2 + x_4)
    x_3 }{U}\, k_n = \frac{Y}{U}\,k_n ,\nonumber\\
   \hspace{-0.3in}b_2 &=& \frac{(x_1 + x_2 + x_3 + x_4)a_2 +
     x_3 a_1}{U} = \frac{x_1 x_3}{U}\, k_n =
   \frac{\bar{Y}}{U}\,k_n,\nonumber\\
   \hspace{-0.3in}Q &=&x_7 M^2_W + \frac{( x_3 + x_5 + x_6 + x_7)
     a^2_1 + 2 x_3 a_1\cdot a_2 + (x_1 + x_2 + x_3 + x_4) a^2_2}{U}
\end{eqnarray}
with $a_1 = - (x_2 + x_3 + x_4) k_n$ and $a_2 = x_3 k_n$. In
Eq.(\ref{eq:C1m.11}) we have neglected the contributions of the terms
proportional to $q$ and $k_e$, appearing in ${\cal F}^{(2)}(k_n, k_e,
q)_{1c}$ in the form of an expansion in powers of $k_n\cdot q/M^2_W$
and $k_n\cdot k_e/M^2_W$ of the order of $10^{-7}$, and the terms of
order $O(m^2_{\pi}/M^2_W) \sim 10^{-6}$. Taking the limit $n \to 4$,
keeping the divergent contributions proportional to $\Gamma(2 - n/2)$
and using the Dirac equations for free fermions, we arrive at the
expression
\begin{eqnarray}\label{eq:C1m.13}
\hspace{-0.30in}&&{\cal F}^{(2)}(k_n, k_e, q)_{1m} = \int^1_0dx_1
\int^1_0dx_2 \int^1_0dx_3 \int^1_0dx_4 \int^1_0dx_5 \int^1_0 dx_6
\int^1_0 dx_7\,\frac{\delta(1 - x_1 - x_2 - x_3 - x_4 - x_5 - x_6 -
  x_7)}{256 \pi^4 U^2}\nonumber\\
\hspace{-0.30in}&& \times\Big\{\Big( - 9 \, \Gamma\Big(2 -
\frac{n}{2}\Big) + f^{(V)}_{1m^{(2)}}(x_1,\ldots, x_7) +
g^{(W)}_{1m^{(2)}}(x_1,\ldots, x_7)\,\frac{m^2_N}{M^2_W}\Big)
\gamma^{\mu} (1 - \gamma^5) \otimes \gamma_{\mu} + \Big(- 9\,
\Gamma\Big(2 - \frac{n}{2}\Big) \nonumber\\
\hspace{-0.30in}&& + f^{(A)}_{1m^{(2)}}(x_1,\ldots, x_7) +
h^{(W)}_{1m^{(2)}}(x_1,\ldots, x_7)\,\frac{m^2_N}{M^2_W}\Big)
\gamma^{\mu} (1 - \gamma^5) \otimes \gamma_{\mu} \gamma^5 +
f^{(W)}_{1m^{(2)}}(x_1,\ldots, x_7)\,\frac{m^2_N}{M^2_W}\,
\frac{\hat{k}_n}{m_N} (1 - \gamma^5) \otimes 1\Big\},\nonumber\\
\hspace{-0.30in}&&
\end{eqnarray}
where we have denoted
\begin{eqnarray}\label{eq:C1m.14}
\hspace{-0.30in}&&f^{(V)}_{1m^{(2)}}(x_1,\ldots, x_7) = 3 + 18 \,
       {\ell n}\frac{M^2_W}{m^2_N} + 18\,{\ell n} \bar{Q}
\end{eqnarray}
and 
\begin{eqnarray}\label{eq:C1m.15}
\hspace{-0.30in}&&g^{(W)}_{1m^{(2)}}(x_1,\ldots, x_7) =
\frac{1}{\bar{Q}}\Big[- \frac{45}{2}\, \frac{Y^2}{U^2} - \frac{7}{2}\,
  \frac{Y\bar{Y}}{U^2} + 17\, \frac{\bar{Y}^2}{U^2}\Big]
\end{eqnarray}
and 
\begin{eqnarray}\label{eq:C1m.16}
\hspace{-0.30in}&&f^{(A)}_{1m^{(2)}}(x_1,\ldots, x_7) = 3 + 18\, {\ell
  n}\frac{M^2_W}{m^2_N} + 18\,{\ell n} \bar{Q}
\end{eqnarray}
and 
\begin{eqnarray}\label{eq:C1m.17}
\hspace{-0.30in}&&h^{(W)}_{1m^{(2)}}(x_1,\ldots, x_7) =
\frac{1}{\bar{Q}}\Big[ \frac{45}{2}\,
  \frac{Y^2}{U^2} - \frac{41}{2}\, \frac{Y\bar{Y}}{U^2} + 19\,
  \frac{\bar{Y}^2}{U^2}\Big]
\end{eqnarray}
and
\begin{eqnarray}\label{eq:C1m.18}
\hspace{-0.30in}&&f^{(W)}_{1m^{(2)}}(x_1,\ldots, x_7) =
\frac{1}{\bar{Q}}\Big[20\, \frac{Y\bar{Y}}{U^2}\Big],
\end{eqnarray}
where for numerical calculation we use
\begin{eqnarray}\label{eq:C1m.19}
  \hspace{-0.30in}\bar{Q} &=& \frac{1}{U}\Big(x_7 U + \big((x_5 + x_6
  + x_7) (x_2 + x_3 + x_4)^2 + x_3 (x_2 + x_4) (x_2 + x_3 + x_4) + x_1
  x^2_3\big)\, \frac{m^2_N}{M^2_W}\Big), \nonumber\\
  \hspace{-0.3in}U &=& (x_1 + x_2 + x_3 + x_4)( x_5 + x_6 + x_7) +
  (x_1 + x_2 + x_4) x_3,\nonumber\\
   \hspace{-0.30in} Y &=& - (x_5 + x_6 + x_7) (x_2 + x_3 + x_4) - (x_2
   + x_4) x_3\;,\; \bar{Y} = x_1 x_3
\end{eqnarray}
with $m_N = (m_n + m_p)/2 = 938.9188\,{\rm MeV}$ and $M_W =
80.379\,{\rm GeV}$ \cite{PDG2020}. After the integration over the
Feynman parameters, we obtain the contribution of the integral ${\cal
  F}^{(2)}(k_n, k_e, q)_{1m}$ to the matrix element $M(n \to p e^-
\bar{\nu}_e)^{(\pi^0)}_{\rm Fig.\,\ref{fig:fig1}m}$: 
\begin{eqnarray}\label{eq:C1m.20}
\hspace{-0.30in}&&M(n \to p e^- \bar{\nu}_e)^{(\pi^0)}_{\rm
  Fig.\,\ref{fig:fig1}m^{(2)}} = \frac{\alpha}{4\pi} \frac{g^2_{\pi
    N}}{16\pi^2} G_V \Big\{A_{1m^{(2)}}\,\Gamma\Big(2 -
\frac{n}{2}\Big) \, \Big[\bar{u}_e \gamma^{\mu} (1 - \gamma^5)
  v_{\bar{\nu}}\Big]\Big[\bar{u}_p \gamma_{\mu} u_n\Big] +
B_{1m^{(2)}} \,\Gamma\Big(2 - \frac{n}{2}\Big)
\nonumber\\ \hspace{-0.30in}&&\times \, \Big[\bar{u}_e \gamma^{\mu} (1
  - \gamma^5) v_{\bar{\nu}}\Big] \Big[\bar{u}_p \gamma_{\mu} \gamma^5
  u_n\Big] + \Big( F^{(V)}_{1m^{(2)}} + G^{(W)}_{1m^{(1)}} \,
\frac{m^2_N}{M^2_W} \Big)\, \Big[\bar{u}_e \gamma^{\mu} (1 - \gamma^5)
  v_{\bar{\nu}}\Big]\Big[\bar{u}_p \gamma_{\mu} u_n\Big]
\nonumber\\ \hspace{-0.30in}&& + \Big( F^{(A)}_{1m^{(2)}} +
H^{(W)}_{1m^{(2)}}\, \frac{m^2_N}{M^2_W}\Big)\, \Big[\bar{u}_e
  \gamma^{\mu} (1 - \gamma^5) v_{\bar{\nu}}\Big] \Big[\bar{u}_p
  \gamma_{\mu} \gamma^5 u_n\Big] + F^{(W)}_{1m^{(2)}} \,
\frac{m^2_N}{M^2_W} \, \Big[\bar{u}_e \frac{\hat{k}_n}{m_N} (1 -
  \gamma^5) v_{\bar{\nu}}\Big] \Big[\bar{u}_p u_n\Big]\Big\}.
\end{eqnarray}
The structure constants are equal to $A_{1m^{(2)}} = - 0.2031$,
$F^{(V)}_{1m^{(2)}} = 2.6738$, $G^{(W)}_{1m^{(2)}} = - 13.6821$,
$B_{1m^{(2)}} = - 0.2031$, $F^{(A)}_{1m^{(2)}} = 2.6738$,
$H^{(W)}_{1m^{(2)}} = 15.8163$ and $F^{(W)}_{1m^{(2)}} = -
1.1188$. The Lorentz structure of the matrix element
Eq.(\ref{eq:C1m.20}) is calculated at the neglect of the contributions
of order $O(k_n \cdot q/M^2_W) \sim O(k_n \cdot k_e/M^2_W) \sim
10^{-7}$ and $O(m^2_{\pi}/M^2_W) \sim 10^{-6}$, respectively.

As a result, the contribution of the Feynman diagram
Fig.\,\ref{fig:fig1}m to amplitude of the neutron beta decay  is
given by
\begin{eqnarray}\label{eq:C1m.21}
\hspace{-0.30in}&&M(n \to p e^- \bar{\nu}_e)^{(\pi^0)}_{\rm
  Fig.\,\ref{fig:fig1}m} = \frac{\alpha}{4\pi} \frac{g^2_{\pi
    N}}{16\pi^2} G_V \Big\{A_{1m}\,\Gamma\Big(2 - \frac{n}{2}\Big) \,
\Big[\bar{u}_e \gamma^{\mu} (1 - \gamma^5)
  v_{\bar{\nu}}\Big]\Big[\bar{u}_p \gamma_{\mu} u_n\Big] +
B_{1m} \,\Gamma\Big(2 - \frac{n}{2}\Big)
\nonumber\\ \hspace{-0.30in}&&\times \, \Big[\bar{u}_e \gamma^{\mu} (1
  - \gamma^5) v_{\bar{\nu}}\Big] \Big[\bar{u}_p \gamma_{\mu} \gamma^5
  u_n\Big] + \Big( F^{(V)}_{1m} + G^{(V)}_{1m} \, \frac{k_n\cdot
  q}{m^2_N} + H^{(V)}_{1m} \, \frac{k_n\cdot k_e}{m^2_N} +
G^{(W)}_{1m}\, \frac{m^2_N}{M^2_W}\Big)\, \Big[\bar{u}_e \gamma^{\mu}
  (1 - \gamma^5) v_{\bar{\nu}}\Big]
\nonumber\\ \hspace{-0.30in}&&\times \Big[\bar{u}_p \gamma_{\mu}
  u_n\Big] + \Big( F^{(A)}_{1m} + G^{(A)}_{1m}\, \frac{k_n\cdot
  q}{m^2_N} + H^{(A)}_{1m}\, \frac{k_n\cdot k_e}{m^2_N} + H^{(W)}\,
\frac{m^2_N}{M^2_W}\Big)\, \Big[\bar{u}_e \gamma^{\mu} (1 - \gamma^5)
  v_{\bar{\nu}}\Big] \Big[\bar{u}_p \gamma_{\mu} \gamma^5 u_n\Big] +
F^{(W)}_{1m} \nonumber\\ \hspace{-0.30in}&&\times \,
\frac{m^2_N}{M^2_W} \, \Big[\bar{u}_e \frac{\hat{k}_n}{m_N} (1 -
  \gamma^5) v_{\bar{\nu}}\Big] \Big[\bar{u}_p u_n\Big]\Big\}.
\end{eqnarray}
The structure constants are equal to $A_{1m} = A_{1m^{(1)}} +
A_{1m^{(2)}} = 0.2136$, $B_{1m} = B_{1m^{(1)}} + B_{1m^{(2)}} =
0.2136$, $F^{(V)}_{1m} = F^{(V)}_{1m^{(1)}} + F^{(V)}_{1m^{(2)}} =
3.4158$, $G^{(V)}_{1m} = G^{(V)}_{1m^{(1)}} = - 1.0198$, $H^{(V)}_{1m}
= H^{(V)}_{1m^{(1)}} = 0.0639$, $G^{(W)}_{1m} = G^{(W)}_{1m^{(2)}} =
-13.6821$, $F^{(A)}_{1m} = F^{(A)}_{1m^{(1)}} + F^{(A)}_{1m^{(2)}} =
3.4158$, $G^{(A)}_{1m} = G^{(A)}_{1m^{(1)}} = - 1.0198$, $H^{(A)} =
H^{(A)}_{1m^{(1)}} = 0.0639$, $H^{(W)}_{1m} = H^{(W)}_{1m^{(2)}} =
15.8163$ and $F^{(W)}_{1m} = F^{(W)}_{1m^{(2)}} = - 1.1188$. The
Lorentz structure of Eq.(\ref{eq:C1m.21}) is calculated at the neglect
of the contributions of order $O(E^2_0/m^2_N)\sim O(m^2_{\pi}/M^2_W)
\sim 10^{-6}$.

\subsection*{C1n. Analytical calculation of the Feynman diagram in
  Fig.\,\ref{fig:fig1}n}
\renewcommand{\theequation}{C1n-\arabic{equation}}
\setcounter{equation}{0}

In the Feynman gauge for the photon propagator the analytical
expression of the Feynman diagram in Fig.\,\ref{fig:fig1}n is given by
(see Eq.(\ref{eq:A.1}))
\begin{eqnarray}\label{eq:C1n.1}
\hspace{-0.30in} &&M(n \to p e^- \bar{\nu}_e)^{(\pi^0)}_{\rm
  Fig.\,\ref{fig:fig1}n} = - 2 e^2 g^2_{\pi N}
G_V\nonumber\\\hspace{-0.30in} &&\times \int
\frac{d^4k}{(2\pi)^4i}\int \frac{d^4p}{(2\pi)^4i}\,\Big[\bar{u}_e
  \gamma^{\mu}(1 - \gamma^5) v_{\bar{\nu}}\Big]\,\Big[\bar{u}_p
  \,\gamma_{\beta} \frac{1}{m_N - \hat{k}_p + \hat{p} -
    i0}\,\gamma^5\, \frac{1}{m_N - \hat{k}_n - \hat{k} -
    i0}\,\gamma^5\, u_n\Big]\nonumber\\ \hspace{-0.30in}&&\times \,
\frac{(2 k + p - q)^{\alpha_2}}{[m^2_{\pi} - k^2 - i0][m^2_{\pi} - (k
    + p - q)^2 - i0]}\, \big[(p - q)_{\mu}\eta^{\beta \alpha_1} - (p -
  2 q)^{\beta} \eta^{\alpha_1}{}_{\mu} - q^{\alpha_1}
  \eta_{\mu}{}^{\beta}\big]\,\frac{1}{p^2 + i0}\,  \nonumber\\ \hspace{-0.30in}&& \times \frac{1}{M^2_W - (p - q)^2 - i0}\Big(\eta_{\alpha_1 \alpha_2} - \frac{(p - q)_{\alpha_1}(p - q)_{\alpha_2}}{M^2_W}\Big),
\end{eqnarray}
where we have made an usual replacement $M^2_W D^{(W)}_{\mu \nu}(- q)
= - \eta_{\mu\nu}$ and used the standard propagator of the electroweak
$W^-$-boson in the physical gauge \cite{Itzykson1980}. Since the
calculation of the matrix element in Eq.(\ref{eq:C1n.1}) is similar to
the calculation of the matrix element in Eq.(\ref{eq:C1a.1}), we skip
standard intermediate calculations and give the r.h.s. of
Eq.(\ref{eq:C1n.1}) in terms of the integrals over the Feynman
parameters. We get
\begin{eqnarray}\label{eq:C1n.2}
\hspace{-0.30in} &&M(n \to p e^- \bar{\nu}_e)^{(\pi^0)}_{\rm
  Fig.\,\ref{fig:fig1}n} = - \frac{\alpha}{2\pi}\frac{g^2_{\pi
    N}}{16\pi^2}\, G_V \int^1_0 \!\! dx_1 \int^1_0\!\! dx_2
\int^1_0\!\! dx_3 \int^1_0 \!\! dx_4 \int^1_0\!\! dx_5 \int^1_0 \!\!
dx_6 \,\frac{\delta(1 - x_1 - x_2 - x_3 - x_4 - x_5 -
  x_6)}{U^2}\nonumber\\
\hspace{-0.30in}&& \times\Big\{\Big(\frac{3}{2}\,\Gamma\Big(2 -
  \frac{n}{2}\Big) + f^{(V)}_{1n}(x_1, \ldots, x_6) +
  g^{(W)}_{1n}(x_1, \ldots, x_6)\, \frac{m^2_N}{M^2_W}\Big)
\Big[\bar{u}_e \gamma^{\mu}(1 - \gamma^5)
  v_{\bar{\nu}}\Big]\Big[\bar{u}_p \gamma_{\mu} u_n\Big] +
f^{(W)}_{1n}(x_1, \ldots, x_6)\, \frac{m^2_N}{M^2_W}\nonumber\\
\hspace{-0.30in}&& \times \Big[\bar{u}_e \frac{\hat{k}_n}{m_N}(1 -
  \gamma^5) v_{\bar{\nu}}\Big]\Big[\bar{u}_p u_n\Big]\Big\},
\end{eqnarray}
where we have taken the limit $n \to 4$, kept the divergent
contribution proportional to $\Gamma(2 - n/2)$ and denoted
\begin{eqnarray}\label{eq:C1n.3}
\hspace{-0.30in} && f^{(V)}_{1n}(x_1, \ldots, x_6) = - 3\, {\ell
  n}\frac{M^2_W}{m^2_N} - 3\, {\ell n} \bar{Q}
\end{eqnarray}
and 
\begin{eqnarray}\label{eq:C1n.4}
\hspace{-0.30in} && g^{(W)}_{1n}(x_1, \ldots, x_6) =
\frac{1}{\bar{Q}}\Big[ \frac{Y^2}{U^2} +
  \frac{\bar{Y}^2}{U^2}\Big]
\end{eqnarray}
and
\begin{eqnarray}\label{eq:C1n.5}
\hspace{-0.30in} && f^{(W)}_{1n}(x_1, \ldots, x_6) =
\frac{1}{\bar{Q}}\Big[- 2\, \frac{Y^2}{U^2} - 2\,
  \frac{\bar{Y}^2}{U^2}\Big]
\end{eqnarray}
with
\begin{eqnarray}\label{eq:C1n.6}
  \hspace{-0.30in}U &=& (x_1 + x_2 + x_4)(x_3 + x_5 + x_6) + (x_1 +
  x_4) x_2,\nonumber\\
  \hspace{-0.30in}\bar{Q} &=& \frac{1}{U}\Big(x_6 U + \big((x_2 + x_3
  + x_5 + x_6) x^2_4 + 2 x_2 x_3 x_4 + (x_1 + x_2 + x_4) x^2_3\big)\,
  \frac{m^2_N}{M^2_W}\Big), \nonumber\\
  \hspace{-0.30in}Y &=& - (x_3 + x_5 + x_6) x_4 - (x_3 + x_4) x_2
  \;,\; \bar{Y} = (x_1 + x_4) x_3 + (x_3 + x_4) x_2.
\end{eqnarray}
After the integration over the Feynman parameters, we obtain the
following contribution of the Feynman diagram in Fig.\,\ref{fig:fig1}n
to the amplitude of the neutron beta decay 
\begin{eqnarray}\label{eq:C1n.7}
\hspace{-0.30in} M(n \to p e^- \bar{\nu}_e)^{(\pi^0)}_{\rm
  Fig.\,\ref{fig:fig1}n} &=& - \frac{\alpha}{2\pi}\frac{g^2_{\pi
    N}}{16\pi^2}\, G_V \Big\{\Big(A_{1n}\,\Gamma\Big(2 -
  \frac{n}{2}\Big) + F^{(V)}_{1n} + G^{(W)}_{1n}\,
  \frac{m^2_N}{M^2_W}\Big) \Big[\bar{u}_e \gamma^{\mu}(1 - \gamma^5)
  v_{\bar{\nu}}\Big]\Big[\bar{u}_p \gamma_{\mu} u_n\Big]\nonumber\\
\hspace{-0.30in}&+& F^{(W)}_{1n}\, \frac{m^2_N}{M^2_W}\Big[\bar{u}_e
  \frac{\hat{k}_n}{m_N}(1 - \gamma^5) v_{\bar{\nu}}\Big]\Big[\bar{u}_p
  u_n\Big]\Big\}.
\end{eqnarray}
The structure constants are equal to $A_{1n} = 1/4$, $F^{(V)}_{1n} = -
3.3958$, $G^{(W)}_{1n} = 3.5834$ and $F^{(W)}_{1n} = - 7.1669$. The
Lorentz structure of the matrix element Eq.(\ref{eq:C1n.7}) is
calculated at the neglect of the contributions of order $O(k_n \cdot
q/M^2_W) \sim 10^{-7}$ and $O(m^2_{\pi}/M^2_W) \sim 10^{-6}$,
respectively.

\subsection*{C1o. Analytical calculation of the Feynman diagram in
  Fig.\,\ref{fig:fig1}o}
\renewcommand{\theequation}{C1o-\arabic{equation}}
\setcounter{equation}{0}

In the Feynman gauge for the photon propagator the analytical
expression of the Feynman diagram in Fig.\,\ref{fig:fig1}o is given by
(see Eq.(\ref{eq:A.1}))
\begin{eqnarray}\label{eq:C1o.1}
 &&M(n \to p e^- \bar{\nu}_e)^{(\pi^0)}_{\rm Fig.\,\ref{fig:fig1}o} =
  + 2 e^2 g^2_{\pi N} G_V \nonumber\\ &&\times \int
  \frac{d^4k}{(2\pi)^4i}\int \frac{d^4p}{(2\pi)^4i}\,\Big[\bar{u}_e
    \gamma^{\beta} \frac{1}{m_e - \hat{k}_e - \hat{p} - i0}\,
    \gamma^{\mu}(1 - \gamma^5) v_{\bar{\nu}}\Big]
  \Big[\bar{u}_p\gamma_{\beta} \frac{1}{m_N - \hat{k}_p + \hat{p} -
      i0} \gamma^5 \frac{1}{m_N - \hat{k}_n - \hat{k} - i0} \gamma^5
    u_n\Big]\nonumber\\ &&\times \, \frac{(2 k + p -
    q)^{\nu}}{[m^2_{\pi} - k^2 - i0][m^2_{\pi} - (k + p - q)^2 -
      i0]}\, \frac{1}{p^2 + i0}\,\Big(\eta_{\mu\nu} + \frac{(p - q)^2
    \eta_{\mu\nu} - (p - q)_{\mu} (p - q)_{\nu}}{M^2_W - (p - q)^2 -
    i0}\Big),
\end{eqnarray}
where we have taken the propagator of the electroweak $W^-$-boson in
the following form
\begin{eqnarray}\label{eq:C1o.2}
  D^{(W)}_{\mu\nu}(p - q) = - \frac{1}{M^2_W}\Big(\eta_{\mu\nu} +
  \frac{(p - q)^2 \eta_{\mu\nu} - (p - q)_{\mu} (p - q)_{\nu}}{M^2_W -
    (p - q)^2 - i0}\Big).
\end{eqnarray}
The calculation of the Feynman diagram in Fig.\,\ref{fig:fig1}o
reduces to the calculation of two integrals ${\cal F}^{(1)}(k_n, k_e,
q)_{1o}$ and ${\cal F}^{(2)}(k_n, k_e, q)_{1o}$, which are defined by
\begin{eqnarray}\label{eq:C1o.3}
 \hspace{-0.30in}&& {\cal F}^{(1)}(k_n, k_e, q)_{1o} = \int
 \frac{d^4k}{(2\pi)^4i}\int \frac{d^4p}{(2\pi)^4i}\,\frac{1}{m^2_{\pi}
   - k^2 - i0}\frac{1}{m^2_{\pi} - (k + p - q)^2 - i0} \frac{1}{m^2_N
   - (k_p - p)^2 - i0} \nonumber\\ \hspace{-0.30in}&&\times
 \frac{1}{m^2_N - (k_n + k)^2 - i0} \frac{1}{m^2_e -(p + k_e)^2 -
   i0}\, \frac{1}{p^2 + i0}\Big(\gamma^{\beta} (2 \hat{p} \hat{k} +
 p^2) (1 - \gamma^5) \otimes \gamma_{\beta} \hat{p} \hat{k}\Big)
\end{eqnarray}
and
\begin{eqnarray}\label{eq:C1o.4}
 \hspace{-0.30in}&& {\cal F}^{(2)}(k_n, k_e, q)_{1o} = \int
 \frac{d^4k}{(2\pi)^4i}\int \frac{d^4p}{(2\pi)^4i}\,\frac{1}{m^2_{\pi}
   - k^2 - i0}\frac{1}{m^2_{\pi} - (k + p - q)^2 - i0} \frac{1}{m^2_N
   - (k_p - p)^2 - i0} \nonumber\\ \hspace{-0.30in} && \times
 \frac{1}{m^2_N - (k_n + k)^2 - i0} \frac{1}{m^2_e -(p + k_e)^2 -
   i0}\, \frac{1}{p^2 + i0} \frac{1}{M^2_W - (p -
   q)^2 - i0} \Big(2 p^2 \gamma^{\beta}\big(\hat{p} \hat{k} - p\cdot
 k\big) (1 - \gamma^5) \otimes \gamma_{\beta} \hat{p} \hat{k}\Big),
 \nonumber\\ \hspace{-0.30in} &&
\end{eqnarray}
where we have kept only leading divergent contributions that
corresponds to the use of the LLA.  The calculation of the integrals
${\cal F}^{(1)}(k_n, k_e, q)_{1o}$ and ${\cal F}^{(2)}(k_n, k_e,
q)_{1o}$ runs as follows.

\subsubsection*{\bf Calculation of the integral
  ${\cal F}^{(1)}(k_n, k_e, q)_{1o}$}

In terms of the integrals over the Feynman parameters and after
integration over the virtual momenta in the $n$-dimensional momentum
space, we obtain for the integral ${\cal F}^{(1)}(k_n, k_e, q)_{1o}$
the following expression
\begin{eqnarray}\label{eq:C1o.5}
 \hspace{-0.21in}&& {\cal F}^{(1)}(k_n, k_e, q)_{1o} = \int^1_0 dx_1
 \int^1_0 dx_2 \int^1_0 dx_3 \int^1_0 dx_4 \int^1_0 dx_5 \int^1_0 dx_6
 \frac{\delta(1 - x_1 - x_2 - x_3 - x_4 - x_5 - x_6)}{2^{2n} \pi^n
   U^{n/2}} \nonumber\\ \hspace{-0.21in}&& \times \Bigg\{
 \frac{n^2}{4}\, \frac{\displaystyle \Gamma\Big(2 -
   \frac{n}{2}\Big)\Gamma\Big(2 - \frac{n - 1}{2}\Big)}{\displaystyle
   2^{n - 3} \Gamma\Big(\frac{1}{2}\Big)}\,
 \Big(\frac{Q}{m^2_N}\Big)^{-4 + n}\,\Big[- \frac{2}{n^2}\,
   \gamma^{\beta} \gamma^{\alpha} \gamma^{\mu} (1 - \gamma^5) \otimes
   \gamma_{\beta} \gamma_{\alpha} \gamma_{\mu}\Big] + \ldots\Bigg\},
\end{eqnarray}
where we have denoted
\begin{eqnarray*}
U&=& (x_1 + x_2 + x_4) (x_3 + x_5 + x_6) + (x_1 + x_4) x_2,
\nonumber\\ b_1 &=& \frac{(x_2 + x_3 + x_5 + x_6) a_1 - x_2 a_2}{U} =
\frac{(x_5 + x_6) x_2 q - ((x_3 + x_5 + x_6) x_4 + (x_3 + x_4)x_2) k_n
  + x_2 x_5 k_e }{U} = \nonumber\\ &=&\frac{X q + Y k_n + Z
  k_e}{U},\nonumber\\ b_2 &=& \frac{(x_1 + x_2 + x_4) a_2 - x_2
  a_1}{U} = \nonumber\\ &=& \frac{((x_1 + x_4)(x_2 + x_3) + x_2 x_3) q
  + ((x_1 + x_4) x_3 + (x_3 + x_4) x_2) k_n - (x_1 + x_2 + x_4) x_5
  k_e}{U} = \frac{\bar{X} q + \bar{Y} k_n + \bar{Z} k_e}{U},
\nonumber\\
\end{eqnarray*}
\begin{eqnarray}\label{eq:C1o.6}
Q &=& (x_1 + x_2) m^2_{\pi} - x_2 q^2 + \frac{(x_2 + x_3 +
  x_5 + x_6) a^2_1 - 2 x_2 a_1 \cdot a_2 + (x_1 + x_2 + x_4) a^2_2}{U}
\end{eqnarray}
with $a_1 = x_2 q - x_4 k_n$ and $a_2 = (x_2 + x_3) q + x_3 k_n - x_5
k_e$. Taking the limit $n \to 4$ and keeping the divergent
contribution proportional to $\Gamma(2 - n/2)$, we get
\begin{eqnarray}\label{eq:C1o.7}
 \hspace{-0.30in}&& {\cal F}^{(1)}(k_n, k_e, q)_{1o} = \int^1_0 dx_1
 \int^1_0 dx_2 \int^1_0 dx_3 \int^1_0 dx_4 \int^1_0 dx_5 \int^1_0 dx_6
 \frac{\delta(1 - x_1 - x_2 - x_3 - x_4 - x_5 - x_6)}{256 \pi^4 U^2}
 \nonumber\\ \hspace{-0.30in}&& \times \Big\{ \Big[- \frac{1}{4}\,
   \Gamma\Big(2 - \frac{n}{2}\Big) + \frac{1}{2}\, {\ell
     n}\frac{Q}{m^2_N}\Big] \, \gamma^{\beta} \gamma^{\alpha}
 \gamma^{\mu} (1 - \gamma^5) \otimes \gamma_{\beta} \gamma_{\alpha}
 \gamma_{\mu} + \ldots\Big\}.
\end{eqnarray}
Using the Dirac equations for free fermions and the algebra of the
Dirac $\gamma$-matrices \cite{Itzykson1980}, we obtain the following
Lorentz structure for the integral ${\cal F}^{(1)}(k_n, k_e, q)_{1o}$
\begin{eqnarray}\label{eq:C1o.8}
 \hspace{-0.30in}&& {\cal F}^{(1)}(k_n, k_e, q)_{1o} = \int^1_0 dx_1
 \int^1_0 dx_2 \int^1_0 dx_3 \int^1_0 dx_4 \int^1_0 dx_5 \int^1_0 dx_6
 \frac{\delta(1 - x_1 - x_2 - x_3 - x_4 - x_5 - x_6)}{256 \pi^4 U^2}
 \nonumber\\ \hspace{-0.30in}&& \times \Big\{ \Big(-
 \frac{5}{2}\,\Gamma\Big(2 - \frac{n}{2}\Big) +
 f^{(V)}_{1o^{(1)}}(x_1, \ldots, x_6) + g^{(V)}_{1o^{(1)}}(x_1,
 \ldots, x_6)\, \frac{k_n \cdot q}{m^2_N} + h^{(V)}_{1o^{(1)}}(x_1,
 \ldots, x_6)\, \frac{k_n \cdot k_e}{m^2_N}\Big) \gamma^{\mu} (1 -
 \gamma^5) \otimes \gamma_{\mu} \nonumber\\ \hspace{-0.30in}&& + \Big(
 \frac{3}{2}\,\Gamma\Big(2 - \frac{n}{2}\Big) +
 f^{(A)}_{1o^{(1)}}(x_1, \ldots, x_6) + g^{(A)}_{1o^{(1)}}(x_1,
 \ldots, x_6)\, \frac{k_n \cdot q}{m^2_N} + h^{(A)}_{1o^{(1)}}(x_1,
 \ldots, x_6)\, \frac{k_n \cdot k_e}{m^2_N}\Big) \gamma^{\mu} (1 -
 \gamma^5) \otimes \gamma_{\mu} \gamma^5 + \ldots\Big\},
 \nonumber\\ \hspace{-0.30in}&& 
\end{eqnarray}
where we have denoted
\begin{eqnarray}\label{eq:C1o.9}
 \hspace{-0.30in}f^{(V)}_{1o^{(1)}}(x_1, \ldots, x_6) &=& + 5\, {\ell
   n}\frac{\bar{Q}}{m^2_N} \quad,\quad g^{(V)}_{1o^{(1)}}(x_1, \ldots,
 x_6) = + 10\, \frac{V}{U}\, \frac{m^2_N}{\bar{Q}} \quad,\quad
 h^{(V)}_{1o^{(1)}}(x_1, \ldots, x_6) = + 10\, \frac{\bar{V}}{U}\,
 \frac{m^2_N}{\bar{Q}},\nonumber\\
 \hspace{-0.30in}f^{(A)}_{1o^{(1)}}(x_1, \ldots, x_6) &=& - 3\, {\ell
   n}\frac{\bar{Q}}{m^2_N} \quad,\quad g^{(A)}_{1o^{(1)}}(x_1, \ldots,
 x_6) = - 6\, \frac{V}{U}\, \frac{m^2_N}{\bar{Q}} \quad,\quad
 h^{(A)}_{1o^{(1)}}(x_1, \ldots, x_6) = - 6\, \frac{\bar{V}}{U}\,
 \frac{m^2_N}{\bar{Q}},
\end{eqnarray}
where for the calculation of the integrals over the Feynman parameters
we use
\begin{eqnarray}\label{eq:C1o.10}
\hspace{-0.3in} \frac{\bar{Q}}{m^2_N} &=& \frac{1}{U}\Big((x_1 + x_2)
U \frac{m^2_{\pi}}{m^2_N} + (x_2 + x_3 + x_5 + x_6) x^2_4 + 2 x_2 x_3
x_4 +(x_1 + x_2 + x_4) x^2_3\Big), \nonumber\\
\hspace{-0.30in} V &=& (x_1 + x_4)(x_2 + x_3) x_3 + x_3 x^2_2 - (x_5 +
x_6) x_2 x_4\;,\; \bar{V} = - (x_1 + x_2) x_3 x_5 - (x_2 + x_3) x_4
x_5
\end{eqnarray}
with $\bar{Q} = Q\big|_{q = k_e = 0}$.  A deviation of $Q$ from
$\bar{Q}$ is taken into account in the linear approximation in the
form of an expansion in powers of $k_n\cdot q/m^2_N$ (or $q_0/m_N$)
and $k_n\cdot k_e/m^2_N$ (or $E_e/m_N$), respectively. After the
integration over the Feynman parameters, we obtain the contribution of
the integral ${\cal F}^{(1)}(k_n, k_e, q)_{1o}$ to the matrix element
$M(n \to p e^- \bar{\nu}_e)^{(\pi^0)}_{\rm Fig.\,\ref{fig:fig1}o}$. We
get
\begin{eqnarray}\label{eq:C1o.11}
 \hspace{-0.30in}&&M(n \to p e^- \bar{\nu}_e)^{(\pi^0)}_{\rm
   Fig.\,\ref{fig:fig1}o^{(1)}} = \frac{\alpha}{2\pi}\frac{g^2_{\pi
     N}}{16\pi^2}\, G_V \Big\{\Big(A_{1o^{(1)}}\,\Gamma\Big(2 -
 \frac{n}{2}\Big) + F^{(V)}_{1o^{(1)}} +
 G^{(V)}_{1o^{(1)}}\,\frac{k_n\cdot q}{m^2_N} +
 H^{(V)}_{1o^{(1)}}\,\frac{k_n\cdot k_e}{m^2_N}\Big)
 \nonumber\\ \hspace{-0.30in}&& \times \Big[\bar{u}_e \gamma^{\mu} (1
   - \gamma^5) v_{\bar{\nu}}\Big] \Big[\bar{u}_p \gamma_{\mu} u_n\Big]
 + \Big(B_{1o^{(1)}}\,\Gamma\Big(2 - \frac{n}{2}\Big) +
 F^{(A)}_{1o^{(1)}} + G^{(A)}_{1o^{(1)}}\,\frac{k_n\cdot q}{m^2_N} +
 H^{(A)}_{1o^{(1)}}\,\frac{k_n\cdot k_e}{m^2_N}\Big)
 \nonumber\\ \hspace{-0.30in}&& \times \Big[\bar{u}_e \gamma^{\mu} (1
   - \gamma^5) v_{\bar{\nu}}\Big]\Big[\bar{u}_p \gamma_{\mu} \gamma^5
   u_n\Big]\Big\}.
\end{eqnarray}
The structure constants are equal to $A_{1o^{(1)}} = - 5/12$,
$B_{1o^{(1)}} = 1/4$, $F^{(V)}_{1o^{(1)}} = - 2.3865$,
$G^{(V)}_{1o^{(1)}} = 47.4131$, $H^{(V)}_{1o^{(1)}} = - 45.8712$,
$F^{(A)}_{1o^{(1)}} = 1.4319$, $G^{(A)}_{1o^{(1)}} = - 28.4479$ and
$H^{(A)}_{1o^{(1)}} = 27.5227$. The Lorentz structure of
Eq.(\ref{eq:C1o.11}) is obtained at the neglect of the contributions
of order $O(E^2_0/m^2_N) \sim 10^{-6}$.

\subsubsection*{\bf Calculation of the integral
  ${\cal F}^{(2)}(k_n, k_e, q)_{1o}$}

In terms of the integrals over the Feynman parameters and after
integration over the virtual momenta in the $n$-dimensional momentum
space, we obtain for the integral ${\cal F}^{(2)}(k_n, k_e, q)_{1o}$
the following expression
\begin{eqnarray*}
\hspace{-0.3in}&&{\cal F}^{(2)}(k_n, k_e, q)_{1o} = \int^1_0 \!\!
dx_1 \int^1_0 \!\! dx_2 \int^1_0 \!\! dx_3 \int^1_0 \!\! dx_4 \int^1_0
\!\! dx_5 \int^1_0 \!\! dx_6 \int^1_0 \!\! dx_7\,\frac{\delta(1 - x_1
  - x_2 - x_3 - x_4 - x_5 - x_6 - x_7)}{2^{2n} \pi^n
  U^{n/2}} \nonumber\\
\end{eqnarray*}
\begin{eqnarray}\label{eq:C1o.12}
\hspace{-0.3in}&& \times \Bigg\{ - \frac{n^2 (n +
  2)}{8}\,\frac{\displaystyle \Gamma\Big(2 -
  \frac{n}{2}\Big)\Gamma\Big(2 - \frac{n - 1}{2}\Big)}{\displaystyle
  2^{n - 3}\Gamma\Big(\frac{1}{2}\Big)}\,\Big(\frac{Q}{m^2_N}\Big)^{-4
  + n}\,\Big[\frac{2}{n}\, \gamma^{\mu} (1 - \gamma^5) \otimes
  \gamma_{\mu}  - \frac{2}{n^2}\, \gamma^{\beta}
  \gamma^{\alpha} \gamma^{\mu} (1 - \gamma^5) \otimes \gamma_{\beta}
  \gamma_{\alpha} \gamma_{\mu}\Big] \nonumber\\
\hspace{-0.3in}&& + \frac{n^2}{4} \Gamma(5 - n)\,Q^{-5+n} m^{2(4-n)}_N
\Big[\frac{2}{n^2}\,b^2_2 \gamma^{\beta} \gamma^{\alpha} \gamma^{\mu}
  (1 - \gamma^5) \otimes \gamma_{\beta} \gamma_{\alpha} \gamma_{\mu} +
  \Big(- 2\, \frac{n + 2}{n^2}\, b^2_1 - 4\, \frac{n + 2}{n^2}\, b^2_2
  \Big) \, \gamma^{\mu} (1 - \gamma^5) \otimes
\gamma_{\mu} \nonumber\\
\hspace{-0.3in}&& + 2\, \frac{n + 2}{n^2}\, \gamma^{\beta}
\gamma^{\alpha} \hat{b}_1 (1 - \gamma^5) \otimes \gamma_{\beta}
\gamma_{\alpha} \hat{b}_1 + 2\, \frac{n + 4}{n^2}\, \gamma^{\beta}
\hat{b}_2 \gamma^{\alpha} (1 - \gamma^5) \otimes \gamma_{\beta}
\hat{b}_2 \gamma_{\alpha} \Big]\Bigg\},
\end{eqnarray} 
where we have used the algebra of the Dirac $\gamma$-matrices in
$n$-dimensional space-time  and denoted
\begin{eqnarray}\label{eq:C1o.13}
  \hspace{-0.3in}U &=& (x_1 + x_2 + x_4)( x_3 + x_5 + x_6 + x_7) +
  (x_1 + x_4) x_2,\nonumber\\
  \hspace{-0.3in}b_1 &=& \frac{(x_2 + x_3 + x_5 + x_6 + x_7)a_1 - x_2
    a_2}{U} = - \frac{(x_3 + x_5 + x_6 + x_7)x_4 + (x_3 + x_4)
    x_2 }{U}\, k_n = \frac{Y}{U}\,k_n ,\nonumber\\
   \hspace{-0.3in}b_2 &=& \frac{(x_1 + x_2 + x_4)a_2 - x_2 a_1}{U} =
   \frac{(x_1 + x_4) x_3 + (x_3 + x_4) x_2}{U}\, k_n =
   \frac{\bar{Y}}{U}\,k_n,\nonumber\\
   \hspace{-0.3in}Q &=&x_7 M^2_W + \frac{(x_2 + x_3 + x_5 + x_6 + x_7)
     a^2_1 - 2 x_2 a_1\cdot a_2 + (x_1 + x_2 + x_4) a^2_2}{U}
\end{eqnarray}
with $a_1 = - x_4 k_n$ and $a_2 = x_3 k_n$. In Eq.(\ref{eq:C1o.12}) we
have neglected the contributions of the terms proportional to $q$ and
$k_e$, appearing in ${\cal F}^{(2)}(k_n, k_e, q)_{1c}$ in the form of
an expansion in powers of $k_n\cdot q/M^2_W$ and $k_n\cdot k_e/M^2_W$
of order $10^{-7}$, and the terms of order $m^2_{\pi}/M^2_W \sim
10^{-6}$. Taking the limit $n \to 4$, keeping  the divergent
contributions proportional to $\Gamma(2 - n/2)$ and  the
contributions of order $O(m^2_N/M^2_W)$ in the large electroweak
$W^-$-boson mass $M_W$ expansion, and using the Dirac equations for a
free neutron and a free proton, we arrive at the expression
\begin{eqnarray}\label{eq:C1o.14}
 \hspace{-0.21in}&& {\cal F}^{(2)}(k_n, k_e, q)_{1o} = \int^1_0 dx_1
 \int^1_0 dx_2 \int^1_0 dx_3 \int^1_0 dx_4 \int^1_0 dx_5 \int^1_0 dx_6
 \int^1_0 dx_7 \frac{\delta(1 - x_1 - x_2 - x_3 - x_4 - x_5 - x_6 -
   x_7)}{256 \pi^4 U^2} \nonumber\\ \hspace{-0.21in}&& \times \Big\{
 \Big( \frac{9}{2}\,\Gamma\Big(2 - \frac{n}{2}\Big) +
 f^{(V)}_{1o^{(2)}}(x_1, \ldots, x_7) + g^{(W)}_{1o^{(2)}}(x_1,
 \ldots, x_7)\, \frac{m^2_N}{M^2_W}\Big) \gamma^{\mu} (1 - \gamma^5)
 \otimes \gamma_{\mu} + \Big(- \frac{9}{2}\,\Gamma\Big(2 -
 \frac{n}{2}\Big) \nonumber\\ \hspace{-0.21in}&&+
 f^{(A)}_{1o^{(2)}}(x_1, \ldots, x_7) + h^{(W)}_{1o^{(2)}}(x_1,
 \ldots, x_6)\, \frac{m^2_N}{M^2_W}\Big) \gamma^{\mu} (1 - \gamma^5)
 \otimes \gamma_{\mu} \gamma^5 + f^{(W)}_{1o^{(2)}}(x_1, \ldots,
 x_7)\, \frac{m^2_N}{M^2_W}\, \frac{\hat{k}_n}{m_N} (1 - \gamma^5)
 \otimes 1\Big\}, \nonumber\\ \hspace{-0.21in}&&
\end{eqnarray}
where we have denoted
\begin{eqnarray}\label{eq:C1o.15}
 \hspace{-0.30in}&&f^{(V)}_{1o^{(2)}}(x_1, \ldots, x_7) = - 9\,{\ell
   n}\frac{M^2_W}{m^2_N} - 9\,{\ell n} \bar{Q}
\end{eqnarray}
and
\begin{eqnarray}\label{eq:C1o.16}
 \hspace{-0.30in}g^{(W)}_{1o^{(2)}}(x_1, \ldots, x_7) =
 \frac{1}{\bar{Q}}\Big[ 3\, \frac{Y^2}{U^2} -
   \frac{\bar{Y}^2}{U^2}\Big]
\end{eqnarray}
and
\begin{eqnarray}\label{eq:C1o.17}
 \hspace{-0.30in}f^{(A)}_{1o^{(2)}}(x_1, \ldots, x_7) = + 9\,{\ell
   n}\frac{M^2_W}{m^2_N} + 9\,{\ell n} \bar{Q}
\end{eqnarray}
and
\begin{eqnarray}\label{eq:C1o.18}
 \hspace{-0.30in}h^{(W)}_{1o^{(2)}}(x_1, \ldots, x_7) =
 \frac{1}{\bar{Q}}\Big[ - 6\,\frac{Y^2}{U^2} -11\,
   \frac{\bar{Y}^2}{U^2}\Big]
\end{eqnarray}
and
\begin{eqnarray}\label{eq:C1o.19}
 \hspace{-0.30in}&&f^{(W)}_{1o^{(2)}}(x_1, \ldots, x_7) =
 \frac{1}{\bar{Q}}\Big[6\, \frac{Y^2}{U^2} + 8\,
   \frac{\bar{Y}^2}{U^2}\Big].
\end{eqnarray}
For the calculation of the integrals over the Feynman parameters we
use
\begin{eqnarray}\label{eq:C1o.20}
  \hspace{-0.30in}\bar{Q} &=& \frac{1}{U}\Big(x_7 U + \big((x_2 + x_3
  + x_5 + x_6 + x_7) x^2_4 + 2 x_2 x_3 x_4 + (x_1 + x_2 + x_4)
  x^2_3\big)\, \frac{m^2_N}{M^2_W}\Big), \nonumber\\
  \hspace{-0.30in} U &=& (x_1 + x_2 + x_4)(x_3 + x_5 + x_6 + x_7) +
  (x_1 + x_4) x_2,\nonumber\\
  \hspace{-0.30in} Y &=& - (x_3 + x_5 + x_6 + x_7)x_4 - (x_3 + x_4)
  x_2 \;,\; \bar{Y} = (x_1 + x_4) x_3 + (x_3 + x_4) x_2,
\end{eqnarray}
where $m_N = (m_n + m_p)/2 = 938.9188\,{\rm MeV}$ and $M_W =
80,379\,{\rm GeV}$ \cite{PDG2020}. After the integration over the
Feynman parameters we obtain the contribution of the integral ${\cal
  F}^{(2)}(k_n, k_e, q)_{1o}$ to the matrix element $M(n \to p e^-
\bar{\nu}_e)^{(\pi^0)}_{\rm Fig.\,\ref{fig:fig1}o}$:
\begin{eqnarray}\label{eq:C1o.21}
 \hspace{-0.30in}&&M(n \to p e^- \bar{\nu}_e)^{(\pi^0)}_{\rm
   Fig.\,\ref{fig:fig1}o^{(2)}} = \frac{\alpha}{2\pi}\frac{g^2_{\pi
     N}}{16\pi^2}\, G_V \Big\{\Big(A_{1o^{(2)}}\,\Gamma\Big(2 -
 \frac{n}{2}\Big) + F^{(V)}_{1o^{(2)}} +
 G^{(W)}_{1o^{(2)}}\,\frac{m^2_N}{M^2_W}\Big) \Big[\bar{u}_e
   \gamma^{\mu} (1 - \gamma^5) v_{\bar{\nu}}\Big] \Big[\bar{u}_p
   \gamma_{\mu} u_n\Big] \nonumber\\ \hspace{-0.30in}&& +
 \Big(B_{1o^{(2)}}\,\Gamma\Big(2 - \frac{n}{2}\Big) +
 F^{(A)}_{1o^{(2)}} + H^{(W)}_{1o^{(2)}}\,\frac{m^2_N}{M^2_W}\Big)
 \Big[\bar{u}_e \gamma^{\mu} (1 - \gamma^5)
   v_{\bar{\nu}}\Big]\Big[\bar{u}_p \gamma_{\mu} \gamma^5 u_n\Big] +
 F^{(W)}_{1o^{(2)}}\,\frac{m^2_N}{M^2_W}\Big) \Big[\bar{u}_e
   \frac{\hat{k}_n}{m_N} (1 - \gamma^5)
   v_{\bar{\nu}}\Big]\nonumber\\
  \hspace{-0.30in}&& \times \,\Big[\bar{u}_p u_n\Big]\Big\}.
\end{eqnarray}
The structure constants are equal to $A_{1o^{(2)}} = 0.1537$,
$B_{1o^{(2)}} = - 0.1537$, $F^{(V)}_{1o^{(2)}} = - 2.0511$,
$G^{(W)}_{1o^{(2)}} = 0.8606$, $F^{(A)}_{1o^{(2)}} = 2.0511$,
$H^{(W)}_{1o^{(2)}} = - 4.5405$ and $F^{(W)}_{1o^{(2)}} = -
3.8898$. The Lorentz structure of the matrix element
Eq.(\ref{eq:C1o.21}) is calculated at the neglect of the contributions
of order $O(k_n\cdot q/M^2_W) \sim O(k_n \cdot k_e/M^2_W) \sim
10^{-7}$ and $O(m^2_{\pi}/M^2_W) \sim 10^{-6}$, respectively.

Summing up the contributions of the integrals ${\cal F}^{(1)}(k_n,
k_e, q)_{1o}$ and ${\cal F}^{(2)}(k_n, k_e, q)_{1o}$, we obtain the
contribution of the Feynman diagram in Fig.\,\ref{fig:fig1}o to the
amplitude of the neutron beta decay 
\begin{eqnarray}\label{eq:C1o.22}
 \hspace{-0.30in}&&M(n \to p e^- \bar{\nu}_e)^{(\pi^0)}_{\rm
   Fig.\,\ref{fig:fig1}o} = \frac{\alpha}{2\pi}\frac{g^2_{\pi
     N}}{16\pi^2}\, G_V \Big\{\Big(A_{1o}\,\Gamma\Big(2 -
 \frac{n}{2}\Big) + F^{(V)}_{1o} + G^{(V)}_{1o}\,\frac{k_n\cdot
   q}{m^2_N} + H^{(V)}_{1o}\,\frac{k_n\cdot k_e}{m^2_N} +
 G^{(W)}_{1o}\,\frac{m^2_N}{M^2_W}\Big) \nonumber\\ \hspace{-0.30in}&&
 \times \Big[\bar{u}_e \gamma^{\mu} (1 - \gamma^5) v_{\bar{\nu}}\Big]
 \Big[\bar{u}_p \gamma_{\mu} u_n\Big] + \Big(B_{1o}\,\Gamma\Big(2 -
 \frac{n}{2}\Big) + F^{(A)}_{1o} + G^{(A)}_{1o}\,\frac{k_n\cdot
   q}{m^2_N} + H^{(A)}_{1o}\,\frac{k_n\cdot k_e}{m^2_N} +
 H^{(W)}_{1o}\,\frac{m^2_N}{M^2_W}\Big) \nonumber\\ \hspace{-0.30in}&&
 \times \Big[\bar{u}_e \gamma^{\mu} (1 - \gamma^5)
   v_{\bar{\nu}}\Big]\Big[\bar{u}_p \gamma_{\mu} \gamma^5 u_n\Big] +
 F^{(W)}_{1o}\,\frac{m^2_N}{M^2_W} \Big[\bar{u}_e
   \frac{\hat{k}_n}{m_N} (1 - \gamma^5)
   v_{\bar{\nu}}\Big] \Big[\bar{u}_p u_n\Big] \Big\}.
\end{eqnarray}
The structure constants are equal to $A_{1o} = A_{1o^{(1)}} +
A_{1o^{(2)}} = - 0.2630$, $B_{1o} = B_{1o^{(1)}} + B_{1o^{(2)}} =
0.0963$, $F^{(V)}_{1o} = F^{(V)}_{1o^{(1)}} + F^{(V)}_{1o^{(2)}} = -
4.4376$, $G^{(V)}_{1o} = G^{(V)}_{1o^{(1)}} = 47.4131$, $H^{(V)}_{1o}
= H^{(V)}_{1o^{(1)}} = - 45.8712 $, $G^{(W)}_{1o}= G^{(W)}_{1o^{(2)}}
= 0.8606$, $F^{(A)}_{1o} = F^{(A)}_{1o^{(1)}} + F^{(A)}_{1o^{(2)}} =
3.4830$, $G^{(A)}_{1o} = G^{(A)}_{1o^{(1)}} = - 28.4479$,
$H^{(A)}_{1o} = H^{(A)}_{1o^{(1)}} = 27.5227$, $H^{(W)}_{1o}=
H^{(W)}_{1o^{(2)}} = - 4.5405$ and $F^{(W)}_{1o} = F^{(W)}_{1o^{(2)}}
= - 3.8898$. The Lorentz structure of the matrix element
Eq.(\ref{eq:C1o.22}) is calculated at the neglect of the contributions
of order $O(E^2_0/m^2_N) \sim O(m^2_{\pi}/M^2_W) \sim 10^{-6}$.

\subsection*{C1p. Analytical calculation of the Feynman diagram in
  Fig.\,\ref{fig:fig1}p}
\renewcommand{\theequation}{C1p-\arabic{equation}}
\setcounter{equation}{0}

In the Feynman gauge for the photon propagator the analytical
expression of the Feynman diagram in Fig.\,\ref{fig:fig1}p is given by
(see Eq.(\ref{eq:A.1}))
\begin{eqnarray}\label{eq:C1p.1}
&&M(n \to p e^- \bar{\nu}_e)^{(\pi^0)}_{\rm Fig.\,\ref{fig:fig1}p} = -
  2 e^2 g^2_{\pi N} M^2_W G_V\nonumber\\ &&\times \int
  \frac{d^4k}{(2\pi)^4i}\int \frac{d^4p}{(2\pi)^4i}\,\Big[\bar{u}_e
    \gamma^{\mu}(1 - \gamma^5)
    v_{\bar{\nu}}\Big]\,\Big[\bar{u}_p\,\gamma_{\beta} \frac{1}{m_N -
      \hat{k}_p + \hat{p} - i0}\,\gamma^5\, \frac{1}{m_N - \hat{k}_n -
      \hat{k} - i0}\,\gamma^5\, u_n\Big]
  \nonumber\\ \hspace{-0.30in}&& \times \, \frac{(2 k + p -
    q)^{\alpha_2}}{[m^2_{\pi} - k^2 - i0][m^2_{\pi} - (k + p - q)^2 -
      i0]}\, \big[(p - q)^{\nu}\eta^{\beta \alpha_1} - (p - 2
    q)^{\beta} \eta^{\alpha_1\nu} - q^{\alpha_1}
    \eta^{\nu\beta}\big]\nonumber\\ &&\times\,
 \frac{1}{p^2 + i0}\, D^{(W)}_{\alpha_1 \alpha_2}(p -
  q) \, D^{(W)}_{\mu\nu}(- q).
\end{eqnarray}
The contribution of the Feynman diagram in Fig.\,\ref{fig:fig1}p is
equal to the contribution of the Feynman diagram in
Fig.\,\ref{fig:fig1}n.
\vspace{-0.15in}
\subsection*{C1q. Analytical calculation of the Feynman diagram in
  Fig.\,\ref{fig:fig1}q}
\renewcommand{\theequation}{C1q-\arabic{equation}}
\setcounter{equation}{0}

In the Feynman gauge for the photon propagator the analytical
expression of the Feynman diagram in Fig.\,\ref{fig:fig1}q is
\begin{eqnarray}\label{eq:C1q.1}
\hspace{-0.30in}&&M(n \to p e^- \bar{\nu}_e)^{(\pi^0)}_{\rm
  Fig.\,\ref{fig:fig1}q} = + 2 e^2 g^2_{\pi N}G_V \int
\frac{d^4k}{(2\pi)^4i}\int
\frac{d^4p}{(2\pi)^4i}\,\Big[\bar{u}_e\,\gamma^{\beta}\,\frac{1}{m_e -
    \hat{k}_e - \hat{p} - i0}\, \gamma^{\mu}(1 - \gamma^5)
  v_{\bar{\nu}}\Big]\nonumber\\ \hspace{-0.30in}&&\times\,\Big[\bar{u}_p\,\gamma_{\beta}
  \frac{1}{m_N - \hat{k}_p + \hat{p} - i0} \,\gamma^5\, \frac{1}{m_N -
    \hat{k}_n - \hat{k} - i0}\,\gamma^5\, u_n\Big]\, \frac{(2 k + p -
  q)^{\nu}}{[m^2_{\pi} - k^2 - i0][m^2_{\pi} - (k + p - q)^2 - i0]}\,
\frac{1}{p^2 + i0}\nonumber\\ \hspace{-0.30in}&&\times \,
\Big(\eta_{\mu\nu} + \frac{(p - q)^2 \eta_{\mu\nu} - (p - q)_{\mu} (p
  - q)_{\nu}}{M^2_W - (p - q)^2 - i0}\Big)
\end{eqnarray}
where we have taken the propagator of the electroweak $W^-$-boson in
the following form
\begin{eqnarray}\label{eq:C1q.2}
  D^{(W)}_{\mu\nu}(p - q) = - \frac{1}{M^2_W} \Big(\eta_{\mu\nu} +
  \frac{(p - q)^2 \eta_{\mu\nu} - (p - q)_{\mu} (p - q)_{\nu}}{M^2_W -
    (p - q)^2 - i0}\Big).
\end{eqnarray}
The result of the calculation of the Feynman diagram in
Fig.\,\ref{fig:fig1}q is equal to the result of the calculation of the
Feynman diagram in Fig.\,\ref{fig:fig1}o.

\subsection*{C1r. Analytical calculation of the Feynman diagram in
  Fig.\,\ref{fig:fig1}r}
\renewcommand{\theequation}{C1r-\arabic{equation}}
\setcounter{equation}{0}

In the Feynman gauge for the photon propagator the analytical
expression of the Feynman diagram in Fig.\,\ref{fig:fig1}r is given by
(see Eq.(\ref{eq:A.1}))
\begin{eqnarray}\label{eq:C1r.1}
\hspace{-0.30in}&&M(n \to p e^- \bar{\nu}_e)^{(\pi^0)}_{\rm
  Fig.\,\ref{fig:fig1}r} = - e^2 g^2_{\pi N} M^2_W G_V \int
\frac{d^4k}{(2\pi)^4i}\int \frac{d^4p}{(2\pi)^4i}\,\Big[\bar{u}_e
  \gamma^{\mu}(1 - \gamma^5)
  v_{\bar{\nu}}\Big]\nonumber\\ \hspace{-0.30in}&&\times
\,\Big[\bar{u}_p \,\gamma_{\beta}\, \frac{1}{m_N - \hat{k}_p + \hat{p}
    - i0}\, \gamma^5 \frac{1}{m_N - \hat{k}_p - \hat{k} + \hat{p} -
    i0}\,\gamma^{\alpha_2}(1 - \gamma^5)\,\frac{1}{m_N - \hat{k}_n
    -\hat{k} - i0}\,\gamma^5 u_n \Big]
\nonumber\\ \hspace{-0.30in}&&\times \,\frac{1}{m^2_{\pi} - k^2 -
  i0}\,\big[(p - q)^{\nu}\,\eta^{\beta \alpha_1} - (p -
  2q)^{\beta} \eta^{\alpha_1 \nu} - q^{\alpha_1} \eta^{\nu
    \beta}\big]\, \frac{1}{p^2 + i0}\, D^{(W)}_{\alpha_1\alpha_2}(p
- q) D^{(W)}_{\mu\nu}(- q).
\end{eqnarray}
Since the calculation of the Feynman diagram in Fig.\,\ref{fig:fig1}r
is similar to the calculation of the Feynman diagrams in
Fig.\,\ref{fig:fig1}a and Fig.\,\ref{fig:fig1}n, we give at once the
expression in terms of the integrals over the Feynman parameters. We
get
\begin{eqnarray}\label{eq:C1r.2}
\hspace{-0.30in}&&M(n \to p e^- \bar{\nu}_e)^{(\pi^0)}_{\rm
  Fig.\,\ref{fig:fig1}r} = - \frac{\alpha}{4\pi} \frac{g^2_{\pi
    N}}{16\pi^2} G_V \int^1_0 \!\!  dx_1 \int^1_0\!\! dx_2 \int^1_0
\!\!dx_3 \int^1_0 \!\! dx_4 \int^1_0 \!\! dx_5 \int^1_0\!\! dx_6
\frac{\delta(1 - x_1 - x_2 - x_3 - x_4 - x_5 - x_6)}{U^2}
\nonumber\\ \hspace{-0.30in}&& \times \Big\{\Big(- \frac{3}{2}\,
\Gamma\Big(2 - \frac{n}{2}\Big) + f^{(V)}_{1r}(x_1, \ldots, x_6) +
g^{(W)}_{1r}(x_1, \ldots, x_6)\,
\frac{m^2_N}{M^2_W}\Big)\Big[\bar{u}_e \gamma^{\mu}(1 - \gamma^5)
  v_{\bar{\nu}}\Big]\Big[\bar{u}_p \gamma_{\mu} u_n\Big] + \Big( -
\frac{3}{2}\, \Gamma\Big(2 - \frac{n}{2}\Big)
\nonumber\\ \hspace{-0.30in}&& + f^{(A)}_{1r}(x_1, \ldots, x_6) +
h^{(W)}_{1r}(x_1, \ldots, x_6) \frac{m^2_N}{M^2_W}\Big)\Big[\bar{u}_e
  \gamma^{\mu}(1 - \gamma^5) v_{\bar{\nu}}\Big]\Big[\bar{u}_p
  \gamma_{\mu}\gamma^5 u_n\Big] + f^{(W)}_{1r}(x_1, \ldots, x_6)
\frac{m^2_N}{M^2_W} \Big[\bar{u}_e \frac{\hat{k}_n}{m_N}(1 - \gamma^5)
  v_{\bar{\nu}}\Big] \nonumber\\ \hspace{-0.30in}&& \times
\,\Big[\bar{u}_p u_n\Big]\Big\}
\end{eqnarray}
where we have taken the limit $n \to 4$, kept the divergent
contributions proportional to $\Gamma(2 - n/2)$, used the Dirac
equations for free fermions and denoted
\begin{eqnarray}\label{eq:C1r.3}
\hspace{-0.30in} f^{(V)}_{1r}(x_1, \ldots, x_6) = 3\, {\ell
  n}\frac{M^2_W}{m^2_N} + 3\, {\ell n} \bar{Q}
\end{eqnarray}
and 
\begin{eqnarray}\label{eq:C1r.4}
\hspace{-0.30in} g^{(W)}_{1r}(x_1, \ldots, x_6) =
\frac{1}{\bar{Q}}\Big[2\, \frac{Y^2}{U^2} - \frac{Y\bar{Y}}{U^2} +
  \frac{\bar{Y}^2}{U^2}\Big]
\end{eqnarray}
and
\begin{eqnarray}\label{eq:C1r.5}
\hspace{-0.30in} f^{(A)}_{1r}(x_1, \ldots, x_6) = 3\, {\ell
  n}\frac{M^2_W}{m^2_N} + 3\, {\ell n} \bar{Q}
\end{eqnarray}
and 
\begin{eqnarray}\label{eq:C1r.6}
\hspace{-0.30in} h^{(W)}_{1r}(x_1, \ldots, x_6) = 
\frac{1}{\bar{Q}}\Big[-  2\,
  \frac{Y^2}{U^2} + \frac{Y\bar{Y}}{U^2} +
  \frac{\bar{Y}^2}{U^2}\Big]
\end{eqnarray}
and
\begin{eqnarray}\label{eq:C1r.7}
\hspace{-0.30in} f^{(W)}_{1r}(x_1, \ldots, x_6) =
\frac{1}{\bar{Q}}\Big[2\, \frac{Y^2}{U^2} - 12\, \frac{Y\bar{Y}}{U^2}
  \Big]
\end{eqnarray}
with
\begin{eqnarray}\label{eq:C1r.8}
  \hspace{-0.30in}\bar{Q} &=& \frac{1}{U}\Big(x_6 U + \big((x_5 +
  x_6)(x_3 + x_4)^2 + (x_2 + x_3)(x_3 + x_4)(x_2 + x_4) + x_1 (x_2 +
  x_3)^2\big)\, \frac{m^2_N}{M^2_W}\Big), \nonumber\\
  \hspace{-0.30in}U &=& (x_1 + x_3 + x_4)(x_2 + x_5 + x_6) + (x_1 +
  x_4) x_3,\nonumber\\
  \hspace{-0.30in} Y &=& - (x_5 + x_6) (x_3 + x_4) - (x_2 + x_3)
  x_4\;,\; \bar{Y} = (x_1 + x_4) x_2 + (x_1 + x_2) x_3.
\end{eqnarray}
After the integration over the Feynman parameters, we obtain the
following contribution of the Feynman diagram in Fig.\,\ref{fig:fig1}r
to the amplitude of the neutron beta decay 
\begin{eqnarray}\label{eq:C1r.9}
\hspace{-0.30in} &&M(n \to p e^- \bar{\nu}_e)^{(\pi^0)}_{\rm
  Fig.\,\ref{fig:fig1}r} = \frac{\alpha}{4\pi}\frac{g^2_{\pi
    N}}{16\pi^2}\, G_V \Big\{\Big(A_{1r}\,\Gamma\Big(2 -
\frac{n}{2}\Big) + F^{(V)}_{1r} + G^{(W)}_{1r}\,
\frac{m^2_N}{M^2_W}\Big) \Big[\bar{u}_e \gamma^{\mu}(1 - \gamma^5)
  v_{\bar{\nu}}\Big]\Big[\bar{u}_p \gamma_{\mu} u_n\Big] +
\Big(B_{1r}\nonumber\\
\hspace{-0.30in}&& \times \,\Gamma\Big(2 - \frac{n}{2}\Big) +
F^{(A)}_{1r} + H^{(W)}_{1r}\, \frac{m^2_N}{M^2_W}\Big) \Big[\bar{u}_e
  \gamma^{\mu}(1 - \gamma^5) v_{\bar{\nu}}\Big]\Big[\bar{u}_p
  \gamma_{\mu} \gamma^5 u_n\Big]+ F^{(W)}_{1r}\,
\frac{m^2_N}{M^2_W}\Big[\bar{u}_e \frac{\hat{k}_n}{m_N}(1 - \gamma^5)
  v_{\bar{\nu}}\Big]\Big[\bar{u}_p u_n\Big]\Big\}.
\end{eqnarray}
The structure constants are equal to $A_{1r} = - 1/4$, $B_{1r} = -
1/4$, $F^{(V)}_{1r} = 3.3962$, $G^{(W)}_{1r} = 6.6760$, $F^{(A)}_{1r}
= 3.3962$, $H^{(W)}_{1r} = - 3.3144$, and $F^{(W)}_{1r} = - 18.6744$.
The Lorentz structure of Eq.(\ref{eq:C1r.9}) is calculated at the
neglect of the contributions of order $O(k_n \cdot q/M^2_W) \sim
10^{-7}$ and $O(m^2_{\pi}/M^2_W) \sim 10^{-6}$, respectively.

\subsection*{C1s. Analytical calculation of the Feynman diagram in
  Fig.\,\ref{fig:fig1}s}
\renewcommand{\theequation}{C1s-\arabic{equation}}
\setcounter{equation}{0}

In the Feynman gauge for the photon propagator the analytical
expression of the Feynman diagram in Fig.\,\ref{fig:fig1}s is 
\begin{eqnarray}\label{eq:C1s.1}
&&M(n \to p e^- \bar{\nu}_e)^{(\pi^0)}_{\rm Fig.\,\ref{fig:fig1}s} =
  e^2 g^2_{\pi N} M^2_W G_V \int \frac{d^4k}{(2\pi)^4i}\int
  \frac{d^4p}{(2\pi)^4i}\,\Big[\bar{u}_e
    \,\gamma^{\beta}\,\frac{1}{m_e - \hat{k}_e - \hat{p} -
      i0}\,\gamma^{\mu}(1 -
    \gamma^5)\,v_{\bar{\nu}}\Big] \nonumber\\ &&\times \Big[\bar{u}_p\,
    \gamma_{\beta}\, \frac{1}{m_N - \hat{k}_p + \hat{p} - i0}
    \gamma^5\, \frac{1}{m_N - \hat{k}_p - \hat{k} + \hat{p} -
      i0}\,\gamma^{\nu} (1 - \gamma^5)\,\frac{1}{m_N - \hat{k}_n -
      \hat{k} - i0}\,\gamma^5 u_n\Big] \nonumber\\ &&\times \,
  \frac{1}{m^2_{\pi} - k^2 - i0}\,\frac{1}{p^2 +
    i0}\,\Big(\eta_{\mu\nu} + \frac{(p - q)^2 \eta_{\mu\nu} - (p -
    q)_{\mu} (p - q)_{\nu}}{M^2_W - (p - q)^2 - i0}\Big),
\end{eqnarray}
where we have taken the propagator of the electroweak $W^-$-boson in
the following form
\begin{eqnarray}\label{eq:C1s.2}
  D^{(W)}_{\mu\nu}(p - q) = - \frac{1}{M^2_W}\Big(\eta_{\mu\nu} +
  \frac{(p - q)^2 \eta_{\mu\nu} - (p - q)_{\mu} (p - q)_{\nu}}{M^2_W -
    (p - q)^2 - i0}\Big).
\end{eqnarray}
The calculation of the Feynman diagram in Fig.\,\ref{fig:fig1}s
reduces to the calculation of two integrals ${\cal F}^{(1)}(k_n, k_e,
q)_{1s}$ and ${\cal F}^{(2)}(k_n, k_e, q)_{1s}$, which are defined by
\begin{eqnarray}\label{eq:C1s.3}
 \hspace{-0.30in}&& {\cal F}^{(1)}(k_n, k_e, q)_{1s} = - \int
 \frac{d^4k}{(2\pi)^4i}\int \frac{d^4p}{(2\pi)^4i}\,\frac{1}{m^2_N -
   (k_p - p)^2 - i0}\frac{1}{m^2_N - (k_p + k - p)^2 - i0}
 \frac{1}{m^2_N - (k_n + k)^2 - i0} \nonumber\\ \hspace{-0.30in} &&
 \times \frac{1}{m^2_e - (p + k_e)^2 - i0} \frac{1}{m^2_{\pi} - k^2 -
   i0}\, \frac{1}{p^2 + i0}\Big( \gamma^{\beta} \hat{p}\gamma^{\mu}(1
 - \gamma^5) \otimes \gamma_{\beta}\big(p^2 - \hat{p}
 \hat{k}\big)\gamma_{\mu} \hat{k} (1 + \gamma^5)\Big)
\end{eqnarray}
and
\begin{eqnarray}\label{eq:C1s.4}
 \hspace{-0.30in}&& {\cal F}^{(2)}(k_n, k_e, q)_{1s} = - \int
 \frac{d^4k}{(2\pi)^4i}\int \frac{d^4p}{(2\pi)^4i}\,\frac{1}{m^2_N -
   (k_p - p)^2 - i0}\frac{1}{m^2_N - (k_p + k - p)^2 - i0}
 \frac{1}{m^2_N - (k_n + k)^2 - i0} \nonumber\\ \hspace{-0.30in} &&
 \times \frac{1}{m^2_e - (p + k_e)^2 - i0} \frac{1}{m^2_{\pi} - k^2 -
   i0}\, \frac{1}{p^2 + i0}\, \frac{1}{M^2_W - (p - q)^2 - i0}\,
 \Big(p^2 \gamma^{\beta} \hat{p}\gamma^{\mu}(1 - \gamma^5) \otimes
 \gamma_{\beta}\big(p^2 - \hat{p} \hat{k}\big)\gamma_{\mu} \hat{k} (1
 + \gamma^5) \nonumber\\ \hspace{-0.30in} && + p^2 \gamma^{\mu} (1 -
 \gamma^5) \otimes \gamma_{\mu} \big(2 (p \cdot k) \hat{p} \hat{k} -
 p^2 \hat{p} \hat{k} - p^2 k^2\big) (1 + \gamma^5)\Big),
\end{eqnarray}
where we have kept only leading divergent contributions that
corresponds to the use of the LLA.  The calculation of the integrals
${\cal F}^{(1)}(k_n, k_e, q)_{1s}$ and ${\cal F}^{(2)}(k_n, k_e,
q)_{1s}$ we give below.

\subsubsection*{\bf Calculation of the integral
  ${\cal F}^{(1)}(k_n, k_e, q)_{1s}$}

Skipping standard intermediate transformations of the integrand of
Eq.(\ref{eq:C1s.3}) and having integrated over the virtual momenta in
the $n$-dimensional momentum space, the integral ${\cal F}^{(1)}(k_n,
k_e, q)_{1s}$ can be expressed in terms of the integrals over the
Feynman parameters. We get
\begin{eqnarray}\label{eq:C1s.5}
 \hspace{-0.30in}&& {\cal F}^{(1)}(k_n, k_e, q)_{1s} = \int^1_0 dx_1
 \int^1_0 dx_2 \int^1_0 dx_3 \int^1_0 dx_4 \int^1_0 dx_5 \int^1_0 dx_6
 \, \frac{\delta(1 - x_1 - x_2 - x_3 - x_4 - x_5 - x_6)}{256 \pi^4
   U^2} \nonumber\\ \hspace{-0.30in}&& \times \Big\{ \Big(
 \frac{1}{4}\,\Gamma\Big(2 - \frac{n}{2}\Big) - \frac{1}{2}\, {\ell
   n}\frac{Q}{m^2_N}\Big)\,\gamma^{\beta} \gamma^{\alpha} \gamma^{\mu}
 (1 - \gamma^5) \otimes \gamma_{\beta} \gamma_{\alpha} \gamma_{\mu} (1
 + \gamma^5) + \ldots \Big\},
\end{eqnarray}
where we have  taken  the limit $n \to 4$, kept the divergent
contribution in the form $\Gamma(2 - n/2)$ and denoted
\begin{eqnarray}\label{eq:C1s.6}
\hspace{-0.15in}U &=& (x_1 + x_3 + x_4) (x_2 + x_5 + x_6) + (x_1 + x_4)
x_3, \nonumber\\
\hspace{-0.15in}b_1 &=& \frac{(x_2 + x_3 + x_5 + x_6) a_1 + x_3
  a_2}{U} =  \nonumber\\
\hspace{-0.15in} &=&- \frac{x_3 (x_5 + x_6) q + ((x_5 + x_6)(x_3 +
  x_4) + (x_2 + x_3) x_4) k_n + x_3 x_5 k_e }{U} = \frac{X q + Y k_n +
  Z k_e}{U}, \nonumber\\
\hspace{-0.15in}b_2 &=& \frac{(x_1 + x_3 + x_4) a_2 + x_3
  a_1}{U} = \nonumber\\
\hspace{-0.15in} &=&\frac{((x_1 + x_4)(x_2 + x_3) + x_2 x_3) q + ((x_1
  + x_2) x_3 + (x_1 + x_4) x_2) k_n - (x_1 + x_3 + x_4) x_5 k_e}{U} =
\frac{\bar{X} q + \bar{Y} k_n + \bar{Z} k_e}{U}, \nonumber\\
\hspace{-0.15in}Q &=& x_1 m^2_{\pi} + \frac{(x_2 + x_3 + x_5 + x_6)
  a^2_1 + 2 x_3 a_1 \cdot a_2 + (x_1 + x_3 + x_4) a^2_2}{U}
\end{eqnarray}
with $a_1 = - x_3 q - (x_3 + x_4) k_n$ and $a_2 = (x_2 + x_3) q + (x_2
+ x_3) k_n - x_5 k_e$. Using the Dirac equations for free fermions and
the algebra of the Dirac $\gamma$-matrices \cite{Itzykson1980}, we
obtain an irreducible Lorentz structure for  the integral ${\cal
  F}^{(1)}(k_n, k_e, q)_{1s}$. We get
\begin{eqnarray}\label{eq:C1s.7}
 \hspace{-0.30in}&& {\cal F}^{(1)}(k_n, k_e, q)_{1s} = \int^1_0 dx_1
 \int^1_0 dx_2 \int^1_0 dx_3 \int^1_0 dx_4 \int^1_0 dx_5 \int^1_0 dx_6
 \frac{\delta(1 - x_1 - x_2 - x_3 - x_4 - x_5 - x_6)}{256 \pi^4 U^2}
 \nonumber\\ \hspace{-0.30in}&& \times \Big\{ \Big(\Gamma\Big(2 -
 \frac{n}{2}\Big) + f^{(V)}_{1s^{(1)}}(x_1, \ldots, x_6) +
 g^{(V)}_{1s^{(1)}}(x_1, \ldots, x_6)\, \frac{k_n \cdot q}{m^2_N} +
 h^{(V)}_{1s^{(1)}}(x_1, \ldots, x_6)\, \frac{k_n \cdot
   k_e}{m^2_N}\Big) \gamma^{\mu} (1 - \gamma^5) \otimes \gamma_{\mu}
 \nonumber\\ \hspace{-0.30in}&& + \Big(\Gamma\Big(2 -
 \frac{n}{2}\Big) + f^{(A)}_{1s^{(1)}}(x_1, \ldots, x_6) +
 g^{(A)}_{1s^{(1)}}(x_1, \ldots, x_6)\, \frac{k_n \cdot q}{m^2_N} +
 h^{(A)}_{1s^{(1)}}(x_1, \ldots, x_6)\, \frac{k_n \cdot
   k_e}{m^2_N}\Big) \gamma^{\mu} (1 - \gamma^5) \otimes \gamma_{\mu}
 \gamma^5 \Big\}, \nonumber\\ \hspace{-0.30in}&&
\end{eqnarray}
where we have denoted
\begin{eqnarray}\label{eq:C1s.8}
 \hspace{-0.30in}f^{(V)}_{1s^{(1)}}(x_1, \ldots, x_6) = - 2\,{\ell
   n}\frac{\bar{Q}}{m^2_N} \quad,\quad g^{(V)}_{1s^{(1)}}(x_1, \ldots,
 x_6) = - 4\, \frac{V}{U}\,\frac{m^2_N}{\bar{Q}}\quad,\quad
 h^{(V)}_{1s^{(1)}}(x_1, \ldots, x_6) = - 4\,
 \frac{\bar{V}}{U}\,\frac{m^2_N}{\bar{Q}},\nonumber\\
 \hspace{-0.30in}f^{(A)}_{1s^{(1)}}(x_1, \ldots, x_6) = - 2\,{\ell
   n}\frac{\bar{Q}}{m^2_N} \quad,\quad g^{(A)}_{1s^{(1)}}(x_1, \ldots,
 x_6) = - 4\, \frac{V}{U}\,\frac{m^2_N}{\bar{Q}}\quad,\quad
 h^{(A)}_{1s^{(1)}}(x_1, \ldots, x_6) = - 4\,
 \frac{\bar{V}}{U}\,\frac{m^2_N}{\bar{Q}}.
\end{eqnarray}
For the calculation of the integrals over the Feynman parameters we
use
\begin{eqnarray}\label{eq:C1s.9}
\hspace{-0.30in} \frac{\bar{Q}}{m^2_N} &=& \frac{1}{U}\Big(x_1 U
\frac{m^2_{\pi}}{m^2_N} + (x_5 + x_6) (x_3 + x_4)^2 + (x_2 + x_3)(x_2
+ x_4)(x_3 + x_4) + x_1 (x_2 + x_3)^2\Big),
\nonumber\\ \hspace{-0.30in} V &=& x_3 (x_3 + x_4) (x_5 + x_6) + x_2
(x_2 + x_3) x_3 + (x_1 + x_4) (x_2 + x_3)^2 \;,\; \bar{V} = - (x_1 +
x_4) x_2 x_5 - (x_1 + x_2) x_3 x_5,\nonumber\\ \hspace{-0.30in}&&
\end{eqnarray}
where $\bar{Q} = Q\big|_{q = k_e = 0}$.  A deviation of $Q$ from
$\bar{Q}$ is taken into account in the linear approximation in the
form of an expansion in powers of $k_n\cdot q/m^2_N$ (or $q_0/m_N$)
and $k_n\cdot k_e/m^2_N$ (or $E_e/m_N$), respectively. After the
integration over the Feynman parameters, the contribution of the
integral ${\cal F}^{(1)}(k_n, k_e, q)_{1s}$ to the matrix element $M(n
\to p e^- \bar{\nu}_e)^{(\pi^0)}_{\rm Fig.\,\ref{fig:fig1}s}$ is given
by
\begin{eqnarray}\label{eq:C1s.10}
 \hspace{-0.30in}&& M(n \to p e^- \bar{\nu}_e)^{(\pi^0)}_{\rm
   Fig.\,\ref{fig:fig1}s^{(1)}}=  \frac{\alpha}{4\pi} \frac{g^2_{\pi
     N}}{16\pi^2}\,G_V \Big\{ \Big(A_{1s^{(1)}} \Gamma\Big(2 -
 \frac{n}{2}\Big) + F^{(V)}_{1s^{(1)}} + G^{(V)}_{1s^{(1)}}\,
 \frac{k_n \cdot q}{m^2_N} + H^{(V)}_{1s^{(1)}}\, \frac{k_n \cdot
   k_e}{m^2_N}\Big) \nonumber\\ \hspace{-0.30in}&& \times
 \Big[\bar{u}_e \gamma^{\mu} (1 - \gamma^5)
   v_{\bar{\nu}}\Big]\Big[\bar{u}_p \gamma_{\mu} u_n\Big] +
 \Big(B_{1s^{(1)}} \Gamma\Big(2 - \frac{n}{2}\Big) +
 F^{(A)}_{1s^{(1)}} + G^{(A)}_{1s^{(1)}} \, \frac{k_n \cdot q}{m^2_N}
 + H^{(A)}_{1s^{(1)}}\, \frac{k_n \cdot k_e}{m^2_N}\Big)
 \Big[\bar{u}_e\gamma^{\mu} (1 - \gamma^5)
   v_{\bar{\nu}}\Big]\nonumber\\ \hspace{-0.30in}&& \times
 \Big[\bar{u}_p \gamma_{\mu} \gamma^5 u_n\Big]\Big\}.
\end{eqnarray}
The structure constants are equal to $A_{1s^{(1)}} = 1/6$,
$B_{1s^{(1)}} = 1/6$, $F^{(V)}_{1s^{(1)}} = 0.5016$,
$G^{(V)}_{1s^{(1)}} = - 0.4546$, $H^{(V)}_{1s^{(1)}} = 0.2488$,
$F^{(A)}_{1s^{(1)}} = 0.5016$, $G^{(A)}_{1s^{(1)}} = - 0.4546$ and
$H^{(A)}_{1s^{(1)}} = 0.2488$. The Lorentz structure of
Eq.(\ref{eq:C1s.10}) is obtained at the neglect of the contributions
of order $O(E^2_0/m^2_N) \sim 10^{-6}$.

\subsubsection*{\bf Calculation of the integral
  ${\cal F}^{(2)}(k_n, k_e, q)_{1s}$}

In terms of the integrals over the Feynman parameters and after
integration over the virtual momenta in the $n$-dimensional momentum
space, we obtain for the integral ${\cal F}^{(2)}(k_n, k_e, q)_{1s}$
the following expression
\begin{eqnarray}\label{eq:C1s.11}
 \hspace{-0.30in}&& {\cal F}^{(2)}(k_n, k_e, q)_{1s} = \int^1_0 dx_1
 \int^1_0 dx_2 \int^1_0 dx_3 \int^1_0 dx_4 \int^1_0 dx_5 \int^1_0 dx_6
 \int^1_0 dx_7 \frac{\delta(1 - x_1 - x_2 - x_3 - x_4 - x_5 - x_6 -
   x_7)}{2^{2n} \pi^n U^{n/2}} \nonumber\\ \hspace{-0.30in}&& \times
 \Bigg\{\frac{n^2 (n + 2)}{8}\,\frac{\displaystyle \Gamma\Big(2 -
   \frac{n}{2}\Big)\Gamma\Big(2 - \frac{n - 1}{2}\Big)}{\displaystyle
   2^{n -
     3}\Gamma\Big(\frac{1}{2}\Big)}\,\Big(\frac{Q}{m^2_N}\Big)^{-4 +
   n}\,\Big[\frac{n - 2}{n}\, \gamma^{\mu} (1 - \gamma^5) \otimes
   \gamma_{\mu} - \frac{n - 2}{n^2}\, \gamma^{\beta} \gamma^{\alpha}
   \gamma^{\mu} (1 - \gamma^5) \otimes \gamma_{\beta} \gamma_{\alpha}
   \gamma_{\mu} \Big] \nonumber\\
\hspace{-0.3in}&& \times (1 + \gamma^5) + \frac{n^2}{4} \Gamma(5 -
n)\,Q^{-5+n} m^{2(4-n)}_N \Big[\Big(2\, \frac{n + 2}{n^2}\, b^2_1 -
  2\, \frac{n^2 - 2 n - 4}{n^2}\, b^2_2\Big)\, \gamma^{\mu}(1 -
  \gamma^5) \otimes \gamma_{\mu} (1 + \gamma^5) + \frac{n - 2}{n^2}\,
  b^2_2 \nonumber\\ \hspace{-0.30in}&& \times \gamma^{\beta}
  \gamma^{\alpha} \gamma^{\mu} (1 - \gamma^5) \otimes \gamma_{\beta}
  \gamma_{\alpha} \gamma_{\mu} (1 + \gamma^5) - \frac{n +
    2}{n^2}\,\gamma^{\beta} \gamma^{\alpha} \gamma^{\mu} (1 -
  \gamma^5) \otimes \gamma_{\beta} \gamma_{\alpha} \hat{b}_1
  \gamma_{\mu} \hat{b}_1 (1 + \gamma^5) - \frac{n^2 + 4 n + 4}{n^2}
  \gamma^{\mu}(1 - \gamma^5) \nonumber\\
\hspace{-0.3in}&& \otimes \gamma_{\mu} \hat{b}_2 \hat{b}_1 (1 +
\gamma^5) + \gamma^{\beta} \hat{b}_2 \gamma^{\alpha} (1 - \gamma^5)
\otimes \Big(4\, \frac{n + 2}{n^2}\, \gamma_{\beta} \gamma_{\alpha}
\hat{b}_1 (1 + \gamma^5) + \frac{n^2 - 4}{n^2}\, \gamma_{\beta}
\hat{b}_2 \gamma_{\alpha} (1 + \gamma^5)\Big) \Big]\Bigg\},
\end{eqnarray}
where we have denoted
\begin{eqnarray}\label{eq:C1s.12}
U&=& (x_1 + x_3 + x_4) (x_2 + x_5 + x_6 + x_7) + (x_1 + x_4) x_3,
\nonumber\\ b_1 &=& \frac{(x_2 + x_3 + x_5 + x_6 + x_7) a_1 + x_3
  a_2}{U} = - \frac{(x_5 + x_6 + x_7) x_3 + (x_2 + x_3) x_4 }{U}\, k_n
= \frac{Y}{U}\, k_n,\nonumber\\ b_2 &=& \frac{(x_1 + x_3 + x_4) a_2 +
  x_3 a_1}{U} = \frac{(x_1 + x_4) x_2 + (x_1 + x_2) x_3}{U}\, k_n =
\frac{ \bar{Y}}{U}\, k_n, \nonumber\\ Q &=& x_1 m^2_{\pi} + \frac{(x_2
  + x_3 + x_5 + x_6 + x_7) a^2_1 + 2 x_3 a_1 \cdot a_2 + (x_1 + x_3 +
  x_4) a^2_2}{U}
\end{eqnarray}
with $a_1 = - (x_3 + x_4) k_n$ and $a_2 = (x_2 + x_3) k_n $. For the
derivation of Eq.(\ref{eq:C1s.11}) we have neglected the contributions
of order $O(k_n\cdot q/M^2_W) \sim O(k_n \cdot k_e/M^2_W) \sim
10^{-7}$. Taking the limit $n \to 4$ and using the algebra of the
Dirac $\gamma$-matrices and Dirac equations for free fermions, we
arrive at the expression
\begin{eqnarray}\label{eq:C1s.13}
 \hspace{-0.30in}&& {\cal F}^{(2)}(k_n, k_e, q)_{1s} = \int^1_0 dx_1
 \int^1_0 dx_2 \int^1_0 dx_3 \int^1_0 dx_4 \int^1_0 dx_5 \int^1_0 dx_6
 \int^1_0 dx_7 \frac{\delta(1 - x_1 - x_2 - x_3 - x_4 - x_5 - x_6 -
   x_7)}{256 \pi^4 U^2} \nonumber\\ \hspace{-0.30in}&& \times \Big\{
 \Big(
 f^{(V)}_{1s^{(2)}}(x_1, \ldots, x_7) + g^{(W)}_{1s^{(2)}}(x_1,
 \ldots, x_7)\, \frac{m^2_N}{M^2_W}\Big) \gamma^{\mu} (1 - \gamma^5)
 \otimes \gamma_{\mu} + \Big( f^{(A)}_{1s^{(2)}}(x_1, \ldots, x_7) +
 h^{(W)}_{1s^{(2)}}(x_1, \ldots, x_6)\, \frac{m^2_N}{M^2_W}\Big)
 \nonumber\\ \hspace{-0.30in}&& \times \gamma^{\mu} (1 - \gamma^5)
 \otimes \gamma_{\mu} \gamma^5 + f^{(W)}_{1s^{(2)}}(x_1, \ldots,
 x_7)\, \frac{m^2_N}{M^2_W}\, \frac{\hat{k}_n}{m_N} (1 - \gamma^5)
 \otimes 1\Big\},
\end{eqnarray}
where we have denoted
\begin{eqnarray}\label{eq:C1s.14}
 \hspace{-0.30in}&&f^{(V)}_{1s^{(2)}}(x_1, \ldots, x_7) = 
 - \frac{3}{2}
\end{eqnarray}
and
\begin{eqnarray}\label{eq:C1s.15}
 \hspace{-0.30in}&&g^{(W)}_{1s^{(2)}}(x_1, \ldots, x_7) =
 \frac{1}{\bar{Q}}\Big[9\,
   \frac{Y^2}{U^2}+ 3\, \frac{Y\bar{Y}}{U^2}\Big]
\end{eqnarray}
and
\begin{eqnarray}\label{eq:C1s.16}
 \hspace{-0.30in}&&f^{(A)}_{1s^{(2)}}(x_1, \ldots, x_7) = -
 \frac{3}{2}
\end{eqnarray}
and
\begin{eqnarray}\label{eq:C1s.17}
 \hspace{-0.30in}&& h^{(W)}_{1s^{(2)}}(x_1, \ldots, x_7) =
 \frac{1}{\bar{Q}}\Big[9 \, \frac{\bar{Y}}{U} - 6\,
   \frac{Y^2}{U^2} + 3\, \frac{Y\bar{Y}}{U^2}\Big]
\end{eqnarray}
and
\begin{eqnarray}\label{eq:C1s.18}
 \hspace{-0.30in}&&f^{(W)}_{1s^{(2)}}(x_1, \ldots, x_7) =
 \frac{1}{\bar{Q}}\Big[
  - 12\, \frac{Y^2}{U^2} - 24\, \frac{Y\bar{Y}}{U^2} + 12 \,
   \frac{\bar{Y}^2}{U^2}\Big].
\end{eqnarray}
For the calculation of the integrals over the Feynman parameters we
use
\begin{eqnarray}\label{eq:C1s.19}
  \hspace{-0.21in}\bar{Q} &=& \frac{1}{U}\Big(x_7 U + \big((x_5 +
  x_6 + x_7) (x_3 + x_4)^2 + (x_2 + x_3)(x_2 + x_4) (x_3 + x_4) + x_1 (x_2 + x_3)^2\big)\, \frac{m^2_N}{M^2_W}\Big), \nonumber\\
  \hspace{-0.21in} U &=& (x_1 + x_3 + x_4)(x_2 + x_5 + x_6 + x_7) +
  (x_1 + x_4) x_3,\nonumber\\ \hspace{-0.21in} Y &=& - (x_5 + x_6 +
  x_7) x_3 - (x_2 + x_3) x_4\;,\; \bar{Y} = (x_1 + x_2) x_3 + (x_1 +
  x_4) x_2,
\end{eqnarray}
where $m_N = (m_n + m_p)/2 = 938.9188\,{\rm MeV}$ and $M_W =
80,379\,{\rm GeV}$ \cite{PDG2020}. After the integration over the
Feynman parameters, the contribution of the integral ${\cal
  F}^{(2)}(k_n, k_e,q)_{1s}$ to the matrix element $M(n \to p e^-
\bar{\nu}_e)^{(\pi^0)}_{\rm Fig.\,\ref{fig:fig1}s}$ is given by
\begin{eqnarray}\label{eq:C1s.20}
 \hspace{-0.21in}&& M(n \to p e^- \bar{\nu}_e)^{(\pi^0)}_{\rm
   Fig.\,\ref{fig:fig1}s^{(2)}}= \frac{\alpha}{4\pi} \frac{g^2_{\pi
     N}}{16\pi^2}\,G_V \Big\{ \Big(F^{(V)}_{1s^{(2)}} +
 G^{(W)}_{1s^{(2)}}\, \frac{m^2_N}{M^2_w}\Big) \Big[\bar{u}_e
   \gamma^{\mu} (1 - \gamma^5) v_{\bar{\nu}}\Big]\Big[\bar{u}_p
   \gamma_{\mu} u_n\Big] \nonumber\\ \hspace{-0.21in}&& +
 \Big( F^{(A)}_{1s^{(2)}} + H^{(W)}_{1s^{(2)}}\,
 \frac{m^2_N}{M^2_W}\Big) \Big[\bar{u}_e\gamma^{\mu} (1 - \gamma^5)
   v_{\bar{\nu}}\Big] \Big[\bar{u}_p \gamma_{\mu} \gamma^5 u_n\Big] +
 F^{(W)}_{1s^{(2)}}\, \frac{m^2_N}{M^2_W} \Big[\bar{u}_e
   \frac{\hat{k}_n}{m_N} (1 - \gamma^5) v_{\bar{\nu}}\Big]
 \Big[\bar{u}_p \gamma^5 u_n\Big]\Big\}.\nonumber\\ \hspace{-0.21in}&&
\end{eqnarray}
The structure constants are equal to $F^{(V)}_{1s^{(2)}} = - 0.0512$,
$G^{(W)}_{1s^{(2)}} = 3.7070$, $F^{(A)}_{1s^{(2)}} = - 0.0512$,
$H^{(W)}_{1s^{(2)}} = 3.7070$ and $F^{(W)}_{1s^{(2)}} = 1.9580$. The
Lorentz structure of the matrix element Eq.(\ref{eq:C1s.20}) is
calculated at the neglect of the contributions of order $O(k_n\cdot
q/M^2_W) \sim O(k_n \cdot k_e/M^2_W) \sim 10^{-7}$ and
$O(m^2_{\pi}/M^2_W) \sim 10^{-6}$, respectively.

As a result, the contribution of the Feynman diagram in
Fig.\,\ref{fig:fig1}s to the amplitude of the neutron beta decay 
is equal to
\begin{eqnarray}\label{eq:C1s.21}
 \hspace{-0.30in}&&~ M(n \to p e^- \bar{\nu}_e)^{(\pi^0)}_{\rm
   Fig.\,\ref{fig:fig1}s}= \frac{\alpha}{4\pi} \frac{g^2_{\pi
     N}}{16\pi^2}\,G_V \Big\{ \Big(A_{1s}\Gamma\Big(2 -
 \frac{n}{2}\Big) + F^{(V)}_{1s}+ G^{(V)}_{1s}\, \frac{k_n \cdot
   q}{m^2_N} + H^{(V)}_{1s}\, \frac{k_n \cdot k_e}{m^2_N} +
 G^{(W)}_{1s}\, \frac{m^2_N}{M^2_w}\Big)
 \nonumber\\ \hspace{-0.30in}&&\times \Big[\bar{u}_e \gamma^{\mu} (1 -
   \gamma^5) v_{\bar{\nu}}\Big]\Big[\bar{u}_p \gamma_{\mu} u_n\Big] +
 \Big(B_{1s}\Gamma\Big(2 - \frac{n}{2}\Big) + F^{(A)}_{1s}+
 G^{(A)}_{1s}\, \frac{k_n \cdot q}{m^2_N} + H^{(A)}_{1s}\, \frac{k_n
   \cdot k_e}{m^2_N} + H^{(W)}_{1s}\, \frac{m^2_N}{M^2_W}\Big)
 \Big[\bar{u}_e\gamma^{\mu} (1 - \gamma^5)
   v_{\bar{\nu}}\Big]\nonumber\\ \hspace{-0.30in}&& \times
 \Big[\bar{u}_p \gamma_{\mu} \gamma^5 u_n\Big] + F^{(W)}_{1s}\,
 \frac{m^2_N}{M^2_W} \Big[\bar{u}_e \frac{\hat{k}_n}{m_N} (1 -
   \gamma^5) v_{\bar{\nu}}\Big] \Big[\bar{u}_p \gamma^5
   u_n\Big]\Big\}.
\end{eqnarray}
The structure constants are equal to $A_{1s} = A_{1s^{(1)}} = 1/6$,
$B_{1s} = B_{1s^{(1)}} = 1/6$, $F^{(V)}_{1s} = F^{(V)}_{1s^{(1)}} +
F^{(V)}_{1s^{(2)}} = 0.4504$, $G^{(V)}_{1s} = G^{(V)}_{1s^{(1)}} = -
0.4546$, $H^{(V)}_{1s} = H^{(V)}_{1s^{(1)}} = 0.2488$, $G^{(W)}_{1s} =
G^{(W)}_{1s^{(2)}} = 3.7070$, $F^{(A)}_{1s} = F^{(A)}_{1s^{(1)}} +
F^{(A)}_{1s^{(2)}} = 0.4504$, $G^{(A)}_{1s} = G^{(A)}_{1s^{(1)}} = -
0.4546$, $H^{(A)}_{1s} = H^{(A)}_{1s^{(1)}} = 0.2488$, $H^{(W)}_{1s} =
H^{(W)}_{1s^{(2)}} = 3.7070$ and $F^{(W)}_{1s} = F^{(W)}_{1s^{(2)}} =
1.9580$. The Lorentz structure of Eq.(\ref{eq:C1s.21}) is obtained at
the neglect of the contributions of order $O(E^2_0/m^2_N) \sim
O(m^2_{\pi}/M^2_W) \sim 10^{-6}$.

\newpage

\subsection*{Fig.1. The contribution  of the  the Feynman diagram in
  Fig.\,\ref{fig:fig1} to the amplitude of the neutron
 beta decay }
\renewcommand{\theequation}{Fig.1-\arabic{equation}}
\setcounter{equation}{0}

Summing up the results of the analytical calculation of the Feynman
diagrams in Fig.\,\ref{fig:fig1}, given in Appendix C, we obtain the
contribution of the Feynman diagrams in Fig.\,\ref{fig:fig1} to the
amplitude of the neutron beta decay. We get
\begin{eqnarray}\label{eq:S1.1}
 \hspace{-0.30in}&& M(n \to p e^- \bar{\nu}_e)^{(\pi^0)}_{\rm
   Fig.\,\ref{fig:fig1}}= \frac{\alpha}{4\pi} \frac{g^2_{\pi
     N}}{16\pi^2}\,G_V \Big\{ \Big(A_1\Gamma\Big(2 - \frac{n}{2}\Big)
 + F^{(V)}_1+ G^{(V)}_1\, \frac{k_n \cdot q}{m^2_N} + H^{(V)}_1\,
 \frac{k_n \cdot k_e}{m^2_N} + G^{(W)}_1\, \frac{m^2_N}{M^2_w}\Big)
 \nonumber\\ \hspace{-0.30in}&& \times \Big[\bar{u}_e \gamma^{\mu} (1
   - \gamma^5) v_{\bar{\nu}}\Big]\Big[\bar{u}_p \gamma_{\mu} u_n\Big]
 + \Big(B_1\Gamma\Big(2 - \frac{n}{2}\Big) + F^{(A)}_1+ G^{(A)}_1\,
 \frac{k_n \cdot q}{m^2_N} + H^{(A)}_1\, \frac{k_n \cdot k_e}{m^2_N} +
 H^{(W)}_1\, \frac{m^2_N}{M^2_W}\Big) \Big[\bar{u}_e\gamma^{\mu} (1 -
   \gamma^5) v_{\bar{\nu}}\Big] \nonumber\\ \hspace{-0.30in}&& \times
 \Big[\bar{u}_p \gamma_{\mu} \gamma^5 u_n\Big] + F^{(W)}_1\,
 \frac{m^2_N}{M^2_W} \Big[\bar{u}_e \frac{\hat{k}_n}{m_N} (1 -
   \gamma^5) v_{\bar{\nu}}\Big] \Big[\bar{u}_p u_n\Big]\Big\}.
\end{eqnarray}
The matrix element Eq.(\ref{eq:S1.1}) is calculated to the LLA.  The
structure constants are equal to $A_1 = 3.8802$, $B_1 = - 0.3074$,
$F^{(V)}_1 = 5.6791$, $G^{(V)}_1 = - 93.2739$, $H^{(V)}_1 = 91.5861$,
$G^{(W)}_1 = 17.3590$, $F^{(A)}_1 = - 13.3792$, $G^{(A)}_1 = 56.3788$,
$H^{(A)}_1 = - 55.2018$, $H^{(W)}_1 = 8.3579$ and $F^{(M)}_1= -
18.5441$. The Lorentz structure of Eq.(\ref{eq:S1.1}) is obtained at
the neglect of the contributions of order $O(E^2_0/m^2_N) \sim
O(m^2_{\pi}/M^2_W) \sim 10^{-6}$. We would like to emphasize that all
structure constants in the matrix element Eq.(\ref{eq:S1.1}) are
induced by the contributions of the {\it first class} currents
\cite{Weinberg1958}, which are $G$-even \cite{Lee1956a} (see also
\cite{Ivanov2018}).

The calculation of the integrals over the Feynman parameters runs as
follows
\begin{eqnarray}\label{eq:S1.2}
 \hspace{-0.30in}&& \int^1_0dx_1\int^1_0dx_2 \ldots \int^1_0 dx_{n -
   1} \int^1_0dx_n \delta\big(1 - \sum^n_{k = 1}x_k\big)\frac{ f(x_1,
   x_2,\ldots, x_{n-1}, x_n)}{U^2(x_1, x_2, \ldots, x_{n - 1}, x_n)} =
 \nonumber\\
 \hspace{-0.30in}&& = \int^1_0dx_1 \int^{1-x_1}_0dx_2 \ldots \int^{1-
   \sum^{n-2}_{k = 1} x_k}_0 dx_{n - 1}\,\frac{f(x_1, x_2, \ldots,
   x_{n-1}, 1 - \sum^{n-1}_{k=1}x_k)}{U^2(x_1, x_2, \ldots, x_{n - 1},
    1 - \sum^{n-1}_{k=1}x_k)}\,,
\end{eqnarray}   
where $n$ is equal to either $n = 6$ or $n = 7$. The determinant
$U(x_1, x_2, \ldots, x_{n - 1}, x_n)$ of the integration over the
momentum space is defined for every diagram Fig.\,\ref{fig:fig1}a -
Fig.\,\ref{fig:fig1}s. Then, the functions $f(x_1, x_2,\ldots,
x_{n-1}, x_n)$ are given in terms of $X/U$, $Y/U$, $Z/U$, $\bar{X}/U$,
$\bar{Y}/U$ and $\bar{Z}/U$, which are functions of the Feynman
parameters $x_j$ for $j = 1,2,\ldots, n-1,n$ in dependence of the
structure of the Feynman diagram.

\newpage

\subsection*{C2. Analytical calculation of the Feynman diagrams in
  Fig.\,\ref{fig:fig2} to the amplitude of the neutron beta decay}
\renewcommand{\theequation}{C2-\arabic{equation}}
\setcounter{equation}{0}

The calculation of the Feynman diagrams in Fig.\,\ref{fig:fig2}
demands intermediate renormalization of masses and wave functions of
virtual hadrons, caused by virtual photon-exchanges \cite{BS1984}. A
direct application of the standard procedure \cite{Smirnov2004,
  Smirnov2006, Smirnov2012} for the calculation of two-loop Feynman
diagrams leads to the Feynman parametrization of the Feynman diagrams
in Fig.\,\ref{fig:fig2} with divergent integrals over the Feynman
parameters. Such an appearance of divergent integrals over the Feynman
parameters is conditioned by the necessity to carry out intermediate
renormalization of masses and wave functions of virtual hadrons,
caused by virtual photon-exchanges. Such a renormalization can be
performed by using the technique proposed by Cvitanovi\'c and
Kinoshita \cite{Kinoshita1974b}. However, as we have found, the
procedure, proposed by Johns \cite{Johns1974}, allows, even for
two-loop Feynman diagrams which we deal with, to avoid the problem of
intermediate renormalization of masses and wave functions of virtual
hadrons, caused by virtual photon-exchanges. Thus, following Johns
\cite{Johns1974} we multiply integrands of the Feynman momentum
integrals by unity $(p^2 + i0)/(p^2 + i0)$, where $p$ is a 4-momentum
of a virtual photon. Then, after the direct application of the
standard procedure to the calculation of two-loop Feynman diagrams
\cite{Smirnov2004, Smirnov2006, Smirnov2012} and in the limit $n \to
4$, we arrive at the Feynman parametrization of the Feynman diagrams
in Fig.\,\ref{fig:fig2} with contributions proportional to $(\Gamma(2
- n/2) - 2\,{\ell n}Q/m^2_N)/U^2$ and multiplied by convergent
integrals over the Feynman parameters. After the integration over the
Feynman parameters the terms, induced by $(\Gamma(2 - n/2) - 2\,{\ell
  n}Q/m^2_N)/U^2\big|_{q = k_e = 0}$, can be removed by
renormalization of the Fermi $G_V$ and axial $g_A$ couping constants
in agreement with Sirlin's procedure \cite{Sirlin1967, Sirlin1978}. No
other renormalization is needed for the extraction of the
contributions of the Feynman diagrams in Fig.\,\ref{fig:fig2} to the
radiative corrections $O(\alpha E_e/m_N)$, calculated to the amplitude
of the neutron beta decay as NLO terms in the large nucleon mass $m_N$
expansion.

\subsection*{C2a. Analytical calculation of the Feynman diagram in
  Fig.\,\ref{fig:fig2}a}
\renewcommand{\theequation}{C2a-\arabic{equation}}
\setcounter{equation}{0}

In the Feynman gauge for the photon propagator the Feynman diagram in
Fig.\,\ref{fig:fig2}a is defined by the following momentum integrals
\begin{eqnarray}\label{eq:C2a.1}
&& M(n \to p e^- \bar{\nu}_e)^{(\pi^0)}_{\rm Fig.\,\ref{fig:fig2}a} =
  - 2 e^2 g^2_{\pi N} G_V \int \frac{d^4k}{(2\pi)^4i}\int
  \frac{d^4p}{(2\pi)^4i}\,\Big[\bar{u}_e \gamma^{\mu}(1 - \gamma^5)
    v_{\bar{\nu}}\Big]\,\,\Big[\bar{u}_p\,\gamma^5\, \frac{1}{m_N -
      \hat{k}_n - \hat{k} - i0}\,\gamma^5 u_n\Big]
  \nonumber\\ &&\times \frac{(2 k - q)_{\mu}(p + 2k - 2 q)^{\beta}(p +
    2k - 2 q)_{\beta}}{[m^2_{\pi} - k^2 - i0][m^2_{\pi} - (k - q)^2 -
      i0]^2[m^2_{\pi} - (k + p - q)^2 - i0]}\, \frac{1}{p^2 + i0},
\end{eqnarray}
where we have made the replacement $M^2_W D_{\mu\nu} (- q) \to -
\eta_{\mu\nu}$, which is valid at the neglect of the
contributions of order $O(m_em_N/M^2_W) \sim 10^{-7}$.  The
calculation of the matrix element Eq.(\ref{eq:C2a.1}) we reduce to the
calculation of the integral ${\cal F}(k_n, k_e, q)_{2a}$, which is
defined by
\begin{eqnarray}\label{eq:C2a.2}
\hspace{-0.3in}&&{\cal F}(k_n, k_e, q)_{2a} = - \int
\frac{d^4k}{(2\pi)^4i}\int \frac{d^4p}{(2\pi)^4i}\,\frac{1}{m^2_N -
  (k_n + k)^2 - i0}\, \frac{1}{m^2_{\pi} - k^2 -
  i0}\,\frac{1}{[m^2_{\pi} - (k - q)^2 -
    i0]^2}\nonumber\\ \hspace{-0.3in}&&\times \,\frac{1}{m^2_{\pi} -
  (k + p - q)^2 - i0}\,\frac{1}{p^2 + i0}\,\Big(2 \hat{k} (1 -
\gamma^5) \otimes \big(p^2 + 4 k^2 + 4 k\cdot p \big)\, \hat{k}\Big),
\end{eqnarray}
where we have kept only leading divergent contributions. Then, we
multiply the integrand by unity $(p^2 + i0)/(p^2 + i0)$
\cite{Johns1974}, merge the denominators by using the Feynman
representation \cite{Feynman1950}, diagonalize the obtained
denominator and integrate in the $n$-dimensional momentum space over
the virtual momenta \cite{Smirnov2004, Smirnov2006, Smirnov2012}. As a
result, we arrive at the following expression
\begin{eqnarray}\label{eq:C2a.3}
 \hspace{-0.30in}&& {\cal F}(k_n, k_e, q)_{2a} = \int^1_0 dx_1
 \int^1_0 dx_2 \int^1_0 dx_3 \int^1_0 dx_4 \int^1_0 dx_5\, x_2 x_5\,
 \frac{\delta(1 - x_1 - x_2 - x_3 - x_4 - x_5)}{2^{2n} \pi^n U^{n/2}}
 \Bigg\{ - 5\,\frac{n(n + 2)}{4} \nonumber\\ \hspace{-0.30in}&& \times
 \, \frac{\displaystyle \Gamma\Big(2 - \frac{n}{2}\Big)\Gamma\Big(2 -
   \frac{n - 1}{2}\Big)}{\displaystyle 2^{n - 3}
   \Gamma\Big(\frac{1}{2}\Big)}\, \Big(\frac{Q}{m^2_N}\Big)^{-4 + n}\,
 \gamma^{\mu} (1 - \gamma^5) \otimes \gamma_{\mu} + \ldots \Bigg\},
\end{eqnarray}
where we have  denoted
\begin{eqnarray*}
\hspace{-0.30in}U &=& (x_1 + x_2 + x_4) (x_3 + x_5) + x_3 x_5,
\nonumber\\
\hspace{-0.30in}b_1 &=& \frac{(x_3 + x_5) a_1 - x_3 a_2}{U} =
\frac{((x_3 + x_5) x_2 + x_3 x_5) q - (x_3 + x_5) x_4 k_n }{U} =
\frac{X q + Y k_n}{U}, \nonumber\\
\end{eqnarray*}
\begin{eqnarray}\label{eq:C2a.4}
\hspace{-0.30in}b_2 &=& \frac{(x_1 + x_2 + x_3 + x_4) a_2 - x_3
  a_1}{U} = \frac{(x_1 + x_4) x_3 q + x_3 x_4 k_n}{U} = \frac{\bar{X}
  q + \bar{Y} k_n}{U}, \nonumber\\
\hspace{-0.30in}Q &=& (x_1 + x_2 + x_3) m^2_{\pi} -(x_2 + x_3) q^2 +
\frac{(x_3 + x_5) a^2_1 - 2 x_3 a_1 \cdot a_2 + (x_1 + x_2 + x_3 +
  x_4) a^2_2}{U}
\end{eqnarray}
with $a_1 = (x_2 + x_3) q - x_4 k_n$ and $a_2 = x_3 q$. Taking the
limit $n \to 4$ and keeping the divergent contribution proportional to
$\Gamma(2 - n/2)$, we transcribe the r.h.s. of Eq.(\ref{eq:C2a.3})
into the form
\begin{eqnarray}\label{eq:C2a.5}
 \hspace{-0.30in}&& {\cal F}(k_n, k_e, q)_{2a} = \int^1_0 dx_1
 \int^1_0 dx_2 \int^1_0 dx_3 \int^1_0 dx_4 \int^1_0 dx_5 \, x_2 x_5\,
 \frac{\delta(1 - x_1 - x_2 - x_3 - x_4 - x_5)}{256 \pi^4 U^2}
 \nonumber\\ \hspace{-0.30in}&& \times \Big\{ \Big(- 15\,\Gamma\Big(2 -
 \frac{n}{2}\Big) + 30\, {\ell n}\frac{Q}{m^2_N}\Big) \gamma^{\mu} (1
 - \gamma^5) \otimes \gamma_{\mu} + \ldots \Big\}.
\end{eqnarray}
Using the Dirac equations for free fermions and the algebra of the
Dirac $\gamma$-matrices \cite{Itzykson1980} we obtain an irreducible
Lorentz structure for the integral ${\cal F}(k_n, k_e, q)_{2a}$. We get
\begin{eqnarray}\label{eq:C2a.6}
 \hspace{-0.30in}&& {\cal F}(k_n, k_e, q)_{2a} = \int^1_0 dx_1
 \int^1_0 dx_2 \int^1_0 dx_3 \int^1_0 dx_4 \int^1_0 dx_5 \,x_2 x_5
 \frac{\delta(1 - x_1 - x_2 - x_3 - x_4 - x_5 )}{256 \pi^4 U^2}
 \nonumber\\ \hspace{-0.30in}&& \times \Big\{ \Big(15\, \Gamma\Big(2 -
 \frac{n}{2}\Big) + f^{(V)}_{2a}(x_1, \ldots, x_5) + g^{(V)}_{2a}(x_1,
 \ldots, x_5)\, \frac{k_n \cdot q}{m^2_N}\Big) \gamma^{\mu} (1 -
 \gamma^5) \otimes \gamma_{\mu} + \ldots \Big\},
\end{eqnarray}
where we have denoted
\begin{eqnarray}\label{eq:C2a.7}
 \hspace{-0.30in}&&f^{(V)}_{2a}(x_1, \ldots, x_5) =  30\,{\ell
   n}\frac{\bar{Q}}{m^2_N} \quad,\quad g^{(V)}_{2a}(x_1, \ldots, x_5)
 =  60\, \frac{V}{U}\,\frac{m^2_N}{\bar{Q}}
\end{eqnarray}
For the calculation of the integrals over the Feynman parameters we
use
\begin{eqnarray}\label{eq:C2a.8}
\hspace{-0.30in} \frac{\bar{Q}}{m^2_N} &=& \frac{1}{U}\Big((x_1 + x_2
+ x_3 )U \frac{m^2_{\pi}}{m^2_N} + (x_3 + x_5) x^2_4\Big),
\nonumber\\ \hspace{-0.30in} V &=& - (x_3 + x_5) x_2 x_4 - x_3 x_4
x_5,
\end{eqnarray}
where $\bar{Q} = Q\big|_{q = 0}$.  A deviation of $Q$ from $\bar{Q}$
is taken into account in the linear approximation in the form of an
expansion in powers of $k_n\cdot q/m^2_N$ (or $q_0/m_N$). After the
integration over the Feynman parameters, the contribution of the
integral ${\cal F}(k_n, k_e, q)_{2a}$ to the matrix element $M(n \to p
e^- \bar{\nu}_e)^{(\pi^0)}_{\rm Fig.\,\ref{fig:fig2}a}$ is given by
\begin{eqnarray}\label{eq:C2a.9}
 \hspace{-0.30in} M(n \to p e^- \bar{\nu}_e)^{(\pi^0)}_{\rm
   Fig.\,\ref{fig:fig2}a} =  - \frac{\alpha}{2\pi} \frac{g^2_{\pi
     N}}{16\pi^2}\,G_V \Big\{ \Big(A_{2a} \Gamma\Big(2 -
 \frac{n}{2}\Big) + F^{(V)}_{2a} + G^{(V)}_{2a}\, \frac{k_n \cdot
   q}{m^2_N}\Big) \Big[\bar{u}_e \gamma^{\mu} (1 - \gamma^5)
   v_{\bar{\nu}}\Big]\Big[\bar{u}_p \gamma_{\mu} u_n\Big]\Big\}.
\end{eqnarray}
The structure constants are equal to $A_{2a} = - 0.5122$, $F^{(V)}_{2a}
= - 3.0916$ and $G^{(V)}_{2a} = - 2.7696$. The Lorentz structure of
Eq.(\ref{eq:C2a.9}) is obtained at the neglect of the contributions
of order $O(E^2_0/m^2_N) \sim 10^{-6}$.

\subsection*{C2b. Analytical calculation of the Feynman diagram in
  Fig.\,\ref{fig:fig2}b}
\renewcommand{\theequation}{C2b-\arabic{equation}}
\setcounter{equation}{0}

In the Feynman gauge for the photon propagator the Feynman diagram in
Fig.\,\ref{fig:fig2}b is defined by the following momentum integrals
\begin{eqnarray}\label{eq:C2b.1}
 && M(n \to p e^- \bar{\nu}_e)^{(\pi^0)}_{\rm Fig.\,\ref{fig:fig2}b} =
  - 2 e^2 g^2_{\pi N}  G_V \int \frac{d^4k}{(2\pi)^4i}\int
  \frac{d^4p}{(2\pi)^4i}\,\Big[\bar{u}_e\, \gamma^{\mu}(1 - \gamma^5)
    v_{\bar{\nu}}\Big]\,\Big[\bar{u}_p\, \gamma^5\,\frac{1}{m_N -
      \hat{k}_n - \hat{k} - i0}\,\gamma^5
    u_n\Big]\nonumber\\ &&\times\, \frac{(2 k - q)_{\mu}(2 k +
    p)^{\beta}(2 k + p)_{\beta}}{[m^2_{\pi} - k^2 - i0]^2[m^2_{\pi} -
      (k + p)^2 - i0][m^2_{\pi} - (k - q)^2 - i0]}\, \frac{1}{p^2 +
    i0},
\end{eqnarray}
where we have made a change of variables $p \to - p$ and the usual
replacement $M^2_W D_{\mu\nu} (- q) \to - \eta_{\mu\nu}$, which is
valid at the neglect of the contributions of order $O(m_em_N/M^2_W)
\sim 10^{-7}$.  The calculation of the matrix element
Eq.(\ref{eq:C2b.1}) we reduce to the calculation of the integral
${\cal F}(k_n, k_e, q)_{2b}$, which is defined by
\begin{eqnarray}\label{eq:C2b.2}
\hspace{-0.3in}&&{\cal F}(k_n, k_e, q)_{2b} = - \int
\frac{d^4k}{(2\pi)^4i}\int \frac{d^4p}{(2\pi)^4i}\,\frac{1}{m^2_N -
  (k_n + k)^2 - i0}\, \frac{1}{m^2_{\pi} - (k - q)^2 - i0}\,
\frac{1}{[m^2_{\pi} - k^2 - i0]^2} \nonumber\\ \hspace{-0.3in}&&\times
\,\frac{1}{m^2_{\pi} - (k + p )^2 - i0}\,\frac{1}{p^2 + i0}\,\Big(2
\hat{k} (1 - \gamma^5) \otimes \big(p^2 + 4 k^2 + 4 k\cdot p\big)\,
\hat{k}\Big).
\end{eqnarray}
The procedure of the calculation of the integral ${\cal F}(k_n, k_e,
q)_{2b}$ is similar to that we have used in subsection C2a. Thus,
multiplying the integrand by unity $(p^2 + i0)/(p^2 + i0)$
\cite{Johns1974}, skipping standard intermediate calculations we
define the integral ${\cal F}(k_n, k_e, q)_{2b}$ in terms of the
integrals over the Feynman parameters. We get
\begin{eqnarray}\label{eq:C2b.3}
 \hspace{-0.30in}&& {\cal F}(k_n, k_e, q)_{2b} = \int^1_0 dx_1
 \int^1_0 dx_2 \int^1_0 dx_3 \int^1_0 dx_4 \int^1_0 dx_5\, x_2 x_5\,
 \frac{\delta(1 - x_1 - x_2 - x_3 - x_4 - x_5)}{2^{2n} \pi^n U^{n/2}}
 \Bigg\{- 5\,\frac{n(n + 2)}{4} \nonumber\\ \hspace{-0.30in}&& \times
 \, \frac{\displaystyle \Gamma\Big(2 - \frac{n}{2}\Big)\Gamma\Big(2 -
   \frac{n - 1}{2}\Big)}{\displaystyle 2^{n - 3}
   \Gamma\Big(\frac{1}{2}\Big)}\, \Big(\frac{Q}{m^2_N}\Big)^{-4 + n}\,
 \gamma^{\mu} (1 - \gamma^5) \otimes \gamma_{\mu} + \ldots \Bigg\},
\end{eqnarray}
where we have denoted
\begin{eqnarray}\label{eq:C2b.4}
 \hspace{-0.30in}U &=& (x_1 + x_2 + x_4) (x_3 + x_5) + x_3 x_5,
\nonumber\\
\hspace{-0.30in}b_1 &=& \frac{(x_3 + x_5) a_1 - x_3 a_2}{U} =
\frac{(x_3 + x_5) x_1  q - (x_3 + x_5) x_4 k_n }{U} =
\frac{X q + Y k_n}{U}, \nonumber\\
\hspace{-0.30in}b_2 &=& \frac{(x_1 + x_2 + x_3 + x_4) a_2 - x_3
  a_1}{U} = \frac{- x_1 x_3 q + x_3 x_4 k_n}{U} = \frac{\bar{X} q +
  \bar{Y} k_n}{U}, \nonumber\\
\hspace{-0.30in}Q &=& (x_1 + x_2 + x_3) m^2_{\pi} - x_1 q^2 +
\frac{(x_3 + x_5) a^2_1 - 2 x_3 a_1 \cdot a_2 + (x_1 + x_2 + x_3 +
  x_4) a^2_2}{U}
\end{eqnarray}
with $a_1 = x_1 q - x_4 k_n$ and $a_2 = 0$. Taking the limit $n \to 4$
and keeping the divergent contribution proportional to $\Gamma(2 -
n/2)$, we transcribe the r.h.s. of Eq.(\ref{eq:C2b.3}) into the form
\begin{eqnarray}\label{eq:C2b.5}
 \hspace{-0.21in}&& {\cal F}(k_n, k_e, q)_{2b} = \int^1_0 dx_1
 \int^1_0 dx_2 \int^1_0 dx_3 \int^1_0 dx_4 \int^1_0 dx_5 \, x_2 x_5\,
 \frac{\delta(1 - x_1 - x_2 - x_3 - x_4 - x_5)}{256 \pi^4 U^2}
 \nonumber\\ \hspace{-0.30in}&& \times \Big\{ \Big(- 15\,\Gamma\Big(2 -
 \frac{n}{2}\Big) + 30\, {\ell n}\frac{Q}{m^2_N}\Big) \gamma^{\mu} (1
 - \gamma^5) \otimes \gamma_{\mu} + \ldots\Big\}.
\end{eqnarray}
Using the Dirac equations for free fermions and the algebra of the
Dirac $\gamma$-matrices \cite{Itzykson1980}, we obtain an irreducible
Lorentz structure for the integral ${\cal F}(k_n, k_e, q)_{2b}$. We get
\begin{eqnarray}\label{eq:C2b.6}
 \hspace{-0.30in}&& {\cal F}(k_n, k_e, q)_{2b} = \int^1_0 dx_1
 \int^1_0 dx_2 \int^1_0 dx_3 \int^1_0 dx_4 \int^1_0 dx_5\, x_2 x_5
 \frac{\delta(1 - x_1 - x_2 - x_3 - x_4 - x_5)}{256 \pi^4 U^2}
 \nonumber\\ \hspace{-0.30in}&& \times \Big\{ \Big(15\, \Gamma\Big(2 -
 \frac{n}{2}\Big) + f^{(V)}_{2b}(x_1, \ldots, x_6) + g^{(V)}_{2b}(x_1,
 \ldots, x_6)\, \frac{k_n \cdot q}{m^2_N}\Big) \gamma^{\mu} (1 -
 \gamma^5) \otimes \gamma_{\mu} + \ldots\Big\},
\end{eqnarray}
where we have denoted
\begin{eqnarray}\label{eq:C2b.7}
 \hspace{-0.30in}&&f^{(V)}_{2b}(x_1, \ldots, x_5) = 30\,{\ell
   n}\frac{\bar{Q}}{m^2_N}\quad,\quad g^{(V)}_{2b}(x_1, \ldots, x_5) =
 60\, \frac{V}{U}\, \frac{m^2_N}{\bar{Q}}.
\end{eqnarray}
For the calculation of the integrals over the Feynman parameters we
use
\begin{eqnarray}\label{eq:C2b.8}
\hspace{-0.30in} \frac{\bar{Q}}{m^2_N} &=& \frac{1}{U}\Big((x_1 + x_2
+x_3) U \frac{m^2_{\pi}}{m^2_N} + (x_3 + x_5) x^2_4\Big),
\nonumber\\ \hspace{-0.30in} V &=& - (x_3 + x_5) x_1 x_4,
\end{eqnarray}
where $\bar{Q} = Q\big|_{q = 0}$.  A deviation of $Q$ from $\bar{Q}$
is taken into account in the linear approximation in the form of an
expansion in powers of $k_n\cdot q/m^2_N$ (or $q_0/m_N$). After the
integration over the Feynman parameters, the contribution of the
integral ${\cal F}(k_n, k_e, q)_{2b}$ to the matrix element $M(n \to p
e^- \bar{\nu}_e)^{(\pi^0)}_{\rm Fig.\,\ref{fig:fig2}b}$ is given by
\begin{eqnarray}\label{eq:C2b.9}
 \hspace{-0.30in}M(n \to p e^- \bar{\nu}_e)^{(\pi^0)}_{\rm
   Fig.\,\ref{fig:fig2}b}= - \frac{\alpha}{2\pi} \frac{g^2_{\pi
     N}}{16\pi^2}\,G_V \Big\{ \Big(A_{2b} \Gamma\Big(2 -
 \frac{n}{2}\Big) + F^{(V)}_{2b} + G^{(V)}_{2b}\, \frac{k_n \cdot
   q}{m^2_N}\Big) \Big[\bar{u}_e \gamma^{\mu} (1 - \gamma^5)
   v_{\bar{\nu}}\Big]\Big[\bar{u}_p \gamma_{\mu} u_n\Big]
 \Big\}.
\end{eqnarray}
The structure constants are equal to $A_{2b} = - 0.5122$, $F^{(V)}_{2b}
= - 3.0916$ and $G^{(V)}_{2b} = - 1.2025$. The Lorentz structure of
Eq.(\ref{eq:C2b.9}) is obtained at the neglect of the contributions
of order $O(E^2_0/m^2_N) \sim 10^{-6}$.

\subsection*{C2c. Analytical calculation of the Feynman diagram in
  Fig.\,\ref{fig:fig2}c}
\renewcommand{\theequation}{C2c-\arabic{equation}}
\setcounter{equation}{0}

In the Feynman gauge for the photon propagator the Feynman diagram in
Fig.\,\ref{fig:fig2}c is defined by the following momentum integrals
\begin{eqnarray}\label{eq:C2c.1}
 \hspace{-0.30in} && M(n \to p e^- \bar{\nu}_e)^{(\pi^0)}_{\rm
   Fig.\,\ref{fig:fig2}c} = - 2 e^2 g^2_{\pi N} G_V \int
 \frac{d^4k}{(2\pi)^4i}\int \frac{d^4p}{(2\pi)^4i}\,\Big[\bar{u}_e\,
   \gamma^{\mu}(1 - \gamma^5) v_{\bar{\nu}}\Big]
 \nonumber\\ \hspace{-0.30in}&& \times \,\Big[\bar{u}_p\,
   \gamma^5\,\frac{1}{m_N - \hat{k}_n - \hat{k} - i0}\,\gamma^{\beta}
   \frac{1}{m_N - \hat{k}_n - \hat{k} - \hat{p} - i0}
   \,\gamma_{\beta}\,\frac{1}{m_N - \hat{k}_n - \hat{k} -
     i0}\,\gamma^5 u_n \Big] \nonumber\\ \hspace{-0.30in}&& \times
 \,\frac{(2 k - q)_{\mu}}{[m^2_{\pi} - k^2 - i0][m^2_{\pi} - (k - q)^2
     - i0]}\,\frac{1}{p^2 + i0},
\end{eqnarray}
where we have made a change of variables $p \to - p$ and a replacement
$M^2_W D_{\mu\nu} (- q) \to - \eta_{\mu\nu}$, which is valid at the
neglect of the contributions of order $O(m_em_N/M^2_W) \sim 10^{-7}$.
The calculation of the matrix element Eq.(\ref{eq:C2c.1}) we reduce to
the calculation of the integral ${\cal F}(k_n, k_e, q)_{2c}$, which is
defined by
\begin{eqnarray}\label{eq:C2c.2}
\hspace{-0.3in}&&{\cal F}(k_n, k_e, q)_{2c} = \int
\frac{d^4k}{(2\pi)^4i}\int \frac{d^4p}{(2\pi)^4i}\,\frac{1}{m^2_N -
  (k_n + k + p)^2 - i0}\frac{1}{[m^2_N - (k_n + k)^2 - i0]^2}\,
\frac{1}{m^2_{\pi} - k^2 - i0}\nonumber\\ \hspace{-0.3in}&&\times
\,\frac{1}{m^2_{\pi} - (k - q)^2 - i0}\, \frac{1}{p^2 + i0}\,\Big(4
\hat{k} (1 - \gamma^5) \otimes \big(k^2 \hat{k} - k^2 \hat{p} + 2
(k\cdot p) \hat{k}\big) \Big).
\end{eqnarray}
Multiplying the integrand by unity $(p^2 + i0)/(p^2 + i0)$
\cite{Johns1974} and following the technique, which we have used in
subsections C2a and C2b, we represent the integral ${\cal F}(k_n, k_e,
q)_{2c}$ as follows
\begin{eqnarray}\label{eq:C2c.3}
 \hspace{-0.30in}&& {\cal F}(k_n, k_e, q)_{2c} = \int^1_0 dx_1
 \int^1_0 dx_2 \int^1_0 dx_3 \int^1_0 dx_4 \int^1_0 dx_5\, x_2 x_5\,
 \frac{\delta(1 - x_1 - x_2 - x_3 - x_4 - x_5)}{2^{2n} \pi^n U^{n/2}}
 \nonumber\\ \hspace{-0.30in}&& \times \Bigg\{\frac{n (n + 2)}{2}
 \, \frac{\displaystyle \Gamma\Big(2 - \frac{n}{2}\Big)\Gamma\Big(2 -
   \frac{n - 1}{2}\Big)}{\displaystyle 2^{n - 3}
   \Gamma\Big(\frac{1}{2}\Big)}\, \Big(\frac{Q}{m^2_N}\Big)^{-4 + n}\,
 \gamma^{\mu} (1 - \gamma^5) \otimes \gamma_{\mu} + \ldots \Bigg\},
\end{eqnarray}
where we have  denoted
\begin{eqnarray}\label{eq:C2c.4}
\hspace{-0.30in}U &=& (x_1 + x_2 + x_4 ) (x_3 + x_5) + x_3 x_5,
\nonumber\\
\hspace{-0.30in}b_1 &=& \frac{(x_3 + x_5) a_1 - x_3 a_2}{U} =
\frac{(x_3 + x_5)x_4 q - ((x_3 + x_5) x_2 + x_3 x_5) k_n }{U} =
\frac{X q + Y k_n}{U}, \nonumber\\
\hspace{-0.30in}b_2 &=& \frac{(x_1 + x_2 + x_3 + x_4) a_2 - x_3
  a_1}{U} = - \frac{x_3 x_4 q + (x_1 + x_4) x_3 k_n}{U} = \frac{\bar{X}
  q + \bar{Y} k_n}{U}, \nonumber\\
\hspace{-0.30in}Q &=& (x_1 + x_4 ) m^2_{\pi} - x_4 q^2 + \frac{(x_3 +
  x_5) a^2_1 - 2 x_3 a_1 \cdot a_2 + (x_1 + x_2 + x_3 + x_4) a^2_2}{U}
\end{eqnarray}
with $a_1 = x_4 q - (x_2 + x_3) k_n$ and $a_2 = - x_3 k_n$. Taking the
limit $n \to 4$ and keeping a divergent contribution proportional to
$\Gamma(2 - n/2)$ we transcribe the r.h.s. of Eq.(\ref{eq:C2c.3}) into
the form
\begin{eqnarray}\label{eq:C2c.5}
 \hspace{-0.30in}&& {\cal F}(k_n, k_e, q)_{2c} = \int^1_0 dx_1
 \int^1_0 dx_2 \int^1_0 dx_3 \int^1_0 dx_4 \int^1_0 dx_5 \, x_2 x_5\,
 \frac{\delta(1 - x_1 - x_2 - x_3 - x_4 - x_5)}{256 \pi^4 U^2}
 \nonumber\\ \hspace{-0.30in}&& \times \Big\{ \Big(6\,\Gamma\Big(2 -
 \frac{n}{2}\Big) - 12\, {\ell n}\frac{Q}{m^2_N}\Big) \gamma^{\mu} (1
 - \gamma^5) \otimes \gamma_{\mu} + \ldots \Big\}.
\end{eqnarray}
Using the Dirac equations for free fermions and the algebra of the
Dirac $\gamma$-matrices \cite{Itzykson1980}, we obtain the irreducible
Lorentz structure for the integral ${\cal F}(k_n, k_e, q)_{2c}$. We get
\begin{eqnarray}\label{eq:C2c.6}
 \hspace{-0.30in}&& {\cal F}(k_n, k_e, q)_{2c} = \int^1_0 dx_1
 \int^1_0 dx_2 \int^1_0 dx_3 \int^1_0 dx_4 \int^1_0 dx_5\, x_2 x_5
 \frac{\delta(1 - x_1 - x_2 - x_3 - x_4 - x_5)}{256 \pi^4 U^2}
 \nonumber\\ \hspace{-0.30in}&& \times \Big\{ \Big(6\, \Gamma\Big(2 -
 \frac{n}{2}\Big) + f^{(V)}_{2c}(x_1, \ldots, x_6) + g^{(V)}_{2c}(x_1,
 \ldots, x_6)\, \frac{k_n \cdot q}{m^2_N}\Big) \gamma^{\mu} (1 -
 \gamma^5) \otimes \gamma_{\mu} + \ldots\Big\},
\end{eqnarray}
where we have denoted
\begin{eqnarray}\label{eq:C2c.7}
 \hspace{-0.30in}f^{(V)}_{2c}(x_1, \ldots, x_5) = - 12\,{\ell
   n}\frac{\bar{Q}}{m^2_N} \quad,\quad &g^{(V)}_{2c}(x_1, \ldots, x_5)
 = - 24\, \frac{V}{U}\,\frac{m^2_N}{\bar{Q}}.
\end{eqnarray}
For the calculation of the integrals over the Feynman parameters we
use
\begin{eqnarray}\label{eq:C2c.8}
\hspace{-0.30in} \frac{\bar{Q}}{m^2_N} &=& \frac{1}{U}\Big((x_1 + x_4)
U \frac{m^2_{\pi}}{m^2_N} + x_5 (x_2 + x_3)^2 + (x_1 + x_2 + x_4)
x^2_3 + x_3 x^2_2\Big),\nonumber\\ \hspace{-0.30in} X &=& (x_3 +
x_5)x_4 \;,\; Y = - (x_3 + x_5) x_2 - x_3 x_5,
\nonumber\\ \hspace{-0.30in} \bar{X} &=& - x_3 x_4 \;,\; \bar{Y} = -
(x_1 + x_4) x_3, \nonumber\\ \hspace{-0.30in} V &=& - (x_3 + x_5) x_2
x_4 - x_3 x_4 x_5,
\end{eqnarray}
where $\bar{Q} = Q\big|_{q = 0}$.  A deviation of $Q$ from $\bar{Q}$
is taken into account in the linear approximation in the form of an
expansion in powers of $k_n\cdot q/m^2_N$ (or $q_0/m_N$). After the
integration over the Feynman parameters, the contribution of the
integral ${\cal F}(k_n, k_e, q)_{2c}$ to the matrix element $M(n \to p
e^- \bar{\nu}_e)^{(\pi^0)}_{\rm Fig.\,\ref{fig:fig2}c}$ is given by
\begin{eqnarray}\label{eq:C2c.9}
 \hspace{-0.30in} M(n \to p e^- \bar{\nu}_e)^{(\pi^0)}_{\rm
   Fig.\,\ref{fig:fig2}c}= \frac{\alpha}{2\pi} \frac{g^2_{\pi
     N}}{16\pi^2}\,G_V \Big\{ \Big(A_{2c} \Gamma\Big(2 -
 \frac{n}{2}\Big) + F^{(V)}_{2c} + G^{(V)}_{2c}\, \frac{k_n \cdot
   q}{m^2_N}\Big) \Big[\bar{u}_e \gamma^{\mu} (1 - \gamma^5)
   v_{\bar{\nu}}\Big]\Big[\bar{u}_p \gamma_{\mu} u_n\Big]\Big\}.
\end{eqnarray}
The structure constants are equal to $A_{2c} = 0.2049$, $F^{(V)}_{2c}
= 0.6001$ and $G^{(V)}_{2c} = 0.3566$.  The Lorentz structure of
Eq.(\ref{eq:C2c.9}) is obtained at the neglect of the contributions of
order $O(E^2_0/m^2_N) \sim 10^{-6}$.

\subsection*{C2d. Analytical calculation of the Feynman diagram in
  Fig.\,\ref{fig:fig2}d}
\renewcommand{\theequation}{C2d-\arabic{equation}}
\setcounter{equation}{0}

In the Feynman gauge for the photon propagator the Feynman diagram in
Fig.\,\ref{fig:fig2}d is defined by the following momentum integrals
\begin{eqnarray}\label{eq:C2d.1}
 \hspace{-0.30in} && M(n \to p e^- \bar{\nu}_e)^{(\pi^0)}_{\rm
   Fig.\,\ref{fig:fig2}d} = - e^2 g^2_{\pi N} G_V \int
 \frac{d^4k}{(2\pi)^4i}\int \frac{d^4p}{(2\pi)^4i}\,\Big[\bar{u}_e\,
   \gamma^{\mu}(1 - \gamma^5) v_{\bar{\nu}} \Big]
 \nonumber\\ \hspace{-0.30in}&& \times \,\Big[\bar{u}_p\,
   \gamma^5\,\frac{1}{m_N - \hat{k}_p - \hat{k} - i0}\,\gamma^{\beta}
   \,\frac{1}{m_N - \hat{k}_p - \hat{k} - \hat{p} -
     i0}\,\gamma_{\beta} \,\frac{1}{m_N - \hat{k}_p - \hat{k} -
     i0}\,\gamma_{\mu} (1 - \gamma^5)\,\frac{1}{m_N - \hat{k}_n -
     \hat{k} - i0} \,\gamma^5\, u_n\Big]
 \nonumber\\ \hspace{-0.30in}&& \times \, \frac{1}{m^2_{\pi} - k^2 -
   i0}\, \frac{1}{p^2 + i0},
\end{eqnarray}
where we have made a change of variables $p \to - p$ and a replacement
$M^2_W D_{\mu\nu} (- q) \to - \eta_{\mu\nu}$, which is valid at
the neglect of the contributions of order $O(m_em_N/M^2_W) \sim
10^{-7}$.  The calculation of the matrix element Eq.(\ref{eq:C2d.1})
we reduce to the calculation of the integral ${\cal F}(k_n, k_e,
q)_{2d}$, which is defined by
\begin{eqnarray}\label{eq:C2d.2}
\hspace{-0.3in}&&{\cal F}(k_n, k_e, q)_{2d} =\int
\frac{d^4k}{(2\pi)^4i}\int \frac{d^4p}{(2\pi)^4i}\,\frac{1}{[m^2_N -
    (k_p + k)^2 - i0]^2}\,\frac{1}{m^2_N - (k_p + k + p)^2 - i0}\,
\frac{1}{m^2_N - (k_n + k)^2 - i0} \nonumber\\ \hspace{-0.3in}&&
\times \, \frac{1}{m^2_{\pi} - k^2 - i0}\, \frac{1}{p^2 +
  i0}\,\Big(\gamma^{\mu} (1 - \gamma^5) \otimes \big( 2 k^2 \hat{k} -
2 k^2 \hat{p}+ 4 (p \cdot k)\, \hat{k}\big) \, \gamma_{\mu} \hat{k} (1
+ \gamma^5)\Big).
\end{eqnarray}
The procedure of the calculation of the integral ${\cal F}(k_n, k_e,
q)_{2d}$ is similar to that we have used in subsections C2a, C2b and
C2c. Thus, multiplying the integrand by unity $(p^2 + i0)/(p^2 + i0)$
\cite{Johns1974} and skipping standard intermediate calculations, we
arrive at the expression
\begin{eqnarray}\label{eq:C2d.3}
 \hspace{-0.30in}&& {\cal F}(k_n, k_e, q)_{2d} = \int^1_0 dx_1
 \int^1_0 dx_2 \int^1_0 dx_3 \int^1_0 dx_4 \int^1_0 dx_5\, x_2 x_5
 \frac{\delta(1 - x_1 - x_2 - x_3 - x_4 - x_5)}{2^{2n} \pi^n U^{n/2}}
 \Bigg\{ - \frac{n(n^2 - 4)}{4} \nonumber\\ \hspace{-0.30in}&& \times
 \, \frac{\displaystyle \Gamma\Big(2 - \frac{n}{2}\Big)\Gamma\Big(2 -
   \frac{n - 1}{2}\Big)}{\displaystyle 2^{n - 3}
   \Gamma\Big(\frac{1}{2}\Big)}\, \Big(\frac{Q}{m^2_N}\Big)^{-4 + n}\,
 \gamma^{\mu} (1 - \gamma^5) \otimes \gamma_{\mu} (1 + \gamma^5) +
 \ldots \Bigg\},
\end{eqnarray}
where we have  denoted
\begin{eqnarray}\label{eq:C2d.4}
\hspace{-0.21in}U &=& (x_1 + x_2 + x_4) (x_3 + x_5) + x_3 x_5,\nonumber\\
\hspace{-0.21in}b_1 &=& \frac{(x_3 + x_5) a_1 - x_3 a_2}{U} = -
\frac{((x_3 + x_5)x_2 + x_3 x_5) q + ((x_2 + x_3)(x_3 + x_5) + x_3
  x_5) k_n }{U} = \frac{X q + Y k_n}{U}, \nonumber\\
\hspace{-0.21in}b_2 &=& \frac{(x_1 + x_2 + x_3) a_2 - x_3 a_1}{U} = -
\frac{(x_1 + x_4) x_3 q + x_1 x_3 k_n}{U} = \frac{\bar{X} q + \bar{Y}
  k_n}{U}, \nonumber\\
\hspace{-0.21in}Q &=& x_1 m^2_{\pi} + \frac{(x_3 + x_5) a^2_1 - 2 x_3
  a_1 \cdot a_2 + (x_1 + x_2 + x_3 + x_4) a^2_2}{U}
\end{eqnarray}
with $a_1 = - (x_2 + x_3) q - (x_2 + x_3 + x_4) k_n$ and $a_2 = - x_3
q - x_3 k_n$. Taking the limit $n \to 4$ and keeping a divergent
contribution proportional $\Gamma(2 - n/2)$, we transcribe the
r.h.s. of Eq.(\ref{eq:C2d.3}) into the form
\begin{eqnarray}\label{eq:C2d.5}
 \hspace{-0.30in}&& {\cal F}(k_n, k_e, q)_{2d} = \int^1_0 dx_1
 \int^1_0 dx_2 \int^1_0 dx_3 \int^1_0 dx_4 \int^1_0 dx_5\, x_2 x_5
 \frac{\delta(1 - x_1 - x_2 - x_3 - x_4 - x_5)}{256 \pi^4 U^2}
 \nonumber\\ \hspace{-0.30in}&& \times \Big\{ \Big(- 6\,\Gamma\Big(2 -
 \frac{n}{2}\Big) + 12\, {\ell n}\frac{Q}{m^2_N}\Big) \gamma^{\mu} (1
 - \gamma^5) \otimes \gamma_{\mu} (1 + \gamma^5) + \ldots \Big\}.
\end{eqnarray}
The integral ${\cal F}(k_n, k_e, q)_{2d}$ has the following Lorentz
structure
\begin{eqnarray}\label{eq:C2d.6}
 \hspace{-0.30in}&& {\cal F}(k_n, k_e, q)_{2d} = \int^1_0 dx_1
 \int^1_0 dx_2 \int^1_0 dx_3 \int^1_0 dx_4 \int^1_0 dx_5\,x_2 x_5
 \frac{\delta(1 - x_1 - x_2 - x_3 - x_4 - x_5)}{256 \pi^4 U^2}
 \nonumber\\ \hspace{-0.30in}&& \times \Big\{ \Big(- 6\, \Gamma\Big(2
 - \frac{n}{2}\Big) + f^{(V)}_{2d}(x_1, \ldots, x_5) +
 g^{(V)}_{2d}(x_1, \ldots, x_5)\, \frac{k_n \cdot q}{m^2_N}\Big)
 \gamma^{\mu} (1 - \gamma^5) \otimes \gamma_{\mu} + \Big(- 6\,
 \Gamma\Big(2 - \frac{n}{2}\Big) \nonumber\\ \hspace{-0.30in}&&+
 f^{(A)}_{2d}(x_1, \ldots, x_5) + g^{(A)}_{2d}(x_1, \ldots, x_5)\,
 \frac{k_n \cdot q}{m^2_N}\Big) \gamma^{\mu} (1 - \gamma^5) \otimes
 \gamma_{\mu} \gamma^5 + \ldots \Big\},
\end{eqnarray}
where we have denoted
\begin{eqnarray}\label{eq:C2d.7}
 \hspace{-0.30in}f^{(V)}_{2d}(x_1, \ldots, x_5) &=& 12\,{\ell
   n}\frac{\bar{Q}}{m^2_N} \quad,\quad g^{(V)}_{2d}(x_1, \ldots, x_5) = 24\, \frac{V}{U}\, \frac{m^2_N}{\bar{Q}},\nonumber\\
 \hspace{-0.30in}f^{(A)}_{2d}(x_1, \ldots, x_5) &=& 12\,{\ell
   n}\frac{\bar{Q}}{m^2_N} \quad,\quad g^{(A)}_{2d}(x_1, \ldots, x_5) = 24\, \frac{V}{U}\, \frac{m^2_N}{\bar{Q}}.
\end{eqnarray}
For the calculation of the integrals over the Feynman parameters we
use
\begin{eqnarray}\label{eq:C2d.8}
\hspace{-0.30in} \frac{\bar{Q}}{m^2_N} &=& \frac{1}{U}\Big(x_1 U
\frac{m^2_{\pi}}{m^2_N} + x_5 (x_2 + x_3 + x_4 )^2 + x_3 (x_2 + x_4)^2
+ (x_1 + x_2 + x_4) x^2_3\Big), \nonumber\\ \hspace{-0.30in} V &=& (x_2 +
x_3)(x_2 + x_4) x_3 + (x_2 + x_3)(x_2 + x_3 + x_4) x_5 + x_1 x^2_3.
\end{eqnarray}
where $\bar{Q} = Q\big|_{q = 0}$.  A deviation of $Q$ from $\bar{Q}$
is taken into account in the linear approximation in the form of an
expansion in powers of $k_n\cdot q/m^2_N$ (or $q_0/m_N$). After the
integration over the Feynman parameters the contribution of the
integral ${\cal F}(k_n, k_e, q)_{2d}$ to the matrix element $M(n \to p
e^- \bar{\nu}_e)^{(\pi^0)}_{\rm Fig.\,\ref{fig:fig2}d}$ is given by
\begin{eqnarray}\label{eq:C2d.9}
 \hspace{-0.30in}&& M(n \to p e^- \bar{\nu}_e)^{(\pi^0)}_{\rm
   Fig.\,\ref{fig:fig2}d} = \frac{\alpha}{4\pi} \frac{g^2_{\pi
     N}}{16\pi^2}\,G_V \Big\{ \Big(A_{2d} \Gamma\Big(2 -
 \frac{n}{2}\Big) + F^{(V)}_{2d} + G^{(V)}_{2d}\, \frac{k_n \cdot
   q}{m^2_N}\Big) \Big[\bar{u}_e \gamma^{\mu} (1 - \gamma^5)
   v_{\bar{\nu}}\Big]\Big[\bar{u}_p \gamma_{\mu} u_n\Big]
 \nonumber\\ \hspace{-0.30in}&& + \Big(B_{2d} \Gamma\Big(2 -
 \frac{n}{2}\Big) + F^{(A)}_{2d} + G^{(A)}_{2d}\, \frac{k_n \cdot
   q}{m^2_N}\Big) \Big[\bar{u}_e \gamma^{\mu} (1 - \gamma^5)
   v_{\bar{\nu}}\Big]\Big[\bar{u}_p \gamma_{\mu} \gamma^5 u_n\Big]
 \Big\}.
\end{eqnarray}
The structure constants are equal to $A_{2d} = - 0.2049$,
$F^{(V)}_{2d} = - 0.3224$, $G^{(V)}_{2d} = 0.5828$, $B_{2d} = -
0.2049$, $F^{(A)}_{2d} = - 0.3224$ and $G^{(A)}_{2d} = 0.5828$. The
Lorentz structure of Eq.(\ref{eq:C2d.9}) is obtained at the neglect of
the contributions of order $O(E^2_0/m^2_N) \sim 10^{-6}$.

\subsection*{Fig.2. The contribution  of the  the Feynman diagram in
  Fig.\,\ref{fig:fig2} to the amplitude of the neutron
 beta decay }
\renewcommand{\theequation}{Fig.2-\arabic{equation}}
\setcounter{equation}{0}

Summing up the results of the analytical calculation of the Feynman
diagrams in Fig.\,\ref{fig:fig2}, given in Appendix C, we obtain the
contribution of the Feynman diagrams in Fig.\,\ref{fig:fig2} to the
amplitude of the neutron beta decay. We get
\begin{eqnarray}\label{eq:S2.1}
 \hspace{-0.30in}&& M(n \to p e^- \bar{\nu}_e)^{(\pi^0)}_{\rm
   Fig.\,\ref{fig:fig2}} = - \frac{\alpha}{2\pi} \frac{g^2_{\pi
     N}}{16\pi^2}\,G_V \Big\{ \Big(A_2\Gamma\Big(2 -
 \frac{n}{2}\Big) + F^{(V)}_2 + G^{(V)}_2\,
 \frac{k_n \cdot q}{m^2_N} \Big) \Big[\bar{u}_e
   \gamma^{\mu} (1 - \gamma^5) v_{\bar{\nu}}\Big] \nonumber\\
\hspace{-0.3in}&& \times \Big[\bar{u}_p \gamma_{\mu} u_n\Big] +
\Big(B_2\Gamma\Big(2 - \frac{n}{2}\Big) + F^{(A)}_2 + G^{(A)}_2\,
\frac{k_n \cdot q}{m^2_N }\Big) \Big[\bar{u}_e \gamma^{\mu} (1 -
  \gamma^5) v_{\bar{\nu}}\Big]\Big[\bar{u}_p \gamma_{\mu}\gamma^5
  u_n\Big] \Big\}.
\end{eqnarray}
The structure constants are equal to $A_2 = -1.1269$, $F^{(V)}_2 = -
6.6221$, $G^{(V)}_2 = - 4.6201$, $B_2 = - 0.2049$, $F^{(A)}_2 =
0.1612$ and $G^{(A)}_2 = - 0.2914$. The Lorentz structure of
Eq.(\ref{eq:S2.1}) is obtained at the neglect of the contributions of
order $O(E^2_0/m^2_N) \sim 10^{-6}$. We would like to emphasize that
all structure constants in the matrix element Eq.(\ref{eq:S2.1}) are
induced by the contributions of the first class currents
\cite{Weinberg1958}, which are $G$-even \cite{Lee1956a} (see also
\cite{Ivanov2018}).

We have to notice that the contributions of the Feynman diagrams in
Fig.\,\ref{fig:fig2} are important for cancellation of gauge
non-invariant contributions of the Feynman diagrams in
Fig.\,\ref{fig:fig1}. However, the contributions of the Feynman
diagrams in Fig.\,\ref{fig:fig2} to the renormalized radiative
corrections, induced by the hadronic structure of the neutron and
defined by the structure constants $G^{(V)}_2 = - 4.6201$ and
$G^{(A)}_2 = - 0.2914$, are at the level of a few parts of
$10^{-6}$. They can be, in principle, neglected. So, we may argue that
the multiplication of the integrands of the analytical expressions of
the Feynman diagrams in Fig.\,\ref{fig:fig2} by unity $(p^2+i0)/(p^2 +
i0)$ allows to simplify both the calculation and renormalization of
these diagrams, while maintaining the role of the self-energy diagrams
in calculating observable quantities \cite{Ivanov1973}.

\newpage

\section*{Appendix D: Analytical calculation of the Feynman diagrams 
in Fig.\,\ref{fig:fig3}}
\renewcommand{\theequation}{D-\arabic{equation}}
\setcounter{equation}{0}

The contributions of the Feynman diagrams in Fig.\,\ref{fig:fig3} such
as Fig.\,\ref{fig:fig3}b and Fig.\,\ref{fig:fig3}g,
Fig.\,\ref{fig:fig3}d and Fig.\,\ref{fig:fig3}i, and
Fig.\,\ref{fig:fig3}c and Fig.\,\ref{fig:fig3}h cancel each other
pairwise. As a result, we are left with the Feynman diagrams in
Fig.\,\ref{fig:fig3}a, Fig.\,\ref{fig:fig3}e and Fig.\,\ref{fig:fig3}f
only. The sum of these diagrams is gauge invariant (see subsection C
in Appendix B) as a consequence of a corresponding Ward-Takahashi-like
identity \cite{Itzykson1980}.

We calculate the Feynman diagrams in Fig.\,\ref{fig:fig3}a,
Fig.\,\ref{fig:fig3}e and Fig.\,\ref{fig:fig3}f in the Feynman gauge
by using the dimensional regularization \cite{Itzykson1980,
  Hooft1972}-\cite{Capper1973} (see also Supplemental Material of
Ref. \cite{Ivanov2019a}) and in the limit $m_{\sigma} \to \infty$ of
the infinite mass of the $\sigma$-meson \cite{Weinberg1967a} (see
also \cite{Ivanov2019a}). This allows to keep the contributions of the
Feynman diagrams with the $\pi$-meson exchanges only. 

The calculation of the Feynman diagrams in Fig.\,\ref{fig:fig3} as
well as of the Feynman diagrams in Fig.\,\ref{fig:fig2} demands
intermediate renormalization of masses and wave functions of virtual
$\pi$-mesons, caused by virtual photon-exchanges \cite{BS1984}. A
direct application of the standard procedure for the calculation of
two-loop Feynman diagrams \cite{Smirnov2004, Smirnov2006, Smirnov2012}
leads to the Feynman parametrization of the Feynman diagrams in
Fig.\,\ref{fig:fig3} with divergent integrals over the Feynman
parameters. In order to avoid a problem of intermediate
renormalization of masses and wave functions of virtual $\pi$-mesons,
caused by virtual photon-exchanges, we follow again Johns
\cite{Johns1974} and multiply integrands of the Feynman momentum
integrals by unity $(p^2 + i0)/(p^2 + i0)$, where $p$ is a 4-momentum
of a virtual photon. Then, after the direct application of the
standard procedure for the calculation of two-loop Feynman diagrams
\cite{Smirnov2004, Smirnov2006, Smirnov2012}, we arrive at the Feynman
parametrization of the Feynman diagrams in Fig.\,\ref{fig:fig3} with
contributions proportional to $(\Gamma(2 - n/2) - 2\,{\ell
  n}Q/m^2_N)/U^2$ and multiplied by convergent integrals over the
Feynman parameters. After the integration over the Feynman parameters,
the terms, induced by $(\Gamma(2 - n/2) - 2\,{\ell
  n}Q/m^2_N)/U^2\big|_{q = k_e=0}$, can be removed by renormalization
of the Fermi $G_V$ and axial $g_A$ coupling constants in agreement
with Sirlin's procedure \cite{Sirlin1967, Sirlin1978}. No other
renormalization is needed for the extraction of the contributions of
the Feynman diagrams in Fig.\,\ref{fig:fig3} to the observable
radiative corrections of order $O(\alpha E_e/m_N)$ to the amplitude of
the neutron beta decay.

In the Feynman gauge the contributions of the Feynman diagrams in
Fig.\,\ref{fig:fig3}a, Fig.\,\ref{fig:fig3}e and Fig.\,\ref{fig:fig3}f
are given by
\begin{eqnarray}\label{eq:D.1}
\hspace{-0.30in}&& M(n \to p e^- \bar{\nu}_e)^{(\pi^0)}_{\rm
  Fig.\,\ref{fig:fig3}a} = - 2 e^2 g^2_{\pi N} G_V \int
\frac{d^4k}{(2\pi)^4i}\int \frac{d^4p}{(2\pi)^4i}\,\Big[\bar{u}_e
  \gamma^{\mu}(1 - \gamma^5) v_{\bar{\nu}}\Big]\,\Big[\bar{u}_p\,
  \gamma^5\,\frac{1}{m_N - \hat{k}_n - \hat{k} - i0}\,\gamma^5
  u_n\Big]\nonumber\\ \hspace{-0.30in} &&\times \,\frac{(p + 2k - 2
  q)_{\mu}}{[m^2_{\pi} - k^2 - i0][m^2_{\pi} - (k - q)^2 -
    i0][m^2_{\pi} - (k + p - q)^2 - i0]}\,\frac{1}{p^2 + i0}
\end{eqnarray}
and 
\begin{eqnarray}\label{eq:D.2}
\hspace{-0.30in}&&M(n \to p e^- \bar{\nu}_e)^{(\pi^0)}_{\rm
  Fig.\,\ref{fig:fig3}e} = - 2 e^2 g^2_{\pi N} G_V \int
\frac{d^4k}{(2\pi)^4i}\int \frac{d^4p}{(2\pi)^4i}\,\Big[\bar{u}_e
  \,\gamma^{\mu}(1 - \gamma^5) v_{\bar{\nu}}\Big]\,\Big[\bar{u}_p\,
  \gamma^5\,\frac{1}{m_N - \hat{k}_n - \hat{k} - i0}\,\gamma^5\,
  u_n\Big] \nonumber\\ \hspace{-0.30in}&&\times \,\frac{(p + 2
  k)_{\mu}}{[m^2_{\pi} - k^2 - i0][m^2_{\pi} - (k + p)^2 -
    i0][m^2_{\pi} - (k - q)^2 - i0]}\, \frac{1}{p^2 + i0}
\end{eqnarray}
and
\begin{eqnarray}\label{eq:D.3}
\hspace{-0.30in}&&M(n \to p e^- \bar{\nu}_e)^{(\pi^0)}_{\rm
  Fig.\,\ref{fig:fig3}f} = - 2 e^2 g^2_{\pi N} 
G_V \nonumber\\ &&\times \int \frac{d^4k}{(2\pi)^4i}\int
\frac{d^4p}{(2\pi)^4i}\,\Big[\bar{u}_e \,\gamma^{\mu}(1 - \gamma^5)
  v_{\bar{\nu}}\Big]\,\Big[\bar{u}_p \, \gamma^5\,\frac{1}{m_N -
    \hat{k}_n - \hat{k} - \hat{p} - i0}\,\gamma_{\mu} \frac{1}{m_N -
    \hat{k}_n - \hat{k} - i0}\,\gamma^5 u_n\Big]
\nonumber\\ \hspace{-0.30in}&& \times \,\frac{1}{[m^2_{\pi} - k^2 -
    i0][m^2_{\pi} - (k + p - q)^2 - i0]}\,\frac{1}{p^2 + i0},
\end{eqnarray}
where we have made a change of variables $p \to - p$ in
Eq.(\ref{eq:D.2}) and the replacement $M^2_W
D^{(W)}_{\mu}{}^{\beta_2}(- q) \to - \eta_{\mu}{}^{\beta_2}$, which is
valid up to the relative contributions of order $O(m_e m_N/M^2_W) \sim
10^{-7}$. The calculation of the matrix elements Eqs.(\ref{eq:D.1}) -
(\ref{eq:D.3}) we perform by calculating the following integrals
\begin{eqnarray}\label{eq:D.4}
\hspace{-0.30in}&& {\cal F}(k_n, k_e, q)_{3a} = - \int
\frac{d^4k}{(2\pi)^4i}\int \frac{d^4p}{(2\pi)^4i}\, \frac{1}{m^2_N -
  (k_n + k)^2 - i0}\,\frac{1}{m^2_{\pi} - k^2 - i0}
\nonumber\\ \hspace{-0.30in}&& \times \, \frac{1}{m^2_{\pi} - (k -
  q)^2 - i0}\,\frac{1}{m^2_{\pi} - (p + k - q)^2 - i0}\,\frac{1}{p^2 +
  i0}\, \Big((\hat{p} + 2 \hat{k}) (1 - \gamma^5) \otimes
\hat{k}\Big)
\end{eqnarray}
and 
\begin{eqnarray*}
\hspace{-0.30in}&& {\cal F}(k_n, k_e, q)_{3e} = - \int
\frac{d^4k}{(2\pi)^4i}\int \frac{d^4p}{(2\pi)^4i}\, \frac{1}{m^2_N -
  (k_n + k)^2 - i0}\,\frac{1}{m^2_{\pi} - k^2 - i0}
\nonumber\\
\end{eqnarray*}
\begin{eqnarray}\label{eq:D.5}
\hspace{-0.30in}&& \times \, \frac{1}{m^2_{\pi} - (k -
  q)^2 - i0}\,\frac{1}{m^2_{\pi} - (p + k)^2 - i0}\,\frac{1}{p^2 +
  i0}\, \Big((\hat{p} + 2 \hat{k}) (1 - \gamma^5) \otimes \hat{k}\Big)
\end{eqnarray}
and
\begin{eqnarray}\label{eq:D.6}
\hspace{-0.30in}&& {\cal F}(k_n, k_e, q)_{3f} = - \int
\frac{d^4k}{(2\pi)^4i}\int \frac{d^4p}{(2\pi)^4i}\,\frac{1}{m^2_N - +
  (k_n + k + p)^2 - i0}\, \frac{1}{m^2_N - (k_n + k)^2 -
  i0}\,\frac{1}{m^2_{\pi} - k^2 - i0} \nonumber\\ \hspace{-0.30in}&&
\times \,\frac{1}{m^2_{\pi} - (k + p - q)^2 - i0}\,\frac{1}{p^2 +
  i0}\, \Big(\gamma^{\mu} (1 - \gamma^5) \otimes (\hat{k} + \hat{p})
\gamma_{\mu} \hat{k} \Big),
\end{eqnarray}
where we have kept only leading divergent contributions. For the
calculation of the integrals in Eqs.(\ref{eq:D.4}) - (\ref{eq:D.6}) we
follow Johns \cite{Johns1974} and multiply the integrands by unity
$(p^2 + i0)/(p^2 + i0)$. Then, we use the standard technique
\cite{Smirnov2004, Smirnov2006, Smirnov2012}.

\subsubsection*{\bf Calculation of the integral
  ${\cal F}(k_n, k_e, q)_{3a}$}
\renewcommand{\theequation}{D3a-\arabic{equation}}
\setcounter{equation}{0}

Merging the denominators in the integral ${\cal F}(k_n, k_e, q)_{3a}$
\cite{Feynman1950, Kinoshita1962, Kinoshita1974a, Kinoshita1974b,
  Smirnov2004, Smirnov2006, Smirnov2012}, we define the integral ${\cal
  F}(k_n, k_e, q)_{3a}$ as follows
\begin{eqnarray}\label{eq:D3a.1}
\hspace{-0.30in}&&{\cal F}(k_n, k_e, q)_{3a} = 5!\int
\frac{d^4k}{(2\pi)^4i}\int \frac{d^4p}{(2\pi)^4i}\int^1_0 dx_1
\int^1_0 dx_2 \int^1_0 dx_3 \int^1_0 dx_4 \int^1_0 dx_5 \, x_5\,
\delta(1 - x_1 - x_2 - x_3 - x_4 - x_5) \nonumber\\ \hspace{-0.30in}&&
\times \,\Big((\hat{p} + 2 \hat{k}) (1 - \gamma^5) \otimes
\hat{k}\, p^2\Big)\Big(x_1(m^2_{\pi} - k^2) + x_2 (m^2_{\pi} - (k -
q)^2) + x_3 (m^2_{\pi} - (p + k - q)^2) \nonumber\\ \hspace{-0.21in}&&
+ x_4(m^2_N - (k_n + k)^2) + x_5(- p^2) - i0\Big)^{-6}.
\end{eqnarray}
After diagonalization and integration over virtual momenta in the
$n$-dimensional space \cite{Smirnov2004, Smirnov2006, Smirnov2012}, we
get
\begin{eqnarray}\label{eq:D3a.2}
\hspace{-0.30in}&& {\cal F}(k_n, k_e, q)_{3a} = \int^1_0 dx_1 \int^1_0
dx_2 \int^1_0 dx_3 \int^1_0 dx_4 \int^1_0 dx_5 \,x_5\,
\frac{\delta(1 - x_1 - x_2 - x_3 - x_4 - x_5)}{2^{2n} \pi^n
  U^{n/2}} \nonumber\\ \hspace{-0.30in}&& \times \,\Bigg\{
\frac{n}{2}\,\frac{\displaystyle \Gamma\Big(2 -
  \frac{n}{2}\Big)\Gamma\Big(2 - \frac{n - 1}{2}\Big)}{\displaystyle
  2^{n - 3}\Gamma\Big(\frac{1}{2}\Big)}\,\Big(\frac{Q}{m^2_N}\Big)^{-4
  + n}\, \gamma^{\mu} (1 - \gamma^5) \otimes \gamma_{\mu} + \ldots\Bigg\},
\end{eqnarray} 
where we have used the algebra of the Dirac $\gamma$-matrices in the
$n$-dimensional space-time and denoted
\begin{eqnarray}\label{eq:D3a.3}
  \hspace{-0.3in}U &=& (x_1 + x_2 + x_4)( x_3 + x_5) + x_3
  x_5,\nonumber\\
  \hspace{-0.3in}b_1 &=& \frac{(x_3 + x_5 )a_1 - x_3 a_2}{U} =
  \frac{((x_3 + x_5) x_2 + x_3 x_5) q - (x_3 + x_5) x_4 k_n }{U}
  = \frac{X q + Y k_n}{U} ,\nonumber\\
   \hspace{-0.3in}b_2 &=& \frac{(x_1 + x_2 + x_3 + x_4)a_2 - x_3
     a_1}{U} = \frac{(x_1 + x_4) x_3 q + x_3 x_4) k_n}{U} =
   \frac{\bar{X} q + \bar{Y}k_n}{U},\nonumber\\
   \hspace{-0.3in}Q &=&(x_1 + x_2 + x_3) m^2_{\pi} - (x_2 + x_3) q^2 +
   \frac{(x_3 + x_5) a^2_1 - 2 x_3 a_1\cdot a_2 + (x_1 + x_2 + x_3 +
     x_4 ) a^2_2}{U}
\end{eqnarray}
with $a_1 = (x_2 + x_3) q - x_4 k_n$ and $a_2 = x_3 q$. Taking the
limit $n \to 4$ and keeping the divergent contribution proportional to
$\Gamma(2 - n/2)$ we arrive at the expression
\begin{eqnarray}\label{eq:D3a.4}
\hspace{-0.30in}&& {\cal F}(k_n, k_e, q)_{3a} = \int^1_0 dx_1 \int^1_0
dx_2 \int^1_0 dx_3 \int^1_0 dx_4 \int^1_0 dx_5 \,x_5\,
\frac{\delta(1 - x_1 - x_2 - x_3 - x_4 - x_5 )}{256 \pi^4 U^2}
\nonumber\\ \hspace{-0.30in}&& \times \,\Big\{ \Big(\Gamma\Big(2 -
\frac{n}{2}\Big) - 2\, {\ell n}\frac{Q}{m^2_N}\Big)\, \gamma^{\mu} (1
- \gamma^5) \otimes \gamma_{\mu} + \ldots\Big\}. 
\end{eqnarray} 
The integral ${\cal F}(k_n, k_e, q)_{3a}$ has the following Lorentz
structure
\begin{eqnarray*}
\hspace{-0.30in}&& {\cal F}(k_n, k_e, q)_{3a} = \int^1_0 dx_1 \int^1_0
dx_2 \int^1_0 dx_3 \int^1_0 dx_4 \int^1_0 dx_5 \,x_5\, \frac{\delta(1
  - x_1 - x_2 - x_3 - x_4 - x_5)}{256 \pi^4 U^2}
\nonumber\\
\end{eqnarray*}
\begin{eqnarray}\label{eq:D3a.5}
\hspace{-0.30in}&& \times \,\Big\{ \Big(\Gamma\Big(2 -
\frac{n}{2}\Big) + f^{(V)}_{3a}(x_1, \ldots, x_5) + g^{(V)}_{3a}(x_1,
\ldots, x_5) \, \frac{k_n\cdot q}{m^2_N}\Big)\, \gamma^{\mu} (1 -
\gamma^5) \otimes \gamma_{\mu} + \ldots \Big\},
\end{eqnarray} 
where we have denoted
\begin{eqnarray}\label{eq:D3a.6}
 \hspace{-0.30in}f^{(V)}_{3a}(x_1, \ldots, x_5) = - 2\, {\ell
   n}\frac{\bar{Q}}{m^2_N} \quad,\quad g^{(V)}_{3a}(x_1, \ldots, x_5)
 = - 4\,\frac{V}{U}\,\frac{m^2_N}{\bar{Q}}.
\end{eqnarray}
For the calculation of the integrals over the Feynman parameters we
use
\begin{eqnarray}\label{eq:D3a.7}
  \hspace{-0.21in}\frac{\bar{Q}}{m^2_N} &=& \frac{1}{U}\Big( (x_1 +
  x_2 + x_3)\, U\, \frac{m^2_{\pi}}{m^2_N} + (x_3 + x_5) x^2_4 \Big),
  \nonumber\\ \hspace{-0.21in} V &=& - (x_3 + x_5) x_2 x_4 - x_3 x_4
  x_5,
\end{eqnarray}
where $m_N = (m_n + m_p)/2 = 938.9188\,{\rm MeV}$ and $m_{\pi} =
139.5706\, {\rm MeV}$ \cite{PDG2020}. After the integration over the
Feynman parameters, the matrix element $M(n \to p
e^-\bar{\nu}_e)^{(\pi^0)}_{\rm Fig.\,\ref{fig:fig3}a}$ takes the form
\begin{eqnarray}\label{eq:D3a.8}
\hspace{-0.3in}&&M(n \to p e^- \bar{\nu}_e)^{(\pi^0)}_{\rm
  Fig.\,\ref{fig:fig3}a} = - \frac{\alpha}{2\pi}\frac{g^2_{\pi
    N}}{16\pi^2}\,G_V \Big\{\Big(A_{3a}\Gamma\Big(2 - \frac{n}{2}\Big)
+ F^{(V)}_{3a} + G^{(V)}_{3a}\,\frac{k_n\cdot q}{m^2_N}\Big)
\Big[\bar{u}_e \gamma^{\mu}(1 - \gamma^5) v_{\bar{\nu}}\Big]\,
\Big[\bar{u}_p\gamma_{\mu}u_n\Big]\Big\},\nonumber\\
\hspace{-0.30in}&&
\end{eqnarray}
where the structure constants are equal to $A_{3a} = 1/6$,
$F^{(V)}_{3a} = 0.9559$ and $G^{(V)}_{3a} = 0.6697$. The Lorentz
structure of Eq.(\ref{eq:D3a.8}) is obtained at the neglect the
contributions of order $O(E^2_0/m^2_N)\sim 10^{-6}$.

\subsubsection*{\bf Calculation of the integral
  ${\cal F}(k_n, k_e, q)_{3e}$}
\renewcommand{\theequation}{D3e-\arabic{equation}}
\setcounter{equation}{0}

Merging the denominators in the integral ${\cal F}^{(1)}(k_n, k_e,
q)_{3a}$ \cite{Feynman1950, Kinoshita1962, Kinoshita1974a, Kinoshita1974b,  Smirnov2004, Smirnov2006,
  Smirnov2012} we arrive at the expression
\begin{eqnarray}\label{eq:D3e.1}
\hspace{-0.30in}&& {\cal F}(k_n, k_e, q)_{3e} = 5!\int
\frac{d^4k}{(2\pi)^4i} \int \frac{d^4p}{(2\pi)^4i} \int^1_0 dx_1
\int^1_0 dx_2 \int^1_0 dx_3 \int^1_0 dx_4 \int^1_0 dx_5 \, x_5 \,
\delta(1 - x_1 - x_2 - x_3 - x_4 - x_5 )
\nonumber\\ \hspace{-0.30in}&& \times \,\Big((\hat{p} + 2 \hat{k}) (1
- \gamma^5) \otimes \hat{k}\,p^2\Big)\Big(x_1(m^2_{\pi} - k^2) + x_2
(m^2_{\pi} - (k - q)^2) + x_3(m^2_{\pi} - (p + k)^2) + x_4(m^2_N -
(k_n + k)^2) \nonumber\\ \hspace{-0.30in}&& + x_5(- p^2) -
i0\Big)^{-6}.
\end{eqnarray}
After diagonalization and integration over virtual momenta in the
$n$-dimensional space \cite{Smirnov2004, Smirnov2006, Smirnov2012}, we
get
\begin{eqnarray}\label{eq:D3e.2}
\hspace{-0.30in}&& {\cal F}(k_n, k_e, q)_{3e} = \int^1_0 dx_1 \int^1_0
dx_2 \int^1_0 dx_3 \int^1_0 dx_4 \int^1_0 dx_5 \, x_5\, \frac{\delta(1
  - x_1 - x_2 - x_3 - x_4 - x_5 )}{2^{2n} \pi^n U^{n/2}}
\nonumber\\ \hspace{-0.30in}&& \times \,\Bigg\{
\frac{n}{2}\,\frac{\displaystyle \Gamma\Big(2 -
  \frac{n}{2}\Big)\Gamma\Big(2 - \frac{n - 1}{2}\Big)}{\displaystyle
  2^{n - 3}\Gamma\Big(\frac{1}{2}\Big)}\,\Big(\frac{Q}{m^2_N}\Big)^{-4
  + n}\, \gamma^{\mu} (1 - \gamma^5) \otimes \gamma_{\mu} + \ldots
\Bigg\},
\end{eqnarray} 
where we have used the algebra of the Dirac $\gamma$-matrices in the
$n$-dimensional space-time and denoted
 \begin{eqnarray}\label{eq:D3e.3}
  \hspace{-0.3in}U &=& (x_1 + x_2 + x_4)( x_3 + x_5 ) + x_3
  x_5,\nonumber\\
  \hspace{-0.3in}b_1 &=& \frac{(x_3 + x_5)a_1 - x_3 a_2}{U} =
  \frac{(x_3 + x_5) x_2 q - (x_3 + x_5) x_4 k_n }{U} = \frac{X q + Y
    k_n}{U} ,\nonumber\\
   \hspace{-0.3in}b_2 &=& \frac{(x_1 + x_2 + x_3 + x_4) a_2 - x_3
     a_1}{U} = \frac{ - x_2 x_3 q + x_3 x_4 k_n}{U} = \frac{\bar{X} q
     + \bar{Y}k_n}{U}, \nonumber\\
   \hspace{-0.3in}Q &=&(x_1 + x_2 + x_3) m^2_{\pi} - x_2 q^2 +
   \frac{(x_3 + x_5 ) a^2_1 - 2 x_3 a_1\cdot a_2 + (x_1 + x_2 + x_3 +
     x_4 ) a^2_2}{U}
\end{eqnarray}
with $a_1 = x_2 q - x_4 k_n$ and $a_2 = 0$. Taking the limit $n \to 4$
and keeping the divergent contribution proportional to $\Gamma(2 -
n/2)$, we transcribe Eq.(\ref{eq:D3e.2}) into the form
\begin{eqnarray}\label{eq:D3e.4}
\hspace{-0.30in}&& {\cal F}(k_n, k_e, q)_{3e} = \int^1_0 dx_1 \int^1_0
dx_2 \int^1_0 dx_3 \int^1_0 dx_4 \int^1_0 dx_5 \int^1_0 dx_6\,
\frac{\delta(1 - x_1 - x_2 - x_3 - x_4 - x_5 - x_6)}{256 \pi^4 U^2}
\nonumber\\ \hspace{-0.30in}&& \times \,\Big\{ \Big(\Gamma\Big(2 -
\frac{n}{2}\Big) - 2\, {\ell n}\frac{Q}{m^2_N}\Big)\, \gamma^{\mu} (1
- \gamma^5) \otimes \gamma_{\mu} + \ldots \Big\}.
\end{eqnarray} 
Using the Dirac equations for free fermions we define the r.h.s. of
Eq.(\ref{eq:D3e.4}) in the following form
\begin{eqnarray}\label{eq:D3e.5}
\hspace{-0.30in}&& {\cal F}(k_n, k_e, q)_{3e} = \int^1_0 dx_1 \int^1_0
dx_2 \int^1_0 dx_3 \int^1_0 dx_4 \int^1_0 dx_5\, x_5 \, \frac{\delta(1
  - x_1 - x_2 - x_3 - x_4 - x_5 )}{256 \pi^4 U^2}
\nonumber\\ \hspace{-0.30in}&& \times \,\Big\{ \Big(\Gamma\Big(2 -
\frac{n}{2}\Big) + f^{(V)}_{3e}(x_1, \ldots, x_5) + g^{(V)}_{3e}(x_1,
\ldots, x_5) \, \frac{k_n\cdot q}{m^2_N}\Big)\, \gamma^{\mu} (1 -
\gamma^5) \otimes \gamma_{\mu} + \ldots \Big\},
\end{eqnarray} 
where we have denoted
\begin{eqnarray}\label{eq:D3e.6}
 \hspace{-0.30in}f^{(V)}_{3e}(x_1, \ldots, x_5) = - 2\, {\ell
   n}\frac{\bar{Q}}{m^2_N}\quad,\quad g^{(V)}_{3e}(x_1, \ldots, x_5) =
 - 4\,\frac{V}{U}\, \frac{m^2_N}{\bar{Q}}.
\end{eqnarray}
For the calculation of the integrals over the Feynman parameters we
use
\begin{eqnarray}\label{eq:D3e.7}
  \hspace{-0.21in} \frac{\bar{Q}}{m^2_N} &=& \frac{1}{U}\Big( (x_1 +
  x_2 + x_3)\, U\, \frac{m^2_{\pi}}{m^2_N} + (x_3 + x_5 ) x^2_4 \Big),
  \nonumber\\ \hspace{-0.21in} V &=& - (x_3 + x_5) x_2 x_4,
\end{eqnarray}
where $m_N = (m_n + m_p)/2 = 938.9188\,{\rm MeV}$ and $m_{\pi} =
139.5706\, {\rm MeV}$ \cite{PDG2020}. After the integration over the
Feynman parameters, the matrix element $M(n \to p
e^-\bar{\nu}_e)^{(\pi^0)}_{\rm Fig.\,\ref{fig:fig3}e}$ takes the form
\begin{eqnarray}\label{eq:D3e.8}
\hspace{-0.3in}&&M(n \to p e^- \bar{\nu}_e)^{(\pi^0)}_{\rm
  Fig.\,\ref{fig:fig3}e} = - \frac{\alpha}{2\pi}\frac{g^2_{\pi
    N}}{16\pi^2}\,G_V \Big\{\Big(A_{3e}\Gamma\Big(2 - \frac{n}{2}\Big)
+ F^{(V)}_{3e} + G^{(V)}_{3e}\,\frac{k_n\cdot q}{m^2_N}\Big)
\Big[\bar{u}_e \gamma^{\mu}(1 - \gamma^5) v_{\bar{\nu}}\Big]\,
\Big[\bar{u}_p\gamma_{\mu}u_n\Big] \Big\}, \nonumber\\
\hspace{-0.3in}&&
\end{eqnarray}
where the structure constants are equal to $A_{3e}= 1/6$,
$F^{(V)}_{3e} = 0.9559$, $G^{(V)}_{3e} = 0.4905$. The Lorentz
structure of Eq.(\ref{eq:D3e.8}) is obtained at the neglect the
contributions of order $O(E^2_0/m^2_N)\sim 10^{-6}$.

\subsubsection*{\bf Calculation of the integral
  ${\cal F}(k_n, k_e, q)_{3f}$}
\renewcommand{\theequation}{D3f-\arabic{equation}}
\setcounter{equation}{0}

Merging the denominators in the integral ${\cal F}^{(1)}(k_n, k_e,
q)_{3a}$ \cite{Feynman1950, Kinoshita1962, Kinoshita1974a, Kinoshita1974b,  Smirnov2004, Smirnov2006,
  Smirnov2012}, we obtain the integral ${\cal F}^{(1)}(k_n, k_e,
q)_{3a}$ in the following form
\begin{eqnarray}\label{eq:D3f.1}
\hspace{-0.30in}&& {\cal F}(k_n, k_e, q)_{3f} = 5!\int
\frac{d^4k}{(2\pi)^4i} \int \frac{d^4p}{(2\pi)^4i} \int^1_0 dx_1
\int^1_0 dx_2 \int^1_0 dx_3 \int^1_0 dx_4 \int^1_0 dx_5 \, x_5\,
\delta(1 - x_1 - x_2 - x_3 - x_4 - x_5 )
\nonumber\\ \hspace{-0.30in}&& \times \,\Big(\gamma^{\mu} (1 -
\gamma^5) \otimes (\hat{p} + \hat{k}) \gamma_{\mu} \hat{k}
\, p^2\Big)\Big(x_1(m^2_{\pi} - k^2) + x_2 (m^2_{\pi} - (k + p - q)^2) +
x_3 (m^2_N - (k_n + p + k)^2) \nonumber\\ \hspace{-0.30in}&& + x_4
(m^2_N - (k_n + k)^2) + x_5(- p^2) - i0\Big)^{-6}.
\end{eqnarray}
After diagonalization and integration over virtual momenta in the
$n$-dimensional space \cite{Smirnov2004, Smirnov2006, Smirnov2012}, we
arrive at the expression
\begin{eqnarray}\label{eq:D3f.2}
\hspace{-0.30in}&& {\cal F}(k_n, k_e, q)_{3f} = \int^1_0 dx_1 \int^1_0
dx_2 \int^1_0 dx_3 \int^1_0 dx_4 \int^1_0 dx_5 \, x_5 \,
\frac{\delta(1 - x_1 - x_2 - x_3 - x_4 - x_5 )}{2^{2n} \pi^n
  U^{n/2}}\Bigg\{ - \frac{n (n - 2)}{4} \nonumber\\ \hspace{-0.30in}&&
\times \,\frac{\displaystyle \Gamma\Big(2 -
  \frac{n}{2}\Big)\Gamma\Big(2 - \frac{n - 1}{2}\Big)}{\displaystyle
  2^{n - 3}\Gamma\Big(\frac{1}{2}\Big)}\,\Big(\frac{Q}{m^2_N}\Big)^{-4
  + n}\, \gamma^{\mu} (1 - \gamma^5) \otimes \gamma_{\mu} + \ldots
\Bigg\},
\end{eqnarray} 
where we have used the algebra of the Dirac $\gamma$-matrices in
$n$-dimensional space-time and denoted
\begin{eqnarray}\label{eq:D3f.3}
  \hspace{-0.3in}U &=& (x_1 + x_4 + x_5)( x_2 + x_3) + (x_1 + x_4)
  x_5,\nonumber\\
  \hspace{-0.3in}b_1 &=& \frac{(x_2 + x_3 + x_5)a_1 - (x_2 + x_3)
    a_2}{U} = \frac{ x_2 x_5 q - ((x_2 + x_3) x_4 + (x_3 + x_4) x_5)
    k_n }{U} = \frac{X q + Y k_n}{U} ,\nonumber\\
   \hspace{-0.3in}b_2 &=& \frac{(x_1 + x_2 + x_3 + x_4 )a_2 - (x_2 +
     x_3) a_1}{U} = \frac{(x_1 + x_4) x_2 q + (x_2 x_4 - x_1 x_3)
     k_n}{U} = \frac{\bar{X} q + \bar{Y}k_n}{U},\nonumber\\
   \hspace{-0.3in}Q &=&(x_1 + x_2) m^2_{\pi} - x_2 q^2 + \frac{(x_2 +
     x_3 + x_5 ) a^2_1 - 2 (x_2 + x_3) a_1\cdot a_2 + (x_1 + x_2 + x_3
     + x_4) a^2_2}{U}
\end{eqnarray}
with $a_1 = x_2 q - (x_3 + x_4) k_n$ and $a_2 = x_2 q - x_3
k_n$. Taking the limit $n \to 4$ and keeping the divergent
contribution proportional to $\Gamma(2 - n/2)$, we get
\begin{eqnarray}\label{eq:D3f.4}
\hspace{-0.30in}&& {\cal F}(k_n, k_e, q)_{3f} = \int^1_0 dx_1 \int^1_0
dx_2 \int^1_0 dx_3 \int^1_0 dx_4 \int^1_0 dx_5 \,x_5 \, \frac{\delta(1
  - x_1 - x_2 - x_3 - x_4 - x_5)}{256 \pi^4 U^2}
\nonumber\\ \hspace{-0.30in}&& \times \,\Big\{ \Big( - \Gamma\Big(2 -
\frac{n}{2}\Big) + 2\, {\ell n}\frac{Q}{m^2_N}\Big)\, \gamma^{\mu} (1
- \gamma^5) \otimes \gamma_{\mu} + \ldots \Big\}.
\end{eqnarray} 
The integral ${\cal F}(k_n, k_e, q)_{3f}$ has the following Lorentz
structure
\begin{eqnarray}\label{eq:D3f.5}
\hspace{-0.30in}&& {\cal F}(k_n, k_e, q)_{3f} = \int^1_0 dx_1 \int^1_0
dx_2 \int^1_0 dx_3 \int^1_0 dx_4 \int^1_0 dx_5 \, x_5 \,
\frac{\delta(1 - x_1 - x_2 - x_3 - x_4 - x_5)}{256 \pi^4 U^2}
\nonumber\\ \hspace{-0.30in}&& \times \,\Big\{ \Big(- \Gamma\Big(2 -
\frac{n}{2}\Big) + f^{(V)}_{3f}(x_1, \ldots, x_5) + g^{(V)}_{3f}(x_1,
\ldots, x_5) \, \frac{k_n\cdot q}{m^2_N}\Big)\, \gamma^{\mu} (1 -
\gamma^5) \otimes \gamma_{\mu} + \ldots\Big\},
\end{eqnarray} 
where we have denoted
\begin{eqnarray}\label{eq:D3f.6}
 \hspace{-0.30in}f^{(V)}_{3f}(x_1, \ldots, x_5) = 2\, {\ell
   n}\frac{\bar{Q}}{m^2_N} \quad,\quad g^{(V)}_{3f}(x_1, \ldots, x_5) = 4\,
 \frac{V}{U}\, \frac{m^2_N}{\bar{Q}}.
\end{eqnarray}
For the calculation of the integrals over the Feynman parameters we
use
\begin{eqnarray}\label{eq:D3f.7}
  \hspace{-0.21in}\frac{\bar{Q}}{m^2_N} &=& \frac{1}{U}\Big( (x_1 +
  x_2)\, U\, \frac{m^2_{\pi}}{m^2_N} + x_5(x_3 + x_4)^2 + (x_2 + x_3)
  x^2_4 + (x_1 + x_4) x^2_3\Big), \nonumber\\ \hspace{-0.21in} V &=& -
  (x_2 + x_5)(x_3 + x_4) x_2 - (x_1 + x_2 + x_4) x_2 x_3,
\end{eqnarray}
where $m_N = (m_n + m_p)/2 = 938.9188\,{\rm MeV}$ and $m_{\pi} =
139.5706\, {\rm MeV}$ \cite{PDG2020}. After the integration over the
Feynman parameters, the matrix element $M(n \to p
e^-\bar{\nu}_e)^{(\pi^0)}_{\rm Fig.\,\ref{fig:fig3}f}$ takes the form
\begin{eqnarray}\label{eq:D3f.8}
\hspace{-0.3in}&&M(n \to p e^- \bar{\nu}_e)^{(\pi^0)}_{\rm
  Fig.\,\ref{fig:fig3}f} = - \frac{\alpha}{2\pi}\frac{g^2_{\pi
    N}}{16\pi^2}\,G_V \Big\{\Big(A_{3f}\Gamma\Big(2 - \frac{n}{2}\Big)
+ F^{(V)}_{3f} + G^{(V)}_{3f}\,\frac{k_n\cdot q}{m^2_N}\Big)
\Big[\bar{u}_e \gamma^{\mu}(1 - \gamma^5) v_{\bar{\nu}}\Big]\,
\Big[\bar{u}_p\gamma_{\mu}u_n\Big] \Big\},\nonumber\\
\hspace{-0.3in}&&
\end{eqnarray}
where the structure constants are equal to $A_{3f} = - 0.1146$,
$F^{(V)}_{3f} = 0.4459$ and $G^{(V)}_{3f} = 0.9876$. The Lorentz
structure of Eq.(\ref{eq:D3f.8}) is obtained at the neglect the
contributions of order $O(E^2_0/m^2_N)\sim 10^{-6}$.

\subsection*{Fig.3. The contribution  of the  the Feynman diagram in
  Fig.\,\ref{fig:fig3} to the amplitude of the neutron
 beta decay }
\renewcommand{\theequation}{Fig.3-\arabic{equation}}
\setcounter{equation}{0}

Summing up the results of the analytical calculation of the Feynman
diagrams in Fig.\,\ref{fig:fig3}, given in Appendix D, we obtain the
contribution of the Feynman diagrams in Fig.\,\ref{fig:fig3} to the
amplitude of the neutron beta decay. We get
\begin{eqnarray}\label{eq:S3.1}
 \hspace{-0.30in}M(n \to p e^- \bar{\nu}_e)^{(\pi^0)}_{\rm
   Fig.\,\ref{fig:fig3}}= - \frac{\alpha}{2\pi} \frac{g^2_{\pi
     N}}{16\pi^2}\,G_V \Big\{ \Big(A_3\Gamma\Big(2 - \frac{n}{2}\Big)
 + F^{(V)}_3 + G^{(V)}_3\, \frac{k_n \cdot q}{m^2_N}\Big)
 \Big[\bar{u}_e \gamma^{\mu} (1 - \gamma^5)
   v_{\bar{\nu}}\Big]\Big[\bar{u}_p \gamma_{\mu} u_n\Big]\Big\}.
\end{eqnarray}
The structure constants are equal to $A_3 = 0.2187$, $F^{(V)}_3 =
2.3577$ and $G^{(V)}_3 = 2.1478$. The Lorentz structure of
Eq.(\ref{eq:S3.1}) is obtained at the neglect of the contributions of
order $O(E^2_0/m^2_N) \sim 10^{-6}$. The calculation of the Feynman
diagrams in Fig.\,\ref{fig:fig3} we have carried out in the LLA. We
would like to emphasize that all structure constants in the matrix
element Eq.(\ref{eq:S3.1}) are induced by the contributions of the
first class currents \cite{Weinberg1958}, which are $G$-even
\cite{Lee1956a} (see also \cite{Ivanov2018}).

It should be noticed that after renormalization of the Fermi $G_V$ and
axial $g_A$ coupling constants the contribution of the Feynman
diagrams in Fig.\,\ref{fig:fig3} to the radiative corrections, induced
by the hadronic structure of the neutron and defined by the structure
constant $G^{(V)}_3 = 2.1478$, is at the level of a few parts of
$10^{-6}$. It can be, in principle, neglected. So, we may argue that
the multiplication of the integrands of the analytical expressions of
the Feynman diagrams in Fig.\,\ref{fig:fig3} by unity $(p^2+i0)/(p^2 +
i0)$ allows to simplify both the calculation and renormalization of
these diagrams.

\newpage

\section*{Appendix E: Analytical calculation of the Feynman diagrams 
in Fig.\,\ref{fig:fig4}}
\renewcommand{\theequation}{E-\arabic{equation}}
\setcounter{equation}{0}

As we have shown in Appendix B that the sum of the Feynman diagrams in
Fig.\,\ref{fig:fig4} is gauge invariant. The calculation of these
Feynman diagrams we carry out in the Feynman gauge by using the
dimensional regularization \cite{Hooft1972}-\cite{Capper1973} (see
also the Supplemental Material of Ref. \cite{Ivanov2019a}), and in the
limit $m_{\sigma} \to \infty$ of the infinite mass of the scalar
isoscalar $\sigma$-meson \cite{Weinberg1967a} (see also
\cite{Ivanov2019a}). This allows to keep the contributions of the
Feynman diagrams with the $\pi$-meson exchanges only.

\subsection*{E4a. Analytical calculation of the Feynman diagram in
  Fig.\,\ref{fig:fig4}a}
\renewcommand{\theequation}{E4a-\arabic{equation}}
\setcounter{equation}{0}

The analytical expression of the Feynman diagram in
Fig.\,\ref{fig:fig4}a, taken in the Feynman gauge, is given by (see
Eq.(\ref{eq:A.1}))
\begin{eqnarray}\label{eq:E4a.1}
\hspace{-0.30in}&& M(n \to p e^- \bar{\nu}_e)_{\rm
  Fig.\,\ref{fig:fig4}a} = + 2 e^2 g^2_{\pi N} G_V \int
\frac{d^4k}{(2\pi)^4i}\int \frac{d^4p}{(2\pi)^4i}\,\Big[\bar{u}_e\,
  \gamma^{\mu}(1 - \gamma^5) v_{\bar{\nu}}\Big]
\nonumber\\ &&\times\,\Big[\bar{u}_p\, \gamma^{\beta}\,\frac{1}{m_N -
    \hat{k}_p - \hat{p} - i0}\,\gamma_{\mu}(1 - \gamma^5) \,
  \frac{1}{m_N - \hat{k}_n - \hat{p} - i0}\,\gamma^5\, \frac{1}{m_N -
    \hat{k}_n - \hat{k} - i0}\,\gamma^5 u_n\Big]
\nonumber\\ \hspace{-0.30in} && \times \,\frac{(2 k -
  p)_{\beta}}{[m^2_{\pi} - k^2 - i0][m^2_{\pi} - (k - p)^2 - i0]}\,
\frac{1}{p^2 + i0},
\end{eqnarray}
where as usual we have made the replacement $M^2_W D^{(W)}_{\mu\nu}(-
q) \to - \eta_{\mu\nu}$, which is valid up to the contributions of
order $O(m_e m_N/M^2_W) \sim 10^{-7}$. Then, we reduce the calculation
of the contribution of the Feynman diagram in Fig.\,\ref{fig:fig4}a to
the calculation of the integral
\begin{eqnarray}\label{eq:E4a.2}
\hspace{-0.21in}&& {\cal F}(k_n, k_e, q)_{4a} = - \int
\frac{d^4k}{(2\pi)^4i}\int \frac{d^4p}{(2\pi)^4i}\,\frac{1}{m^2_N -
  (k_p + p)^2 - i0} \frac{1}{m^2_N - (k_n + p)^2 - i0} \frac{1}{m^2_N
  - (k_n + k)^2 - i0} \frac{1}{m^2_{\pi} - k^2 - i0}
\nonumber\\ \hspace{-0.21in} && \times \,\frac{1}{m^2_{\pi} - (k -
  p)^2 - i0} \frac{1}{p^2 + i0} \Big(\gamma^{\mu} (1 - \gamma^5)
\otimes ( 2 \hat{k} \hat{p} - p^2 ) \gamma_{\mu}\, \hat{p} \hat{k} (1
- \gamma^5) \Big),
\end{eqnarray}
where we have kept only leading divergent contributions that
corresponds to the LLA. Having merged the denominators by using the
Feynman parametrization \cite{Feynman1950, Kinoshita1962,
  Kinoshita1974a, Kinoshita1974b} and after the diagonalization and
integration over the virtual momenta in the $n$-dimensional momentum
space \cite{Smirnov2004, Smirnov2006, Smirnov2012}, we obtain the
expression
\begin{eqnarray}\label{eq:E4a.3}
\hspace{-0.30in}&& {\cal F}(k_n, k_e, q)_{4a} = \int^1_0 dx_1 \int^1_0
dx_2 \int^1_0 dx_3 \int^1_0 dx_4 \int^1_0 dx_5 \int^1_0 dx_6\,
\frac{\delta(1 - x_1 - x_2 - x_3 - x_4 - x_5 - x_6)}{2^{2n} \pi^n
  U^{n/2}} \nonumber\\ \hspace{-0.30in}&& \times \,\Bigg\{\frac{(n -
  2)^2}{2}\,\frac{\displaystyle \Gamma\Big(2 -
  \frac{n}{2}\Big)\Gamma\Big(2 - \frac{n - 1}{2}\Big)}{\displaystyle
  2^{n - 3}\Gamma\Big(\frac{1}{2}\Big)}\,\Big(\frac{Q}{m^2_N}\Big)^{-4
  + n}\, \gamma^{\mu} (1 - \gamma^5) \otimes \gamma_{\mu}(1 -
\gamma^5)  + \ldots \Bigg\},
\end{eqnarray} 
where we have used the algebra of the Dirac $\gamma$-matrices in the
$n$-dimensional space-time and denoted
\begin{eqnarray}\label{eq:E4a.4}
  \hspace{-0.3in}U &=& (x_1 + x_2 + x_4)( x_3 + x_5 + x_6) +
  (x_1 + x_4) x_2,\nonumber\\
  \hspace{-0.3in}b_1 &=& \frac{(x_2 + x_3 + x_5 + x_6)a_1 + x_2
    a_2}{U} = \frac{ - x_2 x_3 q - ((x_3 + x_5)(x_2 + x_4) + (x_2 +
    x_6) x_4) k_n }{U} = \nonumber\\
\hspace{-0.3in}&=&\frac{X q + Y k_n}{U}
  ,\nonumber\\
   \hspace{-0.3in}b_2 &=& \frac{(x_1 + x_2 + x_4)a_2 + x_2 a_1}{U} =
   \frac{-(x_1 + x_2 + x_4) x_3 q - ((x_1 + x_2)(x_3 + x_5) + (x_2 +
     x_3 + x_5) x_4) k_n}{U} = \nonumber\\
\hspace{-0.3in}&=&\frac{\bar{X} q + \bar{Y}k_n}{U},\nonumber\\
   \hspace{-0.3in}Q &=&(x_1 + x_2) m^2_{\pi} +
   \frac{(x_2 + x_3 + x_5 + x_6 ) a^2_1 + 2 x_2 a_1\cdot a_2 + (x_1
     + x_2 + x_4) a^2_2}{U}
\end{eqnarray}
with $a_1 = - x_4 k_n$ and $a_2 = - x_3 q - (x_3 + x_5) k_n$. Taking
the limit $ n \to 4$ and keeping the divergent contributions
proportional to $G(2 - n/2)$, we arrive at the expression
\begin{eqnarray}\label{eq:E4a.5}
\hspace{-0.30in}&& {\cal F}(k_n, k_e, q)_{4a} = \int^1_0 dx_1 \int^1_0
dx_2 \int^1_0 dx_3 \int^1_0 dx_4 \int^1_0 dx_5 \int^1_0 dx_6\,
\frac{\delta(1 - x_1 - x_2 - x_3 - x_4 - x_5 - x_6)}{256 \pi^4 U^2}
\nonumber\\ \hspace{-0.30in}&& \times \,\Big\{\Big(\Gamma\Big(2 -
\frac{n}{2}\Big) - 2 {\ell n}\frac{Q}{m^2_N}\Big)\, \gamma^{\mu} (1 -
\gamma^5) \otimes \gamma_{\mu}(1 - \gamma^5) + \ldots \Big\}.
\end{eqnarray}
The integral ${\cal F}(k_n, k_e, q)_{4a}$ has the following Lorentz
structure
\begin{eqnarray}\label{eq:E4a.6}
\hspace{-0.3in}&&{\cal F}(k_n, k_e, q)_{4a} = \int^1_0dx_1
\int^1_0dx_2 \int^1_0dx_3 \int^1_0dx_4 \int^1_0dx_5 \int^1_0 dx_6 \,
\frac{\delta(1 - x_1 - x_2 - x_3 - x_4 - x_5 - x_6)}{256 \pi^4
  U^2}\nonumber\\
\hspace{-0.3in}&&\times \Big\{\Big( \Gamma\Big(2 - \frac{n}{2}\Big) +
f^{(V)}_{4a}(x_1,\ldots,x_6) +
g^{(V)}_{4a}(x_1,\ldots,x_6)\,\frac{k_n\cdot q}{m^2_N}\Big)
\,\gamma^{\mu}(1 - \gamma^5) \otimes \gamma_{\mu} + \Big(-
\Gamma\Big(2 - \frac{n}{2}\Big) \nonumber\\
\hspace{-0.3in}&& + f^{(A)}_{4a}(x_1,\ldots,x_5) +
g^{(A)}_{4a}(x_1,\ldots,x_6)\,\frac{k_n\cdot q}{m^2_N} \Big)
\,\gamma^{\mu}(1 - \gamma^5) \otimes \gamma_{\mu} \gamma^5 + \ldots
\Big\},
\end{eqnarray}
where we have denoted
\begin{eqnarray}\label{eq:E4a.7}
\hspace{-0.3in}f^{(V)}_{4a}(x_1, \ldots,x_6) &=& - 2\,{\ell
  n}\frac{\bar{Q}}{m^2_N} \quad,\quad g^{(V)}_{4a}(x_1, \ldots,x_6) =
- 4\, \frac{V}{U}\, \frac{m^2_N}{\bar{Q}},\nonumber\\
\hspace{-0.3in}f^{(A)}_{4a}(x_1, \ldots,x_6) &=& + 2\,{\ell
  n}\frac{\bar{Q}}{m^2_N} \quad,\quad g^{(V)}_{4a}(x_1, \ldots,x_6) =
+ 4\, \frac{V}{U}\, \frac{m^2_N}{\bar{Q}}.
\end{eqnarray}
For the calculation of the integrals over the Feynman parameters we
use
\begin{eqnarray}\label{eq:E4a.8}
  \hspace{-0.21in} \frac{\bar{Q}}{m^2_N} &=& \frac{1}{U}\Big( (x_1 +
  x_2)\, U\, \frac{m^2_{\pi}}{m^2_N} + (x_2 + x_3 + x_5 + x_6) x^2_4 +
  2 x_2 x_4 (x_3 + x_5) + (x_1 + x_2 + x_4) (x_3 + x_5)^2\Big),
 \nonumber\\
\hspace{-0.3in} V &=& (x_1 + x_2 + x_4) x^2_3 + (x_1 + x_2) x_3 x_5 +
(x_2 + x_5) x_3 x_4,
\end{eqnarray}
where $m_N = (m_n + m_p)/2 = 938.9188\,{\rm MeV}$ and $m_{\pi} =
139.5706\, {\rm MeV}$ \cite{PDG2020}. After the integration over the
Feynman parameters, the matrix element $M(n \to p
e^-\bar{\nu}_e)^{(\pi^0)}_{\rm Fig.\,\ref{fig:fig4}a}$ takes the form
\begin{eqnarray}\label{eq:E4a.9}
 \hspace{-0.30in}&& M(n \to p e^- \bar{\nu}_e)^{(\pi^0)}_{\rm
   Fig.\,\ref{fig:fig4}a} = \frac{\alpha}{2\pi} \frac{g^2_{\pi
     N}}{16\pi^2}\,G_V \Big\{ \Big(A_{4a}\Gamma\Big(2 - \frac{n}{2}\Big)
 + F^{(V)}_{4a} + G^{(V)}_{4a}\, \frac{k_n \cdot q}{m^2_N}\Big)
 \Big[\bar{u}_e \gamma^{\mu} (1 - \gamma^5)
   v_{\bar{\nu}}\Big]\Big[\bar{u}_p \gamma_{\mu} u_n\Big] \nonumber\\
\hspace{-0.3in}&& + \Big(B_{4a}\Gamma\Big(2 - \frac{n}{2}\Big) +
F^{(A)}_{4a} + G^{(A)}_{4a}\, \frac{k_n \cdot q}{m^2_N}\Big)
\Big[\bar{u}_e \gamma^{\mu} (1 - \gamma^5)
  v_{\bar{\nu}}\Big]\Big[\bar{u}_p \gamma_{\mu}\gamma^5 u_n\Big]\Big\}. 
\end{eqnarray}
The structure constants are equal to $A_{4a} = 1/6$, $F^{(V)}_{4a} =
0.1467$, $G^{(V)}_{4a} = - 0.1843$, $B_{4a} = - 1/6$, $F^{(A)}_{4a} =
- 0.1467$ and $G^{(A)}_{4a} = 0.1843$. The Lorentz structure of
Eq.(\ref{eq:E4a.9}) is obtained at the neglect of the contributions of
order $O(E^2_0/m^2_N) \sim 10^{-6}$. We would like to emphasize that
all structure constants in the matrix element Eq.(\ref{eq:E4a.9}) are
induced by the contributions of the first class currents
\cite{Weinberg1958}, which are $G$-even \cite{Lee1956a} (see also
\cite{Ivanov2018}).

\subsection*{E4b. Analytical calculation of the Feynman diagram in
  Fig.\,\ref{fig:fig4}b}
\renewcommand{\theequation}{E4b-\arabic{equation}}
\setcounter{equation}{0}

The analytical expression of the Feynman diagram in
Fig.\,\ref{fig:fig4}b, taken in the Feynman gauge, is given by (see
Eq.(\ref{eq:A.1}))
\begin{eqnarray}\label{eq:E4b.1}
\hspace{-0.30in}&&M(n \to p e^- \bar{\nu}_e)_{\rm
  Fig.\,\ref{fig:fig4}b} = + 2 e^2 g^2_{\pi N} G_V
\nonumber\\ \hspace{-0.30in} &&\times \int \frac{d^4k}{(2\pi)^4i}\int
\frac{d^4p}{(2\pi)^4i}\,\Big[\bar{u}_e\, \gamma^{\mu}(1 - \gamma^5)
  v_{\bar{\nu}}\Big]\,\Big[\bar{u}_p\, \,\gamma^{\alpha}(1 -
  \gamma^5)\, \frac{1}{m_N - \hat{k}_n - \hat{p} - i0}\, \gamma^5
  \frac{1}{m_N - \hat{k}_n - \hat{k} - i0}\,\gamma^5 u_n\Big]
\nonumber\\
\hspace{-0.3in}&& \times \,\frac{(2 k - p)^{\beta}}{[m^2_{\pi} - k^2 -
    i0][m^2_{\pi} - (k - p)^2 - i0]} \big[(p - q)_{\mu}\eta_{\beta
    \alpha} - (p - 2 q)_{\beta} \eta_{\alpha\mu} - q_{\alpha}
  \eta_{\mu\beta}\big]\,\frac{1}{p^2 + i0} \,\frac{1}{M^2_W - (p -
  q)^2 - i0}, \nonumber\\
\hspace{-0.3in}&&
\end{eqnarray}
where we have taken into account our experience, obtained during the
calculation of the Feynman diagrams in Fig.\,\ref{fig:fig1}a,
Fig.\,\ref{fig:fig1}d and so on. The calculation of the matrix element
Eq.(\ref{eq:E4b.1}) we reduce to the calculation of the following
integral
\begin{eqnarray}\label{eq:E4b.2}
\hspace{-0.3in} &&{\cal F}(k_n, k_e, q)_{4b} = - \int
\frac{d^4k}{(2\pi)^4i}\int \frac{d^4p}{(2\pi)^4i}\,\frac{1}{m^2_N -
  (k_n + p)^2 - i0}\, \frac{1}{m^2_N - (k_n + k)^2 -
  i0}\,\frac{1}{m^2_{\pi} - k^2 - i0}\nonumber\\
\hspace{-0.3in}&& \times \,\frac{1}{m^2_{\pi} - (k - p)^2 -
  i0}\,\frac{1}{p^2 + i0}\, \frac{1}{M^2_W - p^2 - i0}\,\Big(\hat{p}\,
(1 - \gamma^5) \otimes \big(4(k\cdot p) \hat{k} - p^2 \hat{k} - 2 k^2
\hat{p}\big) (1 - \gamma^5) \nonumber\\
\hspace{-0.3in}&& + \gamma^{\mu} (1 - \gamma^5) \otimes \gamma_{\mu}
\big(p^2 - 2 k\cdot p\big) \hat{p} \hat{k} (1 - \gamma^5)\Big),
\end{eqnarray}
where we have neglected the contributions of order $O(k_n\cdot
q/M^2_W) \sim 10^{-7}$. Merging the denominators by using the Feynman
parametrization \cite{Feynman1950, Kinoshita1962, Kinoshita1974a,
  Kinoshita1974b} and after the integration over the virtual momenta
in the $n$-dimensional momentum space \cite{Smirnov2004, Smirnov2006,
  Smirnov2012}, we obtain for the integral ${\cal F}(k_n, k_e, q)_{4b}$
the following expression
\begin{eqnarray}\label{eq:E4b.3}
\hspace{-0.3in} &&{\cal F}(k_n, k_e, q)_{4b} = \int^1_0 dx_1 \int^1_0
dx_2 \int^1_0 dx_3 \int^1_0 dx_4 \int^1_0 dx_5 \int^1_0 dx_6\,
\frac{\delta(1 - x_1 - x_2 - x_3 - x_4 - x_5 - x_6)}{2^{2n} \pi^n
  U^{n/2}} \nonumber\\ \hspace{-0.30in}&& \times \,\Bigg\{- (n -
1)\,\frac{\displaystyle \Gamma\Big(2 - \frac{n}{2}\Big)\Gamma\Big(2 -
  \frac{n - 1}{2}\Big)}{\displaystyle 2^{n -
    3}\Gamma\Big(\frac{1}{2}\Big)}\,\Big(\frac{Q}{m^2_N}\Big)^{-4 +
  n}\, \gamma^{\mu} (1 - \gamma^5) \otimes \gamma_{\mu}(1 - \gamma^5)
- \frac{n}{2}\, \Gamma(5 - n)\,Q^{-5 + n}\,m^{2(4 - n)}_N
\nonumber\\ \hspace{-0.30in}&& \times \, \Big[\Big(- \frac{2}{n}\,
  b^2_2 - \frac{4}{n}\, b^2_1\Big)\, \gamma^{\mu} (1 - \gamma^5)
  \otimes \gamma_{\mu}(1 - \gamma^5) + \, \frac{n + 2}{n}\,
  \gamma^{\mu} (1 - \gamma^5) \otimes \gamma_{\mu} \hat{b}_2 \hat{b}_1
  (1 - \gamma^5) + \frac{4}{n}\, \hat{b}_1 (1 - \gamma^5) \otimes
  \hat{b}_1 (1 - \gamma^5) \nonumber\\ \hspace{-0.30in}&& - \frac{3 n
    + 2}{n}\, \hat{b}_2 (1 - \gamma^5) \otimes \hat{b}_1 (1 -
  \gamma^5) + \frac{4}{n}\, \hat{b}_2 (1 - \gamma^5) \otimes \hat{b}_2
  (1 - \gamma^5)\Big] \Bigg\},
\end{eqnarray}
where we have denoted
\begin{eqnarray}\label{eq:E4b.4}
  \hspace{-0.3in}U&=& (x_1 + x_2 + x_4)(x_3 + x_5 + x_6) + (x_1 + x_4)
  x_2,\nonumber\\
 \hspace{-0.3in}b_1 &=& \frac{(x_2 + x_3 + x_5 + x_6) a_1 + x_2
   a_2}{U} = - \frac{(x_3 + x_5 + x_6) x_4 + (x_3 + x_4) x_2}{U}\, k_n = \frac{Y}{U}\,k_n, \nonumber\\
 \hspace{-0.3in}b_2 &=& \frac{(x_1 + x_2 + x_4) a_2 + x_2 a_1}{U} = -
 \frac{(x_1 + x_4) x_3 + (x_3 + x_4) x_2}{U}\, k_n =
 \frac{\bar{Y}}{U}\,k_n, \nonumber\\
 \hspace{-0.3in} Q &=& x_6\, M^2_W + \frac{(x_2 + x_3 + x_5 + x_6) a^2_1 + 2 x_2 a_1 \cdot a_2 + (x_1 + x_2 + x_4) a^2_2} {U},
\end{eqnarray}
where $a_1 = - x_4 k_n$ and $a_2 = - x_3 k_n$. For the definition of
$Q$ we have neglected the contributions of order $O(m^2_{\pi}/M^2_W)
\sim 10^{-6}$.  Taking the limit $n \to 4$ and keeping the divergent
contributions proportional to $\Gamma(2 - n/2)$, we get
\begin{eqnarray}\label{eq:E4b.5}
\hspace{-0.3in} &&{\cal F}(k_n, k_e, q)_{4b} = \int^1_0
dx_1 \int^1_0 dx_2 \int^1_0 dx_3 \int^1_0 dx_4 \int^1_0 dx_5 \int^1_0
dx_6\, \frac{\delta(1 - x_1 - x_2 - x_3 - x_4 - x_5 - x_6)}{256 \pi^4
  U^2} \nonumber\\ \hspace{-0.3in}&& \times \,\Bigg\{\Big(-
\frac{3}{2}\, \Gamma\Big(2 - \frac{n}{2}\Big) + 3\,{\ell
  n}\frac{Q}{m^2_N}\Big)\, \gamma^{\mu} (1 - \gamma^5) \otimes
\gamma_{\mu}(1 - \gamma^5) + \frac{1}{Q}\, \Big[\big(b^2_2 + 2 b^2_1
  \big)\, \gamma^{\mu} (1 - \gamma^5) - 3\, \gamma^{\mu} (1 -
  \gamma^5) \otimes \gamma_{\mu} \hat{b}_2
  \hat{b}_1\nonumber\\ \hspace{-0.3in}&& \times \, (1 - \gamma^5) -
  2\, \hat{b}_1 (1 - \gamma^5) \otimes \hat{b}_1 (1 - \gamma^5) + 7\,
  \hat{b}_2 (1 - \gamma^5) \otimes \hat{b}_1 (1 - \gamma^5) - 2\,
  \hat{b}_2 (1 - \gamma^5) \otimes \hat{b}_2 (1 - \gamma^5)\Big]
\Bigg\},
\end{eqnarray}
Using the Dirac equation for a free neutron, we obtain for the
integral ${\cal F}(k_n, k_e, q)_{4b}$ the following Lorentz structure
\begin{eqnarray}\label{eq:E4b.6}
\hspace{-0.3in} &&{\cal F}(k_n, k_e, q)_{4b} = \int^1_0
dx_1 \int^1_0 dx_2 \int^1_0 dx_3 \int^1_0 dx_4 \int^1_0 dx_5 \int^1_0
dx_6\, \frac{\delta(1 - x_1 - x_2 - x_3 - x_4 - x_5 - x_6)}{256 \pi^4
  U^2} \nonumber\\ \hspace{-0.30in}&& \times \,\Bigg\{\Big(-
\frac{3}{2}\, \Gamma\Big(2 - \frac{n}{2}\Big) + f^{(V)}_{4b}(x_1,
\ldots, x_6) + g^{(W)}_{4b}(x_1, \ldots, x_6)\,
\frac{m^2_N}{M^2_W}\Big)\, \gamma^{\mu} (1 - \gamma^5) \otimes
\gamma_{\mu} + \Big( \frac{3}{2}\, \Gamma\Big(2 - \frac{n}{2}\Big)
\nonumber\\ \hspace{-0.30in}&& + f^{(A)}_{4b}(x_1, \ldots, x_6) +
h^{(W)}_{4b}(x_1, \ldots, x_6)\, \frac{m^2_N}{M^2_W}\Big)\,
\gamma^{\mu} (1 - \gamma^5) \otimes \gamma_{\mu} \gamma^5 +
f^{(W)}_{4b} (x_1, \ldots, x_6)\,
\frac{m^2_N}{M^2_W}\,\frac{\hat{k}_n}{m_N} (1 - \gamma^5) \otimes
1\Bigg\},~~~~~
\end{eqnarray}
where we have denoted
\begin{eqnarray}\label{eq:E4b.7}
\hspace{-0.3in}f^{(V)}_{4b}(x_1, \ldots,x_6) = 3\,{\ell
  n}\frac{M^2_W}{m^2_N} + 3\, {\ell n}\bar{Q}
\end{eqnarray}
and
\begin{eqnarray}\label{eq:E4b.8}
\hspace{-0.3in}g^{(W)}_{4b}(x_1, \ldots,x_6) = \frac{1}{\bar{Q}} \Big[
  2\,\frac{Y^2}{U^2} - 3\, \frac{Y\bar{Y}}{U^2} +
  \frac{\bar{Y}^2}{U^2}\Big]
\end{eqnarray}
and
\begin{eqnarray}\label{eq:E4b.9}
\hspace{-0.3in}f^{(A)}_{4b}(x_1, \ldots,x_6)  =  - 3\,{\ell
  n}\frac{M^2_W}{m^2_N} - 3\, {\ell n}\bar{Q}
\end{eqnarray}
where $f^{(A)}_{4b}(x_1, \ldots,x_6) = - f^{(V)}_{4b}(x_1,
\ldots,x_6)$, and
\begin{eqnarray}\label{eq:E4b.10}
\hspace{-0.3in}&&h^{(W)}_{4b}(x_1, \ldots,x_6) = - \frac{1}{\bar{Q}}
\Big[ 2\,\frac{Y^2}{U^2} - 3\, \frac{Y\bar{Y}}{U^2} +
  \frac{\bar{Y}^2}{U^2}\Big],
\end{eqnarray}
where $h^{(W)}_{4b}(x_1, \ldots,x_6) = - g^{(W)}_{4b}(x_1,
\ldots,x_6)$, and
\begin{eqnarray}\label{eq:E4b.11}
\hspace{-0.3in}f^{(W)}_{4b}(x_1, \ldots,x_6) = \frac{1}{\bar{Q}} \Big[
  - 2\, \frac{Y^2}{U^2} + 7\, \frac{Y\bar{Y}}{U^2} - 2 \,
  \frac{\bar{Y}^2}{U^2}\Big].
\end{eqnarray}
For the calculation of the integrals over the Feynman parameters we
use
\begin{eqnarray}\label{eq:E4b.12}
 \hspace{-0.3in} \bar{Q} &=& \frac{1}{U}\Big(x_6 U + \big((x_2 + x_3 +
 x_5 + x_6) x^2_4 + 2 x_2 x_3 x_4 + (x_1 + x_2 + x_4) x^2_3\big)\,
 \frac{m^2_N}{M^2_W}\Big),\nonumber\\
  \hspace{-0.3in} Y &=& - (x_3 + x_5 + x_6) x_4 - (x_3 + x_4)
  x_2\;,\; \bar{Y} = - (x_1 + x_4) x_3 - (x_3 + x_4) x_2,
\end{eqnarray}
where $m_N = (m_n + m_p)/2 = 938.9188\, {\rm MeV}$ and $M_W =
80.379\,{\rm GeV}$.  After the integration over the Feynman
parameters, the contribution of the Feynman diagram in
Fig.\,\ref{fig:fig4}b to the amplitude of the neutron beta decay is
equal to
\begin{eqnarray}\label{eq:E4b.13}
\hspace{-0.3in}&& M(n \to p e^- \bar{\nu}_e)^{(\pi^0)}_{\rm
  Fig.\,\ref{fig:fig4}b} = \frac{\alpha}{2\pi} \frac{g^2_{\pi
    N}}{16\pi^2} G_V \Big\{ \Big(A_{4b} \, \Gamma\big(2 -
\frac{n}{2}\Big) + F^{(V)}_{4b} + G^{(W)}_{4b}
\frac{m^2_N}{M^2_W}\Big) \, \Big[\bar{u}_e \gamma^{\mu}(1 - \gamma^5)
  v_{\bar{\nu}}\Big]\Big[\bar{u}_p \gamma_{\mu} u_n\Big]
\nonumber\\ \hspace{-0.3in}&& + \Big(B_{4b} \, \Gamma\big(2 -
\frac{n}{2}\Big)+ F^{(A)}_{4b} + H^{(W)}_{4b}
\frac{m^2_N}{M^2_W}\Big[\bar{u}_e \gamma^{\mu}(1 - \gamma^5)
  v_{\bar{\nu}}\Big]\Big[\bar{u}_p \gamma_{\mu} \gamma^5 u_n\Big] +
F^{(W)}_{4b} \frac{m^2_N}{M^2_W}\Big)\,\Big[\bar{u}_e
  \frac{\hat{k}_n}{m_N}(1 - \gamma^5) v_{\bar{\nu}}\Big]\Big[\bar{u}_p
  u_n\Big]\Big\},\nonumber\\ \hspace{-0.3in}&&
\end{eqnarray}
where the structure constants are equal to $A_{4b} = - 1/4$,
$F^{(V)}_{4b} = 3.3958$, $G^{(W)}_{4b} = 0.6859$, $B_{4b} = 1/4$,
$F^{(A)}_{4b} = - 3.3958$, $H^{(W)}_{4b} = - 0.6859$ and $F^{(W)}_{4b}
= 3.9050$. The Lorentz structure of Eq.(\ref{eq:E4b.13}) is
calculated at the neglect the contributions of order $O(k_n \cdot
q/M^2_W) \sim 10^{-7}$ and $O(m^2_{\pi}/M^2_W) \sim 10^{-6}$,
respectively.

\subsection*{E4c. Analytical calculation of the Feynman diagram in
  Fig.\,\ref{fig:fig4}c}
\renewcommand{\theequation}{E4c-\arabic{equation}}
\setcounter{equation}{0}

In the Feynman gauge for the photon propagator the analytical
expression of the Feynman diagram in Fig.\,\ref{fig:fig4}c is given by
(see Eq.(\ref{eq:A.1}))
\begin{eqnarray}\label{eq:E4c.1}
\hspace{-0.30in}&&M(n \to p e^- \bar{\nu}_e)_{\rm
  Fig.\,\ref{fig:fig4}c} = - 2 e^2 g^2_{\pi N} G_V \int
\frac{d^4k}{(2\pi)^4i}\int \frac{d^4p}{(2\pi)^4i}\,\Big[\bar{u}_e
  \,\gamma^{\beta}\,\frac{1}{m_e - \hat{k}_e - \hat{p} - i0}
  \gamma^{\mu}(1 - \gamma^5) v_{\bar{\nu}}\Big]
\nonumber\\ \hspace{-0.30in}&&\times \,\Big[\bar{u}_p \,
  \gamma^{\nu}(1 - \gamma^5)\, \frac{1}{m_N - \hat{k}_n - \hat{p} -
    i0}\,\gamma^5\, \frac{1}{m_N - \hat{k}_n - \hat{k} - i0}\,\gamma^5
  u_n\Big]\,\frac{(2 k - p)_{\beta}}{[m^2_{\pi} - k^2 - i0][m^2_{\pi}
    - (k - p)^2 - i0]}\nonumber\\ \hspace{-0.30in}&& \times \,
\frac{1}{p^2 + i0}\, \Big(\eta_{\mu\nu} + \frac{(p - q)^2
  \eta_{\mu\nu} - (p - q)_{\mu} (p - q)_{\nu}}{M^2_W - (p - q)^2 -
  i0}\Big),
\end{eqnarray}
where we have taken  the propagator of the electroweak $W^-$-boson
$D^{(W)}_{\mu\nu}(p - q)$ in the form
\begin{eqnarray*}
  D^{(W)}_{\mu\nu}(p - q) = - \frac{1}{M^2_W}\Big(\eta_{\mu\nu} +
  \frac{(p - q)^2 \eta_{\mu\nu} - (p - q)_{\mu} (p - q)_{\nu}}{M^2_W -
    (p - q)^2 - i0}\Big).
\end{eqnarray*}
The calculation of the matrix element Eq.(\ref{eq:E4c.1}) we carry out
by calculating the following two integrals
\begin{eqnarray}\label{eq:E4c.2}
\hspace{-0.30in}&&{\cal F}^{(1)}(k_n, k_e, q)_{4c} = -
\int\frac{d^4k}{(2\pi)^4i}\int \frac{d^4p}{(2\pi)^4i}\, \frac{1}{m^2_e
  - (p + k_e)^2 - i0} \frac{1}{m^2_N - (k_n + p)^2 - i0}
\frac{1}{m^2_N - (k_n + k)^2 - i0}\nonumber\\ &&\times
\frac{1}{m^2_{\pi} - k^2 - i0} \frac{1}{m^2_{\pi} - (k - p )^2 - i0}
\frac{1}{p^2 + i0}\, \Big((2 \hat{k} \hat{p} - p^2) \gamma^{\mu}(1 -
\gamma^5) \otimes \gamma_{\mu} \hat{p} \hat{k}\,(1 -
\gamma^5)\Big)\nonumber\\ \hspace{-0.30in}&&
\end{eqnarray}
and
\begin{eqnarray}\label{eq:E4c.3}
\hspace{-0.30in}&&{\cal F}^{(2)}(k_n, k_e, q)_{4c} = -
\int\frac{d^4k}{(2\pi)^4i}\int \frac{d^4p}{(2\pi)^4i}\, \frac{1}{m^2_e
  - (p + k_e)^2 - i0} \frac{1}{m^2_N - (k_n + p)^2 - i0}
\frac{1}{m^2_N - (k_n + k)^2 - i0}\nonumber\\ &&\times
\frac{1}{m^2_{\pi} - k^2 - i0} \frac{1}{m^2_{\pi} - (k - p )^2 - i0}
\frac{1}{p^2 + i0}\, \frac{1}{M^2_W - (p - q)^2 - i0}\, \Big( p^2\big(
2 \hat{k} \hat{p} - p^2 \big) \gamma^{\mu}(1 - \gamma^5) \otimes
\gamma^{\nu} \hat{p} \hat{k}\,(1 - \gamma^5)
\nonumber\\ \hspace{-0.30in}&& + p^4 \big(\hat{p} - 2 \hat{k}\big) (1
- \gamma^5) \otimes \hat{k} (1 - \gamma^5)\Big),
\end{eqnarray}
where we have used the Dirac equation for a free neutron and have kept
only leading divergent contributions that corresponds to the use of
the LLA.

\subsubsection*{\bf Calculation of the integral
  ${\cal F}^{(1)}(k_n, k_e, q)_{4c}$}

Skipping standard intermediate calculations including the integration
over virtual momenta in the $n$-dimensional momentum space
\cite{Feynman1950, Kinoshita1962, Kinoshita1974a, Kinoshita1974b,
  Smirnov2004, Smirnov2006, Smirnov2012}, we define the integral
${\cal F}^{(1)}(k_n, k_e, q)_{1c}$ in terms of the integrals over the
Feynman parameters. We get
\begin{eqnarray}\label{eq:E4c.4}
\hspace{-0.3in}&&{\cal F}^{(1)}(k_n, k_e, q)_{4c} = \int^1_0dx_1
\int^1_0dx_2 \int^1_0dx_3 \int^1_0dx_4 \int^1_0dx_5 \int^1_0
dx_6\,\frac{\delta(1 - x_1 - x_2 - x_3 - x_4 - x_5 - x_6)}{2^{2n}
  \pi^n U^{n/2}}\nonumber\\
\hspace{-0.3in}&&\times \Bigg\{\frac{1}{2}\,\frac{\displaystyle
  \Gamma\Big(2 - \frac{n}{2}\Big)\Gamma\Big(2 - \frac{n -
    1}{2}\Big)}{\displaystyle 2^{n -
    3}\Gamma\Big(\frac{1}{2}\Big)}\,\Big(\frac{Q}{m^2_N}\Big)^{-4 +
  n}\,\gamma^{\beta} \gamma^{\alpha} \gamma^{\mu} (1 - \gamma^5)
\otimes \gamma_{\mu} \gamma_{\alpha} \gamma_{\beta} (1 - \gamma^5) +
\ldots \Bigg\},
\end{eqnarray}
where we have denoted
\begin{eqnarray}\label{eq:E4c.5}
  \hspace{-0.3in}U &=& (x_1 + x_2 + x_4)(x_3 + x_5 + x_6) + (x_1 +
  x_4)x_2,\nonumber\\
  \hspace{-0.3in}b_1 &=& \frac{(x_2 + x_3 + x_5 + x_6)a_1 + x_2
    a_2}{U} =  - \frac{(x_3 + x_5 + x_6)x_4 + (x_3 + x_4)
    x_2) k_n + x_2 x_5 k_e}{U} =  \nonumber\\
  \hspace{-0.3in} &=&\frac{ Y k_n + Z k_e}{U},\nonumber\\
   \hspace{-0.3in}b_2 &=& \frac{(x_1 + x_2 + x_4)a_2 + x_2 a_1}{U} = -
   \frac{((x_1 + x_4)x_3 + (x_3 + x_4) x_2) k_n + (x_1 + x_2 +
     x_4) x_5 k_e}{U} = \nonumber\\
  \hspace{-0.3in} &=&\frac{ \bar{Y} k_n + \bar{Z} k_e}{U},\nonumber\\
   \hspace{-0.3in}Q &=&(x_1 + x_2) m^2_{\pi} + \frac{(x_2 + x_3 + x_5
     + x_6) a^2_1 + 2 x_2 a_1\cdot a_2 + (x_1 + x_2 + x_4) a^2_2}{U}
\end{eqnarray}
with $a_1 = - x_4 k_n$ and $a_2 = - x_3 k_n - x_5 k_e$.  Taking the
limit $n \to 4$ and keeping the divergent part proportional to
$\Gamma(2 - n/2)$, we transcribe the integral Eq.(\ref{eq:E4c.4}) into
the form
\begin{eqnarray}\label{eq:E4c.6}
\hspace{-0.3in}&&{\cal F}^{(1)}(k_n, k_e, q)_{4c} = \int^1_0dx_1
\int^1_0dx_2 \int^1_0dx_3 \int^1_0dx_4 \int^1_0dx_5 \int^1_0
dx_6\,\frac{\delta(1 - x_1 - x_2 - x_3 - x_4 - x_5 - x_6)}{256 \pi^4
  U^2}\nonumber\\
\hspace{-0.3in}&&\times \Big\{\Big(\frac{1}{4}\, \Gamma(2 -
\frac{n}{2}\Big) - \frac{1}{2}\,{\ell n}
\frac{Q}{m^2_N}\Big)\,\gamma^{\beta} \gamma^{\alpha} \gamma^{\mu} (1 -
\gamma^5) \otimes \gamma_{\mu} \gamma_{\alpha} \gamma_{\beta} (1 -
\gamma^5) + \ldots \Big\}.
\end{eqnarray}
Using the algebra of the Dirac $\gamma$-matrices and having neglected
the contributions of order $O(E^2_0/m^2_N) \sim 10^{-6}$, we arrive at
the expression
\begin{eqnarray}\label{eq:E4c.7}
\hspace{-0.3in}&&{\cal F}^{(1)}(k_n, k_e, q)_{4c} = \int^1_0dx_1
\int^1_0dx_2 \int^1_0dx_3 \int^1_0dx_4 \int^1_0dx_5 \int^1_0
dx_6\,\frac{\delta(1 - x_1 - x_2 - x_3 - x_4 - x_5 - x_6)}{256 \pi^4
  U^2} \nonumber\\
\hspace{-0.3in}&& \times \Big\{\Big(\Gamma\Big(2 - \frac{n}{2}\Big) +
f^{(V)}_{4c^{(1)}}(x_1,\ldots,x_6) + h^{(V)}_{4c^{(1)}}(x_1,\ldots,x_6)
\,\frac{k_n\cdot k_e}{m^2_N}\Big) \,\gamma^{\mu}(1 - \gamma^5) \otimes
\gamma_{\mu} + \Big(- \Gamma\Big(2 - \frac{n}{2}\Big) \nonumber\\
\hspace{-0.3in}&& + f^{(A)}_{4c^{(1)}}(x_1,\ldots,x_6) +
h^{(A)}_{4c^{(1)}}(x_1,\ldots,x_6) \,\frac{k_n\cdot k_e}{m^2_N}\Big)
\,\gamma^{\mu}(1 - \gamma^5) \otimes \gamma_{\mu} \gamma^5 + \ldots
\Big\},
\end{eqnarray}
where we have denoted
\begin{eqnarray}\label{eq:E4c.8} 
\hspace{-0.3in}f^{(V)}_{4c^{(1)}}(x_1,\ldots,x_6) &=& - 2\,{\ell
  n}\frac{\bar{Q}}{m^2_N} \quad,\quad
h^{(V)}_{4c^{(1)}}(x_1,\ldots,x_6) = - 4\, \frac{\bar{V}}{U}\,
\frac{m^2_N}{\bar{Q}},\nonumber\\
\hspace{-0.3in}f^{(A)}_{4c^{(1)}}(x_1,\ldots,x_6) &=& + 2\,{\ell
  n}\frac{\bar{Q}}{m^2_N} \quad,\quad
h^{(A)}_{4c^{(1)}}(x_1,\ldots,x_6) = + 4\, \frac{\bar{V}}{U}\,
\frac{m^2_N}{\bar{Q}}.
\end{eqnarray}
For the calculation of the integrals over the Feynman parameters we
use
\begin{eqnarray}\label{eq:E4c.9}
  \hspace{-0.30in}\frac{\bar{Q}}{m^2_N} &=& \frac{1}{U}\Big( (x_1 +
  x_2)\, U\, \frac{m^2_{\pi}}{m^2_N} + (x_3 + x_5 + x_6) x^2_4 + x_2
  (x_3 + x_4)^2 + (x_1 + x_4) x^2_3\Big), \nonumber\\ \hspace{-0.21in}
  Y &=& - (x_3 + x_5 + x_6) x_4 - (x_3 + x_4) x_2\;,\; Z = - x_2
  x_5,\nonumber\\ \hspace{-0.30in} \bar{Y} &=& - (x_1 + x_4) x_3 -
  (x_3 + x_4) x_2\;,\; \bar{Z} = - (x_1 + x_2 + x_4)
  x_5,\nonumber\\ \hspace{-0.30in} \bar{V} &=& (x_1 + x_4) x_3 x_5 +
  (x_3 + x_4) x_2 x_5,
\end{eqnarray}
where $m_N = (m_n + m_p)/2 = 938.9188\,{\rm MeV}$ and $m_{\pi} =
139.5706\, {\rm MeV}$ \cite{PDG2020}. After the integration over the
Feynman parameters, the contribution of the integral ${\cal
  F}^{(1)}(k_n, k_e, q)_{4c^{(1)}}$ to the matrix element $M(n \to p
e^-\bar{\nu}_e)^{(\pi^0)}_{\rm Fig.\,\ref{fig:fig4}c}$ takes the form
\begin{eqnarray}\label{eq:E4c.10}
 \hspace{-0.30in}&& M(n \to p e^- \bar{\nu}_e)^{(\pi^0)}_{\rm
   Fig.\,\ref{fig:fig4}c^{(1)}}= - \frac{\alpha}{2\pi} \frac{g^2_{\pi
     N}}{16\pi^2}\,G_V \Big\{ \Big(A_{4c^{(1)}}\Gamma\Big(2 -
 \frac{n}{2}\Big) + F^{(V)}_{4c^{(1)}} + H^{(V)}_{4c^{(1)}}\,
 \frac{k_n \cdot k_e}{m^2_N}\Big) \Big[\bar{u}_e \gamma^{\mu} (1 -
   \gamma^5) v_{\bar{\nu}}\Big]\Big[\bar{u}_p \gamma_{\mu} u_n\Big]
 \nonumber\\
\hspace{-0.3in}&& + \Big(B_{4c^{(1)}}\Gamma\Big(2 - \frac{n}{2}\Big) +
F^{(A)}_{4c^{(1)}} + H^{(A)}_{4c^{(1)}}\, \frac{k_n \cdot k_e}{m^2_N
}\Big) \Big[\bar{u}_e \gamma^{\mu} (1 - \gamma^5)
  v_{\bar{\nu}}\Big]\Big[\bar{u}_p \gamma_{\mu}\gamma^5 u_n\Big]\Big\}.
\end{eqnarray}
The structure constants are equal to $A_{4c^{(1)}} = 1/6$,
$F^{(V)}_{4c^{(1)}} = 0.7171$, $H^{(V)}_{4c^{(1)}} = - 0.4455$,
$B_{4c^{(1)}} = - 1/6$, $F^{(A)}_{4c^{(1)}} = - 0.7171$ and
$H^{(A)}_{4c^{(1)}} = 0.4455$. The Lorentz structure of
Eq.(\ref{eq:E4c.10}) is obtained at the neglect of the contributions
of order $O(E^2_0/m^2_N) \sim 10^{-6}$.

\subsubsection*{\bf Calculation of the integral
  ${\cal F}^{(2)}(k_n, k_e, q)_{4c}$}

In terms of the integrals over the Feynman parameters and after the
integration over the virtual momenta in the $n$-dimensional momentum
space, we obtain for the integral ${\cal F}^{(2)}(k_n, k_e, q)_{4c}$
the following expression
\begin{eqnarray}\label{eq:E4c.11}
\hspace{-0.3in}&&{\cal F}^{(2)}(k_n, k_e, q)_{4c} = \int^1_0 \!\!
dx_1 \int^1_0 \!\! dx_2 \int^1_0 \!\! dx_3 \int^1_0 \!\! dx_4 \int^1_0
\!\! dx_5 \int^1_0 \!\! dx_6 \int^1_0 \!\! dx_7\,\frac{\delta(1 - x_1
  - x_2 - x_3 - x_4 - x_5 - x_6 - x_7)}{2^{2n} \pi^n
  U^{n/2}} \nonumber\\
\hspace{-0.3in}&& \times \Bigg\{\frac{n^2 (n +
  2)}{8}\,\frac{\displaystyle \Gamma\Big(2 -
  \frac{n}{2}\Big)\Gamma\Big(2 - \frac{n - 1}{2}\Big)}{\displaystyle
  2^{n - 3}\Gamma\Big(\frac{1}{2}\Big)}\,\Big(\frac{Q}{m^2_N}\Big)^{-4
  + n}\,\Big[\frac{2}{n}\, \gamma^{\mu} (1 - \gamma^5) \otimes
  \gamma_{\mu} (1 - \gamma^5)- \frac{2}{n^2}\, \gamma^{\beta}
  \gamma^{\alpha} \gamma^{\mu} (1 - \gamma^5) \otimes \gamma_{\mu}
  \gamma_{\alpha} \gamma_{\beta} \nonumber\\
\hspace{-0.3in}&& \times (1 - \gamma^5)\Big] + \frac{n^2}{4} \Gamma(5
- n)\,Q^{-5+n} m^{2(4-n)}_N \Big[\frac{2}{n^2}\,b^2_2 \gamma^{\beta}
  \gamma^{\alpha} \gamma^{\mu} (1 - \gamma^5) \otimes \gamma_{\mu}
  \gamma_{\alpha} \gamma_{\beta} (1 - \gamma^5) + \Big(- \frac{2}{n}\,
  b^2_1 + 2\, \frac{n + 4}{n^2}\, b^2_2 \Big) \nonumber\\
\hspace{-0.3in}&& \times \, \gamma^{\mu} (1 - \gamma^5) \otimes
\gamma_{\mu} (1 - \gamma^5) - \frac{(n + 2)(n + 4)}{n^2}\,
\gamma^{\mu} (1 - \gamma^5) \otimes \gamma_{\mu} \hat{b}_2 \hat{b}_1
(1 - \gamma^5) + 2\, \frac{n + 2}{n^2}\, \hat{b}_1
\gamma^{\beta}\gamma^{\alpha} (1 - \gamma^5) \otimes \gamma_{\alpha}
\gamma_{\beta} \hat{b}_1  \nonumber\\
\hspace{-0.3in}&& \times \,(1 - \gamma^5) + 2\, \frac{n +
  4}{n^2}\,\gamma^{\beta} \hat{b}_2 \gamma^{\alpha} (1 - \gamma^5)
\otimes \gamma_{\alpha} \hat{b}_2 \gamma_{\beta} (1 - \gamma^5) - 2\,
\frac{n + 2}{n} \, \hat{b}_1 (1 - \gamma^5) \otimes \hat{b}_1 (1 -
\gamma^5) + \frac{(n + 2)(n + 4)}{n^2} \nonumber\\
\hspace{-0.3in}&& \times\,\hat{b}_2 (1 - \gamma^5)
\otimes \hat{b}_1 (1 - \gamma^5) \Big]\Bigg\},
\end{eqnarray} 
where we have used the algebra of the Dirac $\gamma$-matrices in
$n$-dimensional space-time  and denoted
\begin{eqnarray}\label{eq:E4c.12}
  \hspace{-0.3in}U &=& (x_1 + x_2 + x_4)( x_3 + x_5 + x_6 + x_7) +
  (x_1 + x_4) x_2,\nonumber\\
  \hspace{-0.3in}b_1 &=& \frac{(x_2 + x_3 + x_5 + x_6 + x_7)a_1 + x_2
    a_2}{U} = - \frac{(x_3 + x_5 + x_6 + x_7)x_4 + (x_3 + x_4)
    x_2 }{U}\, k_n = \frac{Y}{U}\,k_n ,\nonumber\\
   \hspace{-0.3in}b_2 &=& \frac{(x_1 + x_2 + x_4)a_2 + x_2 a_1}{U} =
   - \frac{(x_1 + x_4) x_3 + (x_3 + x_4) x_2}{U}\, k_n =
   \frac{\bar{Y}}{U}\,k_n,\nonumber\\
   \hspace{-0.3in}Q &=&x_7 M^2_W + \frac{(x_2 + x_3 + x_5 + x_6 + x_7)
     a^2_1 + 2 x_2 a_1\cdot a_2 + (x_1 + x_2 + x_4) a^2_2}{U}
\end{eqnarray}
with $a_1 = - x_4 k_n$ and $a_2 = - x_3 k_n$. In Eq.(\ref{eq:E4c.12}) we
have neglected the contributions of the terms proportional to $q$ and
$k_e$, appearing in ${\cal F}^{(2)}(k_n, k_e, q)_{4c}$ in the form of
an expansion in powers of $k_n\cdot k_e/M^2_W$ order $O(k_n\cdot
k_e/M^2_W) \sim 10^{-7}$, and the terms of order $O(m^2_{\pi}/M^2_W)
\sim 10^{-6}$. Taking the limit $n \to 4$ and using the Dirac
equations for free fermions, we arrive at the expression
\begin{eqnarray}\label{eq:E4c.13}
 \hspace{-0.21in}&& {\cal F}^{(2)}(k_n, k_e, q)_{4c} = \int^1_0 dx_1
 \int^1_0 dx_2 \int^1_0 dx_3 \int^1_0 dx_4 \int^1_0 dx_5 \int^1_0 dx_6
 \int^1_0 dx_7 \frac{\delta(1 - x_1 - x_2 - x_3 - x_4 - x_5 - x_6 -
   x_7)}{256 \pi^4 U^2} \nonumber\\ \hspace{-0.21in}&& \times
 \Big\{\Big(f^{(V)}_{4c^{(2)}}(x_1, \ldots, x_7) +
 g^{(W)}_{4c^{(2)}}(x_1, \ldots, x_7)\, \frac{m^2_N}{M^2_W}\Big)
 \gamma^{\mu} (1 - \gamma^5) \otimes \gamma_{\mu} +
 \Big(f^{(A)}_{4c^{(2)}}(x_1, \ldots, x_7) + h^{(W)}_{4c^{(2)}}(x_1,
 \ldots, x_6)\, \frac{m^2_N}{M^2_W}\Big)
 \nonumber\\ \hspace{-0.21in}&& \times \gamma^{\mu} (1 - \gamma^5)
 \otimes \gamma_{\mu} \gamma^5 + f^{(W)}_{4c^{(2)}}(x_1, \ldots,
 x_7)\, \frac{m^2_N}{M^2_W}\, \frac{\hat{k}_n}{m_N} (1 - \gamma^5)
 \otimes 1\Big\},
\end{eqnarray}
where we have denoted
\begin{eqnarray}\label{eq:E4c.14}
 \hspace{-0.30in}&&f^{(V)}_{4c^{(2)}}(x_1, \ldots, x_7) = - \frac{3}{2}
\end{eqnarray}
and
\begin{eqnarray}\label{eq:E4c.15}
 \hspace{-0.30in}&&g^{(W)}_{4c^{(2)}}(x_1, \ldots, x_7) =
 \frac{1}{\bar{Q}}\Big[- 2\,
   \frac{Y^2}{U^2} - 12\, \frac{Y\bar{Y}}{U^2} + 6\,
   \frac{\bar{Y}^2}{U^2}\Big]
\end{eqnarray}
and
\begin{eqnarray}\label{eq:E4c.16}
 \hspace{-0.30in}&&f^{(A)}_{4c^{(2)}}(x_1, \ldots, x_7) = \frac{3}{2}
\end{eqnarray}
and
\begin{eqnarray}\label{eq:E4c.17}
 \hspace{-0.30in}h^{(W)}_{4c^{(2)}}(x_1, \ldots, x_7) = -
 \frac{1}{\bar{Q}}\Big[- 2\,
   \frac{Y^2}{U^2} - 12\, \frac{Y\bar{Y}}{U^2} +  6\,
   \frac{\bar{Y}^2}{U^2}\Big]
\end{eqnarray}
and
\begin{eqnarray}\label{eq:E4c.18}
 \hspace{-0.30in}&&f^{(W)}_{4c^{(2)}}(x_1, \ldots, x_7) =
 \frac{1}{\bar{Q}}\Big[ + 12\, \frac{Y\bar{Y}}{U^2} + 16\,
   \frac{\bar{Y}^2}{U^2}\Big].
\end{eqnarray}
For the calculation of the integrals over the Feynman parameters we
use
\begin{eqnarray}\label{eq:E4c.19}
  \hspace{-0.30in}\bar{Q} &=& \frac{1}{U}\Big(x_7 U + \big((x_3 + x_5
  + x_6 + x_7) x^2_4 + x_2 (x_3 + x_4)^2 + (x_1 + x_4) x^2_3\big)\,
  \frac{m^2_N}{M^2_W}\Big), \nonumber\\
  \hspace{-0.30in} U &=& (x_1 + x_2 + x_4)(x_3 + x_5 + x_6 + x_7) +
  (x_1 + x_4) x_2,\nonumber\\
  \hspace{-0.30in} Y &=& - (x_3 + x_5 + x_6 + x_7)x_4 - (x_3 + x_4)
  x_2 \;,\; \bar{Y} = - (x_1 + x_4) x_3 - (x_3 + x_4) x_2,
\end{eqnarray}
where $m_N = (m_n + m_p)/2 = 938.9188\,{\rm MeV}$ and $M_W =
30,379\,{\rm GeV}$ \cite{PDG2020}. After the integration over the
Feynman parameters, we obtain the contribution of the integral ${\cal
  F}^{(2)}(k_n, k_e, q)_{4c}$ to the amplitude of the neutron
beta decay 
\begin{eqnarray}\label{eq:E4c.20}
 \hspace{-0.30in}&&M(n \to p e^- \bar{\nu}_e)^{(\pi^0)}_{\rm
   Fig.\,\ref{fig:fig4}c^{(2)}} = - \frac{\alpha}{2\pi}\frac{g^2_{\pi
     N}}{16\pi^2}\, G_V \Big\{\Big(F^{(V)}_{4c^{(2)}} +
 G^{(W)}_{4c^{(2)}}\,\frac{m^2_N}{M^2_W}\Big) \Big[\bar{u}_e
   \gamma^{\mu} (1 - \gamma^5) v_{\bar{\nu}}\Big] \Big[\bar{u}_p
   \gamma_{\mu} u_n\Big] \nonumber\\ \hspace{-0.30in}&& + \Big(
 F^{(A)}_{4c^{(2)}} + H^{(W)}_{4c^{(2)}}\,\frac{m^2_N}{M^2_W}\Big)
 \Big[\bar{u}_e \gamma^{\mu} (1 - \gamma^5)
   v_{\bar{\nu}}\Big]\Big[\bar{u}_p \gamma_{\mu} \gamma^5 u_n\Big] +
 F^{(W)}_{4c^{(2)}}\,\frac{m^2_N}{M^2_W} \Big[\bar{u}_e
   \frac{\hat{k}_n}{m_N} (1 - \gamma^5)
   v_{\bar{\nu}}\Big]\,\Big[\bar{u}_p u_n\Big]\Big\}.
\end{eqnarray}
The structure constants are equal to $F^{(V)}_{4c^{(2)}} = - 0.0512$,
$G^{(W)}_{4c^{(2)}} = - 2.2306$, $F^{(A)}_{4c^{(2)}} = 0.0512$,
$H^{(W)}_{4c^{(2)}} = 2.2306$ and $F^{(W)}_{4c^{(2)}} = 6.2822$. The
Lorentz structure of the matrix element Eq.(\ref{eq:E4c.20}) is
calculated at the neglect of the contributions of order $O(k_n\cdot
q/M^2_W) \sim O(k_n \cdot k_e/M^2_W) \sim 10^{-7}$ and
$O(m^2_{\pi}/M^2_W) \sim 10^{-6}$, respectively.

The contribution of the Feynman diagram in Fig.\,\ref{fig:fig4}c to
matrix element of thee neutron beta decay  is given by
\begin{eqnarray}\label{eq:E4c.21}
 \hspace{-0.30in}&& M(n \to p e^- \bar{\nu}_e)^{(\pi^0)}_{\rm
   Fig.\,\ref{fig:fig4}c}= - \frac{\alpha}{2\pi} \frac{g^2_{\pi
     N}}{16\pi^2}\,G_V \Big\{ \Big(A_{4c}\Gamma\Big(2 -
 \frac{n}{2}\Big) + F^{(V)}_{4c} + H^{(V)}_{4c}\,
 \frac{k_n \cdot k_e}{m^2_N} +
 G^{(W)}_{4c}\,\frac{m^2_N}{M^2_W}\Big) \Big[\bar{u}_e
   \gamma^{\mu} (1 - \gamma^5) v_{\bar{\nu}}\Big] \nonumber\\
\hspace{-0.3in}&& \times \Big[\bar{u}_p \gamma_{\mu} u_n\Big] +
\Big(B_{4c}\Gamma\Big(2 - \frac{n}{2}\Big) + F^{(A)}_{4c} +
H^{(A)}_{4c}\, \frac{k_n \cdot k_e}{m^2_N } +
H^{(W)}_{4c}\,\frac{m^2_N}{M^2_W}\Big) \Big[\bar{u}_e \gamma^{\mu} (1
  - \gamma^5) v_{\bar{\nu}}\Big]\Big[\bar{u}_p \gamma_{\mu}\gamma^5
  u_n\Big] \nonumber\\
\hspace{-0.3in}&& + F^{(W)}_{4c}\,\frac{m^2_N}{M^2_W} \Big[\bar{u}_e
  \frac{\hat{k}_n}{m_N} (1 - \gamma^5)
  v_{\bar{\nu}}\Big]\,\Big[\bar{u}_p u_n\Big]\Big\}.
\end{eqnarray}
The structure constants are equal to $A_{4c} = A_{4c^{(1)}} = 1/6$,
$F^{(V)}_{4c } = F^{(V)}_{4c^{(1)}}+ F^{(V)}_{4c^{(2)}} = 0.6659$,
$H^{(V)}_{4c } = H^{(V)}_{4c^{(1)}} = - 0.4455$, $G^{(W)}_{4c } =
G^{(W)}_{4c^{(2)}} = - 2.2306$, $B_{4c} = B_{4c^{(1)}} = - 1/6$,
$F^{(A)}_{4c }= F^{(A)}_{4c^{(1)}} + F^{(A)}_{4c^{(2)}} = - 0.6659$,
$H^{(A)}_{4c } = H^{(A)}_{4c^{(1)}} = 0.4455$, $H^{(W)}_{4c } =
H^{(W)}_{4c^{(2)}} = 2.2306$ and $F^{(W)}_{4c } = F^{(W)}_{4c^{(2)}} =
6.2822$. The Lorentz structure of Eq.(\ref{eq:E4c.21}) is obtained at
the neglect of the contributions of order $O(E^2_0/m^2_N) \sim
O(m^2_{\pi}/M^2_W) \sim 10^{-6}$.

\subsection*{E4d. Analytical calculation of the Feynman diagram in
  Fig.\,\ref{fig:fig4}d}
\renewcommand{\theequation}{E4d-\arabic{equation}}
\setcounter{equation}{0}

The analytical expression of the Feynman diagram in
Fig.\,\ref{fig:fig4}d, taken in the Feynman gauge, is given by (see
Eq.(\ref{eq:A.1}))
\begin{eqnarray}\label{eq:E4d.1}
\hspace{-0.30in}&& M(n \to p e^- \bar{\nu}_e)_{\rm
  Fig.\,\ref{fig:fig4}d} = + 2 e^2 g^2_{\pi N} G_V \int
\frac{d^4k}{(2\pi)^4i}\int \frac{d^4p}{(2\pi)^4i}\,\Big[\bar{u}_e \,
  \gamma^{\mu}(1 - \gamma^5) v_{\bar{\nu}}\Big] \Big[\bar{u}_p
  \,\gamma^{\beta}\frac{1}{m_N - \hat{k}_p - \hat{p} - i0}
  \nonumber\\ &&\times \, \gamma_{\mu} (1 - \gamma^5) \, \frac{1}{m_N
    - \hat{k}_n - \hat{p} - i0}\,\gamma^5\,\frac{1}{m_N - \hat{k}_n -
    \hat{k} - \hat{p} - i0}\,\gamma_{\beta} \frac{1}{m_N - \hat{k}_n -
    \hat{k} - i0}\,\gamma^5 u_n \Big]\,\frac{1}{m^2_{\pi} - k^2 - i0}
\, \frac{1}{p^2 + i0}, \nonumber\\ \hspace{-0.30in}&&
\end{eqnarray}
where as usual we have made the replacement $M^2_W D^{(W)}_{\mu\nu}(-
q) \to - \eta_{\mu\nu}$, which is valid up to the relative
contributions of order $O(m_e m_N/M^2_W) \sim 10^{-7}$. We reduce the
calculation of the contribution of the Feynman diagram in
Fig.\,\ref{fig:fig4}d to the calculation of the integral
\begin{eqnarray}\label{eq:E4d.2}
\hspace{-0.30in}&& {\cal F}(k_n, k_e, q)_{4d} = - \int
\frac{d^4k}{(2\pi)^4i}\int \frac{d^4p}{(2\pi)^4i}\,\frac{1}{m^2_N -
  (k_p + p)^2 - i0} \frac{1}{m^2_N - (k_n + p)^2 - i0} \frac{1}{m^2_N
  - (k_n + p + k)^2 - i0} \nonumber\\
\hspace{-0.3in}&& \times \frac{1}{m^2_N - (k_n + k)^2 - i0}
\frac{1}{m^2_{\pi} - k^2 - i0}\, \frac{1}{p^2 + i0} \, \Big(\big(8
(k\cdot p) \hat{p} + 4 p^2 \hat{p} - 4 p^2 \hat{k}\big)(1 - \gamma^5)
\otimes \hat{k}(1 - \gamma^5)\Big). 
\end{eqnarray}
Merging the denominators by using the Feynman parametrization
\cite{Feynman1950, Kinoshita1962, Kinoshita1974a, Kinoshita1974b} and
after the diagonalization and integration over the virtual momenta in
the $n$-dimensional momentum space \cite{Smirnov2004, Smirnov2006,
  Smirnov2012}, we arrive at the expression
\begin{eqnarray}\label{eq:E4d.3}
\hspace{-0.30in}&& {\cal F}(k_n, k_e, q)_{4d} = \int^1_0 dx_1 \int^1_0
dx_2 \int^1_0 dx_3 \int^1_0 dx_4 \int^1_0 dx_5 \int^1_0 dx_6\,
\frac{\delta(1 - x_1 - x_2 - x_3 - x_4 - x_5 - x_6)}{2^{2n} \pi^n
  U^{n/2}} \nonumber\\ \hspace{-0.30in}&& \times \,\Bigg\{ - (n - 2)\,
\frac{\displaystyle \Gamma(2 - \frac{n}{2}\Big) \Gamma\Big(2 - \frac{n
    - 1}{2}\Big)}{\displaystyle 2^{n -
    3}\Gamma\Big(\frac{1}{2}\Big)}\,\Big(\frac{Q}{m^2_N}\Big)^{-4 +
  n}\, \gamma^{\mu} (1 - \gamma^5) \otimes \gamma_{\mu}(1 - \gamma^5)
+ \ldots \Bigg\},
\end{eqnarray} 
where we have used the algebra of the Dirac $\gamma$-matrices in
$n$-dimensional space-time $\gamma^{\alpha} \hat{a} \gamma_{\alpha} =
- (n - 2)\, \hat{a}$, $\gamma^{\alpha} \hat{a} \hat{b}\gamma_{\alpha}
= 4 a\cdot b + (n - 4)\,\hat{a} \hat{b}$, $\gamma^{\alpha} \hat{a}
\hat{b} \hat{c} \gamma_{\alpha} = - 2 \hat{c} \hat{b} \hat{a} - (n -
4) \hat{a} \hat{b} \hat{c}$ and so on at $\gamma^{\alpha}
\gamma_{\alpha} = n$ and $\gamma^{\alpha} \gamma^{\beta} +
\gamma^{\beta} \gamma^{\alpha} = 2\, \eta^{\alpha \beta}$
\cite{Itzykson1980} and denoted
\begin{eqnarray*}
  \hspace{-0.3in}U &=& (x_1 + x_4 + x_5)( x_2 + x_3 + x_6) +
  (x_1 + x_4) x_5,\nonumber\\
  \hspace{-0.3in}b_1 &=& \frac{(x_2 + x_3 + x_5 + x_6)a_1 - x_5
    a_2}{U} = \frac{ x_2 x_5 q - ((x_2 + x_3 + x_5) x_4 + (x_4 + x_5)
    x_6) k_n }{U} = \nonumber\\
\hspace{-0.3in}&=&\frac{X q + Y k_n}{U}
  ,\nonumber\\
   \hspace{-0.3in}b_2 &=& \frac{(x_1 + x_4 + x_5)a_2 - x_5 a_1}{U} =
   - \frac{ (x_1 + x_4 + x_5) x_2 q + ((x_1 + x_4)(x_2 + x_3) + (x_1 +
     x_2 + x_3) x_5) k_n}{U} = \nonumber\\
  \end{eqnarray*}
\begin{eqnarray}\label{eq:E4d.4}  
   \hspace{-0.3in}&=&\frac{\bar{X} q + \bar{Y}k_n}{U},\nonumber\\  
   \hspace{-0.3in}Q &=& x_1 m^2_{\pi} + \frac{(x_2 + x_3 + x_5 + x_6 )
     a^2_1 - 2 x_5 a_1\cdot a_2 + (x_1 + x_4 + x_5) a^2_2}{U}
\end{eqnarray}
with $a_1 = - (x_4 + x_5) k_n$ and $a_2 = - x_2 q - (x_2 + x_3 + x_5)
k_n$. Taking the limit $ n \to 4$ and keeping the divergent
contributions proportional to $\Gamma(2 - n/2)$, we get
\begin{eqnarray}\label{eq:E4d.5}
\hspace{-0.30in}&& {\cal F}(k_n, k_e, q)_{4d} = \int^1_0 dx_1 \int^1_0
dx_2 \int^1_0 dx_3 \int^1_0 dx_4 \int^1_0 dx_5 \int^1_0 dx_6\,
\frac{\delta(1 - x_1 - x_2 - x_3 - x_4 - x_5 - x_6)}{256 \pi^4 U^2}
\nonumber\\ \hspace{-0.30in}&& \times \,\Big\{ \Big(- \Gamma\Big(2 -
\frac{n}{2}\Big) + 2\, {\ell n}\frac{Q}{m^2_N}\Big)\, \gamma^{\mu} (1
- \gamma^5) \otimes \gamma_{\mu}(1 - \gamma^5) + \ldots \Big\}.
\end{eqnarray}
Using the Dirac equations for free fermions and having neglected the
contributions of order $O(E^2_0/m^2_N) \sim 10^{-6}$, we arrive at the
expression
\begin{eqnarray}\label{eq:E4d.6}
\hspace{-0.3in}&&{\cal F}(k_n, k_e, q)_{4d} = \int^1_0dx_1
\int^1_0dx_2 \int^1_0dx_3 \int^1_0dx_4 \int^1_0dx_5 \int^1_0
dx_6\,\frac{\delta(1 - x_1 - x_2 - x_3 - x_4 - x_5 - x_6)}{256 \pi^4
  U^2} \nonumber\\
 \hspace{-0.3in}&&\times \Big\{\Big( - \Gamma\Big(2 - \frac{n}{2}\Big)
 + f^{(V)}_{4d}(x_1,\ldots,x_6) +
 g^{(V)}_{4d}(x_1,\ldots,x_6)\,\frac{k_n\cdot q}{m^2_N}\Big)
 \,\gamma^{\mu}(1 - \gamma^5) \otimes \gamma_{\mu} + \Big(
 \Gamma\Big(2 - \frac{n}{2}\Big) \nonumber\\
\hspace{-0.3in}&& + f^{(A)}_{4d}(x_1,\ldots,x_6) +
g^{(A)}_{4d}(x_1,\ldots,x_6)\,\frac{k_n\cdot q}{m^2_N} \Big)
\,\gamma^{\mu}(1 - \gamma^5) \otimes \gamma_{\mu} \gamma^5 + \ldots
\Big\},
\end{eqnarray}
where we have denoted
\begin{eqnarray}\label{eq:E4d.7}
\hspace{-0.3in}f^{(V)}_{4d}(x_1, \ldots,x_6) &=& + 2\,{\ell
  n}\frac{\bar{Q}}{m^2_N} \quad,\quad g^{(V)}_{4d}(x_1, \ldots,x_6) = + 
4\, \frac{V}{U}\, \frac{m^2_N}{\bar{Q}}, \nonumber\\
\hspace{-0.3in}f^{(A)}_{4d}(x_1, \ldots,x_6) &=& - 2\,{\ell
  n}\frac{\bar{Q}}{m^2_N} \quad,\quad g^{(A)}_{4d}(x_1, \ldots,x_6) = - 
4\, \frac{V}{U}\, \frac{m^2_N}{\bar{Q}}.
\end{eqnarray}
For the calculation of the integrals over the Feynman parameters we
use
\begin{eqnarray}\label{eq:E4d.8}
  \hspace{-0.21in} \frac{\bar{Q}}{m^2_N} &=& \frac{1}{U}\Big( x_1\,
  U\, \frac{m^2_{\pi}}{m^2_N} + x_6 (x_4 + x_5)^2 + (x_2 + x_3 + x_4)
  (x_4 + x_5) (x_2 + x_3 + x_5) + x_1 (x_2 + x_3 + x_5)^2\Big),
  \nonumber\\ \hspace{-0.21in} V &=& (x_2 + x_3) (x_4 + x_5) x_2 +
  (x_2 + x_3 + x_5) x_1 x_2,
\end{eqnarray}
where $m_N = (m_n + m_p)/2 = 938.9188\,{\rm MeV}$ and $m_{\pi} =
139.5706\, {\rm MeV}$ \cite{PDG2020}. After the integration over the
Feynman parameters, the matrix element $M(n \to p
e^-\bar{\nu}_e)^{(\pi^0)}_{\rm Fig.\,\ref{fig:fig4}d}$ takes the form
\begin{eqnarray}\label{eq:E4d.9}
 \hspace{-0.30in}&& M(n \to p e^- \bar{\nu}_e)^{(\pi^0)}_{\rm
   Fig.\,\ref{fig:fig4}d}= \frac{\alpha}{2\pi} \frac{g^2_{\pi
     N}}{16\pi^2}\,G_V \Big\{ \Big(A_{4d}\Gamma\Big(2 - \frac{n}{2}\Big)
 + F^{(V)}_{4d} + G^{(V)}_{4d}\, \frac{k_n \cdot q}{m^2_N}\Big)
 \Big[\bar{u}_e \gamma^{\mu} (1 - \gamma^5)
   v_{\bar{\nu}}\Big]\Big[\bar{u}_p \gamma_{\mu} u_n\Big] \nonumber\\
\hspace{-0.3in}&& + \Big(B_{4d}\Gamma\Big(2 - \frac{n}{2}\Big) +
F^{(A)}_{4d} + G^{(A)}_{4d}\, \frac{k_n \cdot q}{m^2_N}\Big)
\Big[\bar{u}_e \gamma^{\mu} (1 - \gamma^5)
  v_{\bar{\nu}}\Big]\Big[\bar{u}_p \gamma_{\mu}\gamma^5 u_n\Big].
\end{eqnarray}
The structure constants are equal to $A_{4d} = 1/6$, $F^{(V)}_{4d} = 
- 0.2729$, $G^{(V)}_{4d} = - 0.1911$, $B_{4d} = - 1/6$, $F^{(A)}_{4d} =
0.2729$ and $G^{(A)}_{4d} = 0.1911$. The Lorentz structure of
Eq.(\ref{eq:E4d.9}) is obtained at the neglect of the contributions of
order $O(E^2_0/m^2_N) \sim 10^{-6}$.

\subsection*{E4e. Analytical calculation of the Feynman diagram in
  Fig.\,\ref{fig:fig4}e}
\renewcommand{\theequation}{E4e-\arabic{equation}}
\setcounter{equation}{0}

The analytical expression of the Feynman diagram in
Fig.\,\ref{fig:fig4}e, taken in the Feynman gauge, is given by (see
Eq.(\ref{eq:A.1}))
\begin{eqnarray}\label{eq:E4e.1}
\hspace{-0.30in}&& M(n \to p e^- \bar{\nu}_e)_{\rm
  Fig.\,\ref{fig:fig4}e} = + 2 e^2 g^2_{\pi N} G_V \int
\frac{d^4k}{(2\pi)^4i}\int \frac{d^4p}{(2\pi)^4i}\,\Big[\bar{u}_e \,
  \gamma^{\mu}(1 - \gamma^5) v_{\bar{\nu}}\Big]\nonumber\\ &&\times
\,\Big[\bar{u}_p \, \gamma^{\alpha}(1 - \gamma^5)\,\frac{1}{m_N -
    \hat{k}_n - \hat{p} - i0}\, \gamma^5 \, \frac{1}{m_N - \hat{k}_n -
    \hat{k}- \hat{p} - i0}\,\gamma^{\beta} \,\frac{1}{m_N -
    \hat{k}_n - \hat{k} - i0}\,\gamma^5 u_n\Big] \nonumber\\
\hspace{-0.3in}&& \times \, \frac{1}{m^2_{\pi} - k^2 - i0}\,
\frac{1}{p^2 + i0}\, \frac{1}{M^2_W - (p - q)^2 - i0}\,\big[(p -
  q)_{\mu}\eta_{\beta\alpha} - (p - 2 q)_{\beta}\eta_{\alpha \mu} -
  q_{\alpha} \eta_{\mu \beta}\big],
\end{eqnarray}
where we have made a replacement
\begin{eqnarray}\label{eq:E4e.2}
  M^2_W D^{(W)}_{\alpha_1 \alpha_2}(p - q) D^{(W)}_{\mu\nu} (- q) \to
  \frac{\eta_{\alpha_1\alpha_2} \eta_{\mu\nu}}{M^2_W - (p - q)^2 -
    i0},
\end{eqnarray}
according to our experience, obtained during the calculation of the
Feynman diagrams in Fig.\,\ref{fig:fig1}a, Fig.\,\ref{fig:fig1}d and
so on.  The calculation of the matrix element Eq.(\ref{eq:E4e.1}) we
reduce to the calculation of the following integrals
\begin{eqnarray}\label{eq:E4e.3}
\hspace{-0.3in} &&{\cal F}(k_n, k_e, q)_{4e} = - \int
\frac{d^4k}{(2\pi)^4i}\int \frac{d^4p}{(2\pi)^4i}\,\frac{1}{m^2_N -
  (k_n + p)^2 - i0}\, \frac{1}{m^2_N - (k_n + k + p)^2 -
  i0}\,\frac{1}{m^2_N - (k_n + k)^2 -
  i0} \nonumber\\
\hspace{-0.3in}&& \times \,\frac{1}{m^2_{\pi} - k^2 - i0} \,
\frac{1}{p^2 + i0}\, \frac{1}{M^2_W - p^2 - i0}\,\Big(4 \hat{p} (1 -
\gamma^5) \otimes \big(\big(p^2 + p \cdot k\big)\, \hat{k}\, (1 -
\gamma^5) + \gamma^{\mu}(1 - \gamma^5) \otimes \gamma_{\mu} (p^2 k^2 -
p^2 \hat{p}\hat{k} \nonumber\\
\hspace{-0.3in}&& - 2 (k\cdot p) \hat{p} \hat{k}\big) \,(1 -
\gamma^5)\Big),
\end{eqnarray}
where we have kept only leading divergent contributions that
corresponds to the use of the LLA.  Merging the denominators by using
the Feynman parametrization \cite{Feynman1950, Kinoshita1962,
  Kinoshita1974a, Kinoshita1974b} and after the diagonalization and
integration over the virtual momenta in the $n$-dimensional momentum
space \cite{Smirnov2004, Smirnov2006, Smirnov2012}, we arrive at the
following expression for the integral
\begin{eqnarray}\label{eq:E4e.4}
\hspace{-0.3in} &&{\cal F}(k_n, k_e, q)_{4e} = \int^1_0 dx_1 \int^1_0
dx_2 \int^1_0 dx_3 \int^1_0 dx_4 \int^1_0 dx_5 \int^1_0 dx_6\,
\frac{\delta(1 - x_1 - x_2 - x_3 - x_4 - x_5 - x_6)}{2^{2n} \pi^n
  U^{n/2}} \nonumber\\ \hspace{-0.30in}&& \times \,\Bigg\{ \frac{n^2 -
  2 n + 4}{4}\,\frac{\displaystyle \Gamma\Big(2 -
  \frac{n}{2}\Big)\Gamma\Big(2 - \frac{n - 1}{2}\Big)}{\displaystyle
  2^{n - 3}\Gamma\Big(\frac{1}{2}\Big)}\,\Big(\frac{Q}{m^2_N}\Big)^{-4
  + n}\, \gamma^{\mu} (1 - \gamma^5) \otimes \gamma_{\mu}(1 -
\gamma^5) - \frac{n}{2}\, \Gamma(5 - n)\,Q^{-5 + n}\,m^{2(4 - n)}_N
\nonumber\\ \hspace{-0.30in}&& \times \, \Big[\Big( \frac{n -
    2}{n}\,b^2_1 + \frac{n - 2}{n}\, b^2_2\Big)\, \gamma^{\mu} (1 -
  \gamma^5) \otimes \gamma_{\mu}(1 - \gamma^5) - \frac{n +
    2}{n}\,\gamma^{\mu} (1 - \gamma^5) \otimes \gamma_{\mu} \hat{b}_2
  \hat{b}_1 + \frac{4}{n}\,\hat{b}_1 (1 - \gamma^5) \otimes \hat{b}_1
  (1 - \gamma^5) \nonumber\\ \hspace{-0.30in}&& + 4\, \frac{n +
    2}{n}\, \hat{b}_2 (1 - \gamma^5) \otimes \hat{b}_1 (1 - \gamma^5)
  + \frac{4}{n}\, \hat{b}_2 (1 - \gamma^5) \otimes \hat{b}_2 (1 -
  \gamma^5) \Big]\Bigg\},
\end{eqnarray}
where we have denoted
\begin{eqnarray}\label{eq:E4e.5}
  \hspace{-0.3in}U&=& (x_1 + x_3 + x_4)(x_2 + x_5 + x_6) + (x_1 + x_4)
  x_3,\nonumber\\
 \hspace{-0.3in}b_1 &=& \frac{(x_2 + x_3 + x_5 + x_6) a_1 - x_3
   a_2}{U} = - \frac{(x_5 + x_6)(x_3 +  x_4) + (x_2 + x_3) x_4}{U}\, k_n = \frac{Y}{U}\,k_n, \nonumber\\
 \hspace{-0.3in}b_2 &=& \frac{(x_1 + x_3 + x_4) a_2 - x_3 a_1}{U} = -
 \frac{(x_3 + x_4) x_2 + (x_2 + x_3) x_1}{U}\, k_n =
 \frac{\bar{Y}}{U}\,k_n, \nonumber\\
 \hspace{-0.3in} Q &=& x_6\, M^2_W + \frac{(x_2 + x_3 + x_5 + x_6)
   a^2_1 - 2 x_3 a_1 \cdot a_2 + (x_1 + x_3 + x_4) a^2_2} {U},
\end{eqnarray}
where $a_1 = - (x_3 + x_4) k_n$ and $a_2 = - (x_2 + x_3) k_n$. For the
definition of $Q$ we have neglected the contributions of order
$O(m^2_{\pi}/M^2_W) \sim 10^{-6}$.  Taking the limit $n \to 4$ and
keeping the divergent contributions proportional to $\Gamma(2 - n/2)$,
we get
\begin{eqnarray}\label{eq:E4e.6}
\hspace{-0.3in} &&{\cal F}(k_n, k_e, q)_{4e} = \int^1_0 dx_1 \int^1_0
dx_2 \int^1_0 dx_3 \int^1_0 dx_4 \int^1_0 dx_5 \int^1_0 dx_6\,
\frac{\delta(1 - x_1 - x_2 - x_3 - x_4 - x_5 - x_6)}{256 \pi^4 U^2}
\nonumber\\ \hspace{-0.30in}&& \times \,\Big\{\Big(\frac{3}{2} \,
\Gamma\Big(2 - \frac{n}{2}\Big) - 3\,{\ell n}\frac{Q}{m^2_N}\Big)\,
\gamma^{\mu} (1 - \gamma^5) \otimes \gamma_{\mu}(1 - \gamma^5) +
\frac{1}{Q}\, \Big[(- b^2_1 - b^2_2)\, \gamma^{\mu} (1 - \gamma^5)
  \otimes \gamma_{\mu}(1 - \gamma^5) \nonumber\\ \hspace{-0.30in}&& +
  3\,\gamma^{\mu} (1 - \gamma^5) \otimes \gamma_{\mu} \hat{b}_2
  \hat{b}_1 - 2\,\hat{b}_1 (1 - \gamma^5) \otimes \hat{b}_1 (1 -
  \gamma^5) - 12\, \hat{b}_2 (1 - \gamma^5) \otimes \hat{b}_1 (1 -
  \gamma^5) - 2\, \hat{b}_2 (1 - \gamma^5) \otimes \hat{b}_2 (1 -
  \gamma^5) \Big]\Big\},\nonumber\\ \hspace{-0.30in}&&
\end{eqnarray}
where we have neglected the contributions of order $O(k_n\cdot
q/M^2_W) \sim 10^{-7}$.  Using the Dirac equation for a free neutron,
we obtain for the integral ${\cal F}(k_n, k_e, q)_{4e}$ the following
Lorentz structure
\begin{eqnarray*}
\hspace{-0.3in} &&{\cal F}(k_n, k_e, q)_{4e} = \int^1_0
dx_1 \int^1_0 dx_2 \int^1_0 dx_3 \int^1_0 dx_4 \int^1_0 dx_5 \int^1_0
dx_6\, \frac{\delta(1 - x_1 - x_2 - x_3 - x_4 - x_5 - x_6)}{256 \pi^4
  U^2} \nonumber\\ \hspace{-0.30in}&& \times \,\Bigg\{\Big(
\frac{3}{2} \, \Gamma\Big(2 - \frac{n}{2}\Big) + f^{(V)}_{4e}(x_1,
\ldots, x_6) + g^{(W)}_{4e}(x_1, \ldots, x_6)\,
\frac{m^2_N}{M^2_W}\Big)\, \gamma^{\mu} (1 - \gamma^5) \otimes
\gamma_{\mu} + \Big( - \frac{3}{2} \, \Gamma\Big(2 - \frac{n}{2}\Big)
\nonumber\\
\end{eqnarray*}
\begin{eqnarray}\label{eq:E4e.7}
\hspace{-0.30in}&& + f^{(A)}_{4e}(x_1, \ldots, x_6) +
h^{(W)}_{4e}(x_1, \ldots, x_6)\, \frac{m^2_N}{M^2_W}\Big)\,
\gamma^{\mu} (1 - \gamma^5) \otimes \gamma_{\mu} \gamma^5 +
f^{(W)}_{4e} (x_1, \ldots, x_6)\,
\frac{m^2_N}{M^2_W}\,\frac{\hat{k}_n}{m_N} (1 - \gamma^5) \otimes
1\Big\},~~~~~
\end{eqnarray}
where we have denoted
\begin{eqnarray}\label{eq:E4e.8}
\hspace{-0.3in}f^{(V)}_{4e}(x_1, \ldots,x_6) = - 3\,{\ell
  n}\frac{M^2_W}{m^2_N} - 3\, {\ell n}\bar{Q}
\end{eqnarray}
and
\begin{eqnarray}\label{eq:E4e.9}
\hspace{-0.3in}&&g^{(W)}_{4e}(x_1, \ldots,x_6) = \frac{1}{\bar{Q}}
\Big[- \frac{Y^2}{U^2} + 3\,
  \frac{Y\bar{Y}}{U^2} - \frac{\bar{Y}^2}{U^2}\Big]
\end{eqnarray}
and
\begin{eqnarray}\label{eq:E4e.10}
\hspace{-0.3in}f^{(A)}_{4e}(x_1, \ldots,x_6) = 3\, {\ell
  n}\frac{M^2_W}{m^2_N} + 3\, {\ell n}\bar{Q},
\end{eqnarray}
where $f^{(A)}_{4e}(x_1, \ldots,x_6) = - f^{(V)}_{4e}(x_1,
\ldots,x_6)$, and
\begin{eqnarray}\label{eq:E4e.11}
\hspace{-0.3in}&&h^{(W)}_{4e}(x_1, \ldots,x_6) = - \frac{1}{\bar{Q}}
\Big[ - \frac{Y^2}{U^2} + 3\,
  \frac{Y\bar{Y}}{U^2} - \frac{\bar{Y}^2}{U^2}\Big],
\end{eqnarray}
where $h^{(W)}_{4e}(x_1, \ldots,x_6) = - g^{(W)}_{4e}(x_1,
\ldots,x_6)$, and
\begin{eqnarray}\label{eq:E4e.12}
\hspace{-0.3in}f^{(W)}_{4e}(x_1, \ldots,x_6) = \frac{1}{\bar{Q}} \Big[
  - 2\, \frac{Y^2}{U^2} - 12\, \frac{Y\bar{Y}}{U^2} - 2 \,
  \frac{\bar{Y}^2}{U^2} \Big].
\end{eqnarray}
For the calculation of the integrals over the Feynman parameters we
use
\begin{eqnarray}\label{eq:E4e.13}
 \hspace{-0.3in} \bar{Q} &=& \frac{1}{U}\Big(x_6 U + \big((x_5 + x_6)
 (x_3 + x_4)^2 + (x_2 + x_3) (x_3 + x_4) (x_2 + x_4) + x_1 (x_2 +
 x_3)^2\big)\, \frac{m^2_N}{M^2_W}\Big),\nonumber\\
  \hspace{-0.3in} Y &=& - (x_5 + x_6)(x_3 + x_4) - (x_2 + x_3) x_4
  \;,\; \bar{Y} = - (x_3 + x_4) x_2 - (x_2 + x_3) x_1,
\end{eqnarray}
where $m_N = (m_n + m_p)/2 = 938.9188\, {\rm MeV}$ and $M_W =
80.379\,{\rm GeV}$.  After the integration over the Feynman parameters,
the contribution of the Feynman diagram in Fig.\,\ref{fig:fig4}e to
the amplitude of the neutron beta decay  is equal to
\begin{eqnarray}\label{eq:E4e.14}
\hspace{-0.3in}&& M(n \to p e^- \bar{\nu}_e)^{(\pi^0)}_{\rm
  Fig.\,\ref{fig:fig4}e} = \frac{\alpha}{2\pi} \frac{g^2_{\pi
    N}}{16\pi^2} G_V \Big\{ \Big(A_{4e}\,\Gamma\big(2 -
\frac{n}{2}\Big) + F^{(V)}_{4e} + G^{(W)}_{4e}
\frac{m^2_N}{M^2_W}\Big)\, \Big[\bar{u}_e \gamma^{\mu}(1 - \gamma^5)
  v_{\bar{\nu}}\Big]\Big[\bar{u}_p \gamma_{\mu} u_n\Big]
\nonumber\\ \hspace{-0.30in}&& + \Big(B_{4e}\,\Gamma\big(2 -
\frac{n}{2}\Big) + F^{(A)}_{4e} + H^{(W)}_{4e}
\frac{m^2_N}{M^2_W}\Big)\, \Big[\bar{u}_e \gamma^{\mu}(1 - \gamma^5)
  v_{\bar{\nu}}\Big]\Big[\bar{u}_p \gamma_{\mu} \gamma^5 u_n\Big] +
F^{(W)}_{4e} \frac{m^2_N}{M^2_W}\,\Big[\bar{u}_e
  \frac{\hat{k}_n}{m_N}(1 - \gamma^5)
  v_{\bar{\nu}}\Big]\nonumber\\ \hspace{-0.30in}&& \times
\Big[\bar{u}_p u_n\Big]\Big\},
\end{eqnarray}
where the coefficients are equal to $A_{4e} = 1/4$, $F^{(V)}_{4e} = -
3.3962$, $G^{(W)}_{4e} = 0.1746$, $B_{4e} = - 1/4$, $F^{(A)}_{4e} =
3.3962$, $H^{(W)}_{4e} = - 0.1746$ and $F^{(W)}_{4e} = - 22.0375$. The
Lorentz structure of Eq.(\ref{eq:E4e.14}) is calculated at the neglect
the contributions of order $O(k_n \cdot q/M^2_W) \sim 10^{-7}$ and
$O(m^2_{\pi}/M^2_W) \sim 10^{-6}$, respectively.

\subsection*{E4f. Analytical calculation of the Feynman diagram in
  Fig.\,\ref{fig:fig4}f}
\renewcommand{\theequation}{E4f-\arabic{equation}}
\setcounter{equation}{0}

In the Feynman gauge for the photon propagator the analytical
expression of the Feynman diagram in Fig.\,\ref{fig:fig4}f is given by
(see Eq.(\ref{eq:A.1}))
\begin{eqnarray}\label{eq:E4f.1}
\hspace{-0.30in}&&M(n \to p e^- \bar{\nu}_e)_{\rm
  Fig.\,\ref{fig:fig4}f} = - 2 e^2 g^2_{\pi N} G_V \int
\frac{d^4k}{(2\pi)^4i}\int
\frac{d^4p}{(2\pi)^4i}\,\Big[\bar{u}_e\,\gamma^{\beta} \,\frac{1}{m_e
    - \hat{k}_e - \hat{p} - i0}\, \gamma^{\mu}(1 - \gamma^5)
  v_{\bar{\nu}}\Big] \nonumber\\ \hspace{-0.30in}&&
\times\,\Big[\bar{u}_p \, \gamma^{\nu}(1 - \gamma^5) \,\frac{1}{m_N -
    \hat{k}_n - \hat{p} - i0}\, \gamma^5 \, \frac{1}{m_N - \hat{k}_n -
    \hat{k}- \hat{p} - i0}\,\gamma_{\beta} \,\frac{1}{m_N - \hat{k}_n
    - \hat{k} - i0}\,\gamma^5 u_n
  \Big]\nonumber\\ \hspace{-0.30in}&&\times\, \frac{1}{m^2_{\pi} - k^2
  - i0}\,\frac{1}{p^2 + i0} \,\Big(\eta_{\mu\nu} + \frac{(p - q)^2
  \eta_{\mu\nu} - (p - q)_{\mu} (p - q)_{\nu}}{M^2_W - (p - q)^2 -
  i0}\Big)
\end{eqnarray}
where we have taken the propagator of the electroweak $W^-$-boson
$D^{(W)}_{\mu\nu}(p - q)$ in the form
\begin{eqnarray*}
  D^{(W)}_{\mu\nu}(p - q) = - \frac{1}{M^2_W}\Big(\eta_{\mu\nu} +
  \frac{(p - q)^2 \eta_{\mu\nu} - (p - q)_{\mu} (p - q)_{\nu}}{M^2_W -
    (p - q)^2 - i0}\Big).
\end{eqnarray*}
The calculation of the matrix element Eq.(\ref{eq:E4f.1}) we carry out
by calculating the following two integrals
\begin{eqnarray}\label{eq:E4f.2}
\hspace{-0.30in}&&{\cal F}^{(1)}(k_n, k_e, q)_{4f} = -
\int\frac{d^4k}{(2\pi)^4i}\int \frac{d^4p}{(2\pi)^4i}\, \frac{1}{m^2_e
  - (p + k_e)^2 - i0} \frac{1}{m^2_N - (k_n + p)^2 - i0}
\frac{1}{m^2_N - (k_n + p + k)^2 - i0}\nonumber\\ &&\times
\frac{1}{m^2_N - (k_n + k)^2 - i0} \frac{1}{m^2_{\pi} - k^2 - i0}
\frac{1}{p^2 + i0}\, \Big(\gamma^{\beta} \hat{p} \gamma^{\mu}(1 -
\gamma^5) \otimes \gamma_{\mu} (p^2 + \hat{p} \hat{k}) \gamma_{\beta}
\hat{k}(1 - \gamma^5)\Big)
\end{eqnarray}
and
\begin{eqnarray}\label{eq:E4f.3}
\hspace{-0.30in}&&{\cal F}^{(2)}(k_n, k_e, q)_{4f} = -
\int\frac{d^4k}{(2\pi)^4i}\int \frac{d^4p}{(2\pi)^4i}\, \frac{1}{m^2_e
  - (p + k_e)^2 - i0} \frac{1}{m^2_N - (k_n + p)^2 - i0}
\frac{1}{m^2_N - (k_n + p + k)^2 -
  i0}\nonumber\\ \hspace{-0.30in}&&\times \frac{1}{m^2_N - (k_n + k)^2
  - i0} \frac{1}{m^2_{\pi} - k^2 - i0} \frac{1}{p^2 +
  i0}\,\frac{1}{M^2_W - (p - q)^2 - i0}\, \Big(p^2\, \gamma^{\beta}
\hat{p}\gamma^{\mu}(1 - \gamma^5) \otimes \gamma^{\nu} ( p^2 + \hat{p}
\hat{k}) \gamma_{\beta} \hat{k} \nonumber\\ \hspace{-0.30in}&& - p^4
\gamma^{\mu} (1 - \gamma^5) \otimes (\hat{p} + \hat{k}) \gamma_{\mu}
\hat{k} (1 - \gamma^5)\Big),
\end{eqnarray}
where we have kept only leading divergent contributions that
corresponds to the use of the LLA.

\subsubsection*{\bf Calculation of the integral
  ${\cal F}^{(1)}(k_n, k_e, q)_{4f}$}

Skipping standard intermediate calculations, we define the integral
${\cal F}^{(1)}(k_n, k_e, q)_{4f}$ in terms of the integrals over the
Feynman parameters. We get
\begin{eqnarray}\label{eq:E4f.4}
\hspace{-0.3in}&&{\cal F}^{(1)}(k_n, k_e, q)_{4f} = \int^1_0dx_1
\int^1_0dx_2 \int^1_0dx_3 \int^1_0dx_4 \int^1_0dx_5 \int^1_0
dx_6\,\frac{\delta(1 - x_1 - x_2 - x_3 - x_4 - x_5 - x_6)}{2^{2n}
  \pi^n U^{n/2}} \nonumber\\
\hspace{-0.3in}&& \times \Bigg\{- \frac{n - 2}{4}\,\frac{\displaystyle
  \Gamma\Big(2 - \frac{n}{2}\Big)\Gamma\Big(2 - \frac{n -
    1}{2}\Big)}{\displaystyle 2^{n -
    3}\Gamma\Big(\frac{1}{2}\Big)}\,\Big(\frac{Q}{m^2_N}\Big)^{-4 +
  n}\,\gamma^{\beta} \gamma^{\alpha} \gamma^{\mu} (1 - \gamma^5)
\otimes \gamma_{\mu} \gamma_{\alpha} \gamma_{\beta} (1 - \gamma^5) +
\ldots \Bigg\},
\end{eqnarray}
where we have denoted
\begin{eqnarray*}
  \hspace{-0.3in}U &=& (x_1 + x_3 + x_4)(x_2 + x_5 + x_6) + (x_1 +
  x_4)x_3,\nonumber\\
  \hspace{-0.3in}b_1 &=& \frac{(x_2 + x_3 + x_5 + x_6)a_1 -  x_3
    a_2}{U} =  \frac{ -((x_5 + x_6)(x_3 + x_4) + (x_2 + x_3)
    x_4) k_n + x_3 x_5 k_e}{U} = \frac{ Y k_n + Z k_e}{U},\nonumber\\
  \end{eqnarray*}
 \begin{eqnarray}\label{eq:E4f.5} 
   \hspace{-0.3in}b_2 &=& \frac{(x_1 + x_3 + x_4)a_2 - x_3 a_1}{U} = -
   \frac{- ((x_3 + x_4)x_2 + (x_2 + x_3) x_1) k_n - (x_1 + x_3 +
     x_4) x_5 k_e}{U} = \frac{ \bar{Y} k_n + \bar{Z} k_e}{U},\nonumber\\
   \hspace{-0.3in}Q &=& x_1 m^2_{\pi} + \frac{(x_2 + x_3 + x_5
     + x_6) a^2_1 - 2 x_3 a_1\cdot a_2 + (x_1 + x_3 + x_4) a^2_2}{U}
\end{eqnarray}
with $a_1 = - (x_3 + x_4) k_n$ and $a_2 = - (x_2 + x_3) k_n - x_5
k_e$.  Taking the limit $n \to 4$ and keeping the divergent part
proportional to $\Gamma(2 - n/2)$, we transcribe the integral
Eq.(\ref{eq:E4f.4}) into the form
\begin{eqnarray}\label{eq:E4f.6}
\hspace{-0.3in}&&{\cal F}^{(1)}(k_n, k_e, q)_{4f} = \int^1_0dx_1
\int^1_0dx_2 \int^1_0dx_3 \int^1_0dx_4 \int^1_0dx_5 \int^1_0
dx_6\,\frac{\delta(1 - x_1 - x_2 - x_3 - x_4 - x_5 - x_6)}{256 \pi^4
  U^2}\nonumber\\
\hspace{-0.3in}&& \times \Big\{\Big(- \Gamma\Big(2 - \frac{n}{2}\Big) +
2\,{\ell n} \frac{Q}{m^2_N}\Big)\, \gamma^{\mu} (1 - \gamma^5) \otimes
\gamma_{\mu} (1 - \gamma^5) + \ldots \Big\},
\end{eqnarray}
where we have denoted
\begin{eqnarray}\label{eq:E4f.7} 
\hspace{-0.3in}f^{(V)}_{4f^{(1)}}(x_1,\ldots,x_6) &=& + 2\,{\ell
  n}\frac{\bar{Q}}{m^2_N} \quad,\quad
h^{(V)}_{4f^{(1)}}(x_1,\ldots,x_6) = + 4\, \frac{\bar{V}}{U}\,
\frac{m^2_N}{\bar{Q}},\nonumber\\
\hspace{-0.3in}f^{(A)}_{4f^{(1)}}(x_1,\ldots,x_6) &=& - 2\,{\ell
  n}\frac{\bar{Q}}{m^2_N} \quad,\quad
h^{(V)}_{4f^{(1)}}(x_1,\ldots,x_6) = - 4\, \frac{\bar{V}}{U}\,
\frac{m^2_N}{\bar{Q}}.
\end{eqnarray}
For the calculation of the integrals over the Feynman parameters we
use
\begin{eqnarray}\label{eq:E4f.8}
  \hspace{-0.30in}\frac{\bar{Q}}{m^2_N} &=& \frac{1}{U}\Big( x_1\, U\,
  \frac{m^2_{\pi}}{m^2_N} + (x_5 + x_6) (x_3 + x_4)^2 + (x_2 + x_3)
  (x_3 + x_4) (x_2 + x_4) + x_1 (x_2 + x_3)^2\Big),
  \nonumber\\ \hspace{-0.21in} Y &=& - (x_5 + x_6)(x_3 + x_4) - (x_2 +
  x_3) x_4\;,\; Z = x_3
  x_5,\nonumber\\ \hspace{-0.30in} \bar{Y} &=& - (x_3 + x_4)x_2 - (x_2
  + x_3) x_1\;,\; \bar{Z} = - (x_1 + x_3 + x_4)
  x_5,\nonumber\\ \hspace{-0.30in} \bar{V} &=& (x_3 + x_4) x_2 x_5 +
  (x_2 + x_3) x_1 x_5,
\end{eqnarray}
where $m_N = (m_n + m_p)/2 = 938.9188\,{\rm MeV}$ and $m_{\pi} =
139.5706\, {\rm MeV}$ \cite{PDG2020}. After the integration over the
Feynman parameters the contribution of the integral ${\cal
  F}^{(1)}(k_n, k_e, q)_{4f^{(1)}}$ to the matrix element $M(n \to p
e^-\bar{\nu}_e)^{(\pi^0)}_{\rm Fig.\,\ref{fig:fig4}c}$ takes the form
\begin{eqnarray}\label{eq:E4f.9}
 \hspace{-0.30in}&& M(n \to p e^- \bar{\nu}_e)^{(\pi^0)}_{\rm
   Fig.\,\ref{fig:fig4}f^{(1)}}= - \frac{\alpha}{2\pi} \frac{g^2_{\pi
     N}}{16\pi^2}\,G_V \Big\{ \Big(A_{4f^{(1)}}\Gamma\Big(2 -
 \frac{n}{2}\Big) + F^{(V)}_{4f^{(1)}} + H^{(V)}_{4f^{(1)}}\,
 \frac{k_n \cdot k_e}{m^2_N}\Big) \Big[\bar{u}_e \gamma^{\mu} (1 -
   \gamma^5) v_{\bar{\nu}}\Big]\Big[\bar{u}_p \gamma_{\mu} u_n\Big]
 \nonumber\\
\hspace{-0.3in}&& + \Big(B_{4f^{(1)}}\Gamma\Big(2 - \frac{n}{2}\Big) +
F^{(A)}_{4f^{(1)}} + H^{(A)}_{4f^{(1)}}\, \frac{k_n \cdot k_e}{m^2_N
}\Big) \Big[\bar{u}_e \gamma^{\mu} (1 - \gamma^5)
  v_{\bar{\nu}}\Big]\Big[\bar{u}_p \gamma_{\mu}\gamma^5
  u_n\Big]\Big\}.
\end{eqnarray}
The structure constants are equal to $A_{4f^{(1)}} = - 1/6$,
$F^{(V)}_{4f^{(1)}} = - 0.5016$, $H^{(V)}_{4f^{(1)}} = 0.2488$,
$B_{4f^{(1)}} = 1/6$, $F^{(A)}_{4f^{(1)}} = 0.5016$ and
$H^{(A)}_{4f^{(1)}} = - 0.2488$. The Lorentz structure of
Eq.(\ref{eq:E4f.9}) is obtained at the neglect of the contributions of
order $O(E^2_0/m^2_N) \sim 10^{-6}$.

\subsubsection*{\bf Calculation of the integral
  ${\cal F}^{(2)}(k_n, k_e, q)_{4f}$}

In terms of the integrals over the Feynman parameters and after
integration over the virtual momenta in the $n$-dimensional momentum
space, we obtain for the integral ${\cal F}^{(2)}(k_n, k_e, q)_{4f}$
the following expression
\begin{eqnarray*}
\hspace{-0.3in}&&{\cal F}^{(2)}(k_n, k_e, q)_{4f} = \int^1_0 \!\!
dx_1 \int^1_0 \!\! dx_2 \int^1_0 \!\! dx_3 \int^1_0 \!\! dx_4 \int^1_0
\!\! dx_5 \int^1_0 \!\! dx_6 \int^1_0 \!\! dx_7\,\frac{\delta(1 - x_1
  - x_2 - x_3 - x_4 - x_5 - x_6 - x_7)}{2^{2n} \pi^n
  U^{n/2}} \nonumber\\
\hspace{-0.3in}&& \times \Bigg\{ - \frac{n^2 (n +
  2)}{8}\,\frac{\displaystyle \Gamma\Big(2 -
  \frac{n}{2}\Big)\Gamma\Big(2 - \frac{n - 1}{2}\Big)}{\displaystyle
  2^{n - 3}\Gamma\Big(\frac{1}{2}\Big)}\,\Big(\frac{Q}{m^2_N}\Big)^{-4
  + n}\,\Big[\frac{n - 2}{n}\, \gamma^{\mu} (1 - \gamma^5) \otimes
  \gamma_{\mu} (1 - \gamma^5) - \frac{n - 2}{n^2}\, \gamma^{\beta}
  \gamma^{\alpha} \gamma^{\mu} (1 - \gamma^5) \nonumber\\
\hspace{-0.3in}&& \otimes \gamma_{\mu} \gamma_{\alpha} \gamma_{\beta}
(1 - \gamma^5)\Big] + \frac{n^2}{4} \Gamma(5 - n)\,Q^{-5+n}
m^{2(4-n)}_N \Big[2\, \frac{n^2 - 4}{n^2}\, b^2_2 \,\gamma^{\mu} (1 -
  \gamma^5) \otimes \gamma_{\mu} (1 - \gamma^5) - \frac{n -
    2}{n^2}\,b^2_2 \gamma^{\beta} \gamma^{\alpha} \gamma^{\mu} (1 -
  \gamma^5) \nonumber\\
\hspace{-0.3in}&& \otimes \gamma_{\mu} \gamma_{\alpha} \gamma_{\beta}
(1 - \gamma^5) + \frac{n + 2}{n^2}\, \gamma^{\beta} \gamma^{\alpha}
\gamma^{\mu} (1 - \gamma^5) \otimes \gamma_{\mu} \gamma_{\alpha}
\hat{b}_1 \gamma_{\beta} \hat{b}_1 (1 - \gamma^5) + \frac{(n + 2)(n +
  4)}{n^2}\, \gamma^{\beta} \hat{b}_2 \gamma^{\alpha} (1 - \gamma^5)
\otimes \, \gamma_{\alpha} \gamma_{\beta} \hat{b}_1 \nonumber\\
  \end{eqnarray*}
\begin{eqnarray}\label{eq:E4f.10}
\hspace{-0.3in}&& \times (1 - \gamma^5) + \gamma^{\mu} (1 - \gamma^5)
\otimes \Big(- \frac{n + 2}{n}\,\hat{b}_1 \gamma_{\mu} \hat{b}_1 -
\frac{(n + 2)(n + 4)}{n^2}\, \hat{b}_2 \gamma_{\mu} \hat{b}_1\Big) \,
(1 - \gamma^5) - \frac{(n - 2)(n + 4)}{n^2}\, \gamma^{\beta} \hat{b}_2
\gamma^{\alpha} (1 - \gamma^5)\nonumber\\
\hspace{-0.3in}&& \otimes \, \gamma_{\alpha} \hat{b}_2
\gamma_{\beta}\, (1 - \gamma^5)\Big]\Bigg\},
\end{eqnarray} 
where we have used the algebra of the Dirac $\gamma$-matrices in
$n$-dimensional space-time  and denoted
\begin{eqnarray*}
  \hspace{-0.3in}U &=& (x_1 + x_3 + x_4)( x_2 + x_5 + x_6 + x_7) +
  (x_1 + x_4) x_3,\nonumber\\
  \hspace{-0.3in}b_1 &=& \frac{(x_2 + x_3 + x_5 + x_6 + x_7)a_1 - x_3
    a_2}{U} = - \frac{(x_5 + x_6 + x_7) (x_3 + x_4) + (x_2 + x_3)
    x_4 }{U}\, k_n = \frac{Y}{U}\,k_n ,\nonumber\\
   \hspace{-0.3in}b_2 &=& \frac{(x_1 + x_3 + x_4)a_2 - x_3 a_1}{U} = -
   \frac{(x_3 + x_4) x_2 + (x_2 + x_3) x_1}{U}\, k_n =
   \frac{\bar{Y}}{U}\,k_n,\nonumber\\
\end{eqnarray*}
  \begin{eqnarray}\label{eq:E4f.11} 
   \hspace{-0.3in}Q &=&x_7 M^2_W + \frac{(x_2 + x_3 + x_5 + x_6 + x_7)
     a^2_1 - 2 x_3 a_1\cdot a_2 + (x_1 + x_3 + x_4) a^2_2}{U}
\end{eqnarray}
with $a_1 = - (x_3 + x_4) k_n$ and $a_2 = - (x_2 + x_3) k_n$. In
Eq.(\ref{eq:E4f.10}) we have neglected the contributions of the terms
proportional to $q$ and $k_e$, appearing in ${\cal F}^{(2)}(k_n, k_e,
q)_{4f}$ in the form of an expansion in powers of $k_n\cdot k_e/M^2_W$
of order $10^{-7}$, and the terms of order $O(m^2_{\pi}/M^2_W) \sim
10^{-6}$. Taking the limit $n \to 4$ and using the Dirac equations for
free fermions, we arrive at the expression
\begin{eqnarray}\label{eq:E4f.12}
 \hspace{-0.3in}&& {\cal F}^{(2)}(k_n, k_e, q)_{4f} = \int^1_0 dx_1
 \int^1_0 dx_2 \int^1_0 dx_3 \int^1_0 dx_4 \int^1_0 dx_5 \int^1_0 dx_6
 \int^1_0 dx_7 \frac{\delta(1 - x_1 - x_2 - x_3 - x_4 - x_5 - x_6 -
   x_7)}{256 \pi^4 U^2} \nonumber\\ \hspace{-0.3in}&& \times
 \Big\{\Big(f^{(V)}_{4f^{(2)}}(x_1, \ldots, x_7) +
 g^{(W)}_{4f^{(2)}}(x_1, \ldots, x_7)\, \frac{m^2_N}{M^2_W}\Big)
 \gamma^{\mu} (1 - \gamma^5) \otimes \gamma_{\mu} +
 \Big(f^{(A)}_{4f^{(2)}}(x_1, \ldots, x_7) + h^{(W)}_{4f^{(2)}}(x_1,
 \ldots, x_6)\, \frac{m^2_N}{M^2_W}\Big) \nonumber\\ \hspace{-0.3in}&&
 \times \gamma^{\mu} (1 - \gamma^5) \otimes \gamma_{\mu} \gamma^5 +
 f^{(W)}_{4f^{(2)}}(x_1, \ldots, x_7)\, \frac{m^2_N}{M^2_W}\,
 \frac{\hat{k}_n}{m_N} (1 - \gamma^5) \otimes 1\Big\},
 \nonumber\\ \hspace{-0.3in}&&
\end{eqnarray}
where we have denoted
\begin{eqnarray}\label{eq:E4f.13}
 \hspace{-0.30in}&&f^{(V)}_{4f^{(2)}}(x_1, \ldots, x_7) = \frac{3}{2}
\end{eqnarray}
and
\begin{eqnarray}\label{eq:E4f.14}
 \hspace{-0.30in}&&g^{(W)}_{4f^{(2)}}(x_1, \ldots, x_7) =
 \frac{1}{\bar{Q}}\Big[- 12\, \frac{Y^2}{U^2} + 24\,
   \frac{Y\bar{Y}}{U^2} - 8\, \frac{\bar{Y}^2}{U^2}\Big]
\end{eqnarray}
and
\begin{eqnarray}\label{eq:E4f.15}
 \hspace{-0.30in}&&f^{(A)}_{4f^{(2)}}(x_1, \ldots, x_7) = -
 \frac{3}{2},
\end{eqnarray}
where $f^{(A)}_{4f^{(2)}}(x_1, \ldots, x_7) = -
f^{(V)}_{4f^{(2)}}(x_1, \ldots, x_7)$, and
\begin{eqnarray}\label{eq:E4f.16}
 \hspace{-0.30in}h^{(W)}_{4f^{(2)}}(x_1, \ldots, x_7) =
 \frac{1}{\bar{Q}}\Big[- 24\, \frac{Y\bar{Y}}{U^2} - 16\,
   \frac{\bar{Y}^2}{U^2}\Big]
\end{eqnarray}
and
\begin{eqnarray}\label{eq:E4f.17}
 \hspace{-0.30in}&&f^{(W)}_{4f^{(2)}}(x_1, \ldots, x_7) =
 \frac{1}{\bar{Q}}\Big[12\, \frac{Y^2}{U^2} - 48\,
   \frac{Y\bar{Y}}{U^2} - 16\, \frac{\bar{Y}^2}{U^2}\Big].
\end{eqnarray}
For the calculation of the integrals over the Feynman parameters we
use
\begin{eqnarray}\label{eq:E4f.18}
  \hspace{-0.30in}\bar{Q} &=& \frac{1}{U}\Big(x_7 U + \big((x_5 +
  x_6 + x_7) (x_3 + x_4)^2 + (x_2 + x_3 )(x_3 + x_4) (x_2 + x_4) + x_1
  (x_2 + x_3)^2\big)\, \frac{m^2_N}{M^2_W}\Big), \nonumber\\
  \hspace{-0.30in} U &=& (x_1 + x_3 + x_4)(x_2 + x_5 + x_6 + x_7) +
  (x_1 + x_4) x_3,\nonumber\\
  \hspace{-0.30in} Y &=& - (x_5 + x_6 + x_7) (x_3 + x_4) - (x_2 + x_3)
    x_4 \;,\; \bar{Y} = - (x_3 + x_4) x_2 - (x_2 + x_3) x_1,
\end{eqnarray}
where $m_N = (m_n + m_p)/2 = 938.9188\,{\rm MeV}$ and $M_W =
30.379\,{\rm GeV}$ \cite{PDG2020}. After the integration over the
Feynman parameters, we obtain the contribution of the integral ${\cal
  F}^{(2)}(k_n, k_e, q)_{4f}$ to the amplitude of the neutron
beta decay 
\begin{eqnarray}\label{eq:E4f.19}
 \hspace{-0.30in}&&M(n \to p e^- \bar{\nu}_e)^{(\pi^0)}_{\rm
   Fig.\,\ref{fig:fig4}f^{(2)}} = - \frac{\alpha}{2\pi}\frac{g^2_{\pi
     N}}{16\pi^2}\, G_V \Big\{\Big(F^{(V)}_{4f^{(2)}} +
 G^{(W)}_{4f^{(2)}}\,\frac{m^2_N}{M^2_W}\Big) \Big[\bar{u}_e
   \gamma^{\mu} (1 - \gamma^5) v_{\bar{\nu}}\Big] \Big[\bar{u}_p
   \gamma_{\mu} u_n\Big] \nonumber\\
\hspace{-0.3in}&& + \Big( F^{(A)}_{4f^{(2)}} +
H^{(W)}_{4f^{(2)}}\,\frac{m^2_N}{M^2_W}\Big) \Big[\bar{u}_e
  \gamma^{\mu} (1 - \gamma^5) v_{\bar{\nu}}\Big]\Big[\bar{u}_p
  \gamma_{\mu} \gamma^5 u_n\Big] +
F^{(W)}_{4f^{(2)}}\,\frac{m^2_N}{M^2_W}\Big[\bar{u}_e \frac{\hat{k}_n}{m_N} (1 -
  \gamma^5) v_{\bar{\nu}}\Big] \Big[\bar{u}_p u_n\Big]\Big\}.
\end{eqnarray}
The structure constants are equal to $F^{(V)}_{4f^{(2)}} = 0.0512$,
$G^{(W)}_{4f^{(2)}} = - 2.2050$, $F^{(A)}_{4f^{(2)}} = - 0.0512$,
$H^{(W)}_{4f^{(2)}} = - 8.6147$ and $F^{(W)}_{4f^{(2)}} = -
8.0761$. The Lorentz structure of the matrix element
Eq.(\ref{eq:E4f.19}) is calculated at the neglect of the contributions
of order $O(k_n\cdot q/M^2_W) \sim O(k_n \cdot k_e/M^2_W) \sim
10^{-7}$ and $O(m^2_{\pi}/M^2_W) \sim 10^{-6}$, respectively.

The contribution of the Feynman diagram in Fig.\,\ref{fig:fig4}f to
the amplitude of the neutron beta decay is given by
\begin{eqnarray}\label{eq:E4f.20}
 \hspace{-0.30in}&& M(n \to p e^- \bar{\nu}_e)^{(\pi^0)}_{\rm
   Fig.\,\ref{fig:fig4}f}= - \frac{\alpha}{2\pi} \frac{g^2_{\pi
     N}}{16\pi^2}\,G_V \Big\{ \Big(A_{4f}\Gamma\Big(2 -
 \frac{n}{2}\Big) + F^{(V)}_{4f} + H^{(V)}_{4f}\,
 \frac{k_n \cdot k_e}{m^2_N} +
 G^{(W)}_{4f}\,\frac{m^2_N}{M^2_W}\Big) \Big[\bar{u}_e
   \gamma^{\mu} (1 - \gamma^5) v_{\bar{\nu}}\Big] \nonumber\\
\hspace{-0.3in}&& \times \Big[\bar{u}_p \gamma_{\mu} u_n\Big] +
\Big(B_{4f}\Gamma\Big(2 - \frac{n}{2}\Big) + F^{(A)}_{4f} +
H^{(A)}_{4f}\, \frac{k_n \cdot k_e}{m^2_N } +
H^{(W)}_{4f}\,\frac{m^2_N}{M^2_W}\Big) \Big[\bar{u}_e \gamma^{\mu} (1
  - \gamma^5) v_{\bar{\nu}}\Big]\Big[\bar{u}_p \gamma_{\mu}\gamma^5
  u_n\Big] + F^{(W)}_{4f}\,\frac{m^2_N}{M^2_W}\nonumber\\
\hspace{-0.3in}&& \times \,\Big[\bar{u}_e \frac{\hat{k}_n}{m_N} (1 -
  \gamma^5) v_{\bar{\nu}}\Big]\Big[\bar{u}_p u_n\Big] \Big\}.
\end{eqnarray}
The structure constants are equal to $A_{4f} = A_{4f^{(1)}} = - 1/6$,
$F^{(V)}_{4f } = F^{(V)}_{4f^{(1)}} + F^{(V)}_{4f^{(2)}} = - 0.4548$,
$H^{(V)}_{4f } = H^{(V)}_{4f^{(1)}} = 0.2488$, $G^{(W)}_{4f } =
G^{(W)}_{4f^{(2)}}$, $B_{4f} = B_{4f^{(1)}} = 1/6$, $F^{(A)}_{4f }=
F^{(A)}_{4f^{(1)}} + F^{(A)}_{4f^{(2)}} = - 0.4548$, $H^{(A)}_{4f } =
H^{(A)}_{4f^{(1)}} = - 0.2488$, $H^{(W)}_{4f } = H^{(W)}_{4f^{(2)}} =
- 8.6147$ and $F^{(W)}_{4f } = F^{(W)}_{4f^{(2)}} = - 8.0761$. The
Lorentz structure of Eq.(\ref{eq:E4f.20}) is obtained at the neglect
of the contributions of order $O(E^2_0/m^2_N) \sim O(m^2_{\pi}/M^2_W)
\sim 10^{-6}$.

\subsection*{Fig.4. The contribution  of the  Feynman diagrams in
  Fig.\,\ref{fig:fig4} to the amplitude of the neutron
 beta decay }
\renewcommand{\theequation}{Fig.4-\arabic{equation}}
\setcounter{equation}{0}

Summing up the results of the  calculation of the Feynman
diagrams in Fig.\,\ref{fig:fig4}, given in Appendix E, we obtain the
contribution of the Feynman diagrams in Fig.\,\ref{fig:fig4} to the
amplitude of the neutron beta decay. We get
\begin{eqnarray}\label{eq:S4.1}
 \hspace{-0.30in}&& M(n \to p e^- \bar{\nu}_e)^{(\pi^0)}_{\rm
   Fig.\,\ref{fig:fig4}}= - \frac{\alpha}{2\pi} \frac{g^2_{\pi
     N}}{16\pi^2}\,G_V \Big\{ \Big(A_4\Gamma\Big(2 - \frac{n}{2}\Big)
 + F^{(V)}_4 + G^{(V)}_4\, \frac{k_n \cdot q}{m^2_N} +  H^{(V)}_4\,
 \frac{k_n \cdot k_e}{m^2_N} + G^{(W)}_4\,\frac{m^2_N}{M^2_W}\Big) 
 \nonumber\\
\hspace{-0.3in}&& \times \Big[\bar{u}_e \gamma^{\mu} (1 - \gamma^5)
  v_{\bar{\nu}}\Big] \Big[\bar{u}_p \gamma_{\mu} u_n\Big] +
\Big(B_4\Gamma\Big(2 - \frac{n}{2}\Big) + F^{(A)}_4 + G^{(A)}_4\,
\frac{k_n \cdot q}{m^2_N}+ H^{(A)}_4\, \frac{k_n \cdot k_e}{m^2_N } +
H^{(W)}_4\,\frac{m^2_N}{M^2_W}\Big) \nonumber\\ \hspace{-0.30in}&&
\Big[\bar{u}_e \gamma^{\mu} (1 - \gamma^5)
  v_{\bar{\nu}}\Big]\Big[\bar{u}_p \gamma_{\mu}\gamma^5 u_n\Big] +
F^{(W)}_4\, \frac{m^2_N}{M^2_W} \, \Big[\bar{u}_e
  \frac{\hat{k}_n}{m_N} (1 - \gamma^5) v_{\bar{\nu}}\Big]\,
\Big[\bar{u}_p u_n\Big]\Big\}.
\end{eqnarray}
The structure constants are equal to $A_4 = - 1/3$, $F^{(V)}_4 =
0.3421$, $G^{(V)}_4 = 0.3754$, $H^{(V)}_4 = - 0.1967$, $G^{(W)}_4 = -
5.2961$, $B_4 = 1/6$, $F^{(A)}_4 = - 0.3421$, $G^{(A)}_4 = - 0.3754$,
$H^{(A)}_4 = 0.1967$, $H^{(W)}_4 = - 5.5236$ and $F^{(W)}_4 =
16.3386$.  The Lorentz structure of Eq.(\ref{eq:S4.1}) is obtained at
the neglect of the contributions of order $O(E^2_0/m^2_N) \sim
10^{-6}$. We would like to emphasize that all structure constants in
the matrix element Eq.(\ref{eq:S4.1}) are induced by the contributions
of the first class currents \cite{Weinberg1958}, which are $G$-even
\cite{Lee1956a} (see also \cite{Ivanov2018}).

\newpage

\section*{Appendix F: The contribution  of the hadronic structure of the
  neutron to the radiative corrections of order $O(\alpha/\pi)$ and
  $O(\alpha E_e/m_N)$ , described by the Feynman diagrams in
  Fig.\,\ref{fig:fig1} - Fig.\,\ref{fig:fig4}, to the amplitude of the
  neutron beta decay} \renewcommand{\theequation}{F-\arabic{equation}}
\setcounter{equation}{0}

Summing up the results of the analytical calculation of the Feynman
diagrams in Fig.\,\ref{fig:fig1} - Fig.\,\ref{fig:fig4}, we obtain the
contribution of the hadronic structure of the neutron to the radiative
corrections of order $O(\alpha/\pi)$ and $O(\alpha E_e/m_N)$,
described by the Feynman diagrams in Fig.\,\ref{fig:fig1} -
Fig.\,\ref{fig:fig4}, to the amplitude of the neutron beta decay. We
get
\begin{eqnarray}\label{eq:F.1}
 \hspace{-0.30in}&& M(n \to p e^- \bar{\nu}_e)^{(\rm LO + NLO)}_{\rm st}= -
 \frac{\alpha}{2\pi} \frac{g^2_{\pi N}}{16\pi^2}\,G_V \Big\{
 \Big(A_{\rm st} \Gamma\Big(2 - \frac{n}{2}\Big) + F^{(V)}_{\rm st}
 + G^{(V)}_{\rm st}\, \frac{k_n \cdot q}{m^2_N} + H^{(V)}_{\rm st}\,
 \frac{k_n \cdot k_e}{m^2_N} + G^{(W)}_{\rm st}
 \,\frac{m^2_N}{M^2_W}\Big) \nonumber\\
\hspace{-0.3in}&& \times \Big[\bar{u}_e \gamma^{\mu} (1 - \gamma^5)
  v_{\bar{\nu}}\Big] \Big[\bar{u}_p \gamma_{\mu} u_n\Big] +
\Big(B_{\rm st} \Gamma\Big(2 - \frac{n}{2}\Big) + F^{(A)}_{\rm st} +
G^{(A)}_{\rm st} \, \frac{k_n \cdot q}{m^2_N } + H^{(A)}_{\rm st} \,
\frac{k_n \cdot k_e}{m^2_N }+ H^{(W)}_{\rm st}
\,\frac{m^2_N}{M^2_W}\Big) \nonumber\\
\hspace{-0.3in}&&\times \Big[\bar{u}_e \gamma^{\mu} (1 - \gamma^5)
  v_{\bar{\nu}}\Big]\Big[\bar{u}_p \gamma_{\mu}\gamma^5 u_n\Big] +
F^{(W)}_{\rm st}\, \frac{m^2_N}{M^2_W} \, \Big[\bar{u}_e
  \frac{\hat{k}_n}{m_N} (1 - \gamma^5) v_{\bar{\nu}}\Big]
\Big[\bar{u}_p u_n\Big]\Big\},
\end{eqnarray}
where the abbreviations ``LO'' and ``NLO'' mean ``Leading-Oder'' and
``Next-to-Leading-Order'' in the large nucleon mass $m_N$ expansion.
The structure constants are equal to $A_{\rm st} = 2.6387$,
$F^{(V)}_{\rm st} = 1.7562$, $G^{(V)}_{\rm st} = - 95.3708$,
$H^{(V)}_{\rm st} = 91.3894$, $G^{(W)}_{\rm st} = 12.0629$, $B_{\rm
  st} = - 0.3456$, $F^{(A)}_{\rm st} = - 13.5601$, $G^{(A)}_{\rm st} =
56.5862$, $H^{(A)}_{\rm st} = - 55.0051$, $H^{(W)}_{\rm st} = 2.8343$
and $F^{(W)}_{\rm st} = - 2.2055$.  The Lorentz structure of
Eq.(\ref{eq:F.1}) is obtained at the neglect of the contributions of
order $O(m_e m_N/M^2_W) \sim O(k_n\cdot q/M^2_W) \sim O(k_n\cdot
k_e/M^2_W) \sim 10^{-7}$ and $O(E^2_0/m^2_N) \sim O(m^2_N/M^2_W) \sim
10^{-6}$, respectively. We would like to emphasize that all structure
constants in the matrix element Eq.(\ref{eq:F.1}) are induced by the
contributions of the first class currents \cite{Weinberg1958}, which
are $G$-even \cite{Lee1956a} (see also \cite{Ivanov2018}).

Including the factor $g^2_{\pi N}/16\pi^2 = 0.7414$ to the definition
of the structure constants we get
\begin{eqnarray}\label{eq:F.2}
 \hspace{-0.30in}&& M(n \to p e^- \bar{\nu}_e))^{(\rm LO + NLO)}_{\rm
   st}= - \frac{\alpha}{2\pi} \,G_V \Big\{ \Big(A_{\rm st}
 \Gamma\Big(2 - \frac{n}{2}\Big) + F^{(V)}_{\rm st} + G^{(V)}_{\rm
   st}\, \frac{k_n \cdot q}{m^2_N} + H^{(V)}_{\rm st}\, \frac{k_n
   \cdot k_e}{m^2_N} + G^{(W)}_{\rm st} \,\frac{m^2_N}{M^2_W}\Big)
 \nonumber\\
\hspace{-0.3in}&& \times \Big[\bar{u}_e \gamma^{\mu} (1 - \gamma^5)
  v_{\bar{\nu}}\Big] \Big[\bar{u}_p \gamma_{\mu} u_n\Big] +
\Big(B_{\rm st} \Gamma\Big(2 - \frac{n}{2}\Big) + F^{(A)}_{\rm st} +
G^{(A)}_{\rm st} \, \frac{k_n \cdot q}{m^2_N } + H^{(A)}_{\rm st} \,
\frac{k_n \cdot k_e}{m^2_N }+ H^{(W)}_{\rm st}
\,\frac{m^2_N}{M^2_W}\Big) \nonumber\\
\hspace{-0.3in}&&\times \Big[\bar{u}_e \gamma^{\mu} (1 - \gamma^5)
  v_{\bar{\nu}}\Big]\Big[\bar{u}_p \gamma_{\mu}\gamma^5 u_n\Big] +
F^{(W)}_{\rm st}\, \frac{m^2_N}{M^2_W} \, \Big[\bar{u}_e
  \frac{\hat{k}_n}{m_N} (1 - \gamma^5) v_{\bar{\nu}}\Big]
\Big[\bar{u}_p u_n\Big]\Big\}
\end{eqnarray}
where the structure constants are equal to $A_{\rm st} = 1.9562$,
$F^{(V)}_{\rm st} = 1.3020$, $G^{(V)}_{\rm st} = - 70.7050$,
$H^{(V)}_{\rm st} = 67.7533$, $G^{(W)}_{\rm st} = 8.9431$, $B_{\rm st}
= - 0.2562$, $F^{(A)}_{\rm st} = - 10.0530$, $G^{(A)}_{\rm st} =
41.9513$, $H^{(A)}_{\rm st} = - 40.7791$, $H^{(W)}_{\rm st} = 2.1013$
and $F^{(W)}_{\rm st} = - 1.6351$.

Following Sirlin \cite{Sirlin1967} we remove all radiative corrections
of order $O(\alpha/\pi)$ defined by the structure constants $(A_{\rm
  st},F^{(V)}_{\rm st})$ and $(B_{\rm st}, F^{(A)}_{\rm st})$ by
renormalization of the vector $G_V$ and axial $g_A$ coupling
constants, respectively. As result, the contribution of the hadronic
structure of the neutron to the radiative corrections of order
$O(\alpha E_e/m_N)$ to the amplitude of the neutron beta decay is
given by
\begin{eqnarray}\label{eq:F.3}
 \hspace{-0.30in}&& M(n \to p e^- \bar{\nu}_e)^{(\rm NLO)}_{\rm st}= -
 \frac{\alpha}{2\pi} \,G_V \Big\{ \Big(G^{(V)}_{\rm st}\, \frac{k_n
   \cdot q}{m^2_N} + H^{(V)}_{\rm st}\, \frac{k_n \cdot k_e}{m^2_N} +
 G^{(W)}_{\rm st} \,\frac{m^2_N}{M^2_W}\Big)\Big[\bar{u}_e
   \gamma^{\mu} (1 - \gamma^5) v_{\bar{\nu}}\Big] \Big[\bar{u}_p
   \gamma_{\mu} u_n\Big] \nonumber\\
\hspace{-0.3in}&& + \Big( G^{(A)}_{\rm st} \, \frac{k_n \cdot q}{m^2_N
} + H^{(A)}_{\rm st} \, \frac{k_n \cdot k_e}{m^2_N }+ H^{(W)}_{\rm st}
\,\frac{m^2_N}{M^2_W}\Big) \Big[\bar{u}_e \gamma^{\mu} (1 - \gamma^5)
  v_{\bar{\nu}}\Big]\Big[\bar{u}_p \gamma_{\mu}\gamma^5 u_n\Big] +
F^{(W)}_{\rm st}\, \frac{m^2_N}{M^2_W} \, \Big[\bar{u}_e
  \frac{\hat{k}_n}{m_N} (1 - \gamma^5) v_{\bar{\nu}}\Big]
\Big[\bar{u}_p u_n\Big]\Big\}\nonumber\\
\hspace{-0.3in}&&
\end{eqnarray}
with the structure constants $G^{(V)}_{\rm st} = - 70.7050$,
$H^{(V)}_{\rm st} = 67.7533$, $G^{(W)}_{\rm st} = 8.9431$,
$G^{(A)}_{\rm st} = 41.9513$, $H^{(A)}_{\rm st} = - 40.7791$,
$H^{(W)}_{\rm st} = 2.1013$ and $F^{(W)}_{\rm st} = - 1.6351$.

Using the algebra of the Dirac $\gamma$-matrices and Dirac equations
for a free neutron $ \hat{k}_nu_n = m_n u_n$ and a free proton
$\bar{u}_p \hat{k}_p = m_p \bar{u}_p$, one may show that the term
  proportional to the structure constant $F^{(W)}_{\rm st}$ does not
  violate the $V - A$ structure of the radiative corrections $O(\alpha
  E_e/m_N)$ to the amplitude of the neutron beta decay:
\begin{eqnarray}\label{eq:F.4}
 &&\Big[\bar{u}_e \frac{\hat{k}_n}{m_N} (1 - \gamma^5)
   v_{\bar{\nu}}\Big] \Big[\bar{u}_p u_n\Big] =
 \frac{1}{m_N}\Big[\bar{u}_e \gamma^{\mu} (1 - \gamma^5)
   v_{\bar{\nu}}\Big] \Big[\bar{u}_pk_{n\mu} u_n\Big] = \frac{1}{2
   m_N}\Big[\bar{u}_e \gamma^{\mu} (1 - \gamma^5) v_{\bar{\nu}}\Big]
 \Big[\bar{u}_p\big(\hat{k}_n \gamma_{\mu} + \gamma_{\mu}
   \hat{k}_n\big)u_n\Big] =  \nonumber\\
\hspace{-0.3in}&&= \frac{1}{2 m_N}\Big[\bar{u}_e \gamma^{\mu} (1 -
  \gamma^5) v_{\bar{\nu}}\Big] \Big[\bar{u}_p\big(\hat{k}_p
  \gamma_{\mu} + \gamma_{\mu} \hat{k}_n\big)u_n\Big] - \frac{1}{2
  m_N}\Big[\bar{u}_e \gamma^{\mu} (1 - \gamma^5)
  v_{\bar{\nu}}\Big]\Big[\bar{u}_p\big(\hat{q}\gamma_{\mu} u_n\Big] =
\Big[\bar{u}_e \gamma^{\mu} (1 - \gamma^5) v_{\bar{\nu}}\Big]
\Big[\bar{u}_p \gamma_{\mu}u_n\Big] \nonumber\\
\hspace{-0.3in}&&- \frac{m_e}{2 m_N}\Big[\bar{u}_e \gamma^{\mu} (1 -
  \gamma^5) v_{\bar{\nu}}\Big]\Big[\bar{u}_p u_n\Big] - \frac{1}{2
  m_N}\Big[\bar{u}_e \gamma^{\mu} (1 - \gamma^5)
  v_{\bar{\nu}}\Big]\Big[\bar{u}_p i \sigma_{\mu \nu} q^{\nu}u_n\Big]
\to \Big[\bar{u}_e \gamma^{\mu} (1 - \gamma^5) v_{\bar{\nu}}\Big]
\Big[\bar{u}_p \gamma_{\mu}u_n\Big],
\end{eqnarray}
where $k_p = k_n + q$ and $m_N = (m_n + m_p)/2$. The accuracy of the
relation Eq.(\ref{eq:F.4}) is of about $O(E_0/m_N) \sim
10^{-3}$. Plugging Eq.(\ref{eq:F.4}) into Eq.(\ref{eq:F.3}) we get
\begin{eqnarray}\label{eq:F.5}
  M(n \to p e^- \bar{\nu}_e)^{(\rm NLO)}_{\rm st} &=& -
  \frac{\alpha}{2\pi} \,G_V \Big\{ \Big(G^{(V)}_{\rm st}\, \frac{k_n
    \cdot q}{m^2_N} + H^{(V)}_{\rm st}\, \frac{k_n \cdot k_e}{m^2_N} +
  \big(G^{(W)}_{\rm st} + F^{(W)}_{\rm st}\big)
  \,\frac{m^2_N}{M^2_W}\Big)\Big[\bar{u}_e \gamma^{\mu} (1 - \gamma^5)
    v_{\bar{\nu}}\Big] \Big[\bar{u}_p \gamma_{\mu} u_n\Big]
  \nonumber\\ && + \Big( G^{(A)}_{\rm st} \, \frac{k_n \cdot q}{m^2_N
  } + H^{(A)}_{\rm st} \, \frac{k_n \cdot k_e}{m^2_N }+ H^{(W)}_{\rm
    st} \,\frac{m^2_N}{M^2_W}\Big) \Big[\bar{u}_e \gamma^{\mu} (1 -
    \gamma^5) v_{\bar{\nu}}\Big]\Big[\bar{u}_p \gamma_{\mu}\gamma^5
    u_n\Big]\Big\}.
\end{eqnarray}
Apart from the factor $\alpha G_V/2\pi$, the uncertainty, induced by
the relation Eq.(\ref{eq:F.4}), is of about $O(m_NE_0/M^2_W) \sim
10^{-7}$. It is of order of magnitude smaller than the accuracy of the
calculation of the Feynman diagrams in Fig.\,\ref{fig:fig1}-
Fig.\,\ref{fig:fig4}, which is of about a few parts of $10^{-6}$.  The
amplitude Eq.(\ref{eq:F.5}), describing the contribution of the
hadronic structure of the neutron to the radiative corrections
$O(\alpha E_e/m_N)$ to the neutron beta decay, has a required $V - A$
structure in agreement with Sirlin's analysis of the contribution of
the hadronic structure of the neutron, carried out within the current
algebra approach \cite{Sirlin1967, Sirlin1978}.

\end{document}